%% file: these.tex
\documentclass[12pt,twoside,openright,a4paper]{report}
\usepackage[T1]{fontenc}
\usepackage[latin1]{inputenc}
\usepackage[francais,english]{babel}

\usepackage{fancyhdr}
\fancypagestyle{core}{%
\fancyhead[LE]{\leftmark}
\fancyhead[RO]{\rightmark}

\fancyfoot[LE]{\thepage}
\fancyfoot[RO]{\thepage}
}

\fancypagestyle{empty}{%
\fancyhead[LE]{}
\fancyhead[RO]{}

\fancyfoot{}
}

\fancypagestyle{plain}{%
\fancyhead[LE]{}
\fancyhead[RO]{}

\fancyfoot[LE]{\thepage}
\fancyfoot[RO]{\thepage}
}

\fancypagestyle{art}{%
\fancyhead[LE]{\leftmark}
\fancyhead[RO]{\leftmark}

\fancyfoot[LE]{\thepage}
\fancyfoot[RO]{\thepage}
}

\usepackage{chapter}

\usepackage{stmaryrd}
\usepackage{epsf}
\usepackage{epsfig}
\usepackage{amsmath,amsfonts,amssymb, amsbsy}
\usepackage{pifont}
\usepackage{latexsym}
\usepackage{epsfig}
\usepackage{rotating}
\usepackage{graphicx}
\usepackage{float}
\usepackage{pstricks, pst-node, pst-text, pst-all, multido}

\input{macro.tex}

\numberwithin{equation}{section}

\begin{document}

\selectlanguage{francais}
\renewcommand\contentsname{{\huge\rmfamily\scshape Table des mati\`eres}}
\renewcommand\bibname{{\huge\rmfamily\scshape R{\'e}f{\'e}rences bibliographiques}}

\pagestyle{plain}
\nopagenumber

\include{titre}

\setcounter{page}{1}
\pageblanche

\romanpagenumber
\include{remerciements}

\renewcommand{\baselinestretch}{1.2} \normalsize

\include{intro}

\setcounter{page}{1}
\nopagenumber

\pageblanche
\tableofcontents

\pageblanche

\arabicpagenumber

\pagestyle{core}

\include{demarche}
\pageblanche
\include{formalisme}
\pageblanche
\include{contrereac}
\pageblanche

\pagestyle{art}
\include{persp}
\pageblanche

\appendix

\include{df}

\pageblanche


\chapter[{\selectlanguage{english}Open strings in relativistic ion traps \\{\protect\small JHEP {\protect\bf 0305}, 035 (2003) (arXiv:hep-th/0302159)}}]{{\selectlanguage{english}Open strings \\in relativistic ion traps\\{\large JHEP {\bf 0305}, 035 (2003)}\\{\large \bf arXiv:hep-th/0302159}\label{a1}}}
\l{annart1}

\cleardoublepage
\addtocounter{page}{44}
\pageblanche

\chapter[{\selectlanguage{english}Closed strings in Misner space: Stringy fuzziness with a twist \\{\protect\small JCAP {\bf 0410} (2004) 002  (arXiv:hep-th/0407216)}}]{{\selectlanguage{english}Closed strings in Misner space: \\Stringy fuzziness with a twist\\{\large JCAP {\bf 0410} (2004) 002}\\ {\large \bf arXiv:hep-th/0407216}\label{a2}}}
\l{annart2}

\cleardoublepage
\addtocounter{page}{30}
\pageblanche

\chapter[{\selectlanguage{english}Aspects of Dirichlet S-branes \\{\protect\small Phys.\ Lett.\ B {\protect\bf 625} (2005) 291  (arXiv:hep-th/0507059)}}]{{\selectlanguage{english}Aspects of Dirichlet S-branes\\{\large Phys.\ Lett.\ B {\bf 625} (2005) 291}\\ {\large \bf arXiv:hep-th/0507059}\label{a3}}}
\l{annart3}

\cleardoublepage
\addtocounter{page}{12}
\pageblanche

\pagestyle{plain}


\include{biblio}

\pagestyle{empty}
\newpage
\phantom{page blanche}

\end{document}

%% file: macro.tex
\newcommand{\pageblanche}{\newpage{\pagestyle{empty} \phantom{page blanche} \cleardoublepage}}

\newcommand{\nopagenumber}{\renewcommand{\thepage}{}}
\newcommand{\arabicpagenumber}{\renewcommand{\thepage}{\arabic{page}}}
\newcommand{\romanpagenumber}{\renewcommand{\thepage}{\roman{page}}}

\newcommand{\gl}{\guillemotleft~}
\newcommand{\gr}{~\guillemotright}

\DeclareMathOperator{\arctanh}{arctanh}
\DeclareMathOperator{\tr}{tr}
\DeclareMathOperator{\im}{Im}
\DeclareMathOperator{\re}{Re}

\def\be{\begin{equation}}
\def\ee{\end{equation}}
\def\begs{\begin{split}}
\def\ends{\end{split}}
\def\ntee{\notag \end{equation}}
\def\bega{\begin{align}}
\def\enda{\end{align}}
\def\begm{\begin{multline}}
\def\endm{\end{multline}}
\def\bi{\begin{itemize}}
\def\ei{\end{itemize}}
\renewcommand{\i}{\item}
\def\ntba{\begin{align*}}
\def\ntea{\end{align*}}
\def\bse{\begin{subequations}}
\def\ese{\end{subequations}}

\renewcommand{\l}{\label}
\def\bv{\begin{verbatim}}
\def\ev{\end{verbatim}}

\def\Z{\mathbb{Z}}
\def\R{\mathbb{R}}
\def\N{\mathbb{N}}
\def\C{\mathbb{C}}

\newcommand{\p}{\partial}
\newcommand{\bp}{\bar \partial}
\def\dpt{\partial_\tau}
\def\dps{\partial_\sigma}

\renewcommand{\t}{\tau}
\newcommand{\w}{\omega}
\newcommand{\s}{\sigma}

\newcommand{\e}{\epsilon}
\newcommand{\om}{\omega}

\newcommand{\apr}{\alpha'}

\renewcommand{\a}{\alpha}
\newcommand{\ta}{\tilde \alpha}
\newcommand{\ap}{\alpha^+}
\newcommand{\tap}{\tilde \alpha^+}
\newcommand{\am}{\alpha^-}
\newcommand{\tam}{\tilde \alpha^-}

\newcommand{\xp}{X^+}
\newcommand{\xm}{X^-}
\newcommand{\xpm}{X^\pm}

\newcommand{\xpf}{X_{>0}}
\newcommand{\xnf}{X_{<0}}
\newcommand{\xpfp}{X_{>0}^+}
\newcommand{\xpfm}{X_{>0}^-}
\newcommand{\xpfpm}{X_{>0}^\pm}
\newcommand{\xpfmp}{X_{>0}^\mp}
\newcommand{\xnfp}{X_{<0}^+}
\newcommand{\xnfm}{X_{<0}^-}
\newcommand{\xnfpm}{X_{<0}^\pm}
\newcommand{\xnfmp}{X_{<0}^\mp}
\newcommand{\xz}{X_0}
\newcommand{\xzp}{X_0^+}
\newcommand{\xzm}{X_0^-}
\newcommand{\xzpm}{X_0^\pm}

\newcommand{\kp}{k^+}
\newcommand{\km}{k^-}
\newcommand{\kpm}{k^\pm}

\newcommand{\eps}{\varepsilon}

\newcommand\beps{{\bar\varepsilon}}
\newcommand{\ep}{\varepsilon}
\newcommand{\bep}{{\bar \varepsilon}}

\newcommand{\tpsi}{\tilde \psi}
\newcommand{\RR}{\text{R-R}}
\newcommand{\NSNS}{\text{NS-NS}}
\newcommand{\Ra}{\text{R}}
\newcommand{\NS}{\text{NS}}

\newcommand{\pbr}{$p$-brane}

\newcommand{\un}{. 1}
\newcommand{\tun}{. {\tilde 1}}

\newcommand{\bw}{\bar w}
\newcommand{\bz}{\bar z}
\newcommand{\bT}{\overline{T}}
\newcommand{\bM}{\bar M}
\newcommand{\tM}{\overline{M}}
\newcommand{\lno}{\,:\!}
\newcommand{\rno}{\!:\,}

\newcommand{\hypF}{{}_2 F_1}

\newcommand{\A}{{\cal A}}
\newcommand{\D}{{\cal D}}
\newcommand{\F}{{\cal F}}
\newcommand{\J}{{\cal J}}
\newcommand{\M}{{\cal M}}
\newcommand{\cN}{{\cal N}}
\renewcommand{\O}{{\cal O}}
\newcommand{\cP}{{\cal P}}
\renewcommand{\S}{{\cal S}}
\newcommand{\V}{{\cal V}}
\newcommand{\cZ}{{\cal Z}}

%% file: titre.tex

\enlargethispage{3cm}
\thispagestyle{empty}
\vspace*{-2cm}
\centerline{\scshape 
\large Laboratoire de Physique Théorique et des Hautes \'Energies}
\vskip 1.5cm
\centerline{\Large \bfseries TH{\`E}SE DE DOCTORAT DE L'UNIVERSIT{\'E} PARIS VI}
\vskip .8cm
\centerline{\large Sp{\'e}cialit{\'e} : \bfseries\scshape PHYSIQUE TH{\'E}ORIQUE}
\vskip 1.7cm
\centerline{pr{\'e}sent{\'e}e par}
\vskip .6cm
\centerline{\Large \bf M. Bruno DURIN}
\vskip 1cm
\centerline{pour obtenir le grade de}
\vskip .6cm
\centerline{\Large \bf Docteur de l'Universit{\'e} Paris VI}
\vskip 1.5cm
\centerline{Sujet :}
\vskip 1cm
\centerline{\LARGE \bfseries \itshape Configurations dépendantes du temps}
\vskip .3cm
\centerline{\LARGE \bfseries \itshape dans le formalisme perturbatif}
\vskip .2cm
\centerline{\LARGE \bfseries \itshape de la théorie des cordes}
\vskip 3cm
\noindent
Soutenue le 31 janvier 2006 devant le jury compos{\'e} de~:
\vskip 1cm
\begin{tabular}{cll}
MM. & Costas Bachas,&\\
    & Micha Berkooz, &\\
    & Elias Kiritsis, &\\
    & Nikita Nekrasov, &rapporteur,\\
    & Boris Pioline, &directeur de th{\`e}se,\\
    & Pierre Vanhove,&rapporteur, \\
\&  & Paul Windey.\\
\end{tabular}
\cleardoublepage

%% file: remerciements.tex

\centerline{\large \scshape Remerciements}
\vskip 1cm
Je tiens tout d'abord à remercier Laurent Baulieu et Olivier Babelon, directeurs successifs du Laboratoire de Physique Théorique et des Hautes \'Energies de m'y avoir accueilli et m'avoir permis d'y mener mes travaux de recherches.

Je suis très reconnaissant à Boris Pioline, qui a accepté de diriger cette thèse pendant plus de trois ans. Il n'a jamais cessé d'être disponible et patient devant mes errances de thésard. J'adresse ma gratitude à Paul Windey d'avoir bien voulu être mon directeur de thèse administratif, en attendant que Boris Pioline puisse l'être.

Je remercie également Micha Berkooz et Doris Reichmann avec qui j'ai collaboré. Je salue avec reconnaissance mes collègues et amis de bureau ainsi que les membres du laboratoire avec qui j'ai discuté et passé de très bons moments, notamment François Arléo, Olivier Babelon, Marc Bellon, Pedro Bordalo, Guillaume Bossard, Kyryl Kazymyrenko, Bruno Machet, Rédamy Perez-Ramos et Jean-Bernard Zuber.

Je remercie vivement Denis Bernia, Sylvie Dalla-Foglia, Marie-Chritine Lévy, Elysée MacKagny, Annie Richard et Valérie Sabouraud pour leur disponibilité et leur gentillesse, ainsi que Marco Picco, qui malgré son travail de chercheur, a toujours été présent pour résoudre les divers problèmes informatiques que j'ai pu rencontrer.

J'exprime toute ma gratitude aux membres du jury d'avoir accepté d'en faire partie, notamment aux rapporteurs pour leur travail de lecture de la thèse.

Enfin je salue mes amis, ma famille et tout spécialement Albina pour les remercier de leur soutien et de la curiosité qu'ils ont manifestée pendant plus de trois ans.

\cleardoublepage

\centerline{\large \scshape Avertissement}
\vskip 1cm
Ce mémoire comporte un préambule qui vise à présenter de manière accessible les idées qui ont mené à ce travail de thèse. Aux lecteurs réfractaires ou étrangers aux symboles mathématiques, je propose de ne lire que ces premières pages.

Ensuite viennent la table des matières et le corps principal de ce mémoire. Les publications auxquelles a donné lieu le travail de cette thèse sont reproduites en appendice, avant les références bibliographiques.

Ce mémoire a été écrit dans une double démarche, celle tout d'abord de rendre compte de mes travaux et celle ensuite de proposer une exposition pédagogique du formalisme que j'ai utilisé.
Loin de prétendre rivaliser avec les différents livres et cours sur le sujet, cette exposition vise à en concilier les différentes approches et à unifier autant que faire se peut les conventions, pour fournir au lecteur un contenu homogène, qui, je l'espère, sera utile à de futurs thésards.
\cleardoublepage

%% file: intro.tex

\centerline{\large \scshape Préambule}
\vskip 1cm

La description du monde physique repose actuellement sur deux modèles standards. Avant d'en exposer les grandes lignes, nous devons expliquer la différence entre le monde classique et le monde quantique. Nous désignerons par le monde classique le monde physique correspondant aux échelles de tailles (de manière équivalente, de durée, d'énergie) humaines, c'est-à-dire entre le micromètre et le milliard de kilomètres. Hormis quelques exceptions, tous les systèmes de cette taille vérifient le même ensemble de lois physiques que nous appelons théories classiques. La physique des échelles inférieures, dites microscopiques, est décrite par des théories quantiques, où les degrés de liberté, les paramètres qui rendent compte de l'évolution d'un système donné, sont fondamentalement différents des degrés de liberté classiques. Comme l'intuition peut nous l'indiquer, les deux mondes se sont pas étrangers l'un à l'autre : la physique microscopique détermine un grand nombre de paramètres qui interviennent dans les modèles classiques. La physique des échelles supérieures, dites macroscopiques, rassemblant la cosmologie, l'astrophysique et l'astronomie, a un statut beaucoup moins clair. La plupart des théories macroscopiques sont en fait des théories classiques (et quantiques) qui sont appliquées à des échelles de distance et de durée plus grandes et qui sont ajustées pour décrire une grande partie des phénomènes observés, mais le domaine de validité de ces extrapolations est encore inconnu.

Le premier modèle, appelé simplement \emph{modèle standard}, est un modèle unifié de la physique microscopique. Il décrit le comportement des particules élémentaires, de leurs composés, de leurs interactions. Il décrit le monde microscopique, aussi bien la matière\footnote{Par exemple, l'électron est une particule élémentaire, le proton et le neutron, qui consituent le noyau des atomes, sont des composés de différents quarks.} que trois interactions fondamentales sur les quatre connues\footnote{L'électromagnétisme, l'interaction nucléaire faible et l'interaction nucléaire forte sont décrits par le modèle standard, mais pas la gravitation.}. Jusqu'à présent il a été validé par toutes les observations expérimentales microscopiques que nous avons pu faire et donnerait entière satisfaction s'il ne comportait pas un grand nombre\footnote{plus d'une centaine} de paramètres, parmi lesquels la masse de l'électron par exemple, et s'il ne supposait pas l'existence d'une particule encore non détectée : le boson de Higgs.

Le second modèle, appelé \emph{modèle standard cosmologique}, est un modèle global de la physique macroscopique, c'est-à-dire qu'il décrit un certain nombre de propriétés de l'univers considéré dans son ensemble, comme par exemple les propriétés du fond de radiation cosmologique, les propriétés de distribution de la matière dans l'univers, les propriétés d'isotropie et d'homogénéité de l'espace-temps. 
À ce stade, nous devons faire la différence entre ce modèle et les théories macroscopiques que nous avons présentées comme des extrapolations classiques et quantiques. Ces théories sont des modèles phénoménologiques qui décrivent l'évolution d'un système particulier, par exemple la formation, la vie et la fin d'une étoile ou la dynamique d'une galaxie. Par contre, le modèle standard cosmologique vise à décrire de \emph{manière unifiée} le comportement à grande échelle de l'univers et constitue ainsi un premier pas vers une théorie cosmologique unifiée. Cependant les réponses qu'il apporte sont beaucoup plus partielles que celles du modèle standard. Il suppose l'existence d'un phénomène d'inflation dont il ne donne pas l'explication, et n'élucide pas le problème de la constante cosmologique par exemple, dont aucune théorie actuelle ne permet de prédire la valeur très faible mais non nulle déterminée par les mesures astronomiques.

Pour comprendre pourquoi le second modèle parait moins efficace, il faut revenir sur la distinction que nous avons faite entre les deux modèles. Nous avons distingué un modèle unifiant la physique microscopique et un modèle visant à unifier la physique macroscopique. Mais la démarche intellectuelle qui a mené à la construction de ces modèles a pour but d'isoler des constituants fondamentaux microscopiques à partir desquels on construit la physique à toutes les échelles supérieures. Comme la gravitation joue un rôle négligeable aux échelles microscopiques auxquelles nous avons accès, le modèle standard n'inclut pas cette interaction. Aux échelles humaines, la théorie classique de la gravitation (la relativité générale), rend compte parfaitement des résultats expérimentaux. La gravitation, bien que très faible, est une interaction auxquels tous les constituants fondamentaux sont sensibles et dont les effets attractifs sont cumulatifs\footnote{En l'état actuel des connaissances, la seule interaction gravitationnelle répulsive provient de la constante cosmologique et elle est extrêmement faible.}. Aux échelles macroscopiques la gravitation joue donc un rôle primordial, comme les observations astronomiques permettent de la constater. Le modèle standard cosmologique décrit la matière et les trois interactions autres que la gravitation à l'aide du modèle standard ou de théories effectives classiques et y incorpore la théorie classique de la gravitation. Si l'on s'en tient à la démarche initiale, il manque donc une théorie microscopique de la gravitation à partir de laquelle on expliquerait le comportement classique de cette interaction et les observations cosmologiques encore non expliquées.

Malheureusement, aucune théorie microscopique de la gravitation n'est encore pleinement satisfaisante. À la difficulté de construire une telle théorie s'ajoute celle de concevoir des expériences permettant d'en tester les prédictions expérimentales. Devant cet obstacle, on peut imaginer de remettre en cause, pour la gravitation, la démarche déductive, qui suppose la physique classique et macroscopique déterminée par la physique microscopique. Ce faisant, on perd \emph{a priori} toute chance d'obtenir une unification au niveau du formalisme utilisé dans les modèles standards, celui de la théorie des champs, classique pour la gravitation, quantique pour le reste. En pratique les modèles construits dans cette perspective ne sont que phénoménologiques et n'offrent pas de réponses meilleures que celles du modèle standard cosmologique. C'est pourquoi les tentatives les plus sérieusement considérées sont celles visant à construire une théorie microscopique de la gravitation. Dans une telle théorie, il existe une échelle microscopique où les effets quantiques se font sentir. Des considérations dimensionnelles indiquent que cette échelle correspond une distance caractéristique, appelée \emph{longueur de Planck}\footnote{Si cette distance était celle entre nos pieds et notre tête, l'électron serait grand comme la longueur de la galaxie, c'est-à-dire environ cent mille années-lumière.}, de l'ordre de $10^{-35}$~mètres, ou une \emph{énergie de Planck}\footnote{Cette énergie correspond à l'énergie nécessaire pour élever de mille mètres une locomotive de cent tonnes. Elle correspond aussi à l'énergie de repos d'une masse de 10 microgrammes, c'est-à-dire l'énergie qu'on obtiendrait si une poussière de 5 microgrammes de matière s'annihilait avec une poussière de 5 microgrammes d'antimatière.} égale à $10^{19}$~GeV. Notons que, même dans l'hypothèse où les effets quantiques de la gravitation sont très importants, le modèle standard n'est pas remis en cause, puisqu'il ne décrit la physique que jusqu'à une échelle maximale de $10^{15}$~GeV et que les expériences ne sondent que des énergies inférieures à $10^4$~GeV.

Si les autres interactions sont ignorées, il est possible de se focaliser sur la gravitation à une échelle d'énergie proche de l'échelle de Planck. Ce choix a été fait pour la théorie de \emph{gravité quantique de boucle}\footnote{\emph{loop quantum gravity} en anglais}. Construite à partir des principes fondamentaux de la relativité générale, elle vise, à travers un formalisme novateur de \emph{réseaux de spins}, entité mathématique censée décrire la structure de l'espace-temps quantique, à donner un modèle cohérent du comportement microscopique de la gravitation. Ce formalisme fait malheureusement la faiblesse de cette théorie, puisqu'il est encore impossible de faire le lien avec le formalisme de la théorie des champs du modèle standard : la limite de grande\footnote{devant la longueur de Planck} échelle de longueur et de faible courbure n'est pas connue. Cette théorie a donc moins de succès que la \emph{théorie des cordes} qui prend ses racines\footnote{et même ses racines historiques, puisque la théorie des cordes a été dans ses débuts un modèle \gl infructueux\gr\ de l'interaction forte.} dans le modèle standard. La théorie des cordes reprend exactement le même formalisme de la théorie des champs, suit la même démarche qui a mené de la mécanique classique à la mécanique quantique puis à la théorie des champs. Mais au lieu de considérer que l'entité fondamentale est une particule ponctuelle à zéro dimension, elle introduit un objet à une dimension appelé \emph{corde}. Comme cette corde est supposée avoir une dimension de l'ordre de la longueur de Planck, à toutes les échelles de la physique connue, on la voit comme une particule ponctuelle. La corde est donc la brique commune à toutes les particules connues ou prédites par les théories précédentes. La théorie fait même un lien plus profond : \emph{une même} corde peut être telle ou telle particules selon la façon dont elle vibre et dont ses extrémités sont fixées. De même qu'une corde de guitare émet plusieurs notes, la corde peut \gl jouer\gr\ différentes particules. Une \emph{corde fermée}, nouée comme un lacet, pourra correspondre au \emph{graviton}, la particule qui véhicule la gravitation, alors qu'une \emph{corde ouverte}, libre comme un bout de ficelle ou attachée à des parois particulières appelées \emph{D-branes}, pourra correspondre à d'autres interactions ou à des particules de matière. La théorie des cordes propose ainsi un modèle unifié du monde physique connu, ce qui en a fait le succès depuis près de quarante ans.

Cependant, deux difficultés majeures atténuent la portée de cette théorie. La première est une difficulté technique. Même si le formalisme de la théorie des champs pour la particule ponctuelle est un guide précieux, choisir la corde comme entité fondamentale introduit de nouveaux problèmes dont les solutions sont loin d'être connues à l'heure actuelle. Par exemple, le formalisme de la théorie des champs permet de décrire une infinité de degrés de liberté, alors qu'en théorie des cordes seul le formalisme permettant de décrire un nombre fini de degrés de liberté a été correctement posé. Le formalisme qui permettrait de décrire la création et l'annihilation des cordes est encore en gestation. Ainsi, tout un ensemble d'effets quantiques est encore impossible à décrire en théorie des cordes. Un autre exemple particulièrement caractéristique est la difficulté à sélectionner l'état de vide. En effet, la théorie des cordes prédit l'existence d'un nombre immense (mais fini \emph{a priori}) de vides, qui forment ce qu'on appelle le \emph{paysage des vides}\footnote{traduction personnelle de \emph{landscape}}. Les critères qui permettraient de choisir un vide parmi toutes les possibilités prédites par la théorie ne sont pas connus, ni ceux qui indiqueraient si le ou les vides sélectionnés sont compatibles avec le modèle standard.

La seconde difficulté, qui n'est pas étrangère à la première, provient de la manière dont la théorie des cordes a été construite. Façonné dans une démarche qui a mené à l'établissement du modèle standard, le formalisme de la théorie des cordes reste éloigné des principes de la relativité générale, notamment \emph{le principe d'indépendance vis-à-vis du fond} selon lequel les lois de la physique s'écrivent de la même façon quels que soient l'observateur ou la géométrie\footnote{La relativité générale établit un lien canonique entre gravitation et géométrie. Un observateur placé dans un champ gravitationnel peut localement se placer dans un système de coordonnées, c'est-à-dire choisir une géométrie, où le champ gravitationnel est nul.} considérés. En l'absence d'un formalisme permettant d'implémenter ce principe, il est difficile de répondre aux questions suscitées par la relativité générale et d'aller plus loin que les effets semi-classiques de la gravitation quantique\footnote{c'est-à-dire l'influence d'un champ gravitationnel classique sur un système quantique}. En effet, la théorie des cordes n'est définie qu'autour de l'espace-temps de Minkowski et pour des configurations géométriques particulières, majoritairement statiques. L'étude des configurations gravitationnelles non triviales, si possible incorporant des ingrédients du modèle standard cosmologique, représente ainsi un enjeu vital dans la construction de la théorie.

C'est pourquoi dans cette thèse nous avons étudié des configurations géométriques dépendantes du temps. Notre objectif, modeste, a consisté à choisir des configurations auxquelles le formalisme de la théorie des cordes dans l'espace de Minkowski pouvait être appliqué et à essayer de mettre en évidence des effets physiques liés à cette dépendance en temps. Nous avons notamment tenté de comprendre comment un effet comme la contre-réaction gravitationnelle pouvait se manifester. La contre-réaction gravitationnelle est une modification de la configuration géométrique suite aux interactions gravitationnelles des fluctuations quantiques présente dans cette configuration. En particulier, dans une configuration présentant une \emph{singularité}, c'est-à-dire un point où les grandeurs qui caractérisent la géométrie de l'espace-temps deviennent infinie, de nombreuses particules peuvent disparaître ou apparaître à cause de l'influence de la variation dans le temps de géométrie sur un champ quantique et produire un champ gravitationnel dont l'effet sur la singularité est encore inconnu.

\cleardoublepage

%% file: demarche.tex
\chapter{Configurations dépendantes du temps}
\l{demarche}

\section{Modèles exacts}

Comme nous l'avons évoqué dans le préambule, la théorie des cordes propose un cadre qui peut permettre de résoudre les problèmes posés par le modèle standard cosmologique actuel, notamment celui de la nature du Big Bang, ou de manière un peu plus générale de la nature des singularités de genre espace.

Cependant, il n'existe pas de formalisme de théorie des cordes permettant d'intégrer pleinement la théorie de la relativité générale et de construire une théorie de la gravitation quantique. Nous sommes actuellement contraints de nous limiter à une approche semi-classique, où la géométrie est posée comme une hypothèse de travail. C'est tout naturellement que le formalisme de première quantification de la corde a été développé d'abord dans l'espace de Minkowski\footnote{Nous invitons le lecteur à se référer aux ouvrages \cite{Green:1987sp, Green:1987mn, Polchinski:1998rq, Polchinski:1998rr}.}, puis dans des géométries courbes, dans le cadre des \emph{modèles Wess-Zumino-Witten} (WZW) \cite{Knizhnik:1984nr} où la géométrie est décrite par une variété de groupe de Lie. Toutefois les premières géométries étudiées étaient statiques.

Puisqu'il n'était pas possible de comprendre le régime quantique de la gravitation, des modèles ont été développé à partir des premiers espaces courbes et ont permis d'étudier la quantification de la corde évoluant dans une géométrie dépendante du temps dans l'espoir de découvrir des indices permettant de compléter le formalisme ou d'apporter les premiers éléments de réponses aux questions posées par les singularités. Ainsi dès 1985\footnote{mais déjà en 1974 un premier article  \cite{Ademollo:1974te} abordait ce genre de problème.} \cite{Boulware:1985wk} s'est d\'evelopp\'ee une importante direction de recherche visant \`a trouver des modèles qui puissent être résolus exactement, c'est-à-dire des modèles pour lesquelles on peut déterminer explicitement les solutions des équations classiques du mouvement de la corde, mener la procédure de quantification canonique, déterminer le spectre de l'opérateur hamiltonien et calculer les amplitudes les plus simples\footnote{mais toujours dans une approche perturbative, c'est-\`a-dire pour un couplage de corde $g_s$ petit}. Ces mod\`eles s'appellent \emph{mod\`eles exacts}.

Ces modèles sont loin d'être réalistes et cette direction de recherche se démarque des tentatives de construire des géométries cosmologiques de type Friedmann-Robertson-Walker ou inflationnaire de type de Sitter \cite{Kachru:2003sx} qui se situent dans une perspective de théorie effective, à basse énergie.
Mais le compromis qu'il faut faire pour étudier de manière \gl exacte\gr\ un modèle dans le cadre de la théorie des cordes n'empêche pas d'étudier des configurations extrêmement intéressantes du point de vue cosmologiques~: les singularités. Comme nous le détaillons un peu plus bas, certaines géométries comportant une singularité donnent lieu à un traitement exact, ce qui permet d'étudier précisément le comportement de la corde dans cette géométrie. Il est même possible d'aborder les problèmes de contre-réaction gravitationnelle ou de régularisation éventuelle de la singularité. Nous reviendrons sur ce point aux chapitres \ref{contrereac} et \ref{persp}.

Cette thèse s'inscrit dans cette démarche. Nous avons étudié trois modèles différents de géométrie dépendante du temps et extrait des calculs un certain nombre d'effets physiques que nous exposons au chapitre \ref{contrereac}. Notre souci a été de construire des géométries exactes. Nous avons pour cela puisé dans deux des trois différents types de modèles exacts et dépendants du temps actuellement disponibles. Nous insistons sur le fait que les modèles exacts sont des modèles exempts de corrections en $\alpha'$, la tension inverse de la corde, mais qu'ils restent des modèles perturbatifs en $g_s$, la constante de couplage de la corde.

\section{Modèles exacts dépendants du temps}

Comme nous l'expliquerons dans le chapitre \ref{formalisme}, la première quantification de la corde est fondée sur l'interprétation du modèle sigma non-linéaire de la corde en terme de théorie conforme des champs à deux dimensions. La recherche de modèles exacts est donc guidée par la nécessité de trouver la théorie conforme des champs correspondante. Nous pouvons distinguer trois manières générales de construire des modèles exacts dépendants du temps.

\subsection{Orbifold de l'espace de Minkowski}
\l{demorb}

Originellement un orbifold est une variété\footnote{\gl orbi-\gr\ pour \emph{orbital}, rotation et \gl -fold\gr\ de \emph{manifold}, variété.} obtenue à partir de l'espace de Minkowski par quotient sous l'action d'un groupe de rotation fini $\Z_N$ ($N$ est un entier naturel) \cite{Dixon:1985jw}. Par extension, ce nom est appliqué à toute variété obtenue à partir de l'espace plat\footnote{Dans tout ce mémoire, nous utiliserons l'expression \gl espace plat\gr\ comme synonyme de l'expression \gl espace de Minkowksi\gr.} ou d'une autre vari\'et\'e (comme un groupe par exemple, dans les mod\`eles WZW, voir sous-section suivante) par identification sous l'action d'un sous-groupe de Lorentz isomorphe à $\Z$ ou $\Z_N$.

Pour obtenir une géométrie dépendante du temps, il suffit que l'action du sous-groupe corresponde à une transformation non triviale impliquant la coordonnée temporelle. Les points invariants sous cette transformation deviennent des singularités topologiques. L'espace est plat loin des singularités.

La première construction d'orbifold dépendant du temps a \'et\'e faite dans \cite{Horowitz:1991ap}.
L'int\'er\^et cosmologique de ces mod\`eles a \'et\'e mis en \'evidence dans \cite{Khoury:2001bz} et l'\'etude d'orbifolds de l'espace de Minkowski a pris un nouveau d\'epart, avec \cite{Balasubramanian:2002ry}, puis avec l'étude de l'\emph{orbifold null} \cite{Simon:2002ma, Fabinger:2002kr, Cornalba:2002fi, Liu:2002kb} et de l'espace de Misner\footnote{Ces deux types d'orbifolds sont expliqués à la sous-section \ref{ssorb}.} \cite{Nekrasov:2002kf, Berkooz:2002je, Pioline:2003bs, Berkooz:2004re}. La stabilité de ces orbifolds a \'et\'e remise en question \cite{Lawrence:2002aj, Horowitz:2002mw, Cornalba:2002nv}, mais nous avons choisi d'être optimiste et de considérer que même si ces orbifolds n'étaient pas stables, la quantification de la corde dans ces géométries fournirait des indices quant-à leur destin.

L'espace de Misner a l'avantage sur l'orbifold null de présenter une géométrie simplifiée de Big Crunch/Big Bang, comme nous aurons l'occasion de l'expliquer en détail à la section \ref{article2}. Il constitue ainsi un cas d'école de singularité cosmologique.

La possibilité d'une transition douce (\emph{bounce} en anglais) a été étudiée d\`es les articles \cite{Khoury:2001bz, Seiberg:2002hr} et un scénario d'univers cyclique dans l'espace de Misner est évoqué dans \cite{Tolley:2002cv}, qui étudie la théorie des champs dans l'espace de Misner. Une étude complémentaire et des références supplémentaires concernant la quantification d'un champ dans l'espace de Misner se trouvent dans \cite{Pioline:2003bs, Berkooz:2004re}. L'étape suivante consiste à étudier cette possibilité de transition douce dans le cadre de la théorie des cordes. Il s'agit notamment de savoir si la singularité peut être gommée par des effets de contre-réaction gravitationnelle.

Plus modestement, nous avons déterminé dans le cadre de cette thèse les opérateurs de vertex et calculé les amplitudes de diffusion à l'ordre des arbres. Ce travail \cite{Berkooz:2004yy} est exposé à la section \ref{article2} et une partie des détails techniques sont abordés dans les sections \ref{opvertex} et \ref{aoa}.

\subsection{Modèles Wess-Zumino-Witten}
\l{wzw}

Les premières géométries de signature minkowskienne ont été obtenues en construisant des modèles WZW à partir du groupe de Lie non compact SL($2,\R$) (qui est aussi AdS${}_3$, l'espace anti-de Sitter à trois dimensions) \cite{Witten:1991yr, Ishibashi:1991wh}. Il s'agit de variations autour du trou noir à deux dimensions décrit dans l'article \cite{Witten:1991yr} et appelé aussi \gl le cigare\gr.

Ces modèles sont en réalité des \emph{modèles de WZW gaugé}. Ils sont construits à partir d'une variété de groupe quotient. Cette notion est proche mais distincte de celle d'orbifold. Si on note $G$ le groupe de base\footnote{Le plus souvent, il s'agit de SL$(2,\R)$, mais on peut avoir par exemple un produit tensoriel SL$(2,\R) \times \text{SU}(2)$, qui correspond à l'espace à six dimensions AdS${}_3 \times S^3$.} et $H$ un sous-groupe de Lie de $G$, le modèle est construit à partir du quotient $G/H$. Cette construction, dans le cas o\`u $H$ est le sous-groupe compact maximal d'un groupe $G$ non compact, permet de remédier au fait que la repr\'esentation adjointe d'un groupe non compact n'est pas unitaire et d'éliminer ainsi des rep\'esentations adjointes de l'alg\`ebre des courants des états de norme négative (voir le traitement de \cite{Dixon:1989cg} dans le cas de SL($2,\R$)). Par contre, un orbifold est un quotient par un groupe discret, comme nous l'avons dit à la sous-section \ref{demorb}, et il ne fait apparaître que de nouveaux secteurs topologiques dans la théorie conforme initiale. Il est tout à fait possible de construire un orbifold à partir d'un modèle WZW, gaugé ou non. Par exemple \cite{Martinec:2002xq} est un orbifold du modèle WZW sur SL$(2,\R)$.

D'autres modèles dépendants du temps ont ensuite été construits. Ils comportent soit une singularité de type orbifold\footnote{Voir sous-section \ref{ssorb} pour de plus amples explications.} $\R^{1,1}/ \Z$ \cite{Kounnas:1992wc, Bars:1992dx, Giveon:1992kb, Gasperini:1992ym}, soit une singularité de type trou noir BTZ \cite{Banados:1992wn, Banados:1992gq, Troost:2002wk, Martinec:2002xq, Hemming:2002kd}. Le modèle Nappi-Witten \cite{Nappi:1992kv} décrit quant-à lui une géométrie d'univers en expansion. 

Les modèles WZW ont permis également l'étude des problèmes liés aux configurations de Big Crunch/Big Bang \cite{Elitzur:2002rt} dans une approche différente de l'orbifold de Misner. \cite{Elitzur:2002vw} montre que dans ce modèle la singularité peut être gommée par la présence de flux.

L'étude de la propagation de la corde dans un modèle WZW avec une singularité cosmologique type $\R^{1,1}/ \Z$ a été faite dans \cite{Craps:2002ii}. Enfin mentionnons \cite{Buchel:2002kj} qui, dans le cadre d'un tour d'horizon des géométries dépendantes du temps en théorie des cordes, explique comment cette singularit\'e cosmologique appara\^it dans certains mod\`eles WZW.

Nous n'avons pas étudié de modèle WZW dans le cadre de cette thèse. Cependant, comme cet aperçu le suggère il existe des liens entre les modèles WZW et les orbifolds. Ainsi, comme cela est évoqué dans la section \ref{article2} et expliqué en détail dans l'article \cite{Berkooz:2004yy}, certaines amplitudes à l'ordre des arbres dans l'espace de  Misner peuvent être obtenues à partir des amplitudes calculées dans le modèle WZW construit et étudié dans \cite{D'Appollonio:2003dr}, grâce aux analogies formelles entre les deux modèles.

De même qu'il existe des analogies avec les orbifolds, certains modèles WZW, comme ceux étudiés dans \cite{Nappi:1993ie, D'Appollonio:2003dr} sont des modèles d'ondes planes, que nous allons exposer dans la sous-section suivante.

\subsection{Ondes planes}

Les géométries d'ondes planes ont été construites dans une approche différente. Au lieu de trouver directement un modèle exact avec une description en terme de théorie conforme des champs, des géométries ont été d'abord considérées comme des approximations d'ordre dominant en $\apr$ des solutions exactes des équations classiques du mouvement. Il a ensuite été démontré qu'il n'y avait pas de corrections en $\apr$ et qu'ainsi ces solutions étaient exactes. On peut distinguer une première phase \cite{Gueven:1987ad, Amati:1988sa, deVega:1988wp, Horowitz:1989bv, Tseytlin:1992va, Jofre:1993hd, Kiritsis:1993jk} d'une seconde plus récente \cite{Gueven:2000ru, Blau:2001ne, Metsaev:2001bj, Berenstein:2002jq, Kiritsis:2002kz, Gimon:2002sf, Maldacena:2002fy, Papadopoulos:2002bg, Blau:2002js} où ont pu être construites des classes complètes de ces géométries.

Ces ondes planes gravitationnelles, comme les sous-sections \ref{cchem}, \ref{spectrecf} permettent de le comprendre,  correspondent à des configurations de cordes fermées. De manière analogue, on peut coupler la corde ouverte à des ondes planes électromagnétiques  (c'est-\`a-dire la somme d'un champ électrique et d'un champ magnétique orthogonaux et de même intensité), ce que nous expliquerons aux sous-sections \ref{cchem} et \ref{spectreco} et obtenir ainsi un modèle exact \cite{Thorlacius:1997zd}. En fait le cas particulier de l'onde plane monochromatique avait été étudié d\`es 1974 \cite{Ademollo:1974te}, mais ce n'est qu'après avoir réalisé cette propriété d'exactitude que des modèles généraux ont été étudiés \cite{Bachas:2002qt, Bachas:2002jg}.

Dans le cadre de cette thèse, nous avons examiné des configurations d'ondes planes électromagnétiques à profil lin\'eaire, généralisant ainsi les configurations de \cite{Bachas:2002jg}. Nous avons exposé les principaux résultats de cette étude à la section \ref{article1}. La publication correspondante \cite{Durin:2005ts} est reproduite en annexe \ref{annart1}.

\subsection{S-brane de Dirichlet}

Il ne s'agit pas d'une quatrième classe de modèle, mais d'un objet défini dans l'espace de Minkowski. Le concept de S-brane regroupe divers objets qui partagent la propriété d'être localisés dans le temps. Nous y reviendrons à la section \ref{article3}. La S-brane de Dirichlet \cite{Gutperle:2002ai} peut être considérée comme un mur infini qui n'existe qu'un instant et sur lequel les extrémités des cordes ouvertes doivent s'accrocher. Nous définirons précisément cet objet à la sous-section \ref{sdbranes}.

Nous avons, au cours de cette thèse, essayé de préciser les propriétés de cet objet assez particulier car il constitue le plus simple exemple de géométrie dépendante du temps, une fonction delta de la dépendante en temps en quelque sorte. Nous avons rassemblé nos observations à la section \ref{article3}. La publication \cite{Durin:2005ts} à laquelle ces travaux ont donné lieu est reproduite en annexe \ref{annart3}.

\section{Organisation du mémoire}

Comme nous l'avons dit au début de ce chapitre, les modèles exacts permettent de suivre les étapes du formalisme perturbatif de première quantification de la corde jusqu'au calcul des amplitudes. En pratique, quelques ajustements sont nécessaires du fait des subtilités introduites par un modèle particulier. Ainsi dans ce mémoire, nous avons choisi de nous concentrer dans le chapitre \ref{formalisme} sur les problèmes liés à la résolution du modèle. Nous suivons ainsi les étapes de l'étude de la corde dans l'espace de Minkowski et nous montrons comment elles se traduisent dans les différents modèles. Ceci nous permet de remplir un double objectif, celui de se débarasser d'un certain nombre de détails techniques qui risqueraient d'occulter l'exposition des phénomènes physiques et celui de proposer une approche pédagogique où les trois modèles étudiés au cours de cette thèse sont une source d'exemples atypiques.

Ainsi, la physique mise en jeu dans les différents modèles dépendants du temps, ainsi que certains résultats n'ayant pas leurs places dans le chapitre \ref{formalisme} sont exposés au chapitre \ref{contrereac}.

Le chapitre \ref{persp} présente les perspectives et les développements ultérieurs des travaux réalisés au cours de cette thèse.

Enfin, les annexes contiennent un bref formulaire (annexe \ref{df}) et les trois articles publiés au cours de la thèse \cite{Durin:2003gj, Berkooz:2004yy, Durin:2005ts} (annexes \ref{annart1}, \ref{annart2} et \ref{annart3}). Un compte-rendu de conférence \cite{Durin:2005ix} a été aussi publié, il n'est cependant pas reproduit.

%% file: formalisme.tex
\chapter{Formalisme perturbatif}
\l{formalisme}

Dans cette partie, nous exposons le formalisme de première quantification et nous l'illustrons  avec les problèmes que nous avons rencontrés lors de l'étude des différentes configurations dépendantes du temps. Nous commençons par établir un certain nombre de notations et de conventions qui seront utilisées dans ce qui suit.

Nous travaillerons dans un espace-temps, une variété noté $\M$, à $D$ dimensions, de signature $(1,d)$, où $d$ est le nombre de dimension spatiale, avec les conventions habituelles de physique théorique ($\hbar = c= 1$). Le choix des signes pour la signature sera conforme à la convention utilisée en cosmologie : $(-,+,\ldots,+)$. Comme nous allons le préciser ci-dessous, la corde définit une surface d'univers notée $\Sigma$, paramétrée par deux coordonnées $\t$ et $\s$. $\s$ décrit un intervalle compact, $[0,\pi]$ ou $[0,2\pi]$ en général et $\t$ parcourt $\R$. L'interprétation de ces paramètres est la suivante : $\s$ désigne un point de la corde à un instant $\t$. Cette corde évolue dans l'espace-temps $\M$, appelé \emph{espace de plongement} ou \emph{espace-cible}. Les lettres majuscules de la fin de l'alphabet, par exemple $X$, désigneront les coordonnées de plongement de la corde, alors que les lettres minuscules de la fin de l'alphabet, par exemple $x$, désigneront les coordonnées de l'espace-temps. Les indices $\mu$, $\nu$ correspondent à des directions d'espace-temps, tandis que les indices $a$, $b$ des directions de feuille d'univers : $\s_0 = \t, \s_1 = \s$. En toute rigueur, la corde est décrite par la courbe paramétrée suivante
\be
x^\mu = X^\mu (\s_a)
\ee
où $X^\mu$ sont les fonctions de plongements, mais nous ne ferons pas cette distinction et nous appellerons $X^\mu$ les coordonnées. Nous désignerons indifféremment par $\perp$ ou $i$, l'ensemble des directions spatiales ou l'ensemble des directions transverses au cône de lumière, défini plus loin, ou l'ensemble des directions spatiales transverses à une hypersurface de genre temps. Notons enfin que $\nu$ aura un sens bien précis dans le cas d'une configuration de champ électrique ou d'orbifold de boost (voir sections \ref{ssorb} et \ref{cchem}), sans risque de confusion possible.

\section{Théorie du point}
\l{thpoint}

L'étude des systèmes ponctuels apporte une aide précieuse lorsque nous étudions la corde. Non seulement les notions et le formalisme sont similaires, mais l'étude du système ponctuel constitué par le centre de masse de la corde permet, notamment dans le cas de l'espace de Misner (voir la section \ref{article2}), d'extraire une partie significative des effets physiques générés par la géométrie dépendante du temps.

Nous rappelons ainsi le formalisme de ligne d'univers pour les systèmes ponctuels, qui laisse naturellement place à un formalisme de surface d'univers pour la corde, puis nous résumons quelques effets quantiques liés à la présence d'un champ électrique ou d'une géométrie non triviale.

\subsection{Formalisme de ligne d'univers}
\l{lu}

Une particule relativiste (de masse non nulle $m$) est décrite par l'action suivante
\be
\S = -m \int\! ds
\ee
où $s$ est l'abscisse curviligne le long de la trajectoire. Définissons celle-ci par la paramétrisation $x^\mu = X^\mu(\t)$, on a alors
\be
\S = -m \int\! d\t \, \sqrt{-g_{\mu\nu}(X)\dpt X^\mu \dpt X^\nu} \l{actpp}
\ee
La forme de cette action suggère qu'on peut considérer la théorie de la particule relativiste comme une théorie des champs comportant $D$ champs scalaires $X^\mu$ définis sur un espace à une dimension, la ligne d'univers. Ce point de vue sera particulièrement utile lorsque nous traiterons le cas de la corde. Une propriété importante de cette action est son invariance par reparamétrisation de la ligne d'univers $\tau' = f(\tau)$, ce qui caractérise la ligne d'univers indépendamment de tout système de coordonnées. On peut dire que la théorie des $D$ champs scalaires est invariante par difféomorphisme de l'espace sur laquelle elle est définie.

Le traitement quantique de cette action, surtout lorsqu'on la généralise au cas de la corde, est difficile. On peut construire une action quadratique plus maniable qui possède l'avantage de donner un sens au cas de la particule de masse nulle. Cette action quadratique est obtenue en introduisant un degré de liberté de jauge $\eta(\t)$, lié à la métrique unidimensionnelle sur la ligne d'univers ($\eta(\t) = \sqrt{-\gamma_{\t\t} (\t)}$) et qui correspond justement à l'invariance par reparamétrisation de la ligne d'univers de l'action. L'introduction d'une notion de métrique sur la ligne d'univers peut sembler artificielle mais nous verrons qu'elle prend tout leur sens dans le cas de la corde.
\be
\S = \frac12 \int\! d\t \, \left( \frac1\eta g_{\mu\nu}(X) \dpt X^\mu \dpt X^\nu - \eta m^2 \right) \l{actqpp}
\ee
L'équation du mouvement pour $\eta$ s'écrit
\be g_{\mu\nu}(X) \dpt X^\mu \dpt X^\nu + \eta^2 m^2 = 0 \l{eometa}\ee
Si l'on remplace la solution pour $\eta$ dans l'action \eqref{actqpp}, on retrouve l'action initiale \eqref{actpp}.
L'action quadratique \eqref{actqpp} est également invariante sous les reparamétrisations de $\t$. Sous la transformation $\t'=f(\t)$ où $f$ est un difféomorphisme, les champs $X$, $g$ et $\eta$ se transforment de la façon suivante
\bse
\begin{align}
{X'}^\mu(\t') &= X^\mu(\t) \l{diffX}\\
g'_{\mu\nu}(X') &= g_{\mu\nu} (X) \\
\eta'(\t') &= \frac1{\dpt f(\t)} \eta(\t) \l{diffm}
\end{align}
\ese
L'expression de l'action de $X'$ et $\eta'$ est identique à l'expression \eqref{actqpp}. Il suffit de choisir $f$ telle que
\be
\dpt f(\t) = \lambda \eta(t) \l{trjauge}
\ee
où $\lambda$ est une constante pour que $\eta'$ soit une fonction constante. On peut ainsi choisir\footnote{Nous omettons le prime à présent.} la jauge $\eta = 1$. 
L'équation \eqref{eometa} s'écrit alors
\be g_{\mu\nu} (X) \dpt X^\mu \dpt X^\nu + m^2 = 0 \l{ctei}\ee
et correspond ainsi à l'annulation du tenseur énergie-impulsion de la théorie de champs à une dimension définie sur la ligne d'univers. 
Pour éviter les complication inhérentes aux espaces courbes (voir sous-section \ref{curve}), considérons le cas où l'espace de plongement est l'espace plat. Les équations du mouvement, dans cette même jauge, s'écrivent
\be \dpt^2 X^\mu = 0 \l{eomjged}\ee
et ont pour solutions
\be X^\mu = q^\mu + p^\mu \t \l{solpr} \ee
On voit alors que la relation \eqref{ctei} peut être interprétée comme une relation de couche de masse
\be p^\mu p_\mu + m^2 = 0 \l{onshell} \ee
La notation $p^\mu$ n'est pas le fruit du hasard puisque l'impulsion canonique $\pi_\mu$ conjuguée à $X^\mu$ s'écrit
\be \pi_\mu = g_{\mu\nu} \dpt X^\nu = \dpt X_\mu \ee 

Nous pouvons quantifier le système dans le formalisme de quantification canonique ou dans le formalisme d'intégrale des chemins. Comme nous utiliserons surtout le premier dans nos calculs en théorie de cordes et que nous utiliserons certains résultats du second, nous présentons les deux formalismes pour un système ponctuel évoluant dans l'espace plat. La quantification canonique covariante consiste à promouvoir au rang d'opérateur $X^\mu$ et $\pi^\mu$ et à appliquer la prescription suivante
\be [.\, ,.]_{\text{P.B.}} \longrightarrow -i[.\, ,.] \ee
où $[.\, ,.]_{\text{P.B.}}$ désigne les crochets de Poisson\footnote{Nous réservons la notation $\{.\, ,.\}$ pour les crochets d'anticommutation, en dépit de la notation couramment utilisée pour le crochet de Poisson.} et $[.\, ,.]$ les crochets de commutation. $X^\mu$ et $\pi^\mu$, qui classiquement vérifient la relation suivante,
\bse
\begin{align}
[X^\mu(\t)\, ,X^\nu(\t)]_{\text{P.B.}} &= [\pi^\mu(\t)\, ,\pi^\nu(\t)]_{\text{P.B.}} = 0 \\
[X^\mu(\t)\, ,\pi^\nu(\t)]_{\text{P.B.}} &= - \eta^{\mu\nu}
\end{align}
\ese
deviennent des opérateurs qui vérifient les relations de commutation suivantes
\bse
\begin{align}
[X^\mu(\t)\, ,X^\nu(\t)] &= [\pi^\mu(\t)\, ,\pi^\nu(\t)] = 0 \\
[X^\mu(\t)\, ,\pi^\nu(\t)] &= i \eta^{\mu\nu}
\end{align}
\ese
On en déduit les relations de commutation que vérifient les modes\footnote{Nous ne distingons pas au niveau de la notation les quantités classiques des opérateurs quantiques correspondants.} $q^\mu$ et $p^\mu$
\bse
\begin{align}
[q^\mu\, ,q^\nu] &= [p^\mu\, ,p^\nu] = 0 \\
[q^\mu\, ,p^\nu] &= i \eta^{\mu\nu}
\end{align}
\ese
Les états quantiques physiques $\lvert \text{phys} \rangle$ doivent vérifier la condition de couche de masse \eqref{onshell}
\be (p^\mu p_\mu + m^2)\lvert \text{phys} \rangle = 0 \ee
Dans la représentation $q$, l'opérateur $p_\mu$ s'écrit $-i\p/\p q^\mu$ et cette condition devient l'équation de Klein-Gordon pour la fonction d'onde $\phi(q)$ associée à l'état $\lvert \text{phys} \rangle$.
Nous verrons à la section \ref{quantcov} comment nous devons adapter ce formalisme au cas de la corde. 

L'évolution dans le formalisme d'intégrale des chemins est décrite par le propagateur entre deux points de coordonnées $x^\mu$ et ${x'}^\mu$. Nous nous plaçons dans l'espace plat et nous supposons que
\be
X^\mu (\t_0) = x^\mu \text{ , } X^\mu (\t_1) = {x'}^\mu
\ee
L'intégrale des chemins pour le propagateur entre $x^\mu$ et ${x'}^\mu$ s'écrit alors
\be
P(x,x') = \cN \int\! [D\eta][DX]\, \exp\left[\frac{i}2 \int_{\t_0}^{\t_1}\! d\t \, \left(\frac1\eta \dpt X^\mu \dpt X_\mu - \eta m^2 \right)\right]
\ee
où $DX$ représente $\prod_{\mu=0 \ldots D-1} DX^\mu$, $[Df]$ représente le produit infini formel des intégrations sur $f(\t)$, $\prod_{\t \in [\t_0, \t_1]} d[f(\t)]$ et où $\cN$ est une constante de normalisation. L'intégration inclut les degrés de liberté de jauge, qu'il faut éliminer. Nous fixons la jauge en introduisant une fonction $\delta$ normalisée à 1
\be
1 =  \int [Df] \Delta_{\text{F.P.}}(\eta^f) \delta (\eta^f - 1) \l{FD}
\ee
où $\eta^f (f(\t)) = \eta(\t)/\dpt f(\t)$ est une simple réécriture de \eqref{diffm}. De même nous nommons $X^f$ l'image de $X$ sous une transformation de jauge (ici une simple reparamétrisation, voir \eqref{diffX}). Nous obtenons alors
\begin{multline}
P(x,x') = \cN \int\! [D\eta][DX][Df]\\
\Delta_{\text{F.P.}}(\eta^f) \delta(\eta^f - 1) \exp\left[\frac{i}2 \int_{\t_0}^{\t_1}\! d\t \, \left(\frac1\eta \dpt X \cdot \dpt X - \eta m^2 \right)\right]
\end{multline}
Comme l'action et la mesure $[D\eta][DX]$ sont invariantes sous transformation de jauge, nous pouvons remplacer $\eta$ et $X$ par $\eta^f$ et $X^f$, à condition de traiter précisément les bornes d'intégration de l'action,
\begin{multline}
P(x,x') = \cN \int\! [Df][D\eta^f][DX^f]\, \Delta_{\text{F.P.}}(\eta^f) \delta(\eta^f - 1)\\ \exp\left[\frac{i}2 \int_{f(\t_0)}^{f(\t_1)}\! d\t' \, \left(\frac1{\eta^f(\t')} \p_{\t'} X^f(\t') \cdot \p_{\t'} X^f(\t') - \eta^f(\t') m^2 \right)\right]
\end{multline}
Nous avons noté $\t'$ la variable d'intégration pour insister sur le fait que l'invariance sous transformation de jauge nécessite un changement de variable dont l'effet est loin d'être négligeable.

Enfin, nous renommons les variables d'intégration $\eta^f$ et $X^f$ en $\rho$ et $Y$, ce qui montre qu'en fait, rien ne dépend de $f$ dans l'intégrand de l'intégrale des chemins, sauf les bornes de l'intégrale de l'action. Nous pouvons intégrer sur $f(\t)$ pour $\t \in ]\t_0, \t_1[$, ce qui donne un terme infini, car c'est le volume du groupe de jauge. Nous définissons $\cN$ comme l'inverse de ce volume pour éliminer cet infini et ne garder ainsi que la partie pertinente de l'intégrale des chemins.
L'intégration sur $f(\t_0)$ et $f(\t_1)$ est plus subtile. Pour comprendre le sens physique de ces paramètres restants, nous intégrons sur $\rho$, ce qui fixe la jauge à 1, c'est-à-dire $\eta^f(\t')=1$ avant qu'on ne renomme les variables d'intégration. Alors, d'après la relation \eqref{diffm}, $f'(\t) = \eta(\t)$ ou encore
\be
f(\t) = \int_{\t_0}^{\t} \eta(\tilde \t) d\tilde \t + c
\ee
À ce stade l'intégrale des chemins a pour expression
\be
P(x,x') = \int\! df(\t_0)df(\t_1)[DY]\, \Delta_{\text{F.P.}}(1) \exp\left(\frac{i}2 \int_{f(\t_0)}^{f(\t_1)}\! d\t \, \dpt Y(\t) \cdot \dpt Y(\t) - m^2 \right)
\ee
La borne inférieure de l'action, $f(\t_0) = c$, correspond à un choix d'origine sur la ligne d'univers. On peut interpréter ceci comme les transformations $f(\t) = \t + c$ qui laissent la jauge $\eta = 1$ invariante. De manière générale, il s'agit de degrés de liberté correspondant à des symétries non fixées par le choix de jauge et qui forme un \emph{groupe de symétries résiduelles}. Dans notre cas nous pouvons éliminer cette symétrie résiduelle. Nous nous ramenons à une borne inférieure égale à 0 par une transformation $\t' = \t-f(\t_0)$ et, en utilisant le fait que la mesure $[DY]$ est invariante sous cette transformation, nous faisons la même manipulation que précédemment en renommant la variable d'intégration $Y'$ en $X$. On peut alors intégrer sur $f(\t_0)$, ce qui produit à nouveau un facteur infini, absorbé dans $\cN$.

La borne supérieure de l'action est alors égale à $f(\t_1)-f(\t_0)$ (mais ne dépend plus de $f(\t_0)$), c'est-à-dire à $\int_{\t_0}^{\t} \eta(\tilde \t) d\tilde \t$, qui correspond à la longueur invariante des chemins $l=\lVert x' - x \rVert$, comme le suggère l'interprétation de $\eta$ comme une métrique de ligne d'univers. Ce paramètre, qui ne peut pas être éliminé, correspond de manière générale à un degré de liberté de jauge qui n'est pas contraint par les symétries. Il s'agit d'un exemple de \emph{paramètre de Teichmüller} ou \emph{module}.

Ces deux types de degrés de liberté résiduels correspondent à la différence entre les symétries globale et les symétries locales sous lesquelles l'action est invariante.

Il reste un facteur dont nous n'avons pas parlé, le déterminant de Fadeev-Popov $\Delta_{\text{F.P.}}(1)$. Il correspond à la mesure d'intégration de la transformation $f$ dans \eqref{FD}. Dans notre cas il est égal à 1, mais en général, il contribue de façon importante à l'intégrale des chemins. Il est mis sous la forme d'une contribution à l'action en terme de champs appelés \emph{fantômes} car ils ne vérifient pas le théorème spin-statistique en théorie quantique des champs. En pratique, leur contribution élimine les degrés de liberté dont la propagation n'est pas physique. Dans nos calculs, y compris lorsque nous nous exposerons le formalisme de la corde, cette contribution sera toujours incluse de manière implicite.

Finalement, l'intégrale des chemins s'écrit
\be
P(x,x') = \int_0^\infty\!\! dl\int [DX]\, \exp\left(\frac{i}2 \int_0^l\! d\t \, \dpt X(\t) \cdot \dpt X(\t) - m^2 \right) \l{propfinal}
\ee
ce qui donne le résultat habituel\footnote{Un dernier changement de variable $\t' = \t/l$ et un prolongement en temps euclidien $\t'' = -i\t'$ permettent de retrouver l'expression correspondante (3.1.19) dans le cours \cite{Kiritsis:1997hj}, qui expose le calcul détaillé et le résultat (équations (3.1.29) et (3.1.31) de ce même cours).}.

Nous reviendrons sur la construction des amplitudes, sur la procédure qui permet de fixer la jauge et sur les paramètres résiduels en théorie des cordes dans la section \ref{ampl}.

\subsubsection{Couplage au champ électromagnétique}

En guise d'introduction à la configuration de corde couplée à des champs électromagnétiques, nous allons exposer comment le couplage minimal entre une particule ponctuelle et un champ électromagnétique s'écrit dans le formalisme de la ligne d'univers. Le champ électromagnétique $F_{\mu\nu}$ est défini à partir d'un potentiel $A_\mu$. À l'action \eqref{actqpp}, nous ajoutons le terme $\S_{\text{em}}$ dont l'expression est la suivante
\be
\l{cchempp}
S_{\text{em}} = \int \! d\t A_\mu(X) \dpt X^\mu
\ee
et qui modifie les équations du mouvement \eqref{eomjged} de $X^\mu$ comme nous nous y attendions
\be
\dpt^2 X^\mu = F_{\mu\nu} \dpt X^\nu
\ee
Nous avons inclus la charge $e$ de la particule dans la définition du potentiel $A_\mu$. Nous verrons que dans le cas de la corde ouverte (sous-section \ref{cchem}), le couplage au champ électromagnétique se fait de la même façon, où chaque extrémité de la corde constitue un système ponctuel.

\subsection{Création de paires dans un champ électrique constant et uniforme : méthode du temps propre de Schwinger}

La création de paires de particules chargées dans un champ électrique est un effet décrit dans le formalisme de la théorie quantique des champs.
Ce phénomène est caractérisé par l'amplitude de probabilité pour que, dans l'état de vide de particules avec un champ électrique non nul, aucune paire ne soit émise. Notons que la probabilité d'émission de paires peut être interprétée comme la partie imaginaire de l'énergie de cet état de vide. Ce que nous allons présenter n'a donc pas de lien direct avec le formalisme décrivant \emph{une} particule libre, éventuellement couplée à un champ électromagnétique.

Dans le formalisme de la corde, nous calculerons également la partie imaginaire de l'énergie du vide, qui correspond à la production de paires de cordes (comme cela a été mis en évidence pour les cordes ouvertes \cite{Bachas:1992bh}). Notons que l'effet de production de paires n'est pas forcément associé à la présence d'un champ électrique, mais nous nous servirons à plusieurs reprises de l'analogie avec un champ d'électron-positron couplé à un champ électrique.

Nous allons donc résumer la méthode de temps propre introduite par Schwinger \cite{Schwinger:1951nm} et le calcul exposé dans \cite{Itzykson:1980rh}, section 4-3. L'amplitude de probabilité $S_0$ pour qu'aucune paire d'électron-positron ne soit émise dans le vide peut être mise sous la forme suivante
\be
\ln S_0 = \tr \ln \left(\frac{/\!\!\! p - e /\!\!\!\! A(X) - m +i\e}{/\!\!\! p -m+i\e} \right)
\ee
où $\tr$ comprend une trace sur les états de spins et une intégrale sur la position $x$, valeur propre de $X$. Le taux de désintégration $w$ est défini à partir de $S_0$ de la façon suivante
\be
\lvert S_0 \rvert^2 = \exp\left( - \int d^4 x \,w(x) \right)
\ee
Une définition équivalente, que nous utiliserons à la section \ref{article3}, relie $w$ à la fonction de partition du vide $\F$
\be
\int d^4 x \,w(x) = -2\im \F
\ee

De façon réductrice, nous pouvons dire que la méthode de temps propre se résume à la représentation suivante du logarithme d'un quotient
\be
\ln \frac{a}b = \int_0^\infty \frac{ds}s \, (e^{is(b+i\e)} - e^{is(a+i\e)})
\ee
où la limite $\e \to 0$ est sous-entendue.

De manière générale, la méthode de temps propre est une représentation du propagateur de Feynman $G_F$. De manière formelle, le propagateur de Feynman est l'inverse de l'opérateur cinétique écrit, pour un champ scalaire chargé par exemple, sous la forme $\Delta=\delta^{(D)} (x-y) \left( -\D_\mu \D^\mu + m^2 - i\e \right)$ où $\D_\mu$ est la dérivée covariante. On peut calculer 
\be
G_F = -\Delta^{-1} = -i \int_0^\infty d\t \, e^{i\Delta \t}
\ee
$G_F$ s'écrit comme l'opérateur d'évolution d'un système dont l'hamiltonien est $\Delta$. Schwinger propose une formulation plus rigoureuse, en définissant le noyau $K(x,y;\t) = \langle x \rvert e^{i\Delta \t} \rvert y \rangle$. Ce noyau vérifie l'équation de Schrödinger suivante
\be
-i\dpt K(x,y;\t) = (\D_\mu \D^\mu - m^2) K(x,y;\t)
\ee
avec pour condition initiale
\be
K(x,y;0^+) = \delta^{(D)} (x-y)
\ee
Le propagateur de Feynman s'écrit alors
\be
G_F (x,y) = \int_0^\infty d\t\, K(x,y;\t)
\ee
Nous pouvons faire le lien avec l'intégrale des chemins dans le formalisme de ligne d'univers : $G_F$ est le propagateur \eqref{propfinal} et $K$ l'intégrand de l'intégrale sur $l$, à condition de renommer $l$ en $\t$ et la variable d'intégration $\t$ en $\tilde \t$.

Revenons au calcul de la probabilité $w$ de créer des paires par unité de volume et de temps. La représentation en temps propre de Schwinger permet de mener un calcul non perturbatif en $E$, intensité du champ électrique constant et uniforme. Le résultat ne dépend finalement pas de $x$
\be
w = \frac{\alpha E^2}{\pi^2} \sum_{n=1}^\infty \frac1{n^2} \exp \left(-\frac{n\pi m^2}{\lvert eE\rvert} \right)
\ee
où $e$ est la charge de l'électron, $m$ sa masse, et $\alpha$ la constante de structure fine $\alpha = \frac{e^2}{4\pi \hbar c}$.

\subsection{Notions de théorie des champs dans les espaces-temps courbes}
\l{curve}

Comme nous l'avons évoqué à la sous-section précédente et comme nous l'expliquerons en détail dans les sections suivantes, la théorie des cordes peut être considérée comme une théorie de champs et si l'espace de plongement n'est pas plat, nous devons tenir compte des effets de la géométrie sur la quantification des champs $X^\mu$. Comme dans le cas de production de paires dans un champ électrique, il s'agit d'une étude semi-classique, c'est-à-dire que nous décrirons les effets provoqués par une configuration classique du champ gravitationnel sur un champ quantique, en nous inspirant du traitement exposé dans l'ouvrage \cite{Birrell:1982ix}.

\subsubsection{Géométrie dépendante du temps, transformation de Bogolioubov et production de particules}

L'action d'un champ scalaire $\phi(x)$ de masse $m$ dans un espace courbe doté d'une métrique $g_{\mu\nu}(x)$ s'écrit
\be
\S= \frac12 \int d^D x \sqrt{-g(x)}\left[g^{\mu\nu(x)} \p_\mu \phi(x) \p_\nu \phi(x) - \left(\xi R(x) + m^2\right) \phi^2 (x) \right]
\ee
où $\xi$ est un facteur numérique, $R$ le scalaire de Ricci et $g$ le déterminant de la métrique. L'équation du mouvement vérifiée par $\phi$ est
\be
\left[\Box + m^2 + \xi R(x)\right] \phi(x) = 0 \l{eomscg}
\ee
où $\Box \phi = g^{\mu\nu} \nabla_\mu \nabla_\nu \phi = \sqrt{-g}\, \p_\mu\left[\sqrt{-g}\, g^{\mu\nu} \p_\nu \phi\right]$. Nous pouvons distinguer deux cas particuliers
\bi
\i le couplage minimal $\xi = 0$
\i le couplage conforme $\xi = \frac{D-2}{4(D-1)}$, ainsi nommé car les équations du mouvement \eqref{eomscg} sont invariantes sous une transformation conforme $g'_{\mu\nu}(x) = \Omega^2(x) g_{\mu\nu}(x)$, à condition que le champ $\phi$ se transforme de la façon suivante $\phi'(x) = \Omega^{(2-D)/2}(x) \phi(x)$.
\ei

Si nous supposons que l'espace-temps est globalement hyperbolique, c'est-à-dire qu'il admet une foliation $\R \times \Sigma$ où $\Sigma$ est une hypersurface de genre espace et $\R$ représente l'axe d'une coordonnée temporelle $x^0$, nous pouvons généraliser le produit scalaire
\be
(\phi,\psi) = -i\int_\Sigma d\Sigma \sqrt{g_\Sigma (x)}\,n^\mu\, \phi(x) \overset{\leftrightarrow}{\p}_\mu \psi(x)
\ee
où $n^\mu$ est un vecteur unitaire dirigé vers les $x^0$ croissants (le futur) et orthogonal à $\Sigma$. De même que dans l'espace plat, nous pouvons trouver une base orthonormale de solutions $u_p(x)$ de \eqref{eomscg} qui vérifient ainsi
\be
(u_p, u_q) = \delta_{pq} \text{ , } (u_p^*,u_q^*) = -\delta_{pq} \text{ , } (u_p, u_q^*) = 0
\ee
Les indices $p$ et $q$ représentent l'ensemble des quantités qui distinguent les solutions. Le champ $\phi$ peut alors être développé selon ces modes $u_p (x)$
\be
\phi(x) = \sum_p a_p u_p(x) + a_p^\dagger u_p^*(x)
\ee
et quantifié en imposant les relations de commutations suivantes
\be [a_p, a_{p'}^\dagger] = \delta_{pp'} \ee
Nous pouvons alors  construire à partir de là un état du vide $\lvert 0 \rangle_u$, un espace de Fock et définir les mêmes quantités que dans l'espace plat. Cependant, l'analogie s'arrête ici. En effet, dans l'espace de Minkowski, le système de coordonnées cartésien $(t,x,y,z)$ tient une place particulière, liée au groupe de Poincaré sous l'action duquel la métrique plate est invariante. À ce système de coordonnées correspond un ensemble de solutions, les ondes planes $e^{i\vec{k}\cdot\vec{x}-i\w t}$, qu'il est facile de diviser en solutions de fréquence positive $\w >0$ et solutions de fréquence négative $\w<0$. Ces diverses définitions sont absolues, c'est-à-dire valable dans tout l'espace plat, quel que soit l'observateur \emph{inertiel}.

Par contre, dans un espace courbe (ou pour un observateur non inertiel), il n'existe aucun système de coordonnées privilégié. Même si l'espace admet un groupe de symétrie, singulariser un système de coordonnées n'a pas de sens en relativité générale. Ainsi, une autre décomposition du champ $\phi$ en terme de modes $b_q$ est tout aussi valable
\be 
\phi(x)= \sum_q b_q v_q(x) + b_q^\dagger v_q^*(x)
\ee
Nous définissons alors un nouveau vide $\lvert 0 \rangle_v$, qui vérifie $b_q \lvert 0 \rangle_v = 0$ pour tout $q$, et un nouvel espace de Fock. Comme les deux ensembles de solutions sont des bases de l'espace des solutions, il existe une transformation linéaire qui les relient. Cette transformation est appelée \emph{transformation de Bogolioubov}. Elle s'écrit
\be
v_q = \sum_r \alpha_{qr} u_r + \beta_{qr} u_r^*
\ee
ou encore, pour les modes,
\be
b_q = \sum_r \alpha^*_{qr} a_r - \beta^*_{qr} a_r^\dagger
\ee
Les coefficients $\alpha$ et $\beta$ peuvent être calculés à partir des formules suivantes
\be
\alpha_{qr} = (v_q, u_r) \text{ , } \beta_{qr} = -(v_q, u_r^*)
\ee
et vérifient les propriétés suivantes
\bse
\begin{align}
\sum_r (\alpha_{pr} \alpha_{qr}^* - \beta_{pr} \beta_{qr}^*) &= \delta_{pq} \\
\sum_r (\alpha_{pr} \beta_{qr}^* - \beta_{pr} \alpha_{qr}^*) &= 0
\end{align}
\ese
Cette ambiguité dans le développement du champ $\phi$ se traduit physiquement par un phénomène de production de particules. En effet la valeur moyenne dans le vide $\lvert 0 \rangle_v$ de l'opérateur $N_p  = a_p^\dagger a_p$ pour le nombre de particules dans le mode $u_p$ vaut
\be
{}_v \langle 0 \rvert N_p \lvert 0 \rangle_v = \sum_q \lvert \beta_{qp} \rvert^2
\ee
Nous voyons que cette production de particules est générique : elle apparaît dès qu'un coefficient $\beta$ est non nul. Si nous considérons cette production, non comme un effet d'une géométrie dépendante du temps, mais comme un choix de référentiel (inertiel ou non), la situation où un observateur mesure la présence de particules alors qu'un autre ne mesurera rien est tout à fait générale. Le concept de particule en relativité générale est donc problématique, puisqu'il est lié à l'observateur, au choix particulier d'une décomposition en modes, ou encore d'un système de coordonnées. En effet, la réponse d'un détecteur de particule lié à un observateur particulier, donnée par l'opérateur $N = a^\dagger a$, dépend de la décomposition en modes du champ $\phi$ et donc de la base de solution qui correspond à cet observateur (adaptée par exemple à sa ligne d'univers). Les modes sont définis sur l'espace-temps complet, ou du moins sur un large domaine de celui-ci, ce qui explique pourquoi le concept de particule est global et ne peut servir de sonde fiable de l'état quantique du champ $\phi$. Pour cela, il faut définir des grandeurs locales, par exemple la valeur moyenne du tenseur énergie-impulsion $\langle \phi \rvert T_{\mu\nu} (x) \lvert \phi \rangle$. Nous ne n'entrerons pas plus dans les détails.

En effet, les configurations que nous avons étudié en théorie des cordes (voir section \ref{article2}) correspondent à des espace-temps asymptotiquement plat, où nous pouvons donner un sens physique plus précis à la notion de particule. À l'infini passé et futur, l'espace est plat et nous pouvons choisir le vide commun à tous les observateurs inertiels\footnote{Ce vide est l'\emph{un} des vides qu'il est possible de définir dans l'espace de Minkowski, mais c'est le seul qui soit commun à tous les observateur inertiels.  Il suffit en fait que l'espace asymptotique admette un vide privilégié et une définition naturelle des états de particules, sans être forcément plat, mais nous n'aurons pas besoin de ce raffinement pour notre étude.}, dans lequel la notion de particule est correctement définie. Nous définissons ainsi un vide $in$, correspondant au vide de l'espace plat dans l'infini passé, et un vide $out$ correspondant au vide de l'espace plat dans l'infini futur. La décomposition du champ $\phi$ s'écrit ainsi
\bse
\begin{align}
\phi(x) &= \sum_p a^{in}_p u^{in}_p(x) + {a^{in}_p}^\dagger (u^{in}_p)^*(x) \\
\phi(x) &= \sum_p a^{out}_p u^{out}_p(x) + {a^{out}_p}^\dagger (u^{out}_p)^*(x)
\end{align}
\ese
où
\bse
\begin{align}
u^{in}_p(x) &\propto e^{-i\w_{in} x^0} \text{\ \ \ } \phantom{{}^t} x^0 \to -\infty \\
u^{out}_p(x) &\propto e^{-i\w_{out} x^0} \text{\ \ \ } x^0 \to +\infty \\
\end{align}
\ese
Les modes $in$ et $out$ sont reliés par une transformation de Bogolioubov, qui encode la production de particules entre l'infini passé et l'infini futur. Ainsi, si nous partons d'un état de vide à l'infini passé, nous obtenons un état contenant généralement des particules à l'infini futur et ces particules ont été produites par la géométrie intermédiaire.

Notons que cette configuration d'espace-temps asymptotiquement plat permet de traiter plusieurs problèmes intéressants, comme la radiation de Hawking des trous noirs ou la production de particules provoquée par l'expansion cosmologique.

\subsubsection{Vide adiabatique et vide conforme}

Dans ce dernier cas de l'expansion cosmologique, l'espace-temps ne comporte pas forcément de régions statiques $in$ et $out$. Si l'on choisit une base de solutions, le vide et les états de Fock correspondants n'ont pas en général de signification physique directe. Cependant, il existe dans ces géométries une classe d'observateurs privilégiés, ceux liés aux \emph{référentiels comobiles}. La notion d'\emph{observateur comobile} est celle qui se rapproche le plus de la notion d'observateur inertiel dans l'espace de Minkowski. C'est pourquoi nous nous attendons à ce que la notion de particule ait un sens pour ces observateurs. Deux approches permettent alors de résoudre ce problème.

La première consiste à considérer le cas où l'expansion est très faible. Dans la limite où l'expansion est nulle, nous retrouvons l'espace plat, où le concept de particule est bien défini. Nous nous attendons alors à ce qu'il soit possible de définir le concept de particule de manière approché lorsque l'expansion (ou de manière générale la variation de la géométrie) est très petite. Pour cela, nous construisons une base de solutions qui minimise la production totale de particules. Précisons un peu cette construction dans le cas d'une géométrie de Robertson-Walker à section spatiale plate avec un champ scalaire qui a un couplage gravitationnel conforme. La métrique s'écrit
\be
ds^2 = dt^2 - a^2(t) d(x^i)^2
\ee
où $x^i$ représente indifféremment l'ensemble des directions spatiales et le vecteur correspondant. En définissant un \emph{temps conforme} $\eta$ de la façon suivante
\be
t = \int^\eta a(\tilde \eta) d \tilde \eta
\ee
on peut mettre la métrique sous la forme
\be
ds^2  = C(\eta) (d\eta^2 - d(x^i)^2) \l{mconf}
\ee
avec $C(\eta) = a^2(\eta)$. Les solutions des équations du mouvement s'écrivent
\be
u_{k^i} = (2\pi)^{(1-D)/2} C^{(2-D)/4}(\eta)\, e^{i k^i x^i} \xi_k(\eta)
\ee
où $k = \lVert k^i \rVert$. $\xi_k$ vérifie l'équation suivante
\be
\p_\eta^2 \xi_k + \w_k^2 \xi_k =0
\ee
avec $\w_k = k^2 + C(\eta) m^2$.
L'approximation adiabatique consiste à négliger les variations de la solution formelle de type WKB (généralisation de l'onde plane) donnée par l'expression suivante
\be
\xi_k = (2 W_k)^{-\frac12} \, \exp \left(-i\int^\eta W_k(\tilde \eta) d\tilde \eta\right)
\ee
où $W_k$ satisfait l'équation non linéaire suivante
\be
W_k^2 = \w_k^2 - \frac12\left(\frac{\ddot{W}_k}{W_k} - \frac32\frac{\dot{W}_k^2}{W_k^2} \right)
\ee
La borne inférieure de l'intégrale, non spécifiée, correspond à un facteur de phase arbitraire.
Sans entrer plus dans les détails, mentionnons qu'on peut mener un développement lorsque $T \to \infty$, où $T$ est un paramètre de faible variation introduit en remplaçant $\eta$ par $\eta' = \eta/T$. Ainsi, quelle que soit la fonction $C(\eta)$, $C(\eta')$ et toutes ses dérivées varient infiniment peu dans la limite $T \to \infty$. Dans ce développement, le terme en $T^{-n}$ est appelé terme d'ordre adiabatique $n$. À partir de l'approximation d'ordre zéro, $W_k^{(0)} = \w_k$, nous voyons que l'ordre adiabatique $n$ correspond au nombre de dérivées de $C$ dans l'expression de $W_k^{(n)}$. Nous pouvons écrire en général la relation entre les solutions exactes et les solutions approchées $u_k^{(n)}$, obtenues à partir la solution d'ordre adiabatique $n$ $\xi_k^{(n)}$
\be
u_k(\eta) = \alpha_k^{(n)}(\eta)\, u_k^{(n)}(\eta) + \beta_k^{(n)}(\eta)\, {u_k^{(n)}}^*(\eta) \l{adiabdev}
\ee
$\alpha_k^{(n)}(\eta)$ et $\beta_k^{(n)}(\eta)$ sont des constantes à l'ordre $n$, car $u_k^{(n)}$ est solution de l'équation du mouvement à ce même ordre.

Nous pouvons à présent construire le vide adiabatique. Partons d'un ensemble de solutions $u_k^{(n)}$ à l'ordre adiabatique $n$. Nous choisissons ensuite un instant $\eta_0$ quelconque et nous posons
\bse
\begin{align}
\alpha_k^{(n)}(\eta_0) &= 1 + \O(T^{-(n+1)}) \\
\beta_k^{(n)}(\eta_0) &= 0 + \O(T^{-(n+1)})
\end{align}
\ese
Nous construisons ensuite $u_k$ à partir des solutions approchées. Tout d'abord, nous définissons $u_k$ par \eqref{adiabdev}. Ainsi $u_k$ est pour l'instant défini seulement en $\eta_0$. Nous déterminons alors $\alpha$ et $\beta$ pour tout $\eta$ de façon à ce que les $u_k$ forment une base de solutions \emph{exactes} de l'équation du mouvement. Ces solutions sont appelées \emph{modes adiabatiques de fréquence positive}. Nous insistons une nouvelle fois sur le fait qu'il s'agit de solutions exactes. Nous pouvons alors tout à fait construire le vide $\lvert 0 \rangle_\text{ad.}$ correspondant. Ce vide constitue une approximation \emph{adiabatique} au vide $in$ des régions statiques et est appelé naturellement \emph{vide adiabatique}.

À ce stade nous pouvons faire deux remarques
\bi
\i Nous pouvons définir une infinité de vide adiabatique pour un ordre $n$ fixé. En effet, à chaque choix  de la valeur $\eta_0$ correspondra un vide adiabatique \emph{a priori} différent des autres.
\i Bien qu'un vide adiabatique soit moins précis qu'un vide $in$ ou $out$, sa représentation des particules physiques (mesurées par des détecteurs de particules dans un référentiel comobile) est la meilleure disponible si l'espace-temps n'a pas de régions $in$ et $out$.
\ei

La seconde approche au problème de la définition du concept de particules consiste à exploiter les symmétries éventuellement présentes et sélectionner ainsi une base de solutions et des états de particules de manière canonique. Nous nous concentrerons sur le cas d'un champ $\phi$ à couplage conforme, ce qui correspond aux configurations que nous étudierons en théorie des cordes. La métrique peut s'écrire sous la forme suivante
\be
g_{\mu\nu} = \Omega^2(x) \eta_{\mu\nu}
\ee
Comme nous avons supposé que le champ était invariant sous les transformations conformes (ce qui suppose qu'il soit de masse nulle), on peut se ramener par une transformation conforme qui change $\phi$ en $\phi'$ à l'équation du mouvement dans l'espace plat pour $\phi'$. Le champ $\phi$ se décompose alors sur une base de solutions d'onde plane modulo un facteur conforme
\be
\phi(x) = \Omega^{(2-D)/2}(x) \sum_k \left(a_k u_k^0(x) + a_k^\dagger (u_k^0)^*(x)\right)
\ee
où $u_k^0$ est proportionnel à $e^{-i k \cdot x}$. Le vide associé à ces modes est appelé \emph{vide conforme}.

Notons que même un détecteur de particules comobile enregistrera en général la présence de particules dans le vide conforme. Par contre, si un champ est à un instant donné dans l'état de vide conforme, il le restera à tous temps et il n'y aura pas de production de particules. Ceci est interprété de la façon suivante dans une configuration où l'expansion cosmologique cesse de façon douce.

Un détecteur mis en marche après l'expansion n'enregistrera aucune particule. L'éventuelle réponse d'un détecteur mis en marche pendant l'expansion n'est alors pas pertinente : elle est l'artefact de la disparition du sens intuitif du concept de particules dans les régions de courbure non nulle.

Nous rencontrerons ces problèmes liés aux espaces courbes dans notre étude de l'espace de Misner en théorie des cordes que nous exposons dans la section \ref{article2}. Les modes zéros de la corde \emph{twistée}\footnote{Nous conservons le terme anglais. Il est défini à la sous-section \ref{ssorb}.} sont décrits par la théorie quantique de la particule chargée dans l'espace de Rindler, mais nous n'en exposerons pas les détails. Nous nous bornerons à citer les résultats utiles dans la section \ref{article2}. Un traitement complet est donné dans \cite{Gabriel:1999yz}.

\section{Cordes et action sur la feuille d'univers}
\l{cfu}

De la même façon que le mouvement classique d'une particule ponctuelle est décrit par une action proportionnelle à la longueur de la ligne d'univers, le mouvement classique d'une corde est décrit par une action proportionnelle à la surface de la ligne d'univers. Cette action est appelée action de Nambu-Goto
\be
\S_{\text{NG}} = - \frac1{2\pi\apr} \int d\t d\s \sqrt{- \det h_{ab}}
\ee
où $h_{ab}$ est la métrique induite sur la surface d'univers par le plongement dans l'espace-cible
\be
h_{ab} = \p_a X^\mu \p_b X_\mu
\ee
et $\frac1{2\pi\apr}$ aussi noté $T$ est la tension de la corde. $\alpha'$ est appelée la pente de Regge, allusion à la tentative de décrire des séries de résonance en physique des particules par des modèles de cordes appelés \emph{modèles duaux}.
Comme pour la particule ponctuelle, on peut obtenir une action quadratique, \emph{l'action de Polyakov}\footnote{Polyakov a développé le formalisme d'intégrale des chemins correspondant à cette action, mais celle-ci a été trouvée par Brink, di Vecchia, Howe, Deser et Zumino.}, en introduisant une métrique $\gamma$ sur la feuille d'univers
\be
\S_{\text{P}} = - \frac1{4\pi\apr} \int d\t d\s \sqrt{- \gamma} \, \gamma^{ab} \p_a X^\mu \p_b X_\mu \l{actP}
\ee

Cette action est invariante sous les symétries suivantes
\bi
\i invariance de Poincaré
\bse
\begin{align}
{X'}^\mu (\t, \s) &= \Lambda^\mu_{\phantom{\mu}\nu} X^\nu (\t, \s) + a^\mu \\
\gamma'_{ab} (\t, \s) &= \gamma_{ab} (\t, \s)
\end{align}
\ese
\i invariance par difféomorphisme de la feuille d'univers
\bse
\begin{align}
{X'}^\mu (\t', \s') &= X^\mu (\t, \s) \\
\frac{\p{\sigma'}^c}{\p\sigma^a} \frac{\p{\sigma'}^d}{\p\sigma^b} \gamma'_{cd} (\t', \s') &= \gamma_{ab} (\t, \s)
\end{align}
\ese
\i invariance de Weyl de la feuille d'univers
\bse
\begin{align}
{X'}^\mu (\t, \s) &= X^\mu (\t, \s) \\
\gamma'_{ab} (\t, \s) &= e^{2\w (\t, \s)} \gamma_{ab} (\t, \s)
\end{align}
\ese
pour $\w(\t,\s)$ arbitraire
\ei

L'action de Nambu-Goto possède les deux premières symétries mais pas la troisième. Cette troisième symétrie provient du fait que deux métriques de feuille d'univers image l'une de l'autre par une transformation de Weyl définissent le même plongement dans l'espace-cible.


Nous pouvons considérer l'action \eqref{actP} comme définissant une théorie de $D$ champs scalaires définis sur un espace courbe à deux dimensions. L'invariance sous les symétries permet de se ramener à une métrique plate $\gamma_{ab} = \eta_{ab}$. C'est la \emph{jauge unitaire}. La symétrie de Weyl permet de se ramener à une jauge dite \emph{conforme} $\gamma_{ab}(\s) = e^{2w(\s)} \eta_{ab}$ pour laquelle le tenseur de Riemann $R_{abcd}$ de la feuille d'univers est nul. Comme nous le verrons plus en détail dans la sous-section \ref{invconforme}\footnote{Ce point est lié à l'identification de symétries et paramètres résiduels, que nous avons expliqué à la sous-section \ref{lu} dans le cas de la particule ponctuelle et que nous exposerons brièvement dans le cas de la corde à la sous-section \ref{agfu}.}, une fois que la jauge unitaire pour la métrique a été choisie, il reste une symétrie résiduelle sous laquelle l'action est invariante : la symétrie conforme (en toute rigueur, il faut prolonger analytiquement la coordonnées de temps propre $\t$ pour se ramener à une surface euclidienne). La théorie définie sur la feuille d'univers est ainsi une théorie conforme des champs, ce qui permet d'utiliser toute la puissance du formalisme associé (voir section \ref{thconf}).

D'autre part, la procédure de Fadeev-Popov qui permet de fixer la jauge introduit des champs de fantômes et la quantification rigoureuse de l'action complète (matière et fantômes) nécessite d'introduire la \emph{symétrie de Becchi, Rouet, Stora et Tyutin} (BRST) et la \emph{quantification BRST} associée. Nous n'exposerons cependant pas ce formalisme et nous nous restreindrons au formalisme plus simple dit d'\emph{ancienne quantification covariante}, qui introduit des règles a priori arbitraires mais justifiées par le formalisme complet et qui suffit à saisir les problèmes que nous nous sommes posés.

Jusqu'à la section \ref{thconf} nous supposerons la métrique de la feuille d'univers fixée par choix de jauge à la métrique plate $\gamma_{ab} = \eta_{ab}$. Nous allons voir que cela simplifie l'action \eqref{actP}. Nous introduisons les \emph{coordonnées du cône de lumière} sur la feuille d'univers
\be
\s^\pm =  \s \pm \t
\ee
La métrique a alors pour composantes
\be
\eta_{++} = \eta_{--} = 0 \text{ , } \eta_{+-} = \eta_{-+} = \frac12
\ee
et on définit
\be
\p_\pm = \frac12 (\dps \pm \dpt)
\ee
L'action \eqref{actP} s'écrit alors
\be
\S_{\text{P}} = - \frac1{2\pi\apr} \int d^2 \s \, \p_+ X^\mu \p_- X_\mu \l{actPsimpl}
\ee
où $d^2 \s = d\s_- d\s_+ = 2d\t d\s$.

\section{Supercordes : action supersymétrique}

Nous introduisons dès maintenant la version supersymétrique de la théorie des cordes. En effet, cela permet de définir des théories ne comportant pas de \emph{tachyons}, particules de masse carrée négative, comme nous le verrons à la section \ref{superstringM}. Nous distinguons ainsi la \emph{corde bosonique} ainsi nommée car elle ne décrit que des degrés de liberté bosoniques, de la \emph{supercorde} ou corde supersymétrique. Par souci de simplicité, nous nous restreindrons souvent à la théorie de la corde bosonique, l'extension à la théorie supersymétrique étant sous-entendue. Ce n'est que pour la configuration de la S-brane (voir section \ref{article3}) dans l'espace de Minkowski que nous utiliserons explicitement le formalisme supersymétrique. Nous allons, conjointement aux différentes configurations de la corde bosonique, exposer successivement les élements du formalisme de la supercorde en supposant toujours que l'espace-cible est l'espace de Minkowski.

Il existe deux manières d'introduire la supersymétrie en théorie des cordes. Ou bien nous construisons une supersymétrie sur la feuille d'univers, c'est le formalisme de \emph{Ramond-Neveu-Schwarz} (RNS), ou bien nous construisons une supersymétrie dans l'espace-cible, c'est le formalisme de \emph{Green-Schwarz} (GS). Nous n'avons utilisé que le premier, c'est pourquoi nous ne décrirons que celui-ci et nous sous-entendrons dans des expressions comme \gl l'action supersymétrique\gr\ qu'il s'agit de l'objet dans le formalisme RNS. Nous invitons le lecteur à se reporter par exemple à \cite{Green:1987sp, Green:1987mn} pour une exposition détaillée du formalisme GS par leurs auteurs.

Nous ne détaillons pas comment fixer la jauge. Après cette opération, l'action supersymétrique s'écrit\footnote{Les conventions correspondent à \cite{Green:1987sp}.}
\be
\S_{\text{sP}} = - \frac1{4\pi\apr} \int d\t d\s \, \left(\eta^{ab} \p_a X^\mu \p_b X_\mu - i \bar\psi^\mu \rho^a \p_a \psi_\mu \right) \l{actsP}
\ee
où $\psi$ est un spineur de Majorana\footnote{Nous supposons que nous sommes dans un espace-cible à dix dimensions où un spineur peut être à la fois de Majorana et de Weyl. Nous verrons à la sous-section \ref{applthc} que cette hypothèse est \emph{a posteriori} justifiée.}
sur la feuille d'univers et les matrices
\be
\rho^0 = \begin{pmatrix} 0 & -i \\ i & 0 \end{pmatrix} \text{ , } \rho^1= \begin{pmatrix} 0 & i \\ i & 0 \end{pmatrix}
\ee
fournissent une représentation de l'algèbre de Clifford à deux dimensions. Dans cette base, le spineur $\psi$ a pour composantes
\be
\psi^\mu = \begin{pmatrix} \psi_-^\mu \\ \psi_+^\mu \end{pmatrix}
\ee
L'action \eqref{actsP} est invariante sous la transformation de supersymétrie suivante, écrite sous forme infinitésimale,
\be
\delta X^\mu = -\bar \e \psi^\mu \text{ , } \delta \psi^\mu = i \rho^a \p_a X^\mu \e  \l{sinf}
\ee
où $\e$ est un spineur de Majorana constant.

Comme pour l'action bosonique \eqref{actP}, on peut mettre cette action sous une forme plus simple en introduisant les coordonnées du cône de lumière $\s_\pm$ sur la feuille d'univers et utilisant les composantes $\psi_\pm$
\be
\S_{\text{sP}} = - \frac1{2\pi\apr} \int d^2\s\, \left(\p_- X^\mu \p_+ X_\mu - \frac{i}2  \psi_-^\mu \p_+ \psi_{-,\mu} + \frac{i}2 \psi_+^\mu \p_- \psi_{+,\,\mu} \right) \l{actsPsimpl}
\ee

\section{\'Equations du mouvement}

\subsection{Pour la corde bosonique}

L'équation du mouvement pour la métrique $\gamma_{ab}$ impose l'annulation du tenseur énergie-impulsion\footnote{Nous définissons le tenseur énergie-impulsion comme dans \cite{Polchinski:1998rq}
\be
T^{ab} = -4\pi \sqrt{-\gamma}\,\frac{\delta S_P}{\delta \gamma_{ab}}\ .
\ee}
associé à la théorie des champs définie sur la feuille d'univers, ce qui, dans la jauge unitaire, s'écrit
\be T_{ab} = -\frac1{\apr}\left(\p_a X^\mu \p_b X_\mu - \frac12 \eta_{ab} \p_c X^\mu \p^c X_\mu\right) = 0 \l{tei}\ee
La trace du tenseur énergie-impulsion est nulle à cause de l'invariance de l'action sous la symétrie de Weyl. Nous pouvons résumer l'équation précédente par les relations suivantes
\be (\dpt X^\mu \pm \dps X^\mu)^2=0 \l{contrainteVir}\ee
Elle est appelée contrainte de Virasoro pour des raisons que nous expliquerons plus loin (voir section \ref{thconf}). Il s'agit de la contrepartie en théorie des cordes de la relation de couche de masse pour la particule relativiste \eqref{onshell}. Nous nous en servirons pour déterminer la masse des états de la corde après quantification (voir section \ref{quantcov}).

Mentionnons les composantes du tenseur énergie-impulsion dans les coordonnées du cône de lumière $\s_\pm$.
\bse
\l{teilinsk}
\begin{align}
T_{++} &= \frac{1}{\apr} \p_+ X^\mu \p_+ X_\mu \\
T_{--} &= \frac{1}{\apr} \p_- X^\mu \p_- X_\mu \\
T_{+-} &= T_{-+} = 0
\end{align}
\ese
Sous cette forme le lien avec le formalisme de la théorie conforme des champs apparaîtra de manière explicite (voir sous-section \ref{applthc}).

L'équation du mouvement pour les coordonnées $X^\mu$ décrit la propagation classique de la corde
\bse
\l{eqmv}
\be (\dpt^2-\dps^2) X^\mu =0 \ee
ou encore
\be \p_- \p_+ X^\mu =0 \l{eqmvlc} \ee
\ese

\subsection{Pour la supercorde}
\l{eqmvsc}

De même, l'équation du mouvement pour la métrique impose que les composantes du tenseur énergie-impulsion
\bse
\l{Tminsk}
\begin{align}
T_{++} &= \frac1{\apr} \left(\p_+ X^\mu \p_+ X_\mu +\frac{i}2 \psi_+^\mu \p_+ \psi_{+,\,\mu}\right) \\
T_{--} &= \frac1{\apr} \left(\p_- X^\mu \p_- X_\mu -\frac{i}2 \psi_-^\mu \p_- \psi_{-,\,\mu} \right)
\end{align}
\ese
soient nulles.
À l'invariance par supersymétrie correspond la conservation du courant de Noether associé, appelé \emph{supercourant}
\be
\p_- J_+ = 0 \text{ , } \p_+ J_- = 0
\ee
où
\bse
\l{Jminsk}
\begin{align}
J_+ &= \frac{\sqrt{2}}{\apr} \psi_+^\mu \p_+ X_\mu \\
J_- &= -\frac{\sqrt{2}}{\apr} \psi_-^\mu \p_- X_\mu
\end{align}
\ese
Notons qu'il ne s'agit pas sous cette forme d'une équation du mouvement, mais il est possible d'introduire un champ auxiliaire noté habituellement $B$ qui permettent de faire apparaître ces équations comme des équations du mouvement et définir ainsi des transformations de supersymétrie hors couche de masse. Nous n'utiliserons pas cette subtilité.

L'annulation de $T_{++}$, $T_{--}$, $J_+$ et $J_-$ constituent la version supersymétrique de la contrainte de Virasoro. Nous verrons à la sous-section \ref{superconf} qu'à l'action supersymétrique \eqref{actsPsimpl} correspond une théorie de champ \emph{superconforme}.

Les équations du mouvement pour les fermions s'écrivent
\bse
\begin{align}
\p_+ \psi_-^\mu &= 0 \\
\p_- \psi_+^\mu &= 0
\end{align}
\ese

\section{Conditions au bord partie 1 : corde libre dans un espace plat et dans un espace compactifié}

Lorsqu'on calcule la variation de l'action pour obtenir les équations du mouvement \eqref{eqmv}, des termes supplémentaires apparaissent. Pour les annuler, il faut imposer des conditions au bord qui vont contraindre le mouvement de la corde. Ces termes s'écrivent

\be \int d\t \left[\dps X^\mu(\t,\s = l) \delta X_\mu(\t,\s = l) - \dps X^\mu(\t,\s = 0) \delta X_\mu(\t,\s = 0)\right]=0 \ee
où $\s \in [0,l]$ décrit la corde.

Il existe deux manières de satisfaire cette équation. Ou bien les coordonnées sont périodiques et dans ce cas l'invariance conforme nous permet de fixer $l=2\pi$,
\be X^\mu (\t, \s+ 2\pi) = X^\mu(\t,\s) \ee
Nous obtenons une \emph{corde fermée}. Ou bien les coordonnées doivent satisfaire les conditions suivantes 
\be \dps X^\mu(\t,\s = \s_a) \delta X_\mu(\t,\s = \s_a) = 0 \qquad a=0,1\ee
avec $\s_0 = 0$ et $\s_1 = \pi$ (nous avons cette fois choisi de fixer $l=\pi$). Il s'agit alors d'une \emph{corde ouverte}.
Pour chaque coordonnée, chacune des extrémités vérifie une condition au bord de Neumann
\be \dps X^\mu(\t,\s = \s_a) = 0 \ee
ou une condition au bord de Dirichlet
\be \delta X^\mu(\t,\s = \s_a) = 0 \ee
qu'on peut écrire aussi
\be X^\mu(\t,\s = \s_a) = q^\mu \ee
Dans ce dernier cas, l'extrémité de la corde est fixée sur un hyperplan perpendiculaire aux directions selon lesquelles sont imposées les conditions de Dirichlet. Cet hyperplan est appelé D-brane si toutes les directions perpendiculaires sont spatiales ou S-brane de Dirichlet si l'une des directions perpendiculaires est la direction temporelle $X^0$. Nous expliquerons plus en détail l'origine de ces objets en théorie des cordes à la section \ref{dbranes}.

Dans toute cette section et la suivante, nous exposons diverses conditions au bord pour la corde bosonique et nous réservons le cas de la supercorde, que nous supposerons libre dans un espace plat, pour la section \ref{cbsuperc}

\subsection{Solutions des équations de mouvement pour la corde libre dans un espace-cible plat}
\l{soleqmv}

\subsubsection{Corde fermée}

Les solutions des équations du mouvement pour la coordonnée $X^\mu$ d'une corde fermée s'écrivent\footnote{Nous reviendrons dans la sous-section \ref{quantcovco}, note de pied de page \ref{facteurs}, sur les facteurs qui apparaissent dans ces expressions et les suivantes.}
\be
X^\mu (\t, \s) = q^\mu + \apr p^\mu \t + i \sqrt{\frac{\apr}2} \sum_{n \in \Z\backslash\{0\}} \frac{\a_n^\mu}n\, e^{-in (\t-\s)} + \frac{\ta_n^\mu}n\, e^{-in (\t+\s)} \l{cf}
\ee

\subsubsection{Corde ouverte}

Les solutions de l'équation du mouvement pour pour la coordonnée $X^\mu$ d'une corde ouverte s'écrivent
\bi
\i pour des conditions de Neumann à chaque extrémité
\be
X^\mu (\t, \s) = q^\mu + 2 \apr p^\mu \t + i \sqrt{2\apr} \sum_{n \in \Z\backslash\{0\}} \frac{a_n^\mu}n\, e^{-in \t} \cos n\s \l{co}
\ee
\i pour des conditions de Dirichlet à chaque extrémité ($X^\mu (\t, \s_a) = q_a^\mu$)
\be
X^\mu = q_0^\mu +  \frac{q_1^\mu - q_0^\mu}\pi \s - \sqrt{2\apr} \sum_{n \in \Z\backslash\{0\}}
\frac{a_n^\mu}n\, e^{-in \t} \sin n\s \l{codd}
\ee
\i pour une condition de Neumann en $\s = 0$ et une condition de Dirichlet $X^\mu(\t,\pi)=x_1^\mu$
\be
X^\mu (\t, \s) = q_1^\mu + i \sqrt{2\apr} \sum_{n \in \Z} \frac{a_n^\mu}{n+\frac12}\, e^{-i(n+\frac12) \t} \cos \left(n+\frac12\right)\s \l{cond}
\ee
\i pour une condition de Dirichlet $X^\mu(\t,0)=x_0^\mu$ et une condition de Neumann en $\s = \pi$
\be
X^\mu (\t, \s) = q_0^\mu - \sqrt{2\apr} \sum_{n \in \Z} \frac{a_n^\mu}{n+\frac12}\, e^{-i(n+\frac12) \t} \sin \left(n+\frac12\right)\s \l{codn}
\ee
\ei

Notons que pour la corde fermée \eqref{cf} et la corde ouverte avec des conditions de Neumann \eqref{co}, les deux premiers modes $q^\mu$ et $p^\mu$ décrivent le mouvement d'un système ponctuel (voir \eqref{solpr}), le centre de masse de la corde. L'interprétation suivante, valable pour les deux premiers types de cordes ouvertes \eqref{co} et \eqref{codd}, nous sera très utile lors de notre étude des configurations d'ondes planes électromagnétiques (voir section \ref{article1}) : si nous annulons les modes $\a$ ou $a$, qui décrivent des oscillations, les modes restants, appelés aussi \emph{modes zéro}, décrivent un \gl segment rigide\gr, soit en mouvement rectiligne uniforme, soit attaché à des D-branes. Nous utiliserons par la suite le terme \gl approximation de la corde rigide\gr.

\subsection{Corde fermée libre dans un espace plat compactifié}
\l{compact}

Bien qu'elle ne corresponde ni à un espace courbe, ni à une géométrie dépendante du temps, la compactification toroïdale joue un rôle fondamental en théorie des cordes\footnote{notamment pour construire des modèles phénomélogiques où sur les dix dimensions que doit comporter l'espace-cible pour des raisons que nous évoquons à la sous-section \ref{applthc}, six peuvent être réduites à un tore dont les dimensions sont inférieures aux bornes expérimentales.}.

Considérons une direction particulière $X^{\mu_0}$. L'indice 0 sera omis dès à présent. Compactifier la direction $X^\mu$ c'est lui donner la topologie du cercle par l'identification
\be
X^\mu \sim X^\mu + 2\pi R^\mu \l{comp}
\ee
où $R^\mu$ est le \emph{rayon de compactification}. Cette terminologie correspond au plongement d'une variété de topologie $\R^n \times S^1$  dans $\R^{n+2}$ où la variété apparaît comme un cylindre (un tore à partir de deux directions compactifiées).

Les solutions \eqref{cf} deviennent alors
\be
X^\mu (\t, \s) = q^\mu + \apr \frac{v^\mu}{R^\mu} \t + \pi\, w^\mu\,R^\mu \s + i \sqrt{\frac{\apr}2} \sum_{n \in \Z\backslash\{0\}} \frac{\a_n^\mu}n\, e^{-in (\t-\s)} + \frac{\ta_n^\mu}n\, e^{-in (\t+\s)} \l{cfcomp}
\ee
$w^\mu$ est le \emph{nombre d'enroulement} de la corde autour de la direction compacte. Le fait que l'impulsion est égale à $\frac{v^\mu}{R^\mu}$ peut être compris de la façon suivante. À partir de la fonction d'onde quantique $e^{ip\cdot x}$ correspondant à une particule libre se propageant dans l'espace plat $\R^{1,d}$, on construit la fonction d'onde invariante sous l'identification \eqref{comp} par superposition des images dans l'espace de recouvrement de la variété compactifiée, ici l'espace de Minkowski,
\be
\phi(x) = e^{i p_\perp \cdot x_\perp} \sum_{n \in \Z} e^{i \eta_{\mu\mu} p^\mu (x^\mu + 2\pi n R^\mu)}
\ee
Une resommation de Poisson nous donne le résultat attendu
\be
\phi(x)  = e^{i p_\perp \cdot x_\perp} e^{i \eta_{\mu\mu} p^\mu x^\mu} \sum_{v^\mu \in Z} \delta(p^\mu R^\mu - v^\mu)\ .
\ee
Nous avons supposé qu'une seule direction, désignée par $\mu$, était compactifiée et nous avons nommé $x_\perp$ les autres directions.

Nous avons profité de cette configuration très simple pour expliquer la construction des fonctions d'onde quantique correspondant aux modes zéros, bien que nous anticipions sur la quantification canonique traitée à la section \ref{quantcov}. Nous l'avons adaptée au cas de la corde dans l'espace de Misner (voir sous-section \ref{ontwist} et section \ref{article2}) et sans détailler les calculs, qui sont similaires à celui présenté ci-dessus, nous indiquerons les modifications à apporter aux expressions des opérateurs de vertex et des fonctions de corrélation.

\section{Conditions au bord partie 2 : géométrie dépendante du temps}
\l{gdt}

\subsection{Orbifold}
\l{ssorb}

\subsubsection{Trois types d'orbifold}
\l{otypes}

Nous construisons une variété quotient par l'identification suivante
\be X \sim e^{2\pi \J} X \ee
où
\be \J = \beta J_{02} + \theta J_{12}\ ,\ee
$J_{01}$ est le générateur d'un boost selon la direction $x^1$ et $J_{12}$ est le générateur d'une rotation dans le plan $(x^1, x^2)$. Le choix des directions 1 et 2 est arbitraire.
On distingue alors trois cas.
\bi
\i Si $\beta < \theta$, il s'agit d'un \emph{orbifold euclidien}\footnote{car il peut être défini à partir d'une variété de signature euclidienne} ou d'un \emph{orbifold de rotation}. En effet, par une transformation de Lorentz, on peut se ramener à
\be \J' =  \theta' J_{12}\ee
où $\theta' = \sqrt{\theta^2-\beta^2}$. Si $\theta'$ est rationnel, c'est un orbifold au sens premier du terme, sinon on obtient un orbifold noté $\R^2/\Z$ (un produit direct avec $\R^{1,d-2}$ est sous-entendu).
\i Si $\beta = \theta$, il s'agit un \emph{orbifold de genre lumière} ou pour reprendre le terme anglais, un \emph{orbifold null}, noté $\R^{1,2}/\Z$.
\i Si $\beta > \theta$, il s'agit d'un \emph{orbifold Lorentzien} ou \emph{orbifold de boost}, noté $\R^{1,1}/\Z$. On peut se ramener à un boost pur
\be \J' = \beta' J_{02} \ee
où $\beta' = \sqrt{\beta^2-\theta^2}$.
L'espace obtenu de cette façon est un espace cosmologique appelé \emph{espace de Misner}. Il présente une singularité de type Big Crunch/Big Bang. Nous détaillerons cela à la sous-section \ref{geommisner}.
\ei

Les solutions de cordes fermées vérifient la condition de périodicité suivante
\be X^\mu (\t, \s +2\pi) = e^{2\pi w \J} X^\mu(\t, \s) \ , \l{corb}\ee
où $w$ est un entier naturel.

Pour une corde dans le \emph{secteur non twisté} $w = 0$, les solutions ont la même forme que dans l'espace plat \eqref{cf}, mais elles sont restreintes par l'invariance de l'orbifold. De même la fonction d'onde quantique associée aux modes zéro de la corde est modifiée et, comme nous le verrons plus loin, cela modifie les amplitudes de diffusion.

D'autres solutions sont compatibles avec la condition de périodicité \eqref{corb}, elles correspondent aux \emph{secteurs twistés} $w \neq 0$.
\begin{multline}
X^\mu(\t, \s) = e^{w\J \sigma} \left[\left(\int^{\t} e^{-w\J u} du\right)^\mu_{\phantom{\mu}\nu} \a_0^\nu 
+ \left(\int^{\t} e^{w\J u} du\right)^\mu_{\phantom{\mu}\nu} \ta_0^\nu\right] \\
 i \sqrt{\frac{\apr}2} \sum_{n \in \Z\backslash\{0\}} \left[\left(\frac{e^{-i(n-iw\J)(\t-\s)}}{n-iw\J}\right)^\mu_{\phantom{\mu}\nu} \a_n^\nu + \left(\frac{e^{-i(n+iw\J)(\t+\s)}}{n+iw\J}\right)^\mu_{\phantom{\mu}\nu} \ta_n^\nu \right]  \l{orbgen}
\end{multline}

Nous appellerons modes zéros les modes $\a_0$ et $\ta_0$, bien qu'à strictement parler ils ne correspondent pas à l'approximation rigide.

\subsubsection{Orbifold de boost}
\l{oboost}

Regardons de plus près le cas de l'orbifold de boost\footnote{Nous invitons le lecteur cherchant une présentation unifiée des trois types d'orbifold à consulter \cite{Berkooz:2005ym}.} Nous pouvons diagonaliser \footnote{Dans le cas de l'orbifold de rotation, en introduisant des coordonnées d'espace-cible complexe $Z=\frac1{\sqrt{2}}(X^1+iX^2)$ et $\bar Z = \frac1{\sqrt{2}}(X^1-iX^2)$, il est possible également de diagonaliser la matrice $\J$.} la matrice $\J$ en utilisant les coordonnées de cône de lumière $X^\pm = \frac1{\sqrt{2}} (X^0 \pm X^2)$. La vari\'et\'e quotient est obtenu par identification $x^\pm \sim e^{\pm 2\pi \beta} x^\pm$. La condition \eqref{corb} s'\'ecrit alors
\be
X^\pm (\t,\s +2\pi) = e^{\pm 2\pi w \beta} X^\pm(\t,\s) \l{corboost}
\ee
Les solutions ont l'expression suivante, o\`u l'on a pos\'e $\nu = -w\beta$,
\be
X^\pm(\t, \s) = i \sqrt{\frac{\apr}2} \sum_{n \in \Z} \left(\frac{1}{n\pm i\nu}  \a_n^\pm  e^{-i(n\pm i\nu)(\t-\s)}+ \frac{1}{n\mp i\nu} \ta_n^\pm e^{-i(n\mp i\nu)(\t+\s)}\right) \l{orboost}
\ee
Nous faisons deux remarques.
\bi
\i Le cas de l'orbifold de boost peut être obtenu par continuation analytique $\theta \to i\beta$, à partir de l'orbifold de rotation (d'angle irrationnel en général), ce qui permet de calculer certaines amplitudes. Nous invitons le lecteur \`a se r\'ef\'erer \`a \cite{Berkooz:2004yy} pour un exemple d'utilisation de cette propri\'et\'e.
\i cette expression présente des analogies formelles avec les solutions de cordes ouvertes couplées à un champ électromagnétique, analogies qui suggèrent la quantification correcte des modes zéros des cordes twistés, comme nous le verrons à la sous-section suivante et à la section \ref{qobche}.
\ei

\subsection{Couplage à deux champs électromagnétiques}
\l{cchem}

\subsubsection{Configuration de champs générale}

Commençons par une courte digression. Jusqu'ici nous avons travaillé dans le cas ou l'espace-cible où évoluent les cordes est plat (ou possède un espace de recouvrement plat). Cependant il est possible de généraliser l'action de Polyakov au cas où l'espace est courbe. Nous introduisons trois champs : la métrique $g_{\mu\nu}(X)$ , un tenseur antisymétrique $b_{\mu\nu}(X)$ et le dilaton (un champ scalaire) $\phi(X)$. La raison de la présence de ces trois champs sera clarifiée à la sous-section \ref{spectrecf}. Ils définissent la géométrie de l'espace-cible (notamment, le champ antisymétrique correspond à une torsion et le dilaton \`a l'intensit\'e des interactions de corde), c'est pourquoi on les appelle \emph{champs de fond}. Nous écrivons l'action sur une feuille d'univers de signature euclidienne, obtenue par continuation analytique $\t = -i \t_{\text{E}}$. L'action s'écrit alors
\be
\S_{\text{P}} = \frac1{4\pi\apr} \int d\t d\s \sqrt{\gamma_{\text{E}}} \, \left[\left(\gamma_{\text{E}}^{ab} g_{\mu\nu}(X)+i\e^{ab} b_{\mu\nu}(X)\right)\p_a X^\mu \p_b X^\nu +\apr R_{\gamma_{\text{E}}} \phi(X) \right]\l{actPcurved}
\ee
où $\e^{ab}$ est un tenseur complètement antisymétrique défini sur la feuille d'univers $\sqrt{\gamma_{\text{E}}}\,\e^{10} = +1$ et $R_{\gamma_{\text{E}}}$ est le tenseur de Riemann associé à la métrique euclidienne $\gamma_{\text{E}}$.

La théorie des champs ainsi définie porte le nom de \emph{modèle sigma non linéaire}.

Pour que cette action soit invariante sous les transformations de Weyl, les trois champs que nous avons introduits doivent vérifier des équations déterminées perturbativement en calculant les fonctions $\beta$ associées à la renormalisation du modèle sigma non linéaire. Le détail des équations ne nous importe peu (voir \cite{Polchinski:1998rq} p.111), ce qu'il faut retenir, c'est qu'une tentative naïve d'introduire une géométrie dépendante du temps suffisamment simple pour qu'on puisse l'étudier en détail a toutes les chances de briser la symétrie de Weyl. En l'absence de cette symétrie, la théorie n'est plus une théorie conforme et les calculs sont encore hors de portée. Les modèles WZW, dont nous avons parlé à la section \ref{wzw}, sont un exemple de modèle où la symétrie de Weyl n'est pas brisée.

Une autre stratégie consiste, dans le cas des cordes ouvertes, à introduire un champ de fond, le potentiel électromagnétique, qui est couplé aux extrémités de la corde, considérées comme des systèmes ponctuels. À l'action de Polyakov est ainsi ajouté deux termes de bord de la forme \eqref{cchempp}, la nouvelle action s'écrit schématiquement $\S_{\text{em}} = \S_{\text{P}} + \S_{\text{bord}}$ avec
\be \S_{\text{bord}} = e \int d\t A_\mu (X) \dpt X^\mu \l{cem}\ee
où $e$ est la charge électrique et $A_\mu (X)$ le potentiel.
Rappelons que si le champ électromagnétique de fond $F_{\mu\nu}$ est une onde plane, la théorie reste conforme.

Précisons à la lumière de ce formalisme comment un couplage à des ondes planes peut définir une géométrie dépendante du temps. Nous distinguons les coordonnées de cône de lumière $x^\pm = \frac1{\sqrt{2}} (x^0 \pm x^1)$ et des \emph{coordonnées transverses} $x^i$, $i=2,\ldots,d$. Si nous définissons un potentiel dont la seule composante non nulle est
\be A_+ (x^i) = \Phi(x^+, x^i) \ee
nous obtenons alors un champ électromagnétique dépendent du temps
\be F_{i+} = \p_i  \Phi(x^+, x^i) \l{nullfield}\ee
qui constitue un fond exact si
\be \p_i \p^i \Phi = 0 \l{cPhi}\ee
Le calcul de la métrique effective pour les cordes ouvertes donne
\be ds^2 = -2 dx^+ dx^- + dx^i dx^i + (2\pi \apr)^2 \left[\p_i \Phi(x^+, x^i)\right]^2 dx^+ dx^+ \ee
Cette métrique correspond à une onde plane gravitationnelle dans les coordonnées de Brinkmann. Ainsi pour les cordes ouvertes du moins, il s'agit bien d'une géométrie dépendante du temps.

Un autre point de vue, adopté dans les articles \cite{Bachas:2002qt, Bachas:2002jg}, consiste à considérer la version T-duale de ces configurations de champ électromagnétique, c'est-à-dire deux D-branes dont les vitesses et orientations relatives varient au cours du temps. La T-dualité est expliquée à la sous-section \ref{tdco}.

Revenons à notre couplage \eqref{cem}. La variation de l'action dans un espace-cible plat
\begin{multline}
\S_{\text{em}} = -\frac1{4\pi\apr} \int d\t d\s \left(-\dpt X^\mu \dpt X_\mu + \dps X^\mu \dps X_\mu \right) \\
+\frac12 e_0 \int d\t A^{(0)}_\mu (X) \dpt X^\mu - \frac12 e_1 \int d\t A^{(1)}_\mu (X) \dpt X^\mu
\end{multline}
par rapport à $X^\mu(\t,\sigma)$ fournit, après intégration par partie, les conditions au bord suivantes
\be
\dps X^\mu + 2\pi\apr e_a (F^{(a)})^\mu_{\phantom{\mu}\nu}(X)\, \dpt X^\nu = 0 \qquad \s = \s_a \l{cbem}
\ee
Les équations du mouvement \eqref{eqmv} restent, elles, inchangées. Par la suite, on intégrera la charge $e_a$ et le facteur $2\pi \apr$ à la définition du champ $F^{(a)}$.
Les équations du mouvement ont pour solution $X(\t,\s) = f(\t+\s) + g(\t-\s)$ où les indices sont sous-entendus. Les conditions au bord s'écrivent alors
\bse
\begin{align}
& f'(\t) - g'(\t) + F^{(0)}(f(\t)+g(\t))\,(f'(\t) + g'(\t)) = 0 \\
& f'(\t+\pi) - g'(\t-\pi) + F^{(1)}(f(\t+\pi)+g(\t-\pi))\,(f'(\t+\pi) + g'(\t-\pi)) = 0 \l{cbemfg}
\end{align}
\ese
On obtient un système d'équations différentielles non lin\'eaires et non locales, ce qui rend compliquée, sinon impossible, la recherche de solutions générales.

\subsubsection{Champs éléctromagnétiques constants et uniformes, cas du champ électrique}

Commençons par étudier le cas où les champs $F^{(a)}$ sont constants et uniformes. On suppose en outre que les matrices correspondantes $(F^{(a)})^\mu_{\phantom{\mu}\nu}$ commutent $[F^{(0)}, F^{(1)}]=0$. Les solutions des équations du mouvement qui vérifient les conditions au bord \eqref{cbem} s'écrivent
\begin{multline}
X(\t,\s) = q_0 + \left(\int^{\t +\s} e^{\Omega u} du + \frac{1+F^{(0)}}{1-F^{(0)}} \int^{\t -\s} e^{\Omega u} du \right) a_0 \\
+ i\sqrt{\frac{\apr}2} \sum_{n \in \Z\backslash\{0\}}\frac1{n+i\Omega}\left(e^{-i(n+i\Omega)(\t +\s)} +
\frac{1+F^{(0)}}{1-F^{(0)}} e^{-i(n+i\Omega)(\t -\s)} \right) a_n \l{emgen}
\end{multline}
où $\Omega$ est défini par
\be e^{2\pi \Omega} = \frac{1- F^{(1)}}{1+F^{(1)}} \frac{1+ F^{(0)}}{1-F^{(0)}} \ee

Nous nous intéresserons plus particulièrement au cas où les deux champs sont des champs électriques\footnote{Les solutions dans le cas de deux champs magnétiques peuvent être obtenues à partir des solutions \eqref{elec} par un simple prolongement analytique de $\nu$.} $F^{(a)}_{+-} = \e_a$. On peut alors diagonaliser $F^{(a)}$ en se plaçant dans les coordonnées du cône de lumière $X^\pm$ et on obtient 
\be
X^\pm = q_0^\pm + i\sqrt{2\apr}\sum_{n \in \Z}  \frac1{n\pm i\nu} e^{-i(n\pm i\nu)\t} \cos\left[(n\pm i\nu) \s \mp i \arctanh(\e_0) \right]\, a_n^\pm \l{elec}
\ee
où 
\be
\nu = \frac1{\pi} (\arctanh(\e_1) - \arctanh(\e_0))
\ee
Cette expression est la même que dans \cite{Bachas:1992bh} à des choix de conventions près. Maintenant, nous pouvons mettre en évidence la ressemblance formelle entre les équations \eqref{orboost} et \eqref{elec}. De même que dans le cas de la corde libre, on peut considérer la corde fermée \eqref{cf} comme étant constituée d'un jeu d'oscillateurs de cordes ouvertes \eqref{co} pour les modes gauches et d'un jeu d'oscillateurs de cordes ouvertes pour les modes droits, on peut voir la corde fermée dans un orbifold lorentzian comme l'assemblage de deux jeux d'oscillateurs de corde ouverte couplée à un champ électrique ($\e_0 = 0$, $\e_1$ est choisi tel quel $\nu_{\text{elec}} = \nu_{\text{orbifold}} = -w \beta$). Nous reviendrons à la section \ref{qobche} sur cette analogie qui est cruciale puisqu'elle permet de quantifier la corde fermée dans l'orbifold lorentzien à partir de la quantification de la corde ouverte dans un champ électrique. Exposée dans \cite{Pioline:2003bs}, elle a été exploitée par la suite dans \cite{Berkooz:2004re} et nous a permis de mener les calculs de \cite{Berkooz:2004yy}.

\subsubsection{Ondes planes électromagnétiques dépendantes du temps}

Considérons maintenant le cas du champ électromagnétique dépendant du temps. Les équations \eqref{cbemfg} peuvent être simplifiée si nous choisissons d'identifier le temps de la feuille d'univers $\tau$ avec la coordonnée du cône de lumière $X^+$
\be X^+ = q^+ + p^+ \tau \ .\ee
Il s'agit en fait d'un choix de jauge qui permet d'éliminer les redondances introduites par la symétrie par difféomorphisme et par transformation de Weyl. La démarche générale que nous suivons dans ce chapitre consiste à quantifier la théorie puis imposer les contraintes appropriées sur les états du spectre pour éliminer les états non physiques. Ici nous fixons la jauge et nous éliminons les degrés de liberté non physique avant de quantifier. Comme les deux démarches sont équivalentes, nous ne détaillerons pas cette \emph{quantification de cône de lumière}.

Les équations \eqref{cbemfg} deviennent linéaires et il devient possible de trouver des solutions si, pour les champs électromagnétiques $F^{(a)}$, on choisit des fonctions $\Phi^{(a)}$ (voir \eqref{nullfield}) linéaires ou quadratiques. Le cas de fonctions $\Phi^{(a)}$ linéaires, traité dans l'article \cite{Bachas:2002jg}, n'entre pas dans le cadre de cette thèse. Nous allons donc présenter le calcul dans le cas de fonctions $\Phi^{(a)}$ quadratiques et indépendantes de $x^+$. Nous reviendrons sur le cas de champs dépendants du temps dans le chapitre \ref{article1}.

Les fonctions $\Phi^{(a)}$ s'écrivent alors
\be \Phi^{(a)} = \frac12 h_{ij}^{(a)} x^i x^j \ee
où $h_{ij}^{(a)}$ sont des matrices constantes choisies de telle sorte que la condition \eqref{cPhi} sur la fonction $\Phi$ soit remplie. On obtient des champs avec un gradient constant
\be F^{(a)}_{i+} = h_{ij}^{(a)} x^j \ee
En raison de problèmes de stabilité que nous expliquerons dans le chapitre \ref{article1}, il est nécessaire d'ajouter un champ magnétique constant et uniforme. Par souci de simplicité, nous le supposerons nul. Notons toutefois que les équations en présence de ce champ magnétique restent linéaires, mais l'expression des solutions correspondantes devient assez lourde.

Les équations \eqref{cbem} s'écrivent sous ces conditions
\begin{align}
\dps X^+ &= 0 \\
\dps X^i + h_{ij}^{(a)} X^j \dpt X^+ &= 0 \qquad \s=\s_a \\
\dps X^- - h_{ij}^{(a)} X^j \dpt X^+ &=0
\end{align}
Dans la jauge du cône de lumière, seule la condition sur $X^i$ reste pertinente
\be \dps X^i + p^+ h_{ij}^{(a)} X^j = 0 \qquad \s=\s_a \l{cbem2}\ee
On considère le cas où les matrices $h_{ij}$ commutent entre elles et on appelle\footnote{Il y a un facteur $\pi$ de différence avec l'article \cite{Durin:2003gj}.} $e_{ij} = p^+ h_{ij}$. Dans ce cas les matrices sont codiagonalisables et on peut choisir un potentiel de la forme $x^2 - y^2$ ($x$ et $y$ sont deux directions spatiales). On se ramène ainsi à deux problèmes à une dimension image l'un de l'autre par $(e_0,e_1) \to (-e_0,-e_1)$.

Détaillons un peu la résolution de ce problème. $X^i$ s'écrit encore une fois comme la somme d'une partie gauche et d'une partie droite $X(\t,\s) = f(\t+\s)+g(\t-\s)$. Les conditions au bord \eqref{cbem2} s'écrivent alors
\be 
\begin{aligned}
f'(\tau) - g'(\tau) + e_0 \left(f(\tau)+g(\tau)\right) 
 & = 0 \\
f'(\tau+2\pi) - g'(\tau) + e_1 \left(f(\tau+2\pi) + g(\tau)\right)
& = 0 
\end{aligned}
\ee
Les modes de Fourier de $f$ et $g$, d\'efinis par
\be
f(\t) = \int \hat{f}(\w)\, e^{-i\w \t} d\w\ ,
\ee
vérifient le système suivant
\be
\begin{pmatrix} -i\w + e_0 & i\w + e_0 \\
(-i\w + e_1) e^{-2i\pi \w} & 
i\w + e_1 \end{pmatrix}
\begin{pmatrix} \hat{f}(\w) \\ \hat{g}(\w) \end{pmatrix} = 0
\ee
Le système admet des solutions non nulles si et seulement si le déterminant de la matrice $2\times 2$ est nul, ce qui détermine les valeurs des modes $\omega$. On peut mettre cette relation de dispersion sous la forme
\be
\tan(\pi \w) = \frac{ (e_1-e_0) \w}{\w^2 + e_0 e_1} \l{disp}
\ee
qui permet de mettre en évidence le fait qu'il existe un seul mode réel $\omega_n$ sur chaque intervalle $[n,n+1[$, $n \in \N\, \backslash \{0\}$. On garde ainsi la structure des modes de la corde libre, avec un décalage qui rappelle le décalage réel présent pour une corde couplée à des champs électriques. La ressemblance toutefois s'arrête là, comme le montre l'expression des solutions.
\be X(\t,\s) = X_{\text{zéro}} (\t,\s) + \sum_{n=1}^\infty X_n \ee
où
\be
X_n = \cN_n \frac{i}{\w_n} \left(e^{-i\w_n(\tau+\sigma)} +
\frac{i\w_n-e_0}{i\w_n+e_0} e^{-i\w_n (\tau-\sigma)} \right) a_n
+ \text{complexe conjugé} \l{moden}
\ee
et $X_{\text{zéro}}$ correspond aux modes zéros. $\cN_n$ est un coefficient de normalisation dont l'expression peut être trouvé dans l'annexe C de \cite{Durin:2003gj} et dont nous expliquerons la détermination à la section \ref{quantcov}. Dans le cas d'une corde neutre, c'est-à-dire lorsque $e_0 = e_1$, $\w_n = n$ et les modes sont données par l'expression \eqref{moden} en remplaçant $\w_n$ par $n$.

\begin{figure}[h]
\hfill\epsfig{file=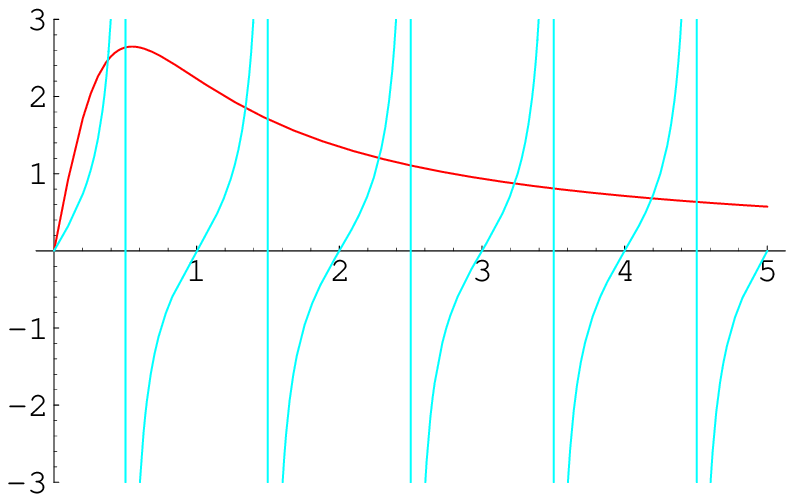,height=3.5cm, width=6cm}
\hfill\epsfig{file=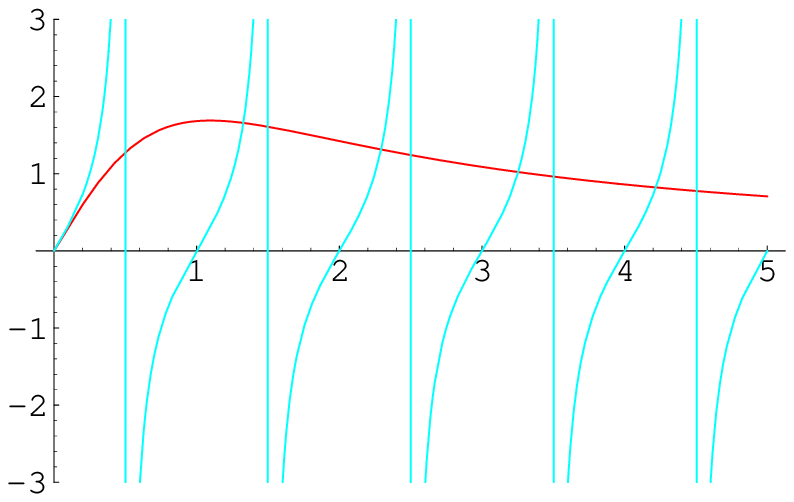,height=3.5cm, width=6cm}
\hfill
\caption{\label{graphsol}\small Solution graphique de la relation de dispersion des cordes ouvertes. \`A gauche : $\delta>0$, $\Delta>0$, toutes les racines sont r\'eelles. \`A droite : $\delta<0$, $\Delta > 0$, deux racines se sont rencontr\'ees \`a $\om=0$ et sont devenues complexes.}
\end{figure}

\begin{figure}[h]
\hfill\epsfig{file=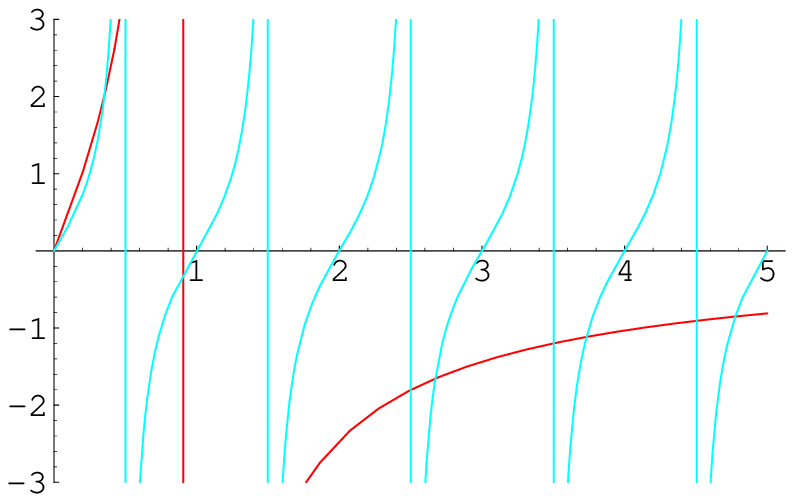,height=3.5cm, width=6cm}
\hfill\epsfig{file=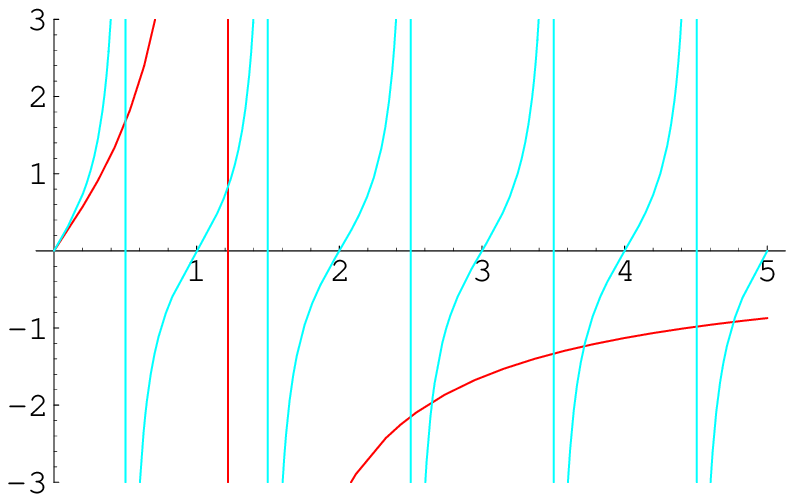,height=3.5cm, width=6cm}
\hfill
\caption{\label{graphsol2}\small Solution graphique de la relation de dispersion des cordes ouvertes. \`A gauche : $\delta<0$, $\Delta<0$, les racines dans l'intervalle $[0, 1[$ correspondent au premier niveau excit\'e. \`A droite : $\delta>0$, $\Delta<0$, deux nouvelles racines se sont rencontr\'ees et sont devenues complexes.}
\end{figure}

Revenons aux modes zéro. Leur expression dépend des valeurs de $e_0$ et $e_1$. Définissons $\delta = e_1 - e_0 - \pi e_0 e_1$ et $\Delta = e_1 - e_0 + \frac2{\pi}$. Cinq cas se présentent
\bi
\i $\delta > 0$ et $\Delta > 0$. La relation de dispersion \eqref{disp} admet une seule racine réelle $\w_0$ dans $]0,1[$ (voir figure \ref{graphsol}). Alors
\be X_{\text{zéro}} = X_{n=0}\ .\ee
\i $\delta=0$ et $\Delta > 0$. Il existe deux modes de fréquence nulle.
\be X_{\text{zéro}}(\t, \s) = \cN^c_0 (1-e_0 \s)(q_0 + p_0 \tau) \l{mode0mac} \ee
\i $\delta<0$ (la courbe $\Delta = 0$ est contenue dans ce domaine). On a deux racines complexes $\w^\pm_0 = \pm i k_0$ (voir figure \ref{graphsol} et \ref{graphsol2})
\be X_{\text{zéro}} = X_+ + X_- \ee
avec
\be X_\pm (\t, \s) = \cN^u_0 \frac{1}{k_0} \left( e^{\pm k_0(\tau+\sigma)} + \frac{\mp k_0-e_0}{\mp k_0+e_0} e^{\pm k_0 (\tau-\sigma)} \right) a_{\pm} \l{mode0inst}
\ee
Cette expression est valable pour une corde neutre, à condition de poser $k_0 = e$ et de redéfinir le mode $a_+$ en posant $a_+ \to \frac{-k_0+e_0}{-k_0-e_0} a_+$. On obtient
\be X_{\text{zéro}}(\t, \s) = \cN^u_0 \frac{1}{e} \left(e^{e(\tau+\sigma)}a_- + e^{-e (\tau-\sigma)} a_+ \right) \ee
\i $\delta = 0$ et $\Delta <0$. Les deux racines $\pm \w_1$ s'annulent (avec les notations données plus haut, il s'agit en fait de fréquence $\w_1$ qui sont devenues $\w_0$ car elles passent de l'intervalle $[1,2[$ à l'intervalle $[0,1[$ ; l'ensemble du spectre s'est décalé). On obtient deux nouveaux modes identiques à \eqref{mode0mac}.
\i $\delta > 0$ et $\Delta < 0$. Les racines $\pm \w_1$ sont devenues complexes (voir figure \ref{graphsol}) et correspondent à deux modes donnés par \eqref{mode0inst}.
\ei

Les phénomènes physiques associés à ces modes seront expliqués au chapitre \ref{article1}.

Cette série de résultats montre qu'une grande variété de configurations peuvent être résolues exactement et permettre ainsi un traitement standard de première quantification.

\section{Conditions au bord partie 3 : fermions}
\l{cbsuperc}

Nous rappelons que pour la corde supersymétrique nous nous plaçons dans l'espace plat et dans la situation où la corde est libre. Nous exposons le calcul des solutions pour les degrés de liberté fermioniques de la corde supersymétrique. Ces résultats sont les éléments constitutifs des expressions d'amplitude que nous écrirons à la section \ref{article3} lorsque nous étudierons la S-brane.

Le terme de bord s'écrit
\be
\int d\t \left[\psi_+ (\pi)\delta \psi_+(\pi) - \psi_-(\pi) \delta \psi_-(\pi) -
\left(\psi_+ (0)\delta \psi_+(0) - \psi_-(0) \delta \psi_-(0)\right)\right] =0
\ee
où la dépendance en $\t$ est sous-entendue.

\subsection{Cordes ouvertes}

Pour les cordes ouvertes, les conditions aux bords sont vérifiées si les spineurs $\psi_\pm^\mu$ vérifient
\bse
\l{scbco}
\begin{align}
\psi_-(0,\tau) &= \eta_1 \psi_+(0,\tau) \\
\psi_-(\pi,\tau) &= \eta_2 \psi_+(\pi,\tau)
\end{align}
\ese
où $\eta_{1,2}$ peuvent prendre indépendamment les valeurs $\pm 1$. Si $\eta_1 = \eta_2$ alors nous sommes dans le \emph{secteur de Ramond}, noté R, de la corde ouverte. Si $\eta_1 = -\eta_2$ alors nous sommes dans le \emph{secteur de Neveu-Schwarz}, noté NS, de la corde ouverte.

Les solutions pour les spineurs s'écrivent alors
\be
\psi_\mp^\mu = \sqrt{\apr}\sum_r b_r^\mu\, e^{-ir(\t\mp\s)}
\ee
où $r\in \Z +\frac12$ dans le secteur NS et $r \in \Z$ dans le secteur R.

\subsection{Cordes fermées}

Dans le cas des cordes fermées, les deux composantes $\psi_+$ et $\psi_-$ sont indépendantes ; elles peuvent être soit périodiques, soit anti-périodiques. Cela s'écrit
\bse
\begin{align}
\psi_-(0,\tau) &= \eta_3 \psi_-(2\pi,\tau) \\
\psi_+(\pi,\tau) &= \eta_4 \psi_+(2\pi,\tau)
\end{align}
\ese

Nous avons alors quatre secteurs différents
\bi
\i le \emph{secteur de Ramond,Ramond}, noté R-R ou RR, qui correspond à $\eta_3 = \eta_4 = +1$,
\i le \emph{secteur de Neveu-Schwarz,Neveu-Schwarz}, noté NS-NS ou NSNS, qui correspond à $\eta_3 = \eta_4 = -1$,
\i le \emph{secteur de Ramond,Neveu-Schwarz}, noté R-NS ou RNS, qui correspond à $\eta_3 = -\eta_4 = +1$,
\i le \emph{secteur de Neveu-Schwarz,Ramond}, noté NS-R ou NSR, qui correspond à $\eta_3 = -\eta_4 = -1$,
\ei

Les solutions pour les spineurs s'écrivent
\bse
\begin{align}
\psi_-^\mu &= \sqrt{\apr}\sum_r \beta_r^\mu\, e^{-ir(\t-\s)} \\
\psi_+^\mu &= \sqrt{\apr}\sum_s \tilde\beta_s^\mu\, e^{-is(\t+\s)}
\end{align}
\ese
où, pour les modes $\beta$, $r\in \Z +\frac12$ dans le secteur NS et $r \in \Z$ dans le secteur R et, pour les modes $\tilde\beta$, $s\in \Z +\frac12$ dans le secteur NS et $s \in \Z$ dans le secteur R.

Avant d'exposer la procédure de première quantification pour la corde, nous allons expliquer rapidement l'intégrale des chemins en théorie de cordes et montrer comment nous aboutissons à une théorie conforme à deux dimensions.

\section{Intégrale des chemins et théorie conforme des champs}
\l{thconf}

Il existe deux formalismes majeurs pour implémenter la procédure de quantification : l'intégrale des chemins et le formalisme d'opérateur (quantification canonique).

Nous avons entrevu l'existence de la quantification du cône de lumière, où les contraintes sont résolues avant la quantification. Ceci brise la covariance manifeste de la théorie, qu'il faut vérifier \emph{a posteriori}. Inversement, il est possible de quantifier puis d'imposer les contraintes sur les états physiques, ce qui permet de préserver explicitement la covariance. Ce programme est réalisé par l'\emph{ancienne quantification covariante} et dans la \emph{quantification BRST}. Comme nous l'avons évoqué à la section \ref{cfu}, la seconde est la version rigoureuse de la première.

Dans cette section, nous introduisons le formalisme d'intégrale des chemins, qui permet de présenter plus aisément le lien avec la théorie conforme des champs et les concepts qui nous seront utiles par la suite, notamment les fonctions de corrélation.

\subsection{Feuille d'univers euclidienne et intégrale des chemins}

Une fonction de corrélation $C= \langle \Phi[X]\rangle$ s'écrit dans le formalisme de l'intégrale des chemins
\be C = \int [DX] [D\gamma]\, \Phi[X]\, e^{i S_{\text{P}} [X,\gamma]}\ . \ee
Pour donner une définition mathématiquement plus acceptable de l'intégrale des chemins, il est d'usage de considérer une intégrale des chemins avec une action euclidienne, qui est ici obtenue par prolongement analytique de la coordonnée $\tau$ de feuille d'univers $\tau = -i\tau_{\text{E}}$. La feuille d'univers devient une surface euclidienne et l'action
\be
\S_{\text{E}} = \frac1{4\pi\apr} \int d\t_{\text{E}} d\s \sqrt{\gamma_{\text{E}}} \,\gamma_{\text{E}}^{ab} \p_a X^\mu \p_b X_\mu \l{actPeucl}
\ee
où le changement de signe devant l'action provient de la convention choisie pour l'intégrale de chemin euclidienne : le poids affecté à chaque trajectoire classique est $e^{-\S_{\text{E}}}$. Les symétries permettent de choisir une jauge pour la métrique qui peut se ramener à la métrique plane euclidienne $\delta_{ab}$. 

Fixer la jauge, comme pour la particule ponctuelle, produit un déterminant de Fadeev-Popov, qui s'exprime comme une contribution en terme de fantômes à l'action de la corde. Nous n'entrerons pas plus dans les détails, mais notons qu'aux fantômes correspond également une théorie conforme des champs. Le formalisme que nous exposons dans cette section s'applique donc à ces champs.

Nous avons fixé la jauge pour métrique. Il reste cependant une classe de transformations combinant une transformation de Weyl et un difféomorphisme qui ne modifie pas la métrique et les degrés de liberté correspondants ne sont donc pas éliminés par le choix de jauge. Il s'agit des transformations conformes. Il faudra donc tenir compte des contraintes imposées par cette symétrie résiduelle lorsque nous procéderons à la quantification de la théorie et au calcul des fonctions de corrélation.

\subsection{Invariance par transformation conforme de l'action avec la jauge de métrique fixée}
\l{invconforme}

Avant d'aborder le formalisme de la théorie conforme des champs à deux dimensions, expliquons brièvement la symétrie conforme. Le prolongement analytique $\tau = -i\tau_{\text{E}}$ changent les coordonnées de cône de lumière $\s_\mp$ en coordonnées complexes $w = \sigma + i\tau_{\text{E}}$, $\bar w = \sigma -i\tau_{\text{E}}$. La métrique de la feuille d'univers, dans la jauge unitaire, s'écrit alors
\be ds_\Sigma^2 = dw d\bw \ee
La métrique du plan est factorisée. Ceci est une particularité de la dimension deux et la métrique ne se factorise pas pour un espace plat de dimension supérieure à deux.

Appliquons à présent le difféomorphisme suivant
\be w' = f(w) \qquad \bw' = \overline{f(\w)} \ee
à la suite de la transformation de Weyl de paramètre $\w (w, \bw)$ que nous déterminerons \emph{a posteriori}. $f$ est une fonction holomorphe.
La métrique devient alors
\be {ds'_\Sigma}^2 = e^{2\w (w, \bw)}\frac1{\lvert\p_w f \rvert^2} dw' d\bw' \ee
Si l'on choisit $\w (w, \bw)= \ln \lvert\p_w f\rvert$, la métrique reste inchangée. Nous voyons donc que les transformations conformes apparaissent comme un sous-ensemble des transformations de Weyl associées à un difféomorphisme et qu'elles ne sont pas fixées par le choix de jauge.

Avant de continuer, faisons deux remarques. La première concerne les transformations conformes. Ces transformations sont définies sur le plan doté de la métrique $\delta_{ab}$. Les transformations conformes n'agissent pas, contrairement aux difféomorphismes ou aux transformations de Weyl, sur une métrique indépendante qu'on peut faire varier ; elles changent les distances entre les points. Il s'agit donc d'une contrainte additionnelle sur la dynamique des champs sur la feuille d'univers. Lorsque nous écrivons $ds'^2$, c'est bien la distance infinitésimale qui change et non la métrique, bien que les notations ne distinguent pas cette nuance.

La seconde remarque porte sur la dimension deux. Nous avons insisté sur le fait qu'en dimension deux la métrique s'écrivait sous forme factorisée et c'est cette forme qui permet de considérer des transformations $w' = f(w)$ avec une fonction $f$ holomorphe quelconque. Par contre, en dimension supérieure à deux, il n'existe qu'un nombre fini de transformations conformes : les translations, les rotations, les dilatations et l'inversion, c'est-à-dire la transformation
\be x^\mu \to \frac1{x^\mu x_\mu} x^\mu \ee
Les contraintes imposées sur les fonctions de corrélations en dimension deux sont donc bien plus puissantes qu'en dimension supérieure. C'est l'une des raison pour lesquelles la définition d'une \gl théorie des membranes\gr\ (la corde est remplacée par une surface dont l'évolution temporelle correspondrait à un volume d'univers) ou autres objets de dimension supérieure n'est pas encore claire.

\subsection{Générateurs classiques des transformations con\-formes}

Nous allons à présent mettre en évidence la structure algébrique des transformations conformes au niveau classique. Considérons une transformation conforme infinitésimale $w' = w+\e(w)$, où $\e$ est une fonction holomorphe (la transformation de $\bw$, sous-entendue, est identique). Notons que le domaine sur lequel la fonction est holomorphe n'est pas précisé. Si nous choisissons le plan complexe tout entier, toute fonction bornée est constante, cette solution ne peut pas être retenue. Nous pouvons alors définir un domaine $\{w, \lvert w \rvert < R\}$ où $R$, alors $\e(w)$ et bornée et admet un dévelopement de Taylor $\e(w) = \sum_{n=-1}^\infty b_n w^{n+1}$. 
Nous pouvons restreindre encore le domaine à un anneau $\{w, r < \lvert w-w_0 \rvert < R\}$, ce qui autorise la fonction $\e(w)$ à avoir des singularités artificielles ou essentielles ou des pôles hors de l'anneau (et notamment en $w_0$). $\e(w)$ admet alors un développement en série de Laurent $\e(w)= \sum_{n=-\infty}^\infty a_n (w-w_0)^{n+1}$. $r$ et $R$ sont choisis de telle sorte à éviter d'éventuelles singularités\footnote{En général, $r$ est fixé à 0 et le domaine est un disque percé du point $w_0$.} et pour qu'il soit possible de borner $\e(w)$ sur l'anneau par une valeur infinitésimale. Par la suite, lorsque nous parlerons de fonction holomorphe infinitésimale, c'est dans le sens de fonction holomorphe infinitésimale définie sur un anneau. Bien entendu, tout ceci s'applique aussi à $\bar\e(\bw)$.

$\e$ admet donc un développement en série de Laurent\footnote{$w_0$ peut être fixé sans perte de généralité à 0 par une translation.}
\be \e(w) = \sum_{n=-\infty}^\infty a_n w^{n+1} \ee
La transformation conforme infinitésimale agit sur un champ scalaire classique $\Phi(w)$ (de même, la dépendance sur $\bw$ est sous-entendue) de la façon suivante
\be \delta_{\e} \Phi = -\sum_{n=-\infty}^\infty a_n w^{n+1} \p_w \Phi \ee
où $\delta_{\e} \Phi = \Phi'(w)-\Phi(w)$ avec  $\Phi'(w)=\Phi(w-\e(w))$.
Les opérateurs $l_n$, définis par
\be l_n = -w^{n+1} \p_w \ee
sont les générateurs des transformations conformes classiques. Ils vérifient l'\emph{algèbre de Wit}\footnote{ ou \emph{algèbre de Virasoro classique}, bien que ce terme prête à confusion.} :
\be [l_n, l_m] = (n-m) l_{n+m} \ee
On a  la même relation de commutation pour les opérateurs $\bar l_n$.

\subsection{Générateurs quantiques des transformations con\-formes}
\l{genquant}

Au niveau quantique, ce n'est plus sur un champ classique $\Phi$ qu'il faut étudier l'effet d'une transformation conforme, mais sur les fonctions de corrélation contenant $\Phi$. L'effet d'une transformation conforme infinitésimale sur le champ $\Phi$ dans une fonction de corrélation $\langle \Phi(w_0) \Phi_1(w_1) \ldots \Phi_N(w_N) \rangle$ est le suivant\footnote{Nous suivons en partie l'exposition faite dans \cite{Dotsenko}. L'article de référence est \cite{Belavin:1984vu}.}
\begin{multline}
\langle \delta_{\e}\Phi(w_0) \Phi_1(w_1) \ldots \Phi_N(w_N) \rangle = \frac1{2\pi} \int_B\!\! d^2 \xi \, \partial^a \e^b (\xi, \bar \xi) \langle T_{ab}(\xi, \bar \xi) \Phi(w_0) \Phi_1 \ldots \Phi_N \rangle =\\
\begin{aligned}
&\frac1{2\pi i} \oint_C\!\! d\xi \left[\e(\xi) \langle T_{ww}(\xi, \bar \xi) \Phi(w_0) \Phi_1 \ldots \Phi_N \rangle
+ \overline{\e(\xi)} \langle T_{\bw w}(\xi, \bar \xi) \Phi(w_0) \Phi_1 \ldots \Phi_N \rangle \right] \\
-&\frac1{2\pi i} \oint_C\!\! d \bar\xi \left[\overline{\e(\xi)} \langle T_{\bw\bw}(\xi, \bar \xi)  \Phi(w_0)\Phi_1 \ldots \Phi_N \rangle
+ \e(\xi) \langle T_{w\bw}(\xi, \bar \xi) \Phi(w_0) \ldots \Phi_N(w_N) \rangle \right]
\end{aligned}\l{Ward}
\end{multline}
où $C$ est un contour qui contient seulement $w_0$ et où, $B$ est le domaine privé de $w_0$ tel que $C = \p B$ et sur lequel $\e$ est défini. Sur la première ligne,  $\partial^a$ agit sur le produit $\e^b T_{ab}$. $T$ est le tenseur énergie-impulsion. Enfin, le facteur $\frac1{2\pi}$ est une convention commode.

En choisissant $\e(w) = \e$ (translation) puis $\e(w) = \lambda (w-w_0)$ (dilatation), on obtient, pour $w \neq \w_i$
\begin{align}
\p_{\bw} T_{ww} = \p_w T_{\bw\bw} &= 0 \l{holo}\\
T_{w \bw} = T_{\bw w} &= 0 \l{trace}
\end{align}

Ces relations sont des relations sur l'\emph{opérateur} $T$, ou autrement dit, elles doivent être comprises en terme de fonction de corrélation, c'est-à-dire que $\p_{\bw} T_{ww}= 0$ signifie en fait
\be \langle \p_{\bw} T_{ww} \Phi (w_0) \Phi_1(w_1) \ldots \Phi_N(w_N) \rangle = 0 \ee
quels que soient les opérateurs $\Phi$ et $\Phi_i$. Ceci est valable pour toutes les équations qui suivent, qu'elles impliquent $T$ ou un autre opérateur. Les équations \eqref{holo} permettent de se ramener à deux fonctions respectivement holomorphe $T_{ww} = T(w)$ et antiholomorphe\footnote{Dans le sens indiqué plus haut de fonctions développables en série de Laurent.} $T_{\bw\bw} = \bT(\bw)$. On peut montrer par ailleurs que $T_{w \bw} = T_{\bw w} = T^a_{\phantom{a}a}$ ; ainsi l'équation \eqref{trace} correspond à l'annulation de la trace de $T$. Les deux fonctions $T$ et $\bar T$ suffisent pour définir le tenseur énergie-impulsion. Cependant, l'équation \eqref{trace} n'est valable que pour une feuille d'univers plate. L'anomalie de Weyl modifie cette équation
\be T_{w \bw} = T_{\bw w} = -\frac{c}{12} R \ee
où $R$ est la courbure de Ricci de la feuille d'univers et $c$ est une constante qui est aussi la \emph{charge centrale} de la théorie conforme considérée, notion que nous définirons un peu plus loin. Toutefois, en ce qui nous concerne, nous travaillerons dans des situations cette anomalie est une constante, ce qui permet de se restreindre aux deux fonctions $T$ et $\bT$.

L'identité de Ward \eqref{Ward} s'écrit à présent
\begin{multline}
\langle \delta_{\e}\Phi(w_0) \Phi_1(w_1) \ldots \Phi_N(w_N) \rangle = \\
\frac1{2\pi i} \oint_C\!\! d\xi \,\e(\xi) \langle T(\xi) \Phi(w_0) \Phi_1 \ldots \Phi_N \rangle
-\frac1{2\pi i} \oint_C\!\! d \bar\xi\; \overline{\e(\xi)}\; \bT(\bar \xi)  \Phi(w_0)\Phi_1 \ldots \Phi_N \rangle
\end{multline}
Cette relation, dans le cadre d'une théorie conforme particulère, permet de calculer la transformation de l'opérateur $\Phi$.

Les générateurs des transformations conformes sont les coefficients de $T$ et de $\bT$ des séries de Laurent suivantes
\be T(w) = \sum_{n=-\infty}^{\infty} w^{-n-2}\, L_n \text{\ ,\ } \bT(w) = \sum_{n=-\infty}^{\infty} \bw^{-n-2}\, \bar L_n \ee
Les opérateurs $L_n$ et $\bar L_n$ vérifient les mêmes relations de commutation\footnote{Rappelons que cette équation doit être comprise comme une relation sur des fonctions de corrélation. Voir \cite{Dotsenko} pour la démonstration dans ce formalisme.}
\be [L_n, L_m] = (n-m) L_{n+m} + \frac{c}{12} n(n^2-1) \delta_{n+m,0} \ee
et constituent ainsi une \emph{algèbre de Virasoro}. $c$ est la \emph{charge centrale}.

\subsection{Représentations irréductibles de l'algèbre des générateurs, champs primaires}

Les représentations irréductibles de cette algèbre, appelées aussi \emph{module de Verma} $V_{c,h}$ sont caractérisées par la charge centrale $c$ et le \emph{poids conforme} $h$, qui est la valeur propre de $L_0$ de l'état de plus haut poids. N'oublions pas la présence de l'algèbre \emph{antiholomorphe} correspondant aux $\bar L_n$, avec des modules de Verma $\overline{V}_{\bar c, \bar h}$ caractérisés par une charge centrale $\bar c$ et un poids conforme $\bar h$. Les états de notre théorie appartiennent donc au produit direct $V_{c,h} \times \overline{V}_{\bar c, \bar h}$. Les notations sont ici trompeuses puisque $\bar c$ n'est pas le conjugué de $c$ ni $\bar h$ celui de $h$. On notera le poids conforme $(h,\bar h)$.

L'état de plus haut poids de la représentation irréductible est appelé aussi \emph{champ primaire}.
Sous une transformation conforme $w \to f(w)$, il se transforme de la façon suivante
\be
\Phi(w,\bw) \to \Phi' (w,\bw) = (\p_w f(w))^h (\overline{\p_w f(w)})^{\bar h} \Phi(f(w),\overline{f(w)}) \l{chprim}
\ee

On peut aussi caractériser un état de plus haut poids par les conditions suivantes
\bse
\begin{align}
L_n \Phi &= 0 \text{\ \ \ \ \ pour tout\ }n >0 \\
L_0 \Phi &= h\, \Phi
\end{align}
\ese
et les conditions correspondantes pour les opérateurs $\bar L_n$.

Ainsi, le champ scalaire que nous avons considéré plus haut est un champ primaire de poids $(0,0)$. Dans ce qui suit, tous les champs que nous considérerons seront des champs primaires, sauf le tenseur énergie-impulsion, qui appartient au module de Verma de l'opérateur identité de poids conforme nul.

Comme cette dernière phrase permet de le constater, nous employons indifféremment \gl champ\gr, \gl opérateur\gr\ ou \gl état\gr\ pour désigner les éléments des modules de Verma. \gl Champ\gr\ fait référence aux fonctions de corrélation de champs, \gl opérateur\gr\ fait référence à l'écriture condensée des relations entre fonctions de corrélation sous forme de relation entre opérateurs (sous-entendu, agissant sur les champs qu'on peut rajouter dans les fonctions de corrélation) et \gl état\gr\ se trouve justifié par l'équivalence entre le formalisme d'intégrale des chemins et le formalisme de quantification canonique, que nous aborderons dans la section \ref{quantcov}.

\subsection{Application à la théorie de la corde bosonique : théorie conforme minimale $c=1$ du boson libre, ordre normal}
\l{applthc}

Revenons maintenant à la théorie des cordes. Au lieu d'utiliser les variables $w$, nous utiliserons les variables $z$ définies par
\be z = e^{-iw} = e^{\t_{\text{E}} -i\s} \qquad \bz = e^{i\bw} = e^{\t_{\text{E}} +i\s} \ee
de sorte que les expressions des solutions des équations du mouvement s'écrivent comme des séries de Laurent. Par exemple, pour la corde fermée libre, après prolongement analytique $\t = -i\t_{\text{E}}$, la solution \eqref{cf} s'écrit
\be
X^\mu (z, \bz) = x^\mu - i \frac{\apr}2 p^\mu \ln (z\bz) + i \sqrt{\frac{\apr}2} \sum_{n \in \Z\backslash\{0\}} \left(\frac{\a_n^\mu}n\, z^{-n} + \frac{\ta_n^\mu}n\, {\bz}^{-n}\right) \l{cfz}
\ee
et peut être décomposée en une partie holomorphe
\be 
X_{\text{h.}}^\mu (z) = \frac12 x^\mu - i \frac{\apr}2 p^\mu \ln (z) + i \sqrt{\frac{\apr}2} \sum_{n \in \Z\backslash\{0\}} \frac{\a_n^\mu}n\, z^{-n} \l{cfzh}
\ee
et une partie antiholomorphe
\be 
X_{\text{a.h.}}^\mu (z) = \frac12 x^\mu - i \frac{\apr}2 p^\mu \ln (\bz) + i \sqrt{\frac{\apr}2} \sum_{n \in \Z\backslash\{0\}} \frac{\ta_n^\mu}n\, {\bz}^{-n} \l{cfzah}
\ee
L'action euclidienne \eqref{actPeucl} s'écrit
\be
\S_{\text{E}} = \frac1{2\pi\apr} \int d^2 w\, \p_w X^\mu \p_{\bw} X_\mu = \frac1{2\pi\apr} \int d^2 z\, \p X^\mu \bp X_\mu \l{actPeuclsimpl}
\ee
où $d^2 w = dw d\bw = 2 d\t_{\text{E}} d\s$, $\p = \p_z$ et $\bp = \p_{\bz}$.
Les composantes non nulles du tenseur énergie-impulsion\footnote{Les expressions \eqref{teitc} comportent un signe $-$ supplémentaire par rapport aux expressions \eqref{teilinsk}. Ce signe provient du prolongement analytique vers la feuille d'univers euclidienne qui change le signe de l'action (comparer \eqref{actPeuclsimpl} et \eqref{actPsimpl}).} sont
\be
T(z) = -\frac1{\apr} \lno \p X^\mu(z) \p X_\mu(z)\rno \text{\ ,\ \ } \bT(\bz) = -\frac1{\apr} \lno\bp X^\mu(\bz) \bp X_\mu(\bz) \rno \l{teitc}
\ee
Les deux points indiquent l'\emph{ordre normal} que l'on va appeler \emph{conforme}, par opposition à l'ordre normal dans le formalisme de la quantification canonique. Les deux ordres normaux produisent le même résultat, mais l'ordre normal conforme est défini différemment. Voyons comment sur l'exemple de $T(z)$.

La fonction de corrélation $\langle X^\mu(z,\bz) X^\nu(z',\bz') \rangle$ peut être calculée sans obstacle dans le formalisme de l'intégrale des chemins puisque $\S_{\text{E}}$ est quadratique. Le résultat s'écrit
\be \langle X^\mu(z,\bz) X^\nu(z', \bz') \rangle = -\frac{\apr}2 \eta^{\mu\nu} \ln (\lvert z-z'\rvert^2 S) \ee
où $S$ est la surface de la feuille d'univers et correspond à un régulateur à grande distance. Remarquons en passant que $X$ n'est pas un champ primaire. La fonction de corrélation $\langle \p X^\mu(z) \p X_\mu(z')\rangle$ est déduite par différentiation (la variable $\bz$ n'intervient plus à présent) :
\be \langle \p X^\mu(z) \p X^\nu(z')\rangle = - \frac{\apr}2 \eta^{\mu\nu} \frac1{(z-z')^2} \ee
On constate alors que la limite $z \to z'$ est divergente, ce qui pose problème pour définir $T$. On régularise cette divergence à courte distance à l'aide de l'ordre normal, défini par
\be
\lno X^\mu(z,\bz) X^\nu(z', \bz')\rno = X^\mu(z,\bz) X^\nu(z', \bz') + \frac{\apr}2 \eta^{\mu\nu} \ln (\lvert z-z'\rvert^2)
\ee

L'ordre normal appliqué à plus de deux champs est défini récursivement par
\be \lno\Phi_1 \Phi_2 \ldots \Phi_N \rno = \Phi_1 \lno\Phi_2 \ldots \Phi_N \rno -  \sum_{i=2}^N \langle \Phi_1 \Phi_i \rangle \lno\Phi_2 \ldots \Phi_{i-1} \Phi_{i+1} \ldots \Phi_N \rno \ee
On vérifie que cette procédure est cohérente en calculant le produit $T(z) T(z')$ (toujours avec des fontions de corrélation sous-entendues)
\be T(z) T(z') = \frac{D}2 \frac1{(z-z')^4} + \frac2{(z-z')^2} T(z') + \frac1{z-z'} \p_{z'} T(z') + \ldots \ee
où $\ldots$ fait l'ellipse des termes non divergents lorsque $z \to z'$.
Ce développement de $T(z) T(z')$ correspond à ce à quoi on s'attend pour une théorie conforme dont la charge centrale serait $c=D$. D'autre part, on peut aussi vérifier que
\be T(z) \p_{z'} X^\rho (z') = \frac1{(z-z')^2} \p_{z'} X^\rho (z') + \frac1{(z-z')} \p_{z'}(\p_{z'} X^\rho) (z') + \ldots
\ee
ce qui correspond à ce qu'on attend pour un champ primaire de poids conforme $(1,0)$. On peut vérifier directement que $\p_z X^\rho (z)$ obéit à la loi de transformation \eqref{chprim} avec $h=1$ et $\bar h =0$.

Ainsi nous sommes en présence d'une théorie conforme avec une charge centrale $c=D$. Cette théorie est la superposition de $D$ bosons libres $X^\mu$ qui définissent chacun une théorie conforme avec une charge centrale $c=1$. Un outil puissant permet de calculer les fonctions de corrélations de cette théorie : la représentation des champs primaires par des opérateurs exponentiels du champ libre $X^\mu$ (chapitre 6 de \cite{Dotsenko}). À un opérateur $\Phi_h(z,\bz)$ de la théorie\footnote{On ne considère que le cas où $h=\bar h$.}, on associe un \emph{opérateur de vertex} $V_p (z,\bz)$
\be V_p (z,\bz) = \lno e^{i p X(z, \bz)}\rno \ee
où $p$ vérifie la relation suivante
\be h = \bar h = \frac{\apr}4 p^2 \ee

Ceci se généralise à la théorie contenant $D$ bosons $X^\mu$. Les fonctions de corrélations feront ainsi intervenir les opérateurs de vertex
\be V_p (z,\bz) = \lno e^{i p_\mu X^\mu(z, \bz)}\rno \ee
où $p$ est l'impulsion de la corde entrante. Ces opérateurs se généralisent aux produits
\be V^{\{\mu_i, \nu_j\}}_p (z,\bz) = \lno \prod_i \p^{m_i} X^{\mu_i} \prod_j \bp^{n_j} X^{\nu_j}  e^{i p_\mu X^\mu(z, \bz)}\rno \ee
qui correspondent à tous les états que nous trouverons en quantification canonique.

Les amplitudes de la théorie des cordes doivent être invariante sous une transformation conforme. Comme la position de l'opérateur de vertex est arbitraire sur la feuille d'univers, les amplitudes font intervenir les opérateurs \emph{intégrés}
\be \V = \int d^2 z \lno e^{i p_\mu X^\mu(z, \bz)} \rno \ee
qui doivent être invariant, ce qui fixe le poids conforme de $V_p (z,\bz)$ à $(1,1)$ et permet de déterminer le spectre dans ce formalisme.

Nous avons ainsi donné un aperçu de l'origine de la condition de quantification, de l'ordre normal et de la forme des opérateurs de vertex que nous introduirons dans le cadre de la quantification canonique sans plus de justification.

Nous ne développerons pas plus les formalismes de l'intégrale des chemins et de la théorie conforme et nous terminerons par deux remarques
\begin{itemize}
\i Dans le cas de la corde ouverte, les conditions aux bords identifient les algèbres de Virasoro holomorphes et antiholomorphes $L_n = \bar L_n$. 
\i La théorie des cordes est bien définie lorsque la charge centrale totale est nulle. Notre théorie conforme à $c=D$ est complétée par la théorie conforme des fantômes de Fadeev-Popov qui ont une charge centrale $c'= - 26$ dans le cas de la corde bosonique. La charge centrale totale est ainsi $D-26$, ce qui fixe la dimension de l'espace-temps à vingt-six dimensions.
\end{itemize}

\subsection{Intégrale des chemins pour les fermions, théorie superconforme}
\l{superconf}

\subsubsection{Intégrale des chemins pour la corde supersymétrique}

L'intégrale des chemins pour l'action supersymétrique fait intervenir, comme en théorie des champs, des variables de Grassmann et les résultats de l'intégrale de Berezin. On peut écrire l'action en terme de \emph{superchamps} définis sur un \emph{superespace}, ce qui rend manifeste l'invariance par supersymétrie, mais nous n'exposerons pas ce formalisme dont nous n'aurons pas besoin par la suite. Les degrés de liberté de symétrie sont éliminés en fixant la \emph{jauge superconforme}. Le déterminant de Fadeev-Popov sera exprimé en terme non seulement de fantômes, mais aussi de \emph{superfantômes}. Comme précédemment, nous ne détaillerons pas les subtilités liés à ces superfantômes et nous en tiendrons compte directement dans le résultat.

\subsubsection{Théorie superconforme des champs}

Nous avons vu à la sous-section \ref{eqmvsc} qu'aux transformations de supersymétrie correspondaient un supercourant $J_\pm$. En procédant de la même manière que dans le cas bosonique, nous pouvons montrer que l'action supersymétrique, une fois la jauge fixée, est invariante sous l'action des transformations \emph{superconformes}
\be
\delta X^\mu (w,\bw) = -\bar \e(w,\bw)\, \psi^\mu (w,\bw)\text{ , } \delta \psi^\mu (w,\bw) = i \rho^a \p_a X^\mu (w,\bw)\, \e(w,\bw) \l{sconf}
\ee
où nous insistons sur la dépendance en $w$, $\bw$ de $\e$. $\bar \e$ n'est pas ici le conjugué de $\e$, mais le conjugué multiplié par la matrice de Pauli $\rho_0$. L'expression \eqref{sconf} est à comparer à celle de la supersymétrie \eqref{sinf} et à celle des transformations conformes (dont l'expression infinitésimale est $\delta X^\mu(w,\bw) = -\e(w,\bw)\, X^\mu(w,\bw)$).

La symétrie résiduelle est ainsi une symétrie superconforme, qui contraint la forme des fonctions de corrélation et permet de travailler avec le formalisme de la théorie superconforme des champs.

De même que les transformations conformes sont générées par des modes $L_n$ de $T$ qui forment l'algèbre de Virasoro, les transformations superconformes sont générées par les modes $L_n$ et les modes $G_r$ du supercourant $J$, fonction holomorphe de $w$ obtenue à partir de $J_-$ lors du prolongement analytique vers la feuille d'univers euclidienne\footnote{$\bar J$, supercourant antiholomorphe, est obtenu à partir de $J_+$ et participe à une seconde superalgèbre de Virasoro.}
\be
J(w) = \sum_{r\in \Z + \nu} w^{-r-\frac32} \, G_r \text{\ ,\ }
\bar J(\bw) = \sum_{r \in \Z + \nu} {\bw}^{-r-\frac32} \, \bar G_r
\ee
où $\nu = 0$ si nous sommes dans le secteur de Ramond et $\nu = \frac12$ si nous sommes dans le secteur de Neveu-Schwarz.

Ces modes forment la \emph{superalgèbre de Virasoro}.
\bse
\begin{align} 
[L_n, L_m] &= (n-m) L_{n+m} + \frac{c}{12} n(n^2-1) \delta_{n+m,0} \\
\{G_r, G_s \} &= 2 L_{r+s} + \frac{c}{12}(4r^2-1) \delta_{r+s,0} \\
[L_n, G_r] &= \frac{n-2r}2 G_{n+r}
\end{align}
\ese
Lorsque $r$ et $s$ sont des entiers, il s'agit de l'\emph{algèbre de Ramond}, lorsque $r$ et $s$ sont demi-entiers, il s'agit de l'\emph{algèbre de Neveu-Schwarz}.

\subsubsection{Application à la théorie de la supercorde : théorie superconforme minimale $c=\frac12$ du fermion libre}

Comme pour la corde bosonique, il faut utiliser les variables $z$ et $\bz$ pour que les solutions des équations du mouvement s'écrivent comme des séries de Laurent. L'action euclidienne s'écrit
\be
\l{susyactioneucl}
\begin{split}
\S_{\text{E}} &= \frac1{2\pi\apr} \int d^2 w\, \left(\p_w X^\mu \p_{\bw} X_\mu - \frac{i}2  \psi_-^\mu \p_{\bw} \psi_{-,\mu} + \frac{i}2 \psi_+^\mu \p_w \psi_{+,\,\mu} \right) \\
 &= \frac1{2\pi\apr} \int d^2 z\, \left(\p X^\mu \bp X_\mu - \frac{i}2  \psi_-^\mu \bp \psi_{-,\mu} + \frac{i}2 \psi_+^\mu \p \psi_{+,\,\mu} \right)
\end{split}
\ee
où, dans la seconde égalité, $\psi_\pm$, fonctions de $z, \bz$, sont définis par
\bse
\begin{align}
\psi_- (w) &= (\p_w z)^{\frac12} \psi_- (z) = (-iz)^{\frac12} \psi_- (z) \\
\psi_+ (\bw) &= (\p_{\bw} \bz)^{\frac12} \psi_+ (\bz) = (i\bz)^{\frac12} \psi_+ (\bz)
\end{align}
\ese
pour que l'action soit invariante sous la transformation conforme. Comme on peut le vérifier en calculant le produit d'opérateurs $T \psi$ à partir des expressions données ci-dessous, ces transformations  correspondent à des opérateurs de poids conformes respectifs $(\frac12,0)$ et $(0,\frac12)$. Nous avons les expressions suivantes pour le tenseur énergie-impulsion et le supercourant\footnote{Les expressions \eqref{TJeucl} comportent un signe $-$ supplémentaire par rapport aux expressions \eqref{Tminsk} et \eqref{Jminsk}. Ce signe provient du prolongement analytique vers la feuille d'univers euclidienne qui change le signe de l'action (comparer \eqref{susyactioneucl} et \eqref{actsPsimpl}).}
\bse
\l{TJeucl}
\begin{align}
T_{--}(z) = T(z) &= -\frac{1}{\apr}\left(\lno \p X^\mu(z) \p X_\mu(z)\rno -\frac{i}2 \!\lno \psi_-(z) \p \psi_-(z)\rno\right) \\
T_{++}(\bz) = \bT(\bz) &= -\frac{1}{\apr}\left(\lno \bp X^\mu(\bz) \bp X_\mu(\bz)\rno +\frac{i}2 \!\lno \psi_+(\bz) \bp \psi_+(\bz)\rno\right) \\
J_-(z) = J(z) &= \frac{\sqrt{2i}}{\apr} \!\lno \psi_-^\mu (z) \p X_\mu(z)\rno \\
J_+(\bz) = \bar J(\bz) &= -\frac{\sqrt{-2i}}{\apr} \!\lno \psi_+^\mu (\bz) \bp X_\mu(\bz)\rno
\end{align}
\ese
Le développement en modes de $\psi_\pm$ s'écrit
\bse
\begin{align}
\psi_-^\mu (z) &= \sqrt{i\apr}\sum_r \beta_r^\mu\, z^{-r-\frac12} \\
\psi_+^\mu (\bz) &= \sqrt{-i\apr}\sum_s \tilde\beta_s^\mu\, {\bz}^{-s-\frac12}
\end{align}
\ese
Notons que $J$ est un opérateur de poids conforme $(\frac32,0)$ et $\bar J$ un opérateur de poids conforme $(0,\frac32)$. Nous rappelons que $T$ et $\bar T$ sont les descendants de l'opérateur identité dans l'algèbre holomorphe et antiholomorphe respectivement et ne se transforment pas comme les champs primaires. Nous n'entrerons pas plus dans les détails.

Les fonctions de corrélation à deux points des champs fermioniques s'écri\-vent
\bse
\begin{align}
\langle \psi_-^\mu(z) \psi_-^\nu(z') \rangle &= i \apr \eta^{\mu\nu} \frac1{z-z'} \\
\langle \psi_+^\mu(\bz) \psi_+^\nu(\bz') \rangle &= -i \apr \eta^{\mu\nu} \frac1{\bz-\bz'}
\end{align}
\ese
et permettent de définir l'ordre normal conforme pour ces champs.

Pour calculer des fonctions de corrélation, des opérateurs de vertex supersymétriques peuvent être construits comme pour la corde bosonique. Comme cette construction fait intervenir des subtilités liés aux superfantômes et que nous n'aurons pas besoin de la forme précise de ces opérateurs de vertex par la suite, nous n'entrerons pas dans les détails.

Enfin, pour que la théorie puisse être correctement définie, il faut à nouveau que la charge centrale totale soit nulle. En tenant compte des fantômes et superfantômes, celle-ci est égale à $D-10$. Le nombre de dimensions de l'espace-temps est donc fixée à dix, contre vingt-six dans le cas de la corde bosonique.

\section{Quantification canonique : ancienne quantification covariante}
\l{quantcov}

Il ne sera pas aussi facile d'appliquer la condition de couche de masse et la sélection des états physiques est obtenue par une règle \emph{ad hoc}, qu'on peut justifier dans le formalisme BRST ou celui d'intégrale des chemins. Nous n'exposerons pas le formalisme BRST, nous nous contenterons d'exposer certains aspects du formalisme d'intégrale des chemins qui nous seront utiles pour l'explication des calculs que nous avons réalisés en théorie des cordes. Nous ne détaillerons pas les autres aspects, notamment le traitement complet des \emph{fantômes} qui sont des degrés de liberté supplémentaires introduits lorsqu'on élimine les degrés de liberté de jauge ; nous nous contenterons de mentionner les effets qu'ils ont sur nos calculs.

\subsection{Quantification canonique de la corde ouverte libre}
\l{quantcovco}

La procédure que nous allons suivre ressemble formellement à ce qu'on appelle \emph{seconde quantification} en théorie des champs. Gardons à l'esprit cependant que pour la corde il s'agit d'une \emph{première quantification}. Nous allons détailler cette procédure dans le cas de la corde libre (conditions au bord de Neumann aux deux extrémités) et ouverte. Nous donnerons seulement les résultats pour les autres cas (orbifold et champ électromagnétique).

Il s'agit de promouvoir au rang d'opérateurs les modes $x^\mu$, $p^\mu$ et $a^\mu_n$ de la corde.
Comme pour le système ponctuel (voir section \ref{thpoint}), la prescription de la quantification canonique s'écrit
\be [.\, ,.]_{\text{P.B.}} \longrightarrow -i[.\, ,.] \ee
Les crochets de Poisson sur $X^\mu$ et son impulsion conjuguée $\pi^\mu = -\frac1{2\pi \apr}\dpt X^\mu$
\bse
\begin{align}
[X^\mu (\t,\s)\, , X^\nu (\t,\s')]_{\text{P.B.}} &= [\pi^\mu (\t,\s)\, , \pi^\nu (\t,\s')]_{\text{P.B.}} =0 \\
[X^\mu (\t,\s)\, ,\pi^\nu (\t,\s')]_{\text{P.B.}} &= -\eta^{\mu\nu}\, \delta(\s-\s')
\end{align}
\ese
permettent de calculer les crochets de Poisson sur les modes de la corde (nous omettrons définitivement d'écrire les crochets de commutation nuls)\footnote{\l{facteurs} Les facteurs qui apparaissent dans les expressions en termes de modes des champs $X$ (cela est valable aussi pour les champs $\psi$) sont déterminés de façon à ce que ces crochets de Poisson soient standards, sachant que $x$ et $p$ sont des modes décrivant le mouvement rectiligne uniforme du centre de masse, que les modes $a$, $\a$ et $\ta$ décrivent les oscillations d'un champ bosonique et les modes $b$, $\beta$ et $\tilde \beta$, celles d'un champ fermionique.}
\bse
\begin{align}
[a_n^\mu\, , a_m^\nu]_{\text{P.B.}} &= i n\, \eta^{\mu\nu}\,\delta_{m+n,0} \\
[x^\mu, p^\nu]_{\text{P.B.}} &= - \eta^{\mu\nu}
\end{align}
\ese
Après quantification , les opérateurs $X^\mu$ et $\pi^\nu$ vérifient les relations de commutation suivantes 
\be [X^\mu (\t,\s)\, ,\pi^\nu (\t,\s')] = i\,\eta^{\mu\nu}\, \delta(\s-\s') \ee
et les modes
\bse
\begin{align}
[a_n^\mu\, , a_m^\nu] &= n\, \eta^{\mu\nu}\,\delta_{m+n,0} \\
[x^\mu, p^\nu] &= i \eta^{\mu\nu}
\end{align}
\ese
Les champs classiques $X^\mu$ sont réels. Cette propriété devient une condition d'hermiticité pour l'opérateur $X^\mu$ et impose la relation suivante $(a_n^\mu)^\dagger = a_{-n}^\mu$.
On reconnaît respectivement les relations de commmutation de l'oscillateur harmonique (à un facteur de normalisation près) et celles de la particule libre. L'espace des états $E$ sera donc le produit direct entre les états de l'oscillateur harmonique et les états propres de l'impulsion $p$
\be E = \left\{\Bigl\lvert \left\{N_n^\mu\right\}_{n\in \N\, \backslash\{0\}},\, k^\mu,\, \mu \in\{0,\ldots,D\} \Bigr\rangle \right\} \ee
où $k^\mu$ est la valeur propre de l'opérateur impulsion $p^\mu$ et $N_n^\mu$ le niveau du $n$-ième oscillateur harmonique de la coordonnée $X^\mu$.

\subsection{Contrainte de Virasoro dans le formalisme de quantification canonique}

L'espace de Hilbert $H$ des états physiques est cependant bien plus petit, car il faut d'une part, imposer l'équivalent des contraintes classiques (voir \eqref{tei}) et d'autre part éliminer les redondances d'états qu'on ne peut pas distinguer entre eux (parce que leur différence est un état physique orthogonal à tous les autres états, lui-même compris, appelé état \emph{null} pour reprendre le terme anglais). L'élimination des redondances se fait par simple quotient de l'espace des états physiques par l'espace des états null. Par la suite, nous parlerons de l'espace des états remplissant les contraintes physiques et l'opération subséquente d'élimination des états null sera implicite.

Au niveau classique, les équations du mouvement sur la métrique de la feuille d'univers \eqref{tei} s'écrivent en terme des modes de $T$
\be L_n = 0 \qquad \text{pour tout } n \ee
Après quantification, cette condition doit être adoucie, comme le formalisme BRST permet de le démontrer. Les états physiques $\lvert \text{phys} \rangle$ vérifient les conditions suivantes
\be L_n \lvert \text{phys} \rangle = 0 \text{ , } n > 0 \qquad (L_0-a) \lvert \text{phys} \rangle = 0 \l{Virquant}\ee
où $a$ est la constante d'ordre normal\footnote{Nous écrirons aussi $L_0$ en y incluant la constante d'ordre normal. La condition devient alors $L_0 \lvert \text{phys}\rangle =0$. Nous utiliserons la convention suivante : si la constante $a$ n'est pas écrite explicitement, elle est supposée incluse dans l'expression de $L_0$. Nous définirons de même $\bar L_0$.}. Avant de voir comment cette constante apparaît en quantification canonique, remarquons que la contrainte sur les états quantiques correspond à la définition d'un champ primaire $\Phi$ en théorie conforme
\be L_n \Phi = 0 \qquad n > 0 \qquad \text{et} \qquad L_0 \Phi = h \Phi \ee
La détermination du spectre de la corde est ainsi équivalente à l'identification des opérateurs primaires de la théorie conforme associée. Nous avons vu que ces opérateurs étaient représentés par des opérateurs de vertex qui devaient avoir un poids conforme\footnote{Dans la théorie conforme associée à la corde ouverte, il n'y a qu'une seule algèbre de Virasoro et nous notons simplement le poids conforme $h=1$.} 1. On peut montrer que dans le formalisme de quantification canonique, il faut aussi que $a= 1$ pour que le spectre soit cohérent.

Revenons à la corde ouverte libre. À partir de l'expression de $T$ en fonction de $X^\mu$ \eqref{teilinsk} ou \eqref{teitc}, nous obtenons l'expression des $L_n$ en fonction des modes quantifiés
\bse
\begin{align}
L_n &= \frac12 \sum_{m \in\Z} a_{-m}^\mu a_{n+m, \, \mu} \\
L_0 &= \apr p^\mu p_\mu + \frac12 \sum_{m \in\Z} a_{-m}^\mu a_{m,\,\mu}
\end{align}
\ese
Mais il subsiste une ambiguité dans l'ordre des opérateurs pour $L_0$. Pour résoudre cette ambiguité, on applique la prescription d'\emph{ordre normal}, cette fois-ci \emph{d'opérateur}, qui impose de placer les opérateurs de création ($a_{-n}^\mu$, $n>0$) à gauche et les opérateurs d'annihilation ($a_n^\mu$, $n>0$) à droite. Cet ordre normal introduit une constante supplémentaire
\be L_0 = \apr p^\mu p_\mu + \sum_{n=1}^{\infty} a_{-n}^\mu a_{n,\,\mu} -\frac{D-2}{24} \ee
La constante d'ordre normal est donc $a = \frac{D-2}{24}$. Elle est égale à 1 lorsque $D=26$.

Dans les autres cas, où la corde ouverte a des conditions aux bords Dirichlet-Dirichlet (DD) \eqref{codd}, Neumann-Dirichlet (ND) \eqref{cond} ou Dirichlet-Neumann (DN) \eqref{codn}, pour une coordonnée $X^{\mu_0}$ donnée, la contribution de cette direction à $L_0$, que nous noterons $L_0^{\mu_0}$
\bse
\begin{align}
L_{0, DD}^{\mu_0} &= \frac{(q_1-q_0)^2}{(2\pi)^2\apr} + \sum_{n=1}^\infty a_{-n} a_n - \frac{1}{24}\\
L_{0, ND}^{\mu_0} = L_{0, DN}^{\mu_0} &= \sum_{n=1}^\infty a_{-n} a_n - \frac{1}{24} 
\end{align}
\ese
où nous avons omis $\mu_0$ pour alléger les expressions et inclus la contribution à la constante d'ordre normal. Les expressions des $L_n$, $n\neq 0$ restent inchangées.

Notons que pour une corde ouverte ayant des conditions DD sur le même nombre de dimensions, $L_0$ s'écrira
\be
L_0 = \apr p_{\sslash}^2 + \frac{(q^\perp_1-q^\perp_0)^2}{(2\pi)^2\apr} + \sum_{n=1}^\infty a_{-n}\cdot a_n -1 \l{Lbddbr}
\ee
où $\sslash$ désigne les directions du volume d'univers des D-branes et $\perp$ les directions transverses. 

\subsection{Détermination du spectre de la corde libre ouverte}
\l{spectreco}

Nous allons classer les états à l'aide du nombre total $N$ d'excitations des oscillateurs harmoniques, valeur propre de l'opérateur
\be
\hat N = \sum_{n=1}^{\infty} a_{-n}^\mu a_{n,\,\mu}
\ee
En effet, si l'on applique cet opérateur à un élément de $E$, on obtient
\be
\hat N \Bigl\lvert \{N_n^\mu\},\, k^\mu \Bigr\rangle = \left(\sum_{n=1}^{\infty} \sum_{\mu=0}^{D} N_n^\mu\right)\Bigl\lvert \{N_n^\mu\},\, k^\mu \Bigr\rangle = N \Bigl\lvert \{N_n^\mu\},\, k^\mu \Bigr\rangle 
\ee

Nous n'entrerons pas dans les détails de l'analyse détaillée du spectre qui est faite dans \cite{Green:1987sp} p.81-86 ou \cite{Polchinski:1998rq} p.121-124 et nous nous contenterons d'en donner les résultats.
\begin{itemize}
\i Le spectre est dépourvu d'états de norme négative quand $D=26$. On retrouve la même condition que celle évoquée dans le cadre de la théorie conforme. La constante d'ordre normal qui apparaît dans l'expression de $L_0$ est donc égale à $-1$.
\i La condition sur $L_0$ correspond à la condition de couche de masse sur les états physiques
\be \l{mass} m^2 = \frac1{\apr} (N-1) \ee
\i Il existe un seul état au niveau $N=0$, noté $\lvert 0, k^\mu \rangle$ de masse carré négative. La théorie comporte donc un tachyon.
\i Au niveau $N=1$, nous avons 24 états de masse nulle, définis par $e_\mu a_{-1}^\mu \lvert 0,\, k^\rho \rangle$ où $e^\mu$ vérifie la condition $k^\mu e_\mu = 0$ et est défini modulo $e_\mu \sim e_\mu + x k_\mu$ avec $x$ réel quelconque. Ces états correspondent aux 24 polarisations possibles pour un boson vecteur de masse nulle évoluant dans un espace-temps à 26 dimensions.
\i Les autres états sont obtenus par applications successives des opérateurs $a_{-n}^\mu$. Nous n'en aurons pas besoin par la suite.
\end{itemize}

Nous allons à présent donner les résultats pertinents pour les autres conditions au bord que nous avons évoquées précédemment.

\subsection{Corde fermée libre dans un espace plat}
\l{spectrecf}

Les modes de la corde fermée libre \eqref{cf} vérifient les mêmes relations de commutation que la corde ouverte
\bse
\begin{align}
[\a_n^\mu\, , \a_m^\nu] &= n\, \eta^{\mu\nu}\,\delta_{m+n,0} \\
[\ta_n^\mu\, , \ta_m^\nu] &= n\, \eta^{\mu\nu}\,\delta_{m+n,0} \\
[x^\mu, p^\nu] &= i\, \eta^{\mu\nu}
\end{align}
\ese
Les états physiques doivent vérifier la condition \eqref{Virquant} pour les $L_n$ et pour les $\bar L_n$, ainsi qu'une condition supplémentaire dite de \emph{level matching}\footnote{Une traduction pourrait être \emph{ajustement de niveau} mais nous préférons garder le terme anglais.}
\be
(L_0-\bar L_0) \lvert\text{phys} \rangle =0
\ee
Notons que $L_0$ dans le cas de la corde ouverte, ou $L_0 + \bar L_0$ dans le cas de la corde fermée, est le générateur des translations par rapport au temps $\tau_{\text{E}}$ de la feuille d'univers euclidienne et consitue ainsi l'hamiltonien du système. De même, $L_0-\bar L_0$ est le générateur des rotations. Les états physiques sont donc invariants sous les rotations de la feuille d'univers. Nous verrons dans la section suivante que ce n'est pas toujours le cas.

Les générateurs des algèbres de Virasoro $L_n$ et $\bar L_n$ sont identiques pour $n \neq 0$ à ceux de la corde ouverte.
\bse
\begin{align}
L_n &= \frac12 \sum_{m \in\Z} \a_{-m}^\mu \a_{n+m, \, \mu} \\
\bar L_n &= \frac12 \sum_{m \in\Z} \ta_{-m}^\mu \ta_{n+m, \, \mu} \l{Lncf}
\end{align}
\ese
Par contre, nous avons
\bse
\begin{align}
L_0 &= \frac{\apr}4 p^\mu p_\mu + \sum_{n=1}^{\infty} \a_{-n}^\mu \a_{n, \, \mu} - 1\\
\bar L_0 &= \frac{\apr}4 p^\mu p_\mu + \sum_{n=1}^{\infty} \ta_{-n}^\mu \ta_{n, \, \mu} - 1 \l{L0cf}
\end{align}
\ese
où nous avons posé $D=26$.

On peut définir comme pour la corde ouverte un niveau $\bar N$. La condition de level matching impose que $N = \bar N$. La masse des états physiques est définie par la relation suivante :
\be
m^2 = \frac4{\apr} (N-1)
\ee
Comme précédemment, l'état $N=0$ est un tachyon. Les états suivants sont obtenus par application successives d'un nombre égal d'opérateur $\a_{-n}^\mu$ et $\ta_{-n}^\nu$.

Les états de masse nulle sont particulièrement intéressants. Il s'agit des états $e_{\mu\nu} \a_{-1}^\mu \ta_{-1}^\nu \lvert 0,\, k^\rho \rangle$ où $e_{\mu\nu}$ vérifie la condition $k^\mu e_{\mu\nu} = k^\nu e_{\mu\nu} = 0$ et est défini modulo $e_{\mu\nu} \sim e_{\mu\nu} + x_\mu k_\nu + y_\nu k_\mu$ avec $x_\mu k^\mu = y_\mu k^\mu = 0$. Nous avons 576 ($24^2$) états de masse nulle qu'on peut organiser en un scalaire $\phi$, le dilaton, un tenseur symétrique de trace nulle $g$, la métrique, et un tenseur antisymétrique $B$. Il est tout à fait légitime de donner de tels noms à ces états car une perturbation de la métrique par exemple autour de l'espace plat est décrit par l'insertion de l'opérateur de vertex du graviton (voir \cite{Polchinski:1998rq} p.108). Ainsi l'extrapolation de cette observation  conduit à inclure un dilaton, une métrique quelconque et un champ antisymétrique à l'action de Polyakov, ce qui donne le résultat que nous avons déjà donné \eqref{actPcurved}. Toutefois, en l'absence de théorie de champ de cordes permettant de décrire des états cohérents de gravitons par exemple, il est encore impossible de déterminer de tels champs et nous sommes contraints de les choisir sans autre guide que la possibilité de définir la théorie perturbative et la faisabilité des calculs.

\subsection{Corde fermée et compactification}

Nous allons supposer qu'une seule direction est compactifiée et nous omettrons l'indice d'espace-temps correspondant. Pour traiter le cas de la corde avec un nombre d'enroulement non nul, il faut distinguer la partie gauche $X_L$, ou partie antiholomorphe \eqref{cfzah} sur la feuille d'univers euclidienne, et la partie droite\footnote{Ces dénominations supposent que $\s$ est orienté de gauche à droite et sont opposées aux conventions habituelles (qui supposent que $\s$ est orienté de droite à gauche) où les parties droite et gauche sont échangées, auquel cas la partie gauche est associée à la partie holomorphe et la partie droite, à la partie antiholomorphe.} $X_R$ de la solution \eqref{cfcomp}, ou partie holomorphe \eqref{cfzh},
\bse
\begin{align}
X_R (\t, \s) &= x_R^\mu + \frac{\apr}2 p_R (\tau-\sigma)  + i \sqrt{\frac{\apr}2} \sum_{n \in \Z\backslash\{0\}} \frac{\a_n^\mu}n\, e^{-in (\t-\s)} \l{cfcompR}\\
X_L (\t, \s) &= x_L^\mu + \frac{\apr}2 p_L (\tau+\sigma)  + i \sqrt{\frac{\apr}2} \sum_{n \in \Z\backslash\{0\}} \frac{\ta_n^\mu}n\, e^{-in (\t+\s)}\l{cfcompL}
\end{align}
\ese
avec
\be p_R = \frac{v}{R} + \frac{wR}{\apr} \qquad p_L =\frac{v}{R} - \frac{wR}{\apr} \ee
Les cordes non enroulées vérifient $p_L = p_R$ et les cordes enroulées sont celles pour lesquelles $p_L \neq p_R$. C'est pourquoi il faut introduire les modes conjugués à $p_L$ et $p_R$ : $x_L$ et $x_R$.

Les relations de commutation sont les mêmes que pour la corde fermée dans un espace plat, mais celle pour les modes zéro s'écrivent pour un nombre d'enroulement non nul :
\be [x_L\, , p_L] = [x_R\, , p_R] = i \ee

On obtient alors, pour $L_0$ et $\bar L_0$
\bse
\begin{align}
L_0 &= \frac{\apr}4 (p_R^2 + p_\perp^2) + \sum_{n=1}^{\infty} \a_{-n}^\mu \a_{n, \, \mu} - 1\\
\bar L_0 &= \frac{\apr}4 (p_L^2+ p_\perp^2) + \sum_{n=1}^{\infty} \ta_{-n}^\mu \ta_{n, \, \mu} - 1
\end{align}
\ese
où $p_\perp^2$ est la norme au carré de l'impulsion dans les directions non compactifiées.

Les conditions de Virasoro imposent :
\bse
\l{spcfcomp}
\begin{gather}
m^2 = \frac{v^2}{R^2} + \frac{w^2 R^2}{\apr} + \frac{2}{\apr}(N+\bar N -2) \\
N-\bar N + vw = 0
\end{gather}
\ese

Nous évoquerons les propriétés de symmétrie de ce spectre et ses conséquences physiques à la section \ref{dbranes}

\subsection{Corde fermée dans un orbifold et corde ouverte dans un champ électromagnétique constant et uniforme}
\l{qorbchem}

Comme nous avons vu à la section \ref{gdt}, il existe une analogie formelle entre la corde fermée dans un orbifold et la corde ouverte dans un champ électromagnétique constant et uniforme et nous allons voir qu'elle reste présente après la quantification. Dans le cas de l'orbifold de boost que nous expliquerons plus en détail à la sous-section suivante, cette analogie est même indispensable pour quantifier correctement les modes zéro.

Dans le cas général de l'orbifold \eqref{orbgen}, les relations de commutation (non nulles) entre les modes $\a_n$ et $\ta_n$ s'écrivent \cite{Berkooz:2005ym}
\bse
\begin{align}
[\a_n^\mu\,, \a_m^\nu] &= (n-iw \J)^\mu_{\phantom{\mu}\rho} \eta^{\rho\nu}\, \delta_{n+m,0} \\
[\ta_n^\mu\,, \ta_m^\nu] &= (n+iw \J)^\mu_{\phantom{\mu}\rho} \eta^{\rho\nu}\, \delta_{n+m,0} \l{rcorb}
\end{align}
\ese
Ces relations sont valables pour $n=0$ sauf pour les directions selon lesquelles $\J$ a une valeur propre nulle, auquel cas, il faut définir\footnote{Nous invitons le lecteur à se référer à la sous-section 2.4 de \cite{Berkooz:2005ym} pour les détails de ce cas particulier. Dans faire le lien entre notre présentation et cet article, il faut échanger les modes $\a$ et $\ta$.} à partir de $\a_0$ et $\ta_0$ des opérateurs de type $x$ et $p$. Cette redéfinition doit être faite pour les cordes non twistées $w=0$ ($\a_0^\mu = \ta_0^\mu = \sqrt{\frac{\apr}2} p^\mu$). Par ailleurs, dans ce cas, on retrouve bien les relations de la corde fermée dans un espace plat.

Ces relations de commutation sont à comparer à celles de la configuration de champ électromagnétique \eqref{emgen}
\be
[a_n^\mu\,, a_m^\nu] = (n+i \Omega)^\mu_{\phantom{\mu}\rho} \eta^{\rho\nu}\, \delta_{n+m,0} 
\ee
valables pour $n=0$ sauf si $\Omega$ a une valeur propre nulle (cas des ondes planes, qui est résolu en adoptant la jauge du cône de lumière). Il faut ajouter les relations de commutation sur les $q_0^\mu$
\be
[q_0^\mu\,, q_0^\nu] = \left(\frac{i}{F^{(0)} - F^{(1)}}\right)^\mu_{\phantom{\mu}\rho} \eta^{\rho\nu}
\ee
La configuration de corde fermée est composée de deux copies des modes de cordes ouvertes, si l'on ne tient pas compte des modes zéro dont le traitement est plus délicat. Nous verrons comment les modes zéro sont reliés dans le cas de l'orbifold de boost et de la configuration de champs électriques.

En ce qui concerne les générateurs des algèbres de Virasoro, pour la corde fermée et pour la corde ouverte, les expressions des $L_n$ ($\bar L_n$) \eqref{Lncf} restent inchangées tandis que dans les expressions de $L_0$ ($\bar L_0$), la constante d'ordre normal change à cause d'une part, du décalage dans les énergies des modes ($\pm iw\J$ ou $+iw\Omega$) et d'autre part, dans le cas d'un orbifold de boost ou du champ électrique, les modes zéros ne sont pas des opérateurs d'oscillateur harmonique et ne doivent donc pas être ordonnés. Dans ces derniers cas, où la quantification est particulièrement délicate, nous détaillerons les expressions des générateurs $L_n$ et la façon dont il faut procéder.

Pour le cas général, nous invitons le lecteur à se reporter à l'annexe B de \cite{Berkooz:2005ym} qui présente des résultats unifiés pour les trois configurations d'orbifold, qui peuvent être transposés aux  configurations de champs électromagnétiques constants et uniformes.

\subsection{Configuration de champs électriques et orbifold de boost}
\l{qobche}

Considérons le cas de la corde ouverte couplée aux champs électriques d'intensité $\e_0$ et $\e_1$. Les relations de commutation s'écrivent
\bse
\begin{align}
[a_n^+ \,, a_m^-] &= -(n+i\nu)\, \delta_{n+m,0} \\
[q_0^+\, , q_0^-] &= \frac{i}{\e_0-\e_1} \l{rcE}
\end{align}
\ese
En particulier, les modes $a_0^\pm$ vérifient la relation de commutation suivante
\be [a_0^+ \,, a_0^-] = -i\nu \ee
qui correspondent à un oscillateur harmonique inversé \cite{Pioline:2003bs}. Si l'on se restreint aux modes $x_0$ et $a_0$, nous avons exactement la même situation qu'une particule dans un champ électrique. Nous en expliquerons les conséquences physiques dans le chapitre \ref{article2}. Alors que pour les autres modes, malgré le décalage $i\nu$ de leur énergie, l'espace des états est un espace de Fock, pour $a_0$, l'espace des états est le continuum d'états correspondants au spectre de l'oscillateur harmonique inversé. La distinction \gl opérateur de création\gr/\gl opérateur d'annihilation\gr\ n'a pas de sens pour $a_0$, il ne faut donc pas l'ordonner. La prescription donnée par \cite{Pioline:2003bs} consiste à symétriser le produit $a_0^+ a_0^-$. Ainsi, $L_0$ s'écrit
\be
L_0 = L_0^\perp -\frac12(a_0^+ a_0^- + a_0^- a_0^+) - \sum_{n=1}^\infty a_{-n}^+ a_n^- + a_{-n}^- a_n^+ + \frac12 \nu^2 -\frac1{12}
\ee
où $L_0^\perp$ correspond aux directions transverses au cône de lumière\footnote{La contribution des fantômes est sous-entendue.}. Notons que $L_0$ n'est pas le prolongement analytique $\theta \to i\beta$ de l'opérateur $L_0$ de la corde ouverte couplée à des champs magnétiques. Un tel prolongement donnerait, à la place du terme $\frac12 \nu^2$, $\frac12 i\nu (1-i\nu)$.

Pour la corde fermée dans un orbifold de boost, les relations de commutation s'écrivent
\bse
\begin{align}
[\a_n^+\,, \a_m^-] &= -(n+i\nu)\, \delta_{n+m,0} \\
[\ta_n^+\,, \ta_m^-] &= -(n-i\nu)\, \delta_{n+m,0} \l{rcorboost}
\end{align}
\ese
Pour les modes zéros, nous avons en particulier
\bse
\begin{align}
[\a_0^+\,, \a_0^-] &= -i\nu\\
[\ta_0^+\,, \ta_0^-] &= i\nu
\end{align}
\ese
Il s'agit de deux copies de l'algèbre de la corde ouverte, les modes droits de la corde fermée $\a_0^\pm$ correspondant à un paramètre $+\nu$, les modes gauches $\ta_0^\pm$ à un paramètre $-\nu$. Il est possible de faire correspondre exactement les quatre modes zéros de la corde fermée avec les quatre modes zéros de la corde ouverte et ainsi transposer directement à la corde fermée l'analyse faite pour la corde ouverte. Il suffit d'identifier
\be
\a_0^\pm = a_0^\pm \qquad \ta_0^\pm = \pm \sqrt{\nu(\e_1-\e_0)} q_0^\pm
\ee
$a_0^\pm$ et $q_0^\pm$ peuvent être représentés par la dérivée covariante dans un champ électrique constant appliquée à la fonction d'onde $\phi(x^+, x^-)$ d'un état propre des modes zéro du cône de lumière
\be
a_0^\pm = i\p_\mp \pm \frac{\nu}2 x^\pm \qquad q_0^\pm = \mp \frac1{\sqrt{\nu(\e_1-\e_0)}} \left(i\p_\mp \mp \frac{\nu}2 x^\pm \right)
\ee
On aura alors
\be
\a_0^\pm = i\nabla_\mp = i\p_\mp \pm \frac{\nu}2 x^\pm \qquad \ta_0^\pm = i\overline{\nabla}_\mp = i\p_\mp \mp \frac{\nu}2 x^\pm \l{repx}
\ee
Nous pouvons à partir de ces définitions nous focaliser sur la physique dans le cône de lumière. Nous allons voir qu'en définissant de façon appropriée la masse, les conditions de Virasoro impliquent que la fonction d'onde $\phi$ vérifie l'équation de Klein-Gordon dans le cône de lumière en présence d'un champ électrique, ce qui justifie \emph{a posteriori} notre interprétation en terme d'oscillateur harmonique inversé.

Les expressions de $L_0$ et $\bar L_0$ sont
\bse 
\begin{align}
L_0 &= L_0^\perp -\frac12(\a_0^+ \a_0^- + \a_0^- \a_0^+) - \sum_{n=1}^\infty \a_{-n}^+ \a_n^- + \a_{-n}^- \a_n^+ + \frac12 \nu^2 -\frac1{12} \\
\bar L_0 &= \bar L_0^\perp -\frac12(\ta_0^+ \ta_0^- + \ta_0^- \ta_0^+) - \sum_{n=1}^\infty \ta_{-n}^+ \ta_n^- + \ta_{-n}^- \ta_n^+ + \frac12 \nu^2 -\frac1{12}
\end{align}
\ese
Le générateur de boost $J$
\be
\begin{split}
J =& \frac1{2\nu}(\a_0^+ \a_0^- + \a_0^- \a_0^+) - \frac1{2\nu}(\ta_0^+ \ta_0^- + \ta_0^- \ta_0^+) \\
&+ i \sum_{n=1}^\infty \left(\frac{\a_{-n}^+ \a_n^- + \a_{-n}^- \a_n^+}{n+i\nu} - \frac{\ta_{-n}^+ \ta_n^- + \ta_{-n}^- \ta_n^+}{n-i\nu} \right) \l{boostgen}
\end{split}
\ee
doit être un multiple de $\frac{1}{\beta}$ à cause de la relation d'équivalence de l'orbifold.

Définissons maintenant une masse \gl droite\gr\ $M$ et une masse \gl gauche\gr\ $\bar M$
\bse 
\begin{align}
M^2 &= -2\sum_{n=1}^\infty \a_{-n}^+ \a_n^- + \a_{-n}^- \a_n^+ + \nu^2 -\frac16 + L_0^\perp\\
\bM^2 &= -2\sum_{n=1}^\infty \ta_{-n}^+ \ta_n^- + \ta_{-n}^- \ta_n^+ + \nu^2 -\frac16 + \bar L_0^\perp
\end{align}
\ese
et les conditions de Virasoro s'écrivent
\be
M^2 = \a_0^+ \a_0^- + \a_0^- \a_0^+ \qquad \bM^2 = \ta_0^+ \ta_0^- + \ta_0^- \ta_0^+
\ee
c'est-à-dire, dans la représentation des $\a_0$ et $\ta_0$ en terme d'opérateurs différentiels agissant sur la fonction d'onde $\phi$, de l'équation de Klein-Gordon dans $\R^{1,1}$ d'une particule de charge $+\nu$
\bse
\be
(\nabla^+ \nabla^- + \nabla^- \nabla^+)\phi = M^2\phi \l{scct1}
\ee
ou d'une particule de charge $-\nu$
\be
(\overline{\nabla}^+ \overline{\nabla}^{\,-} + \overline{\nabla}^{\,-} \overline{\nabla}^+)\phi = \bM^2\phi \l{scct2}
\ee
\ese
La signification physique pour les cordes de ces équations sera précisée \`a la sous-section \ref{ctob}.

Terminons par la définition de la masse de cône de lumière $\mu$
\be
\mu^2 = \frac12 (M^2+\bM^2)
\ee
et de $j$ la partie de mode zéro du générateur de boost
\be
j = \frac1{2\nu} \left(M^2 - \bM^2 \right) \l{impboost}
\ee

\subsection{Corde ouverte dans des ondes planes électromagnétiques \`a profil lin\'eaire}
\l{cocheeq}

Dernière configuration à avoir été étudiée dans la section \ref{gdt}, elle présente la ressemblance suivante avec la configuration de champ électrique : selon la valeur de $\delta = e_1 - e_0 - \pi e_0 e_1$, les modes zéro sont soit de type magnétique ($\delta > 0$), c'est-à-dire des opérateurs d'oscillateur harmonique, soit de type onde plane ($\delta=0$), où ils se réduisent à deux modes de corde libre $x_0$ et $p_0$, soit de type électrique $\delta<0$, c'est-à-dire des opérateurs d'oscillateur harmonique inversé. Nous verrons au chapitre \ref{article1} que le type magnétique correspond à une configuration stable alors que le type électrique, à une instabilité cinématique de la corde.

Nous avons vu à la section \ref{gdt} que cette configuration pouvait se ramener, dans le cas où la corde n'est pas neutre, à deux problèmes unidimensionnels selon deux directions d'espace. Nous avons ainsi les relations de commutation suivantes
\be [a_n\,, a_m^\dagger] = \w_n\, \delta_{n,m} \ ,\ee
pour $n \neq 0$ et
\be [a_0\,, a_0^\dagger] = \w_0 \text{\ \ ou \ \ } [x_0\,,p_0] = i \text{\ \ ou \ \ } [a+\,, a_-] = ik_0 \ee
pour les modes zéro, selon la valeur de $\delta$ (nous nous limitons à $\Delta >0$, $\Delta=e_1 - e_0 +\frac2{\pi}$).
Les normalisations, dont nous n'avons pas expliqué l'origine à la section \ref{gdt}, sont fixés par ces relations de normalisations, de telle sorte que dans le membre de droite apparaissent l'énergie du mode correspondant, comme dans le cas de la corde libre dans un espace plat.

$L_0$ s'écrit alors
\be p^+ L_0 = \sum_{n=1}^\infty a_n^\dagger + E +
\left\{ \begin{matrix} a_0^\dagger a_0 & \text{ si } \delta > 0 \\
                       p_0^2           & \text{ si } \delta = 0 \\
                       a_+ a_-         & \text{ si } \delta < 0
\end{matrix} \right.
\ee
À ce stade nous devons faire trois remarques :
\begin{itemize}
\i Le facteur $p^+$ apparaît car nous avons utilisé le formalisme de la quantification dans la jauge du cône de lumière.
\i $E$ est une constante d'ordre normal dont l'expression peut être trouvée dans \cite{Arutyunov:2001nz}. Il est important de noter qu'elle est réelle.
\i Les modes électriques ont été ordonnés, malgré ce qui a été dit précédemment. En effet, les états peuvent être exprimés comme des états de Fock. La constante $E$ étant réelle même dans le cas $\delta<0$, la condition de Virasoro peut tout à fait être vérifiée. La différence par rapport à la configuration de champs électriques réside dans le type d'instabilité. Dans notre cas, l'instabilité est seulement cinématique, alors que dans l'autre cas, l'instabilité est due à la production de paires. Il serait intéressant d'étudier d'autres configurations de type électrique pour savoir si cette corrélation entre type d'instabilité et nature du spectre peut être généralisée.
\end{itemize}

\subsection{Corde supersymétrique dans l'espace plat}
\l{superstringM}

\subsubsection{Quantification canonique}

Les opérateurs fermioniques doivent vérifier les relations d'anticommutation suivantes
\bse
\begin{align}
\{\psi^\mu_-(\t, \s)\,,\psi^\nu_-(\t,\s')\} &= \frac1{\apr} \eta^{\mu\nu} \delta(\s-\s') \\
\{\psi^\mu_+(\t, \s)\,,\psi^\nu_+(\t,\s')\} &= \frac1{\apr}\eta^{\mu\nu}  \delta(\s-\s') \\
\{\psi^\mu_-(\t, \s)\,,\psi^\nu_+(\t,\s')\} &= 0
\end{align}
\ese
Les modes vérifient alors
\be
\{\beta_r^\mu\,, \beta_s^\nu \} = \{\tilde \beta_r^\mu\,, \tilde\beta_s^\nu\} = \eta^{\mu\nu} \delta_{r+s,0}
\ee

Les générateurs $L_n$ et $G_r$ ont pour expression
\bse
\begin{align}
L_n &=  \frac12 \sum_{m \in\Z} \a_{-m}^\mu \a_{m+n, \, \mu} + \frac12 \sum_s \left(\frac{n}2 + s \right) \beta^\mu_{-s} \beta_{s+n, \, \mu} \qquad n \neq 0 \\
G_r &= \sqrt{2\apr} p^\mu \beta_{r,\,\mu} + \sum_{m \in \Z\backslash \{0\}} \a^\mu_{-m} \beta_{m+r,\,\mu} \\
L_0 &= \frac{\apr}4 p^\mu p_\mu + \sum_{m=1}^\infty \a_{-m}^\mu \a_{m,\,\mu} + \sum_{s>0} s \beta^\mu_{-s} \beta_{s, \, \mu} + a
\end{align}
\ese
avec $a = \frac12$ pour le secteur R et $a=0$ pour le secteur NS. Les expressions pour les quantités antiholomorphes s'écrivent de façon identique avec $\ta$ et $\tilde\beta$. Pour la corde ouverte, $\a$ et $\beta$ sont remplacés par $a$ et $b$ et $\frac{\apr}4 p^\mu p_\mu$ par $\apr p^\mu p_\mu$. 

Il est utile d'introduire un \emph{nombre de spineurs sur la feuille d'univers}, noté $F$, défini modulo 2, pour classer les états du spectre selon leur valeur propre $\pm 1$ pour l'opérateur $e^{i\pi F}$. L'expression de $F$ ne nous est pas utile, elle peut être trouvé dans \cite{Polchinski:1998rr} chapitre 10.

\subsubsection{Corde ouverte}

La condition d'états physiques s'écrit pour la corde ouverte
\bse
\begin{align}
L_n \lvert \text{phys} \rangle &= 0 \text{\ \ } n>0 \\
G_r \lvert \text{phys} \rangle &= 0 \text{\ \ } r\geq 0 \l{Gcond}\\
(L_0-a) \lvert \text{phys}\rangle &=0
\end{align}
\ese
où $a = \frac12$ dans le secteur NS  et $a=0$ dans le secteur R.

Nous obtenons alors la condition de couche de masse suivante, à $D=10$ dimensions et en tenant compte des contributions des fantômes et superfantômes,
\be
m^2 = \frac1{\apr} (N-\frac12)
\ee
pour le secteur NS et
\be
m^2 = \frac1{\apr} N
\ee
pour le secteur R.

Dans le secteur NS, l'état du vide est l'état annihilé par tous les $\beta_r^\mu$, avec $r>0$,  et est noté $\lvert 0;k \rangle$. Le secteur NS contient donc des bosons. L'\'etat $\lvert 0;k \rangle$ est un état tachyonique, noté NS$-$ car il a une valeur propre $-1$ pour l'opérateur\footnote{Ceci signifie qu'il y a un spineur sur la feuille d'univers. En effet, il s'agit d'un spineur superfantôme, qu'il faut inclure dans le décompte.} $e^{i\pi F}$ . Les états excités bosoniques sont construits en appliquant des opérateurs $\beta_{-r}^\mu$, $r>0$, une fois chaque maximum à cause des relations d'anticommutation. On peut montrer qu'il existe 8 états de masse nulle qui représentent les différentes polarisations d'une boson vecteur à 8 dimensions. On les note NS$+$.

Par contre dans le secteur R, il existe plusieurs états du vide, qui ont une masse nulle. Ces états $\lvert u;k \rangle$ (nous anticipons la notation sur le résultat) sont déterminés par la condition \eqref{Gcond} pour $r>0$ et par la condition suivante
\be
G_0 \lvert u;k \rangle = \sqrt{\apr} \lvert s';k \rangle\, k \cdot \Gamma_{s's} u_s = 0
\ee
où $u_s$ est la polarisation du spineur $u$ : $\lvert u;k \rangle = \lvert s;k \rangle u_s$ et la somme sur $s$ est sous-entendue. Nous reconnaissons l'équation de Dirac pour un fermion de masse nulle
\be
k\cdot \Gamma_{s's} u_s = 0
\ee
où $\Gamma^\mu$ sont les matrices réalisant l'algèbre de Cardiff à $D$ dimensions
\be
\Gamma^\mu = \sqrt{2} \beta_0^\mu
\ee
Le secteur R contient donc des fermions. Il est possible de démontrer qu'il y a 16 états du vide dans le secteur R, qui se décomposent en 8 états correspondant aux polarisations d'un spineur de chiralité positive\footnote{Nous définissons $\Gamma = \Gamma^0 \ldots \Gamma_9$. Un état de \emph{chiralité positive} est un état ayant pour valeur propre +1 pour $\Gamma$, un état de \emph{chiralité négative}, un état ayant pour valeur propre $-1$ pour $\Gamma$.
}, notés R$+$, et en 8 états correspondants aux polarisations d'un spineur de chiralité négative, notés R$-$. Les signes $\pm$ font toujours référence à la valeur propre de l'état pour $e^{i\pi F}$.

Ainsi, les états tachyoniques et de masse nulle sont classés dans les secteurs NS$\pm$ et R$\pm$. Trois conditions de cohérence supplémentaires sur la théorie, localité, fermeture de l'algèbre des produits d'opérateurs et invariance modulaires des amplitudes, ainsi que le couplage à la théories de corde fermée dont nous parlerons à la sous-section suivante, sélectionnent deux secteurs : NS$+$ et R$+$. C'est la \emph{théorie de type I}. Elle contient des cordes ouvertes non orientées. Le spectre ne comporte pas de tachyon (qui est dans NS$-$).

\subsubsection{Corde fermée}

Le spectre de la corde fermée est consituté de deux copies des spectres de la corde ouverte. Si on note $a$ et $\bar a$ les constantes d'ordre normal associées respectivement aux parties holomorphe et antiholomorphe, on a la condition de couche de masse suivante

\be
m^2 = \frac4{\apr} (N-a) = \frac4{\apr} (\bar N-\bar a)
\ee
$a$ et $\bar a$ sont déterminés comme pour la corde ouverte selon le secteur dans lequel on se trouve. ainsi les secteurs RR et NSNS contiennent des bosons, alors que les secteurs RNS et NSR, des fermions.

Les trois conditions de cohérence sélectionnent deux théories :
\begin{itemize}
\i la \emph{théorie de type IIB}, qui comporte les secteurs (NS+,NS+), (R+,R+), (R+,NS+), (NS+,R+)
\i la \emph{théorie de type IIA}, qui comporte les secteurs (NS+,NS+), (R+,R$-$), (R+,NS+), (NS+,R$-$)
\end{itemize}
Cette sélection est opérée sur le spectre total par l'intermédiaire de la \emph{projection Gliozzi-Scherk-Olive} (GSO).  De m\^eme cette projection GSO s\'electionne les \'etats de cordes ouvertes. Pour la théorie de type IIB, la projection conserve les secteurs qui vérifient
\be
e^{i\pi F} = e^{i\pi \bar F} = +1
\ee
Pour la théorie de type IIA, la projection conserve les secteurs qui vérifient
\be
\begin{aligned}
e^{i\pi F} &= +1 \\
e^{i\pi \bar F} &= +1 \text{\ si le secteur est NS,\ \ \ } e^{i\pi \bar F} = -1 \text{\ si le secteur est R}
\end{aligned}
\ee
Aucune de ces deux théories ne contient de tachyons, ce qui fait tout l'intérêt des théories supersymétriques.

À partir de la théorie de type IIB, dont les secteurs holomorphes et antiholomorphes ont la même chiralité,  il est possible de définir une théorie de cordes fermées non orientées en ne sélectionnant que les états du spectre dont la valeur propre est +1 pour l'opérateur de parité sur la feuille d'univers, noté $\Omega$. On obtient alors la \emph{théorie de cordes fermées non orientées de type I}. C'est la seule théorie avec laquelle peuvent interagir les cordes ouvertes, ce qui explique pourquoi elles doivent elles aussi être non orientées. C'est dans le cadre de ces théories, de type I pour les cordes ouvertes, de type IIA et IIB pour les cordes fermées, que nous étudierons la S-brane, à la section \ref{dbranes} et \ref{article3}.

\section{D-branes et Dirichlet S-branes}
\l{dbranes}

Nous revenons sur le cas de la corde fermée dans un espace plat compactifié et nous exposons la T-dualité, qui permet d'introduire un objet très important en théorie des cordes, la D-brane ou membrane de Dirichlet.

\subsection{T-dualité pour les cordes fermées}

Comme précédemment nous ne considérons qu'une seule direction (supposée spatiale) et nous omettons l'indice d'espace-temps correspondant. Le spectre de la corde fermée dans un espace plat compactifié, déterminé par les conditions \eqref{spcfcomp}, est invariant sous la transformation
\be
R \to \frac{\apr}{R} \text{ , } v\leftrightarrow w
\ee
En particulier, lorsque $R\to 0$, le spectre de la corde présente un continuum et s'approche de celui dune dimension non compacte, ce qui est totalement différent de la situation en théorie du point.
Cette équivalence est aussi valable pour les interactions\footnote{Cependant, nous ne détaillerons pas cet aspect dans les sections suivantes.}. Sous cette transformation, le mode $p_R$ ne change pas, tandis que $p_L$ est changé en $-p_L$. Si l'on note $X'$ le champ obtenu par T-dualité à partir de $X$, on peut écrire
\be
X' = X_R - X_L
\ee
La relation entre $X'$ et $X$ est donc non-locale sur la feuille d'univers.

Finalement la théorie de la corde compactifiée sur un cercle de rayon $R$ est équivalente à la théorie compactifiée sur un cercle de rayon $\apr/R$. Les rayons définissant des théories différentes appartiennent à l'intervalle $[\sqrt{\apr}, \infty[$. $\sqrt{\apr}$ est le rayon auto-dual et consistue en quelque sorte une distance minimale pour la corde fermée.

De même pour les champs fermioniques, la T-dualité change $\psi_{-,L}$ en $-\psi_{-,L}$ et $\psi_{+,L}$ en $-\psi_{+,L}$. Ceci multiplie par $-1$ la valeur propre pour $e^{i \pi \bar F}$ des états du secteur R antiholomorphe. La théorie IIA compactifiée sur un cercle de rayon $R$ est donc équivalente par T-dualité à la théorie IIB compactifiée sur un cercle de rayon $\apr/R$ et inversement, la théorie IIB est T-duale à la théorie IIA.

\subsection{T-dualité pour les cordes ouvertes et D-branes}
\l{tdco}

Pour les cordes ouvertes, par contre, dont le spectre est identique au spectre \eqref{mass} d'espace plat non compactifié, la limite $R \to 0$ est identique à celle en théorie du point. La corde ouverte, lorsque $R = 0$, est contrainte à se propager dans un espace comportant une direction spatiale de moins. Comme nous l'avons vu pour la corde fermée, la théorie T-duale est décrite par le champ $X' = X_L - X_R$. La condition de Neumann pour $X$ devient une condition de Dirichlet pour $X'$. Les coordonnées $\left. X(\s,\t)\right\rvert_{\s=0, \pi}$ des extrémités de la corde ouverte sont ainsi fixées. Par contre, pour $\psi_\pm$, les conditions aux bords ne sont pas modifiées.

Dans le cas général, où l'on a fait une T-dualité dans une ou plusieurs directions, le lieu où les extrémités de la corde sont contraintes d'évoluer est un hyperplan de l'espace-temps, appelé \emph{D-brane}.

Nous pouvons faire plusieurs remarques
\bi
\i On peut tout à fait définir la transformation de T-dualité par l'opération $X_L \to - X_L$ (et $\psi_{\pm, L}\to -\psi_{\pm, L}$) indépendamment de la compactification, par exemple dans le cas décompactifié où $R=\infty$.
\i Dans la théorie supersymétrique, la T-dualité à partir de la théorie de type I compactifiée fait apparaître en plus des D-branes, des objets qui correspondent au fait que la corde ouverte est non orientée : des orientifolds. Il s'agit de plans qui sont en quelque sorte des miroirs dans lesquels les deux versions orientées de la corde se reflètent. Loin des D-branes, nous avons un théorie de type II, car la projection d'orientation $\Omega$ relie un état de corde avec son image derrière l'orientifold, mais n'impose aucune contrainte locale sur les états. À partir de la théorie de type I (dont les secteurs ont même chiralité), nous obtenons par T-dualité une théorie de type IIA hors des D-branes (et type I sur les D-branes) puis par une T-dualité supplémentaire une théorie de type IIB, etc.
\i La position d'une D-brane est T-duale à la valeur d'un potentiel de jauge. À un système de $n$ D-branes correspond un champ de jauge $U(n)$ qui est couplé à la corde ouverte. Dans ce cas, les positions des D-branes sont encodées comme les valeurs propres d'une matrice  $U(n)$. À une seule D-brane correspond un champ de jauge électromagnétique.
\i Les D-branes sont en réalité des objets ayant une dynamique propre. En effet, les fluctuations des D-branes sont simplement T-duales des fluctuations du champ de jauge (par exemple, ce champ peut dépendre du temps, alors la position de la D-brane aussi. Ce genre de configuration est étudiée par exemple dans \cite{Bachas:2002jg}). Par la suite, nous ne considérerons que des D-branes \gl rigides\gr.
\i Les D-branes sont des objets dont la masse est proportionnelle à l'inverse de la constante de couplage de la théorie et sont à ce titre considérées comme des états non perturbatifs de la théorie.
\i Les D-branes peuvent être définies dans d'autres géométries. Un exemple proche des travaux que nous avons effectués dans cette thèse peut être trouvé dans \cite{Hikida:2005ec}
\i Nous noterons D\pbr\ une D-brane comportant $p$ directions spatiales. Le volume d'univers qu'elle décrit est donc un espace à $p+1$ dimensions. Dans l'espace-plat, il s'agit de $\R^{1,p}$.
\ei

La D\pbr, dans les théories à basse énergie, est une solution de type soliton. Elle est décrite par une action dite de \emph{Born-Infeld} pour la partie bosonique
\be \l{DBIact}
\begin{split}
S_p = - T_p \int\! d^{p+1} \xi\, \Bigl[&\tr e^{-\phi}\, \sqrt{-\det \left(g + b + 2\pi\apr F\right)}\Bigr.\\
& \Bigl. + i \tr \exp (2\pi\apr F + b) \wedge \sum_q C_q \Bigr] 
\end{split}
\ee
où $T_p$ est une constante qui correspond à la tension de la D-brane. $\phi$ est le dilaton $g$ et $b$ sont la métrique et le tenseur anti-symmétrique (états de masse nulle du secteur NSNS de la corde fermée) induit sur la D\pbr
\be
g_{\a\beta} = \frac{\p X^\mu}{\p \xi^\a} \frac{\p X^\nu}{\p \xi^\beta} \, g_{\mu\nu} (X(\xi))
\text{ , } b_{\a\beta} = \frac{\p X^\mu}{\p \xi^\a} \frac{\p X^\nu}{\p \xi^\beta}\, b_{\mu\nu} (X(\xi))
\ee
$F$ est le champ de jauge se propageant dans la D-brane (état de masse nulle du secteur NS de la corde ouverte). Enfin, les champs $C_q$ sont les différents états RR admis dans la théorie de type II, $q$ indiquant le degré de la forme différentielle associée.
Pour la partie fermionique de l'action de la D-brane, nous invitons le lecteur à se reporter à \cite{Polchinski:1998rr} et aux références qui y sont données.

Si l'on considère les D-branes comme des états de la théorie, leur construction par T-dualité montrent que la théorie de type IIA contient les D8-, D6-, D4-, D2-, D0-branes et la théorie de type IIB, les D7-, D5-, D3-, D1-branes, auxquelles il faut ajouter la D9-brane, identifiée dans la théorie à basse énergie (supergravité de type IIB). Ces états sont stables\footnote{Ce sont des états BPS.}.
À ces D-branes stables, on peut adjoindre des D-branes instables\footnote{ou non-BPS}, qui ne font pas à strictement parler partie de la théorie puisqu'elles se désintègre au bout d'un certain temps, mais qui jouent un rôle important dans les configurations dépendantes du temps (nous en redirons quelques mots à la sous-section \ref{Sbranes}). Ainsi en théorie de type IIA, les D9-, D7-, D5-, D3-, D1-branes sont instables, et en théorie de type IIB, ce sont les D8-, D6-, D4-, D2-, D0-branes.

\subsection{S-branes de Dirichlet en théorie de type II${}^\star$}
\l{sdbranes}

Si nous faisons une T-dualité sur la direction $X^0$, nous obtenons une D-brane très particulière : elle n'existe qu'un instant ! Pour insister sur cette particularité, elle est appelée S-brane (de Dirichlet, pour des raisons qui apparaîtront à la sous-section \ref{Sbranes}). Cette construction est décrite en détail dans \cite{Hull:1998vg}. En théorie des cordes bosoniques, elle ne pose aucun problème particulier. Cependant en théorie des supercordes, lorsqu'on se place dans la limite des basses énergies, les S-branes sont des solutions de théories appelées II${}^\star$ et dont la définition est problématique. Le statut de ces S-branes est donc assez imprécis. Nous reparlerons de ce problème à la section \ref{brane}, consacrée aux \emph{états de bord}.

\section{Opérateur de vertex et ordre normal}
\l{opvertex}

Dans cette section, nous parlerons uniquement des opérateurs de vertex pour la théorie bosonique car nous n'avons calculé en théorie supersymétrique que des amplitudes ne faisant intervenir aucun opérateur de vertex (amplitudes du vide, sans états entrant ni sortant).

\subsection{Construction des opérateurs de vertex pour les différents états du spectre}

Les amplitudes autres que les amplitudes du vide correspondent à des fonctions de corrélation à un ou plusieurs opérateurs, dont nous avons expliqué qu'elles pouvaient être calculées dans une représentation de champ libre à l'aide des opérateurs de vertex. Nous allons donner quelques exemples qui nous seront utiles, puis nous étudierons en détail le problème d'ordre normal que pose les opérateurs de vertex dans la configuration d'orbifold de boost.

Considérons pour l'instant le cas de la corde libre dans un espace plat. Nous avons évoqué dans la section \ref{thconf} le fait que les fonctions de corrélation pour le champ libre bosonique s'écrivaient en terme d'opérateur de vertex de forme exponentielle $e^{ikX}$. Cet opérateur de vertex représente l'insertion d'un état tachyonique $\lvert 0;k^\mu \rangle$ dans l'amplitude de diffusion. Pour construire les opérateurs de vertex correspondant aux autres états du spectre, il faut trouver l'équivalent des opérateurs de création $\a_{-n}^\mu$ dans la représentation de champ libre conforme. Rappelons que le champ $\p X$ peut s'écrire comme une série de Laurent
\be
\p X^\mu (z) = - i \frac{\apr}2 p^\mu \frac1z - i \sqrt{\frac{\apr}2} \sum_{n \in \Z\backslash\{0\}} \a_n^\mu\, z^{-n-1}
\ee
pour $z$ appartenant à un anneau $\A$ centré sur 0 (on suppose qu'on veuille insérer un état en 0). Par application du théorème des résidus, on exprime alors $\a_n^\mu$ comme
\be \a_n^\mu = \sqrt{\frac{2}{\apr}} \oint_C \frac{dz}{2\pi} z^{-n} \p X^\mu(z) \ee
où $C$ est un contour contenu dans l'anneau $\A$, ce qui revient à
\be \a_n^\mu \sim \p^n X^\mu (0) \ee
Ainsi à l'action de $\a_n^\mu$ sur un état correspond la multiplication de l'opérateur de vertex par $\p^n X^\mu (0)$. Nous nous permettons d'écrire $\p^n X^\mu (0)$ puisque les éventuelles divergences sont retirées à l'aide de l'ordre normal. Toutefois, nous avons privilégié la clarté à la rigueur ; il est possible de démontrer cette correspondance en utilisant l'équivalence entre le formalisme d'intégrale des chemins et celui de quantification canonique.

En résumé, pour l'état de niveau $N=0$ (état tachyonique pour la corde bosonique libre dans un espace plat), l'opérateur de vertex s'écrit
\be
\V_0 = \lno e^{ik \cdot X (z,\bar z)} \rno
\ee
Pour un état de niveau $N=1$ (état de masse nulle pour la corde bosonique libre dans un espace plat), nous avons
\be
\V_1^\mu = \lno \p X^\mu(z)  e^{ik \cdot X (z,\bar z)} \rno
\ee
dans la cas de la corde ouverte et
\be
\V_1^{\mu\nu} = \lno \p X^\mu(z) \bp  X^\nu(\bar z) e^{ik \cdot X (z,\bar z)} \rno
\ee
dans le cas de la corde fermée.

\subsection{Ordre normal dans une amplitude comportant des états twistés}
\l{ontwist}

Ces expressions sont valables pour les cordes fermées non twistées dans l'orbifold de boost. Par contre, pour les états de cordes twistés, la construction des opérateurs de vertex correspondants n'est pas connue. Nous avons alors choisi d'écrire les amplitudes dans le formalisme des opérateurs, où jusqu'à deux cordes twistées peuvent être insérées sous la forme des états initiaux et finaux, qu'il est possible de déterminer dans ce cas. Nous allons donc revenir sur le problème de l'ordre normal pour les opérateurs de vertex des cordes non twistées dans l'orbifold de boost.

Considérons d'abord l'opérateur de vertex $\V_0$ pour une corde libre dans un espace plat.
Nous écrivons alors $X$ comme la somme d'une partie de mode 0 $X_0$, d'une partie de fréquences positives $\xpf$, qui ne contient que des opérateurs d'annihilation et d'une partie de fréquences négatives $\xnf$, qui ne contient que des opérateurs de création. L'opérateur de vertex
\be
\V_0 = \lno e^{-ik^+ \xm - i k^- \xp}e^{ik^\perp \cdot X^\perp} \rno
\ee
s'écrit, en se restreignant aux coordonnées du cône de lumière,
\be
\V_0 = e^{-ik^+ \xnfm - i k^- \xnfp} e^{-ik^+ \xpfm - i k^- \xpfp} e^{-ik^+ \xzm - i k^- \xzp} \l{noV0}
\ee
où on a placé les opérateurs de création à gauche, les opérateurs d'annihilation à droite, sans se préoccuper des éventuels commutateurs des opérateurs entre eux. En effet, les termes qui proviennent des commutateurs sont singuliers et on peut considérer qu'ils sont retirés par l'ordre normal, ce qui régularise l'expression des opérateurs de vertex. Cependant, nous allons voir que la prescription d'ordre normal opératoriel est naïve dans le cas de l'orbifold de boost, car les termes issus des commutateurs ne sont pas tous singuliers. Il faut revenir à la simple régularisation introduite par l'ordre normal conforme et traduire cette régularisation en terme de prescription sur les opérateurs de création et d'annihilation.
Examinons en détail le cas de la corde libre dans un espace plat. Nous notons $\V^*_0$ l'opérateur de vertex non régularisé et écrivons
\be
\V^*_0 = e^{-ik^+ \xm - i k^- \xp} = e^{(-ik^+ \xnfm - ik^- \xnfp) + (-ik^+ \xpfm - i k^- \xpfp) + (-ik^+ \xzm - i k^- \xzp)}
\ee
Pour simplifier les expressions, nous absorbons le facteur $-ik^\pm$ dans $X^\mp$ ; nous supposerons par ailleurs que $\apr = 2$,
\be
\V^*_0 = e^{\xnf(z,\bz) + \xpf(z,\bz) + \xz(z,\bz)}
\ee
Nous souhaitons utiliser la formule de Baker-Campbell-Hausdorff $e^{A+B} = e^A e^B e^{-\frac12 [A\,,B]}$ pour des opérateurs $A$ et $B$ tels que $[A,B]$ soit un multiple de l'identité\footnote{C'est ainsi que nous désignerons les \emph{c-number}.}, mais le commutateur $[\xnf\,, \xpf]$, qui s'écrit
\be
[\xnf(z,\bz)\,, \xpf(z,\bz)] = -4k^+k^- \sum_{n=1}^\infty \frac1n \l{csd}
\ee
est infini. Pour étudier la singularité, séparons les positions des opérateurs $\xnf$ et $\xpf$ et écrivons
\be
\V^*_0 = \lim_{z' \to z} e^{\xnf(z',\bz') + \xpf(z,\bz) + \xz(z,\bz)}
\ee
La formule de Baker-Campbell-Hausdorff donne alors
\be
\V^*_0 = \lim_{z' \to z} e^{\xz(z,\bz)} e^{\xnf(z',\bz')} e^{\xpf(z,\bz)} e^{-k^+k^-\ln \left\lvert 1-\frac{z'}z \right\rvert^2}
\ee
Nous obtenons alors \eqref{noV0} en supprimant le dernier facteur. Cette régularisation peut être faite sans introduire la limite $z' \to z$, en adoptant la prescription suivante : écrire le commutateur \eqref{csd} sous la forme d'une série divergente et soustraire les termes dominants divergents.

Cette analyse a pu paraître triviale dans le cas de la corde libre, mais examinons maintenant l'opérateur de vertex correspondant à l'insertion d'un état de corde non twistée dans une amplitude de diffusion. Comme nous l'avons expliqué dans le cas de la corde compactifiée, il faut superposer les images de la corde sous l'opération de boost. Nous obtenons un résultat dépendant de la valeur propre du moment cinétique de boost $j$
\be
\V_0 = \lno \int_{-\infty}^\infty dv e^{-ik^+ X^- e^{-v} -ik^- X^+ e^{v} +i k^\perp \cdot X^\perp -ijv} \rno \l{vont}
\ee
Pour ne pas alourdir inutilement les expressions, nous enlevons tout ce qui dépend de $v$, sachant qu'à la fin nous pouvons remettre l'intégrale sur $v$, le facteur $e^{-ijv}$ et faire la substitution $k^\pm \to k^\pm e^{\mp v}$ pour obtenir le résultat correct.

Nous écrivons ainsi l'opérateur de vertex non régularisé de la façon suivante
\be
\V^*_0 = e^{\xnf(z,\bz) + \xpf(z,\bz) + \xz(z,\bz)}
\ee
Pour une amplitude dont les états initiaux et finaux appartiennent au secteur non twisté, $X^\pm$ aura l'expression correspondante au secteur non twisté où $w=0$, ce qui revient au cas étudié ci-dessus. Par contre dès lors que les états initiaux et finaux appartiennent à un secteur twisté $w$, $X^\pm$ a l'expression correspondant à ce secteur $w$. Physiquement, la corde non twistée est émise dans le vide des cordes twistées. C'est dans cette situation que nous nous plaçons maintenant.

En suivant la même démarche que ci-dessus, nous séparons les fréquences positives des fréquences négatives
\be
\V^*_0 = \lim_{z' \to z} e^{\xnf(z',\bz') + \xpf(z,\bz) + \xz(z,\bz)}
\ee
et nous obtenons
\be
\V^*_0 = \lim_{z' \to z} e^{\xz(z,\bz)} e^{\xnf(z',\bz')} e^{\xpf(z,\bz)} e^{\frac12 k^+ k^- C^*(z,z')} \l{V02}
\ee
avec
\be
\begin{split}
C^*(z,z') =& \frac1{1+ i\nu}\left(\frac{z'}{z}\right)^{1+i\nu}
\hypF \left(1+i\nu, 1; 2+i\nu; \frac{z'}z\right) \\
&+\frac1{1-i\nu}\left(\frac{z'}{z}\right)^{1-i\nu}
\hypF \left(1-i\nu, 1; 2-i\nu; \frac{z'}z\right) \\
&+\frac1{1+i\nu}\left(\frac{\bz'}{\bz}\right)^{1+i\nu}
\hypF \left(1+i\nu, 1; 2+i\nu; \frac{\bz'}\bz\right) \\
&+\frac1{1-i\nu}\left(\frac{\bz'}{\bz}\right)^{1-i\nu}
\hypF \left(1-i\nu, 1; 2-i\nu; \frac{\bz'}\bz\right)
\end{split}
\ee
où $\hypF(a,b;c;x)$ est une fonction hypergéométrique définie dans l'annexe~\ref{df}.
Du comportement de cette fonction lorsque $x \to 1$ (voir l'annexe~\ref{df}), on déduit celui de $C^*$
\be
C^*(z,z') = - 2 \ln \left\lvert 1- \frac{z'}z \right\rvert^2 + 2\left[2\psi(1) - \psi(1+i\nu) - \psi(1-i\nu)\right] + o(1) \qquad z' \to z
\ee
où $\psi$ est la fonction digamma. On régularise en éliminant le ln dans $C$. Les termes non-divergents doivent être gardés par contre et on obtient l'opérateur de vertex régularisé
\be 
\V_0 = e^{\xz(z,\bz)} e^{\xnf(z,\bz)} e^{\xpf(z,\bz)} e^{k^+k^- \left[2\psi(1) - \psi(1+i\nu) - \psi(1-i\nu)\right]} 
\l{V0reg}
\ee

En utilisant la prescription que nous avons donnée, le calcul est plus simple
\be
\V^*_0 = e^{\xz(z,\bz)} e^{\xnf(z,\bz)} e^{\xpf(z,\bz)} e^{k^+k^- S^*}
\ee
avec
\be
S^*= \sum_{n=1}^\infty \frac1{n+i\nu} + \frac1{n-i\nu}
\ee
Le terme dominant est $\frac2n$. Pour régulariser, il suffit de remplacer $S^*$ par\footnote{Dans la section \ref{article2} et l'article \cite{Berkooz:2004yy}, annexe \ref{a2}, nous avons utilisé $\Delta$, qui est l'opposé de $S$ : $S(\nu)=-\Delta(\nu)$.}
\be
S= \sum_{n=1}^\infty \frac1{n+i\nu} + \frac1{n-i\nu} - \frac2n
\ee
$S$ est finie et peut être évaluée, ce qui permet de confirmer le résultat obtenu \eqref{V0reg}

Remarquons que $S$ peut être écrite comme la différence de commutateurs
\be
S = [\xnfm\,, \xpfp]_0 + [\xnfp\,, \xpfm]_0 - [\xnfm\,, \xpfp]_\nu - [\xnfp\,, \xpfm]_\nu
\ee
où l'indice $\nu$ indique que le commutateur doit être calculé dans le secteur twisté de paramètre $\nu$ et l'indice $0$ indique que le commutateur doit être calculé dans le secteur non twisté.

Enfin, nous reviendrons dans la sous-section \ref{am3c} sur la signification physique du facteur $e^{k^+k^- S}$.

\subsection{Ordonnement des modes zéro pour les opérateurs de vertex de l'état $N=1$ dans une amplitude comportant des états twistés}

Il existe une seconde subtilité avec les opérateurs de vertex $\V_1$ pour des amplitudes où les états initiaux et finaux appartiennent au secteur twisté. Considérons dans un premier temps un opérateur $\V_1$ de corde ouverte. La définition
\be
\V_1^\mu = \lno \p X^\mu (z)\, e^{ik \cdot X(z,\bz)} \rno
\ee
impose un ordre arbitraire pour les modes zéro. Dans le cas d'un opérateur de vertex dans une amplitude sans états twistés, mettre $\p X^\mu$ à droite ou à gauche de l'exponentielle revient au même une fois qu'on contracte l'opérateur de vertex avec la polarisation de l'état (à cause des contraintes sur cette polarisation). Par contre dans le cas d'une amplitude avec états twistés, il faut symétriser l'opérateur de vertex. Notons que dans l'expression de $\V_0$, les modes zéro ne sont pas ordonnés et que pour $\V_1^\perp$ il n'y a pas de problème de modes zéro. Pour $\V_1^\pm$, il faut donc adopter la définition suivante
\be
\V_1^\pm = \frac12 \left( \lno \p X^\pm (z)\, e^{ik\cdot X(z,\bz)} \rno + \lno  e^{ik \cdot X(z,\bz)}\, \p X^\pm (z) \rno \right)
\ee
où $k\cdot X = - k^+ \xm - k^- \xp + k^\perp \cdot X^\perp$.
Comme pour $\V_0$, la dépendance sur $v$ est sous-entendue, mais à cause du facteur $\p X^\pm$ supplémentaire, il faut ici rajouter un facteur $e^{\pm v}$ pour obtenir un opérateur de vertex invariant sous la transformation de boost.
Le calcul des opérateurs régularisés se fait comme précédemment.
Le premier terme s'écrit comme la limite régularisée quand $z' \to z$ de
\be
\begin{split}
\p\xpm (z')\! \lno e^{ik\cdot X} \rno\!\! (z) &= \p\xnfpm (z')\! \lno e^{ik\cdot X}\rno\!\! (z) + \\
               & +  \lno e^{ik\cdot X}\rno\!\! (z)\, \p\xpfpm (z') + [\p\xpfpm (z')\,,  \lno e^{ik\cdot X}\rno\!\! (z)] +\\
               & + \p\xzpm (z')\!  \lno e^{ik\cdot X}\rno\!\! (z)
\end{split}
\ee
où
\be [\p\xpfpm (z')\,,  \lno e^{ik\cdot X}\rno\!\! (z)] = [\p\xpfpm (w),-i\kpm \xnfmp (z)] \lno e^{ik\cdot X}\rno\!\! (z) \l{com1}\ee
et
\be [\p\xpfpm (z')\,,-i\kpm \xnfmp (z)] = -i \kpm \left(\frac{z}{z'}\right)^{\pm i\nu} \frac1{z'-z} \ee
Le second terme quant-à lui, s'écrit dans cette même limite
\be
\begin{split}
 \lno e^{ik\cdot X}\rno\!\! (z')\, \p\xpm (z) &= \p\xnfpm (z)\!  \lno e^{ik\cdot X}\rno\!\! (z') + [ \lno e^{ik\cdot X}\rno\!\! (z')\,, \p\xnfpm (z)]\\
               & +  \lno e^{ik\cdot X}\rno\!\! (z')\, \p\xpfpm (z) \\
               & +  \lno e^{ik\cdot X}\rno\!\! (z')\, \p\xzpm (z)
\end{split}
\ee
où
\be [ \lno e^{ik\cdot X}\rno\!\! (z')\,, \p\xnfpm (z)] = [-i\kpm \xpfmp (z'), \p\xnfpm (z)]  \lno e^{ik\cdot X}\rno\!\! (z') \l{com2}\ee
et
\be [-i\kpm \xpfmp (z')\,, \p\xnfpm (z)] = i \kpm \left(\frac{z}{z'}\right)^{\mp i\nu} \frac1{z'-z} \ee
On obtient finalement
\be
\begin{split}
{\cal V}^\pm_1 (z) &= \p\xnfpm  \lno e^{ik\cdot X}\rno\!\! (z) +  \lno e^{ik\cdot X}\rno\! \p\xpfpm (z) \\
          & \frac12 \left(\p\xzpm  \lno e^{ik\cdot X}\rno\!\!(z) +  \lno e^{ik\cdot X}\rno\! \p\xzpm (z)\right) \\
          & \mp \frac{\nu \kpm}z  \lno e^{ik\cdot X}\rno\!\! (z) + \frac{i}2 \kpm \p\!\!  \lno e^{ik\cdot X}\rno\!\! (z) \l{ov1ord}
\end{split}
\ee
Le dernier terme peut être éliminé puisqu'il correspond à un courant conservé.

Nous constatons que, d'une part, les modes zéro sont bien symétrisés et que, d'autre part, le terme issu des commutateurs \eqref{com1} et \eqref{com2} peut être incorporé dans la partie de mode 0, à condition de redéfinir $\a_0^\pm \to \a_0^\pm \mp i\nu k^\pm z^{\pm i\nu}$ dans $\p\xzpm$.

Pour l'opérateur de vertex $\V_1^{\mu\nu}$ pour l'insertion d'une corde fermée non twistée dans une amplitude à états twistés, la symmétrisation s'écrit
\be
\begin{split}
\V_1^{\mu\nu} (z,\bz) =& \frac14 \left[ \lno \p X^\mu (z) \bp X^\nu (\bz) e^{ik\cdot X(z,\bz)} \rno + \lno \p X^\mu (z) e^{ik\cdot X(z,\bz)} \bp X^\nu (\bz) \rno \right. \\
 & \left. + \lno \bp X^\nu (\bz) e^{ik\cdot X(z,\bz)} \p X^\mu (z) \rno +  \lno e^{ik\cdot X(z,\bz)} \p X^\nu (z) \bp X^\mu (\bz) \rno \right] \l{ovsym}
\end{split}
\ee
L'effet de cette symmétrisation n'apparaît que lorsque $\mu$ ou $\nu$ est $\pm$. On obtient un résultat similaire à \eqref{ov1ord}, avec les opérateurs de création à gauche, ceux d'annihilation à droite, les modes zéros non ordonnés et la règle suivante\footnote{Il y a un changement de signe par rapport à l'article \cite{Berkooz:2004yy} car la convention pour l'onde plane qui y a été choisie est $e^{-ik\cdot X}$ et non $e^{ik\cdot X}$. On peut simplement passer d'une convention à l'autre en changeant $k^\pm$ en $-k^\pm$ (et $k^\perp$ et $-k^\perp$).}
\be
\a_0^\pm \to \a_0^\pm \mp i\nu\, k^\pm\, z^{\pm i\nu} \text{\ , \ \ } \ta_0^\pm \to \ta_0^\pm \pm i\nu\, k^\pm\, z^{\mp i\nu}
\ee
ces remplacements devant être fait seulement dans $\p\xzpm$ et $\bp\xzpm$.
L'opérateur de vertex pour le graviton, qui nous intéresse plus particulièrement pour les amplitudes dont nous parlerons au chapitre \ref{article2}, est la combinaison symétrique de trace nulle de $\V_1^{\mu\nu}$.

\section{Amplitudes, amplitudes à l'ordre des arbres et fonction de partition}
\l{ampl}

Nous disposons de tous les éléments pour construire les amplitudes de diffusion. Contrairement aux précédentes sections, nous ne détaillerons pas les différents cas correspondants aux diverses configurations que nous avons étudiées car les spécificités de chacune nous ont mené vers des amplitudes différentes. Nous exposerons donc les résultats au chapitre \ref{contrereac} et nous nous bornerons dans cette section à rappeler la définition des amplitudes et certaines de leurs propriétés.

\subsection{Série perturbative en théorie des cordes}

De même qu'en théorie du point, les diagrammes de Feynmann encodent la série perturbative d'une amplitude. Ces diagrammes en théorie des cordes, après continuation analytique vers la feuille d'univers euclidienne, sont des surfaces de Riemann et sont classifiées selon leur genre, qui est aussi la puissance à laquelle est élevée la constante de couplage dans la série perturbative. Ainsi, les deux ordres les plus bas sont les surfaces de genre zéro comme la sphère ou le disque qui correspondent à l'ordre des arbres et les surface de genre un comme le tore qui correspondent à l'ordre d'une boucle. Nous nous limiterons à l'ordre des arbres pour les amplitudes comportant des états externes et à l'ordre d'une boucle pour les amplitudes du vide. Notons que le formalisme de la théorie des cordes ne permet en l'état actuel de ne traiter que des amplitudes sur la couche de masse, ce que nous supposerons tout au long de notre étude.

\subsection{Aspects globaux de la feuille d'univers et symétries}
\l{agfu}

Considérons une amplitude à quatre cordes fermées. Le diagramme à l'ordre des arbres est représenté sur la figure \ref{d4cf}. L'invariance conforme de l'action permet de simplifier ce diagramme, en ramenant la feuille d'univers (euclidienne) à une sphère et chaque patte à un point, qui est la position de l'opérateur de vertex correspondant à l'état entrant correspondant (voir figure \ref{d4cf}).

\begin{figure}[h]
\begin{center}
\epsfig{file=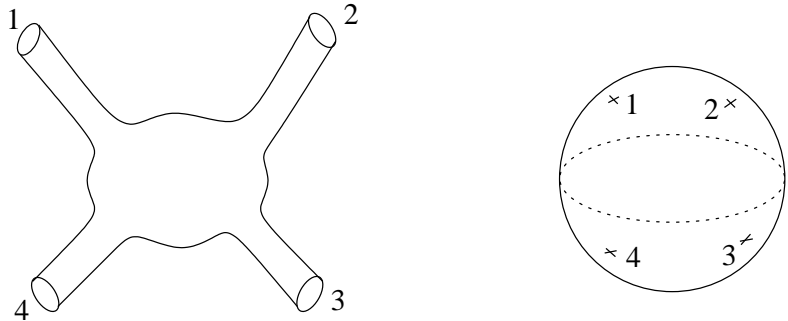, width=10.4cm}
\end{center}
\caption{\l{d4cf}\small Amplitude de diffusion de quatre cordes ferm\'ees. \`A gauche : les cordes ferm\'ees entrante sont bien visibles sur la feuille d'univers (les quatre tubes). \`A droite : apr\`es transformation conforme, on peut se ramener \`a une sph\`ere perc\'ee de quatre trous correspondant aux cordes ferm\'ees.}
\end{figure}

Lorsque nous avons exposé l'action de Polyakov \eqref{actP}, nous avons vu que nous pouvions éliminer par un choix de jauge sur la métrique les degrés de liberté superflus introduits par les symétries de Weyl et de difféomorphisme, sauf ceux liés à l'invariance conforme. Cependant nous avons travaillé au niveau local, sans nous préoccuper des contraintes globales, imposées notamment par la structure différentielle et la topologie de la feuille d'univers. Comme pour la particule ponctuelle, nous aurons un groupe de symétries résiduelles et des paramètres de Teichmüller associés à la feuille d'univers. 

Tout d'abord, la symétrie résiduelle qu'est l'invariance conforme est restreinte : la sphère par exemple n'est pas invariante sous une transformation conforme quelconque $z \to f(z)$, mais seulement sous les transformations conformes dont la forme infinitésimale\footnote{$a_0$, $a_1$, $a_2$, $\alpha$, $\beta$, $\gamma$ et $\eta$ sont des nombres complexes.} est $\delta z = a_0 + a_1 z + a_2 z^2$, c'est-à-dire celles de la forme $z\to \frac{\alpha z+ \beta}{\gamma z + \eta}$ avec $\alpha \eta - \gamma \beta = 1$. En effet on peut montrer que ce sont les seules transformations conformes qui sont définies sur la sphère toute entière. Dans le cas général, on a un nombre fini\footnote{d'après le théorème de Riemann-Roch} de degrés de liberté de symétrie non fixés par le choix de jauge. Le groupe de symétries résiduelles s'appelle \emph{groupe de Killing conforme}. Notons que localement, l'invariance conforme est toujours présente ; ainsi le formalisme de théorie conforme n'est pas disqualifié pour le calcul des amplitudes.

D'autre part, les paramètres de Teichmüller sont dans le cas de la corde les paramètres\footnote{en nombre fini d'après le même théorème} de la métrique qui ne sont pas fixés par les symétries. Par exemple, les métriques ne sont pas toutes dans la même orbite du groupe de jauge et la famille des métriques non images l'une de l'autre par une transformation de jauge est paramétrée par un module. Par exemple, le tore a deux modules\footnote{On compte les paramètres réels.} écrit sous la forme d'un nombre complexe $\t$, notation traditionnelle qui peut prêter à confusion -- nous choisirons $\rho$. L'espace des modules est noté $F_0$ et il correspond à une classe d'équivalence du demi-plan supérieur sous l'action du groupe PSL(2,$\Z$) qui est le groupe des transformations de coordonnées globales du tore.

Enfin, il reste une dernière catégorie de paramètres : les positions des opérateurs de vertex, qu'on appelle aussi par exenstion modules. On peut jauger les symétries du groupe de Killing conforme en fixant autant de positions d'opérateur de vertex.

Considérons une amplitude générale dans le formalisme d'intégrale des chemins. L'intégration sur les métriques et les positions des opérateurs de vertex est mise sous la forme d'une intégration sur le groupe de jauge, les modules et les positions libres des opérateurs de vertex. La mesure d'intégration est donnée par le jacobien de cette transformation, jacobien qui se met sous la forme d'une contribution de fantômes à l'action de Polyakov. Comme nous l'avons déjà évoqué, l'amplitude peut se calculer en terme de fonctions de corrélations d'une théorie conforme de boson, de fermions et de fantômes. Lorsque nous calculerons une amplitude dans les différentes configurations que nous avons étudiés, nous utiliserons implicitement les résultats de la procédure d'intégrale des chemins pour écrire l'amplitude sous la forme d'une fonction de corrélation de théorie conforme dépendante des différents paramètres sur lesquelles on intègre. Nous n'écrirons pas les contributions des fantômes, nous en tiendrons compte directement dans le résultat final. Enfin, nous calculerons les fonctions de corrélation dans le formalisme des opérateurs.

\subsection{Amplitudes à l'ordre des arbres}
\l{aoa}

Appliquons ce qui a été dit plus haut au cas de l'amplitude de diffusion à l'ordre des arbres de trois états tachyoniques de cordes fermées libres dans l'espace plat. La feuille d'univers à considérer est une sphère $S^2$. Elle n'a aucun module, par contre elle a six paramètres dans le groupe de Killing conforme et ces six paramètres réels peuvent être éliminés si on fixe les positions des trois opérateurs de vertex. Le choix standard est $z_1=0$, $z_2=1$ et $z_3=\infty$. L'amplitude s'écrit
\be
\A_3 = \lim_{z_3 \to \infty} \langle \lno e^{ik_1 \cdot X(0)} \rno  \lno e^{ik_2 \cdot X(1)} \rno \lno e^{ik_3 \cdot X(z_3, \bz_3)} \rno \rangle
\ee
où, dans l'équation précédente et toutes les équations concernant les amplitudes, le symbole $=$ signifie égal à une constante de proportionnalité près (constante de couplage incluse).
Dans le formalisme des opérateurs, cette amplitude s'écrit
\be
\A_3 = \langle 0;k_1 \rvert \!\! \lno e^{ik_2 \cdot X(1)} \rno\!\!\lvert 0;k_3 \rangle
\ee
On obtient le résultat suivant
\be
\A_3 = \delta (k_1+k_2+k_3)
\ee

L'amplitude suivante
\be
\A^{\mu\nu}_3 = \langle \lno e^{ik_1 \cdot X(0)} \rno  \lno \p X^\mu(1) \bp X^\nu(1) e^{ik_2 \cdot X(1)} \rno \lno e^{ik_3 \cdot X(z_3, \bz_3)} \rno \rangle
\ee
que nous noterons
\be
\A^{\mu\nu}_3 = \langle 0;k_1 \lvert\, T^{\mu\nu}\, \vert 0;k_3 \rangle
\ee
vaut
\be
\A^{\mu\nu}_3 = (k_1 - k_3)^\mu (k_1 - k_3)^\nu \delta (k_1+k_2+k_3) \l{a3g}
\ee
Notons qu'en toute rigueur, l'amplitude complète est $e_{2\mu\nu} \A^{\mu\nu}_3$ où $e_2$ est la polarisation de l'état de masse nulle.

Nous étudierons en détail le cas où les états initiaux et finaux sont des états twistés et nous noterons simplement $\langle 1 \rvert\, T^{\mu\nu}\, \lvert 3 \rangle$.

Citons enfin l'amplitude de diffusion de quatre états tachyoniques
\be
\begin{split}
\A_4 &= \int_\C d^2 z_3 \lim_{z_4 \to \infty} \langle \lno e^{ik_1 \cdot X(0)} \rno  \lno e^{ik_2 \cdot X(1)} \rno \lno e^{ik_3 \cdot X(z_3, \bz_3)} \rno \lno e^{ik_4 \cdot X(z_4, \bz_4)} \rno \rangle \\
     &= \int d^2 z_3 \, \langle 0;k_1 \rvert\, \text{{\bfseries T}}\!\! \lno e^{ik_2 \cdot X(1)} \rno \lno e^{ik_3 \cdot X(z_3, \bz_3)} \rno\!\! \lvert 0;k_4 \rangle
\end{split}
\ee
où {\bf T} désigne le produit ordonné dans le temps. Après calculs, on obtient, dans un premier temps, l'expression suivante
\be
\A_4 = \delta({\textstyle \sum_{i=1}^4} k_i) \int_\C d^2 z_3 \lvert z_3 \rvert^{-\apr u/2-4} \lvert 1-z_3 \rvert^{-\apr t/2-4}
\ee
puis, en intégrant, l'amplitude de Virasoro-Shapiro
\be
\A_4 = \delta({\textstyle\sum_{i=1}^4} k_i) \frac{\Gamma\left(-1-\frac{\apr s}4\right) \Gamma\left(-1-\frac{\apr t}4\right) \Gamma\left(-1-\frac{\apr u}4\right)}{\Gamma\left(2+\frac{\apr s}4\right) \Gamma\left(2+\frac{\apr t}4\right) \Gamma\left(2+\frac{\apr u}4\right)}
\ee
où $s$, $t$ et $u$ sont les variables de Mandelstam telles qu'elles sont définies en théorie des champs ($s=(k_1+k_2)^2$, $t=(k_1+k_3)^2$, $u=(k_1+k_4)^2$).

Nous n'exposons pas les résultats concernant les amplitudes de cordes ouvertes à l'ordre des arbres, car cela ne nous sera pas utile par la suite. D'autre part, nous n'étudierons que la version bosonique de ce type d'amplitude.

\subsection{Fonctions de partition}
\l{oneloop}

Comme en théorie des champs, les \emph{fonctions de partition} sont des amplitudes de corde sans pattes externes. Comme nous l'avons dit plus haut il s'agira toujours d'amplitude à une boucle. Pour une fonction de partition de cordes ouvertes, la feuille d'univers est un anneau, ou, ce qui est équivalent du point de vue topologique, un cylindre (voir figure\ref{oponeloop}). Il y a un module, que nous noterons $t$ ($0 < t < \infty$). $2\pi t$ est la circonférence du cylindre\footnote{Le facteur $2\pi$ est une convention commode qui permet de passer du tore au cylindre en posant $\rho = it$ et faisant le quotient par la relation d'équivalence $w \sim -\bw$.}. La fonction de corrélation correspondant à la fonction de partition est $\langle 1 \rangle$ et s'écrit, dans le formalisme d'opérateur,
\be
\cZ = \int_0^\infty \frac{dt}{2t} \, \tr e^{-2\pi t L_0} \l{fpco}
\ee
où la trace porte sur l'espace des états de la corde ouverte et inclut l'intégration sur les impulsions. Le dénominateur $2t$ élimine la redondance due au module et au choix d'orientation de la boucle. $L_0$ constitue l'hamitonien du système et les contributions des fantômes sont sous-entendues. Pour faire le lien avec l'intégrale des chemins correspondante, il faut considérer la feuille d'univers annulaire, comme une bande de longueur $2\pi t$ le long de laquelle les états de cordes se propagent (la translation selon $\t_{\text{E}}$ est générée par $L_0$) et dont on colle les deux extrémités.

\begin{figure}[h]
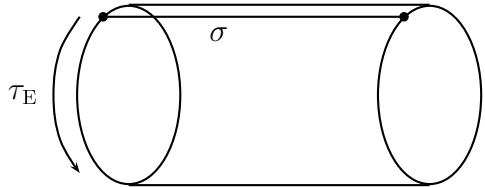

\vspace{1.5cm}
\begin{center}
\rput{0}(-1.8,0){%
  \psset{dotsep=2pt}
  \psellipse[linecolor=black](0,0)(.7,1.2)
  \psellipse[linecolor=black](4,0)(.7,1.2)
  \psline[linecolor=black](0,1.2)(4,1.2)
  \psline[linecolor=black](0,-1.2)(4,-1.2)
  \pscurve[linecolor=black]{->}(-.65,1.04)(-.91,.6)(-1,0)(-.91,-.6)(-.65,-1.04)
  \rput[r]{0}(-1.2,0){\textcolor{black}{$\t_{\text{E}}$}}
  \psline[linecolor=black]{*-*}(-.34,1.04)(3.66,1.04)
  \rput[t]{0}(1.2,.9){\textcolor{black}{$\s$}}
}
\end{center}
\vspace{.8cm}
\caption{\l{oponeloop}\small La corde ouverte d\'ecrit une bande dont les extr\'emit\'es sont coll\'ees pour former un cylindre (ou un anneau). $\t_E \in [0, 2\pi t]$, $\s \in [0,\pi]$.}
\end{figure}

Dans le cas des cordes fermées, la feuille d'univers est un tore ; nous en avons parlé un peu plus haut. Nous écrirons $\rho=\rho_1+i\rho_2$ où $\rho_1$ et $\rho_2$ sont les deux modules du tore. On peut considérer que le tore est formée par l'évolution d'un état de corde pendant un temps euclidien $2\pi \rho_2$ et la translation du même état d'une distance $2 \pi \rho_1$. Dans le formalisme d'opérateur, ceci s'écrit
\be
\cZ = \int_{F_0} \frac{d\rho d\bar\rho}{4 \rho_2}\, \tr e^{2\pi i \rho_1 (L_0 -\bar L_0) - 2\pi \rho_2 (L_0+\bar L_0}
\ee
La trace porte cette fois sur l'espace des états de la corde fermée et inclut de même l'intégration sur les impulsions. $L_0 - \bar L_0$ est le générateur des translations selon $\s$, $L_0+\bar L_0$ est le générateur des translations\footnote{$L_0+\bar L_0$ sans la constante d'ordre normal est l'hamiltonien dans le système de coordonnées $z, \bz$ mais avec nos conventions, elle est sous-entendue dans l'expression donn\'ee et nous avons bien le g\'en\'erateur des translations selon $\t_{\text{E}}$ correspond au fait que le tenseur énergie-impulsion n'est pas un opérateur primaire} selon $\t_{\text{E}}$. Le dénominateur $4\rho_2$ inclut la contribution des modes excités des fantômes pour $d\rho d\bar\rho$, la redondance due au groupe de Killing conforme et un facteur 2 dû au choix d'orientation du tore.

Par commodité, nous appellerons aussi fonction de partition ou amplitude du vide les amplitudes $\cZ(t)$ ou $\cZ(\tau)$, sans intégration sur les modules.

\subsubsection{Cas de la corde supersymétrique}

Dans l'équation \eqref{fpco} la trace comprend aussi une somme sur les différents secteurs. Il est plus pratique pour la suite de définir des fonctions de partition par secteurs, que nous noterons également $\cZ$
\be
\cZ = \int_0^\infty \frac{dt}{2t} \, \tr \left(\cP\, e^{-2\pi t L_0}\right)
\ee
où $\cP$ est un opérateur de projection et où la trace porte sur un secteur particulier. Nous verrons des exemples de cela à la section suivante \eqref{brane}.
Si la partie de mode zéro des fonctions de partition supersymétriques est identique à celle des fonctions de partition bosonique, la contribution des états des spineurs et des superfantômes modifient le facteur correspondant aux états excités. Nous ne donnerons pas plus de détails et nous invitons le lecteur à se reporter à \cite{Polchinski:1998rr}, \cite{Kiritsis:1997hj} ou \cite{DiVecchia:1999rh}. 

\subsection{Dualité de feuille d'univers}
\l{dualco}

Revenons à la fonction de partition de corde ouverte. Sous sa forme d'anneau, la feuille d'univers évoque bien une boucle pour cordes ouvertes, mais si nous la considèrons comme un cylindre, il s'agit aussi de la feuille d'univers correspondant à la propagation d'une corde fermée. Ce changement de point de vue est illustré sur la figure \ref{doc}. Nous considérons une feuille d'univers décrivant la propagation d'une corde ouverte de longueur $\pi$ sur un intervalle de temps $2\pi t$. Si nous  définissons de nouvelles coordonnées de feuille d'univers $\t'_{\text{E}}, \s'$ de la façon suivante
\be
\t'_{\text{E}} = \frac1t \s \text{\ \ , \ \ } \s' = \frac1t \t_{\text{E}}
\ee
alors la feuille d'univers décrit la propagation d'une corde fermée d'une longueur standard $2\pi$ sur un invervalle de temps $\frac{\pi}t$ (voir figure \ref{doc}). Ainsi l'amplitude de propagation d'une corde fermée\footnote{souvent notée $\cZ_{\text{c.f.}}$ bien que ce ne soit pas à proprement parler une fonction de partition} $\A_{\text{c.f.}}$ est égale à l'amplitude du vide des cordes ouvertes à une boucle $\cZ_{\text{op}}$. C'est la \emph{dualité de feuille d'univers}. Comme le changement de coordonnées est une transformation conforme, cette dualité est la conséquence de la symétrie conforme présente dans la théorie.

\begin{figure}[h]
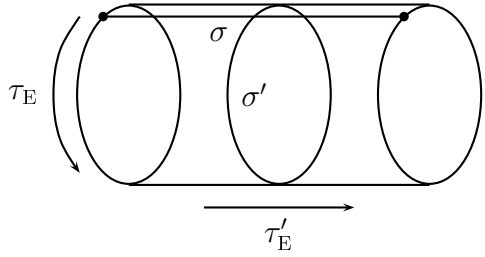

\vspace{1.5cm}
\begin{center}
\rput{0}(-1.8,0){%
  \psset{dotsep=2pt}
  \psellipse[linecolor=black](0,0)(.7,1.2)
  \psellipse[linecolor=black](4,0)(.7,1.2)
  \psline[linecolor=black](0,1.2)(4,1.2)
  \psline[linecolor=black](0,-1.2)(4,-1.2)
  \pscurve[linecolor=black]{->}(-.65,1.04)(-.91,.6)(-1,0)(-.91,-.6)(-.65,-1.04)
  \rput[r]{0}(-1.2,0){\textcolor{black}{$\t_{\text{E}}$}}
  \psline[linecolor=black]{*-*}(-.34,1.04)(3.66,1.04)
  \rput[t]{0}(1.2,.9){\textcolor{black}{$\s$}}
  \psellipse[linecolor=black](2,0)(.7,1.2)
  \rput[l]{0}(1.5,0){\textcolor{black}{$\s'$}}
  \psline[linecolor=black]{->}(1,-1.5)(3,-1.5)
  \rput[t]{0}(2,-1.65){\textcolor{black}{$\t'_{\text{E}}$}}
}
\vspace{1.5cm}
\end{center}
\caption{\l{doc}\small $\t_E \in [0, 2\pi t]$, $\s \in [0,\pi]$ sont les coordonn\'ees pour la corde ouverte. $\t'_E \in [0,\frac{\pi}t] $, $\s' \in [0, 2\pi]$ sont les coordonn\'ees pour la corde ferm\'ee.}
\end{figure}

Nous introduisons à la section suivante \ref{brane} le formalisme d'états de bord qui permet de calculer les amplitudes de propagation de la corde fermée et de vérifier cette dualité. Lorsque nous considérons une amplitude d'anneau comme une amplitude d'anneau de corde ouverte à une boucle, nous dirons plus simplement que nous nous plaçons dans le \emph{canal de corde ouverte} ; de même, si nous considérons la même amplitude comme une amplitude de cylindre de corde fermée à l'ordre des arbres, alors nous parlerons de \emph{canal de corde fermée}.

\subsubsection{Remarque à propos des configurations de D-branes}

De même que la fonction de partition du vide électromagnétique constitué par deux plaques conductrices qui imposent des conditions aux bords spécifiques est égale au produit entre une durée infinie d'interaction et l'énergie d'interaction par effet Casimir entre ces deux plaques, la fonction de partition des cordes ouvertes dans le vide constitué par deux D-branes est égale au produit entre une durée infinie d'interaction et l'énergie d'interaction entre ces deux D-branes. Dans le cas de deux S-branes, on parlera respectivement de longueur d'interaction $L$ et d'impulsion d'interaction. Le calcul de la fonction de partition, déjà motivé par la signification physique que celle-ci revêt, est de plus indispensable à la définition complète des états de bord.

En effet, dans une configuration de D-branes, la dualité de feuille d'univers fait correspondre à notre amplitude à une boucle de corde ouverte (avec conditions aux bords de Dirichlet) une amplitude correspondant à la propagation d'une corde fermée émise par une des deux membranes et absorbée par l'autre.

\section{Membranes de Dirichlet et formalisme d'états de bord}
\l{brane}

\subsection{Formalisme d'états de bord}

Nous utilisons dans cette section le terme \emph{membranes de Dirichlet} pour désigner indifférement les D-branes et les S-branes de Dirichlet, c'est-à-dire les objets correpondants à des conditions de Dirichlet sur la coordonnées $X^0$. Nous avons vu à la sous-section \ref{sdbranes} qu'elles pouvaient être obtenues par T-dualité sur la direction temporelle, mais ici, nous supposerons seulement que des conditions de Dirichlet ont été imposées sur la coordonnée $X^0$ dans une configuration d'espace plat non compactifiée.

Dans le canal de cordes fermées, l'amplitude de propagation peut s'écrire dans le formalisme d'opérateurs, à condition de définir des \emph{états de bords} décrivant l'interaction entre une membrane et une corde fermée.
\be
\A_{\text{c.f.}}(t) = \langle B_1 \rvert P \lvert B_2 \rangle
\ee
où $P$ est le propagateur de la corde fermée, c'est-à-dire
\be
P= \frac{\apr}{4\pi} \int_{\lvert z \rvert \leq 1} \frac{d^2 z}{\lvert z \rvert^2}\, z^{L_0} \bz^{\bar L_0}
\ee
Les notations $L_0$ et $\bar L_0$ incluent la constante d'ordre normal.
Ces états de bords sont déterminés en appliquant la transformation conforme correspondant à la dualité sur les conditions aux bords que vérifient les coordonnées bosoniques $X$ et fermioniques $\psi$. Les conditions au bord que vérifie une corde ouverte attachée à une \pbr\ en $\s=0$ (ou de façon similaire en $\s = \pi$) s'écrivent pour les coordonnées bosoniques
\be
\p_\s X^\mu = 0 \text{\ \ , \ \ } X^\nu = y^\nu \qquad \qquad \s = 0
\ee
où $\mu$ décrit le volume d'univers de la \pbr\ et $\nu$ désigne les directions perpendiculaires à la \pbr. Pour une D\pbr, $\mu = 0, \ldots\,p$ et $\nu=p+1, \ldots, d$. Pour une S\pbr\ de Dirichlet, $\mu = 1, \ldots, p$, $\nu = 0, p+1, \ldots, d$.
Les conditions que doivent vérifier l'état de bord $\lvert B \rangle$ correspondant à cette \pbr\ sont donc
\be
\p_\t X^\mu \lvert B \rangle = 0 \text{\ \ , \ \ } (X^\nu- y^\nu) \lvert B \rangle = 0 \qquad \qquad \t = 0
\ee
Il est utile d'écrire ces conditions en terme de modes
\begin{align}
(\a_n^\mu + \ta_{-n}^\mu) \lvert B \rangle &= 0 & p^\mu \lvert B \rangle &=0 \\
(\a_n^\nu - \ta_{-n}^\nu) \lvert B \rangle &= 0 & (x^\nu - y^\nu) \lvert B \rangle &= 0 \l{bsbc}
\end{align}

De même, les conditions aux bords pour les fermions \eqref{scbco} imposent que la partie fermionique de l'état de bord ne comportent que les secteurs R-R et NS-NS. Elles imposent en outre,
\begin{align}
(\psi_r^\mu + i\eta\tpsi_{-r}^\mu) \lvert B \rangle &= 0  \\
(\psi_r^\nu - i\eta\tpsi_{-r}^\nu) \lvert B \rangle &= 0 \l{bsfc}
\end{align}
où $r$ est entier dans le secteur R-R et demi-entier dans le secteur NS-NS. Le facteur $i$ provient de la transformation conforme et $\eta$ vaut  $\pm 1$. Nous avons ainsi deux types d'états pour chacun des secteurs R-R et NS-NS, notés en général $\lvert B, \eta \rangle$ et $\lvert B, + \rangle$ ou $\lvert B, - \rangle$ si la valeur de $\eta$ est fixée. Ces états permettent de construire des états de bord invariants sous la projection GSO et qui donc se couplent bien aux secteurs physiques de la théorie des supercordes considéré. Pour le secteur R-R, nous noterons cet état physique $\lvert B \rangle_\RR$ ; pour le secteur NS-NS, $\lvert B \rangle_\NSNS$. Dans notre étude, nous nous plaçons en théorie II, mais nous donnerons les expressions correspondantes un peu plus loin et nous exposons  l'utilisation de la dualité de feuille d'univers en toute généralité.

De la même manière, les contraintes imposées par les fantômes peuvent être exprimés en terme de modes, mais nous n'entrerons pas dans les détails, nous tiendrons seulement compte de la contributation des fantômes dans le résultat final.

Les solutions aux équations \eqref{bsbc} et \eqref{bsfc} s'écrivent
\begin{align}
\lvert B, \eta \rangle_\NSNS &=  {\cal N}_\NSNS \, \delta^{(d_\perp)} (x-y)\, e^{-\sum_{n=1}^\infty \alpha_{-n}^\mu S_{\mu\nu} \tilde \alpha_{-n}^\nu}\, e^{i \eta\sum_{r=1/2}^\infty \psi_{-r}^\mu S_{\mu\nu} \tpsi_{-r}^\nu }\, \lvert 0; k=0 \rangle \\
\lvert B, \eta \rangle_\RR &= {\cal N}_{R-R}\, \delta^{(d_\perp)} (q-y)\, e^{-\sum_{n=1}^\infty \alpha_{-n}^\mu S_{\mu\nu} \tilde \alpha_{-n}^\nu}\, e^{i \eta\sum_{n=1}^\infty \psi_{-n}^\mu S_{\mu\nu} \tilde \psi_{-n}^\nu}\, {\cal M}_{AB}  \lvert A \rangle \lvert \tilde B\rangle
\end{align}
où $\lvert 0; k=0 \rangle$ est le vide NS-NS, 
$\lvert A\rangle \lvert \tilde B \rangle$ est le vide R-R
avec des indices de spineur de SO$(1,9)$ A et B. La matrice $S^\mu_{\phantom{\mu}\nu}$
a des valeurs propres $+1$ dans les directions du volume d'univers de la \pbr\ et
$-1$ dans les directions perpendiculaires à la \pbr. La matrice ${\cal M}_{AB}$ est une solution de l'équation
\be
(\Gamma^\mu)^{\text{T}} {\cal M} - i\eta S^\mu_{\phantom{\mu}\nu} \Gamma_{11} {\cal M} \Gamma^\nu = 0 \ . \l{eqM}
\ee
Les solutions s'écrivent
\be
\l{MDbr}
{\cal M} = C \Gamma^0 \Gamma^1 \ldots \Gamma^{p} \frac{1+ i\eta \Gamma_{11}}{1 + i\eta}\ ,
\ee
où $C$ est la matrice de conjugaison de charge. Une éventuelle constante de proportionnalité est en fait intégrée dans la normalisation de l'état. Pour la S-brane, nous distinguerons cependant le cas où nous avons un facteur $i$ supplémentaire.

La dualité de feuille d'univers identifie la propagation d'une corde ouverte le long d'une boucle de longueur $2\pi t$ avec la propagation d'une corde fermée sur une durée $\frac{\pi}t$. D'autre part nous  obtenons par calcul direct\footnote{en faisant notamment le changement de variable $s = \frac1t$} les identités suivantes 
\begin{gather}
\int_0^\infty ds \, {}_\NSNS\langle B, \eta \rvert e^{-\pi s (L_0 + \bar L_0)} \lvert B, \eta \rangle_\NSNS
=  \frac{8\cN_\NSNS^2}{\pi\apr T_p^2} \int_0^\infty \frac{dt}{2t} \, \tr_\NS e^{-2\pi t L_0} \\ 
\int_0^\infty ds \, {}_\NSNS\langle B, \eta \rvert e^{-\pi s (L_0 + \bar L_0)} \lvert B, -\eta \rangle_\NSNS
= \frac{8\cN_\NSNS^2}{\pi\apr T_p^2} \int_0^\infty \frac{dt}{2t} \, \tr_\Ra e^{-2\pi t L_0} \\ 
\int_0^\infty ds \, {}_\RR\langle B, \eta \rvert e^{-\pi s (L_0 + \bar L_0)} \lvert B, \eta \rangle_\RR
= - \frac{\cN_\RR^2}{2\pi\apr T_p^2} \int_0^\infty \frac{dt}{2t} \, \tr_\NS (-1)^F e^{-2\pi t L_0} \\ 
\int_0^\infty ds \, {}_\RR\langle B, \eta \rvert e^{-\pi s (L_0 + \bar L_0)} \lvert B, -\eta \rangle_\RR
= -\frac{\cN_\RR^2}{2\pi\apr T_p^2} \int_0^\infty \frac{dt}{2t} \, \tr_\Ra (-1)^F e^{-2\pi t L_0}
\end{gather}
où 
\be T_p = \pi^{7/2 -p} 2^{3-p} {\apr}^{3/2-p/2} \ee
est la tension de la \pbr. $\tr_\NS$ désigne une trace sur le secteur NS de la corde ouverte seulement, $\tr_\Ra$ sur le secteur R seulement. Nous avons écrit le propagateur, après changement de variable $z=e^{-\pi s}$,
\be
P = \frac{\pi \apr}2 \int_0^\infty ds\, e^{-\pi s(L_0 + \bar L_0)}\ .
\ee

À partir de ces identités, on construit des états de bords pour les D\pbr\ qui soient invariants sous projection GSO et qui correspondent à une amplitude de corde ouverte où se propagent aussi des états invariants par projection GSO. En théorie de type II, les combinaisons détats de bords invariantes sous projection GSO sont
\begin{align}
\lvert B \rangle_\NSNS &= \lvert B,+ \rangle_\NSNS - \lvert B,- \rangle_\NSNS \\
\lvert B \rangle_\RR &= \lvert B,+ \rangle_\RR + \lvert B,- \rangle_\RR
\end{align}
L'état $\lvert B \rangle_\RR$ n'est invariant sous projection GSO que si $p$ est pair en théorie IIA et impair en théorie IIB. On construit alors deux types d'états de bords de D-branes : un état de bord de D-brane \emph{BPS}, ainsi nommé car il est supersymétrique et conduit à une interaction nulle,
\be \lvert D \rangle_{\text{BPS}} = \lvert B \rangle_\NSNS + \lvert B \rangle_\RR \ee
et un état de bord de D-brane \emph{instable}\footnote{car il n'est pas protégé par la supersymétrie} 
\be \lvert D \rangle_{\text{inst.}} = \lvert B \rangle_\NSNS \ee
qui est un état indépendant pour $p$ impair (en type IIA) ou pair (en type IIB).

La dualité impose, pour les D-branes BPS,
\be
\cN_\NSNS^2 = -\frac1{16} \cN_\RR^2
\ee

Les états de bord, qui représentent les conditions au bord induite par une D-brane dans le secteur de corde fermée, permettent de calculer les champs du spectre de corde fermée émis par la membrane. Pour cela, il faut projeter l'état de bord auquel on a appliqué le propagateur sur les états de masse nulle adéquats. Ces projecteurs sont construits à partir des opérateurs de vertex de la corde fermée supersymétrique. Ainsi, le champ NS-NS $J^{\mu\nu} (k)$ émis par une D-brane est donné par\footnote{Dans les deux équations qui suivent, nous suivons les conventions de \cite{DiVecchia:1997pr}. Les indices des opérateurs de vertex indiquent la représentation (\emph{picture} en anglais) des superfantômes.}
\be J^{\mu\nu} (k) = \lim_{z, \bz \to \infty} \langle 0 \rvert \!\lno \V_{-1}^\mu (k/2; z)  \tilde \V_{-1}^\nu (k/2; \bz)\rno\!  \lvert B\rangle_\NSNS \l{chNSNS}\ee
Le champ R-R est donné par
\be J^{\dot{a}\dot{b}} (k) = \lim_{z, \bz \to \infty} \langle 0 \rvert \!\lno \V_{-3/2}^{\dot{a}} (k/2; z)  \tilde \V_{-1/2}^{\dot{b}} (k/2; \bz)\rno\! \lvert B\rangle_\RR \l{chRR}\ee
où ${\dot{a}}$ et ${\dot{b}}$ sont des indices correspondant à la représentation conjuguée de Weyl des spineurs de SO(1,9).

Après calcul, la comparaison de ces expressions \eqref{chNSNS} et \eqref{chRR} avec les champs émis par la solution classique de supergravité correspondant à la D-brane achève de fixer les normalisations
\be
\cN_\NSNS = i\frac{T_p}4 \text{\ \ \ et\ \ \ } \cN_\RR = \pm 4 T_p\ .
\ee
pour les D-branes BPS
et
\be
\cN_\NSNS = i\frac{T_p}{2\sqrt{2}}
\ee
pour les D-branes instables.


\subsection{\'Etats de bord pour les S-branes}
\l{sbr}

Pour la S-brane, la situation est moins claire. Il existe deux prescriptions possibles pour construire les états $\lvert B, \eta \rangle_\RR$. La première consiste à prolonger analytiquement la solution de D-brane \eqref{MDbr} et à écrire
\be \M_{\text{S-brane}} = i C \Gamma^1 \ldots \Gamma^{p} \Gamma^{p+1} \frac{1+ i\eta \Gamma_{11}}{1 + i\eta} \ee
Cette prescription conduit à des champs NS-NS \eqref{chNSNS} et R-R \eqref{chRR} qui ne sont pas tous les deux réels. On peut vérifier que du point de vie des cordes ouvertes, les états qui se propagent dans la boucle sont des états ayant une valeur propre $-1$ pour l'opérateur de projection GSO, en particulier le tachyon de corde ouverte qui est habituellement éliminé par une projection sur les états de valeur propre $+1$. La seconde prescription consiste à garder la même matrice $\M$ que dans le cas de la D-brane. Les champs émis sont maintenant réels (si l'on choisit correctement les normalisations), les états de cordes ouvertes qui se propagent dans la boucle ont une valeur propre $+1$ pour l'opérateur de projection GSO. Cependant, ces états ne vérifient pas les équations de supergravité de type II, mais de type II${}^\star$.

Ainsi, les S-branes de Dirichlet émettant des champs NS-NS et R-R physiques n'existent pas dans une théorie de type II. Par contre, nous pouvons considérer des S-branes émettant seulement un champ NS-NS, qui sont la contrepartie des D-branes instables. Nous allons, dans la section \ref{article3} du chapitre suivant \ref{contrereac}, étudier plus en détail l'amplitude à une boucle de cordes ouvertes tendues entre deux S-branes de Dirichlet. Il s'agira alors de S-branes \gl instables\gr.

\subsection{\'Etats de bord dans une configuration non triviale}

Dans le cas de la configuration d'ondes planes \`a profil lin\'eaire que nous avons décrite à la sous-section \ref{cchem}, les conditions aux bords sont linéaires\footnote{Nous n'avons étudié cette configuration que dans le cadre de la théorie bosonique.}. Nous pouvons alors résoudre les conditions imposées sur les états de bords et calculer l'amplitude de propagation de la corde fermée en présence des champs. Nous avons d'autre part calculé l'amplitude à une boucle de la corde ouverte. Nous avons ensuite vérifié explicitement la dualité de feuille d'univers. Nous ne détaillons pas ici ces calculs, qui sont semblables à ceux exposés ci-dessus, et nous invitons le lecteur à se réferer à l'article \cite{Durin:2003gj} (annexe \ref{a1}) et notamment à son annexe A. Ceci constitue un exemple non trivial de cette dualité.

\vfill

{\it Conclusion du chapitre}

\vspace{2mm}
À travers cet exposé du formalisme de première quantification de la théorie des cordes, nous avons mis en place les outils nécessaires à la discussion que nous allons mener au chapitre suivant \ref{contrereac}. Nous avons essayé de montrer comment divers problèmes techniques issus de configurations en apparence très différentes pouvaient être traités de manière similaire. Dans le chapitre suivant, nous traiterons les aspects physiques, plus spécifiques, de chacune des configurations.

\clearpage

%% file: contrereac.tex

\chapter{Effets physiques produits par les configurations dépendantes du temps}
\l{contrereac}

\section{Corde ouverte dans une onde plane \`a profil lin\'eaire}
\l{article1}

\subsection{Configuration statique, configuration dépendante du temps}

Avant de nous attaquer à la configuration dépendante du temps, nous avons étudié le cas statique.
Nous avons mis en évidence l'existence d'un gradient critique, qui correspond à la limite entre un domaine de stabilité et un domaine d'instabilité. En première analyse, les extrémités de la corde ont des domaines de stabilité disjoints, la configuration est donc globalement instable. Dans un premier temps, nous avons cherché un mécanisme permettant de stabiliser la configuration statique. Dans un second temps, nous avons essayé de résoudre les équations du mouvement dans une configuration dépendante du temps.

La solution au premier problème est la transposition directe des mécanismes de stabilisation des pièges à ions. Le piège de Penning, obtenu par l'ajout d'un champ magnétique constant et uniforme élargit la zone de stabilité pour chaque extrémité de la corde et permet l'existence d'une configuration stable. Le mécanisme du \emph{piège de Paul}, qui confine les ions en imposant une variation périodique du champ électrique (ici, nous faisons varier le gradient des champs \'electromagn\'etiques), fonctionne aussi.

Le second problème n'a pas pu être résolu sans faire certaines approximation, qui nous ont permis de mener les calculs à leur terme. Ceci nous a notamment permis de mettre en évidence un degré de liberté que nous avons interprété comme la contre-réaction due à la dépendance en temps.

\subsection{Particule chargée dans une onde plane \'electromagn\'etique}
\l{preq}

Nous commen\c cons notre étude par le cas d'une particule chargée. En effet, comme nous le constaterons également à la section \ref{article2}, l'étude de modèles ponctuels reproduisant le comportement des modes zéros de la corde permet de mettre en évidence une partie non négligeable des effets physiques engendrés par la configuration choisie. Ici, la particule chargée est loin d'être un modèle fidèle des modes zéros, mais elle permet de comprendre facilement la source de l'instabilité de la configuration.

Nous considérons donc une particule relativiste de masse $m$ et de charge unité, plongée dans une configuration composée d'une onde plane \`a profil lin\'eaire de potentiel $A = \Phi(x^+,x^i) dx^+$ (vérifiant la condition d'harmonicité \eqref{cPhi}) et d'un champ magnétique constant et uniforme $B$. Les équations du mouvement correspondant à l'action donnée dans la section \ref{lu} dans le formalisme de la ligne d'univers, conduisent à la relation de couche de masse suivante
\be \frac12 (p_i)^2 + p^+ p^- + p^+ \Phi(x^+,x^i) + m^2 = 0 \ee
où $p_i = \dpt x_i- B_{ij} x^j$ et $p^- = \dpt x^- - \Phi$ sont les impulsions canoniques conjuguées à $x^i$ et $x^+$. $p^-$ est l'Hamiltonien pour l'évolution sur la ligne d'univers.

Le mouvement de la particule relativiste se réduit donc à celui d'une particule non relativiste dans un potentiel \gl \'electrostatique\gr\ qui dépend du temps $\Phi(x^+,x^i)$. L'apparition d'une cinématique non-relativiste est une caractéristique habituelle de la quantification dans la jauge du c\^one de lumière, qui est naturellement imposée ici par l'équation du mouvement sur $x^+$ : $x^+ = x_0^+ + p^+ \t$.

En l'absence de champ magnétique, les points critiques du potentiel harmonique ont la forme suivante à deux dimensions (spatiales) $\Phi = \re(z^n/n)$ avec $z=x + iy$. Pour $n=2$, le cas habituel, $\Phi = \frac12 (x^2-y^2)$ et le mouvement le long de $y$ est instable (il correspond à un oscillateur harmonique inversé. Mais pour des points critiques d'ordre supérieur, le mouvement converge vers $z=0$ lorsque $\t \to \infty$. En effet $z(\t) = [n(n-2)\t]^{1/(2-n)}$. Si l'on revient aux équations du mouvement pour la corde \eqref{cbem}, on est confronté à des équations non linéaires, que nous n'avons pas su résoudre\footnote{En perdant la linéarité des équations du mouvement, on perd aussi la possibilité de superposer les sources et leurs solutions particulières.}. Néanmoins, l'instabilité du cas $n=2$ n'est pas rédhibitoire ; il existe plusieurs fa\c cons de stabiliser un piège électrostatique quadrupolaire.
\begin{itemize}
\i La première possibilité, anecdotique en ce qui concerne notre problème, consiste à considérer un piège pour  des molécules de polarisabilité négative. Les molécules sont alors attirées vers le centre du piège, où le champ électrostatique $E$ est nul. Une polarisabilité négative peut être obtenue pour des états excités dégénérés lorsque $E=0$.
\i La seconde possibilité consiste à rajouter un champ magnétique constant et uniforme pour confiner les direction instables. Le piège s'appelle \emph{piège de Penning}. Le cas le plus simple est le piège à trois dimensions, où le potentiel quadrupolaire s'écrit $V(x)=-\frac12 \, e \, (x^2 + y^2 - 2 z^2)$. En ajoutant un champ magnétique dans la direction $z$, on peut confiner les ions dans le plan $(x,y)$. En effet, les équations du mouvement
\be
\begin{pmatrix} \ddot{x} \\ \ddot{y} \\ \ddot{z} \end{pmatrix}
+ 2b  \begin{pmatrix} -\dot{y} \\ \dot{x} \\ 0 \end{pmatrix}
- e \begin{pmatrix} x \\ y \\ -2z \end{pmatrix}
= \begin{pmatrix} 0 \\ 0 \\ 0 \end{pmatrix}
\ee
conduisent à la relation de dispersion suivante
\be
(\w^2-2 \w b+e)(\w^2+2 \w b+e)(2e-\w^2) = 0 
\ee
où les 2 premiers facteurs correspondent  aux polarisations gauche et droite dans le plan $(x,y)$ et le dernier, à celle dans la direction $z$. Le mouvement est stable si toutes les racines $\w$ sont réelles, c'est-à-dire si $b^2 > e$ et si la charge $e$ est strictement positive. Ainsi, seules les particules chargées positivement sont piégées, et ce, seulement si le champ magnétique est suffisamment intense.
\i La troisième possibilité consiste à faire varier le potentiel $\Phi$ à une fréquence donnée de fa\c con à obtenir une résonance paramétrique avec les modes propres du champ quadrupolaire. Il s'agit du piège de Paul. À deux dimensions, le potentiel s'écrit
$\Phi(x^+)=\frac12(e + f \sin(\omega \tau)(x^2-y^2)$. Les équations du mouvement s'écrivent alors
\be
\ddot{x} + \left[ e + f \sin(\omega \tau) \right] x = 0
\ee
Cette équation s'appelle \emph{équation de Mathieu} et présente qualitativement le même diagramme de stabilité que pour une modulation rectangulaire, plus simple à calculer (voir figure \ref{dsPaul}).
\end{itemize}

Avant de voir comment ces mécanismes s'appliquent aux cas de la corde ouverte dans une onde plane \`a profil lin\'eaire, nous allons étudier le comportement d'un dip\^ole élastique dans une telle configuration de champ. En effet, ce système reproduit une grande partie de la physique de la corde à basse énergie.

\subsection{Dip\^ole élastique dans une onde plane \'electromagn\'etique \`a profil lin\'eaire}
\l{decheq}

Nous commen\c cons par expliquer comment nous nous ramenons au modèle du dip\^ole élastique. Dans la jauge du cone de lumière $X^+ = x^+ + p^+ \t$, les conditions de Virasoro pour la corde ouverte s'écrivent
\begin{subequations}
\begin{align}
p^+ \p_\sigma X^- &+ \p_\tau X^i \p_\sigma X^i = 0 \l{v1} \\ 
- p^+ \p_\tau X^- &+ \frac12 \left[ (\p_\tau X^i)^2 + (\p_\sigma X^i)^2 \right]
=0 \l{v2}
\end{align}
\end{subequations}
En réécrivant la contrainte Hamiltonienne \eqref{v2} en terme des impulsions canoniques (la dépendance en $\t$ est sous-entendue)
\be
\begin{split}
\pi_- &= -\p_\tau X^+ = -p^+\ ,\quad \pi_i = \p_\tau X^i \\
\pi_+ &= -\p_\tau X^- +\, \Phi^{(0)}( X(0)) \delta(\s)\, -\, \Phi^{(1)}( X(\pi)) \delta(\s-\pi)
\end{split}
\ee
et en intégrant sur $\s$ entre $0$ et $\pi$ nous obtenons l'expression de l'Hamiltonien de c\^one de lumière $p^-$
\be
p^- = {\cal H}_{\text{c.l.}} = \frac1{2 p^+} \int_0^\pi  \left[(\pi_i)^2 + (\dps X^i)^2 \right] d\s
\,-\, \Phi^{(0)}(X(0)) \,+\, \Phi^{(1)}(X(\pi)) \l{lchamil}
\ee
Cet Hamiltonien est conservé si les potentiels $\Phi^{(a)}$ sont indépendants de $x^+$.
De même que dans le cas de la particule relativiste (voir sous-section \ref{preq}), ${\cal H}_{\text{c.l.}}$ peut être interprété comme l'Hamiltonien de deux particules non relativistes évoluant sous l'effet d'un potentiel dépendant du temps égal à $p^+ \Phi^{(1)}(X(\pi)) - p^+ \Phi^{(0)}(X(\pi))$ et liées par un potentiel élastique $\int_0^\pi \frac12 ( \dps X^i)^2$. L'énergie de liaison cro\^it comme la longueur de la corde au carré, alors que dans le cas relativiste, cette énergie augmente de fa\c con linéaire. Cette différence est cruciale, comme nous allons le constater par la suite.

Le comportement à basse énergie des cordes ouvertes peut être modélisé par le modèle simplifié du dip\^ole élastique, qui correspond à l'approximation de corde rigide. Notre définition du dip\^ole est plus large que celle du dip\^ole habituel, puisque nous autorisons la charge totale à prendre n'importe quelle valeur (autrement dit notre dip\^ole est un \gl vrai\gr\ dip\^ole seulement dans le cas où $\Phi^{(0)}(x^+,x^i) = \Phi^{(1)}(x^+,x^i)$). Notre dip\^ole est ainsi constitué de deux charges dont les positions respectives sont $x^i_L (\t)$ et $x^i_R(\t)$ et qui sont reliées par une corde \gl rigide\gr
\be
X^i(\t,\s) = \frac1{\pi} \left[\s\, x^i_R + (\pi-\s)\,x^i_L\right] \l{sstr}
\ee

L'énergie de c\^one de lumière de ce dip\^ole est donnée par la relation suivante
\be
\l{ms2}
-p^+ p^- + \frac12 (p_L^2 + p_R^2) + V(x_L,x_R) = 0
\ee
où $V$ est l'énergie potentielle
\be
\l{pote}
V(x_L,x_R)= \frac{1}{2\pi} (x_L^i-x_R^i)^2 + p^+ \left[ \Phi^{(1)}(x_R) - \Phi^{(0)}(x_L) \right]
\ee
Remarquons qu'une solution \emph{statique} de ce problème est une solution exacte des équations pour la corde puisque l'équation \eqref{sstr} satisfaira les équations du mouvement \eqref{eqmv} et les conditions aux bords suivantes
\begin{subequations}
\l{cbc}
\begin{align}
x_R^i-x_L^i\, +\,  \pi p^+\,\p_i \Phi^{(0)}(x_L)  &= 0   &&\text{at } \sigma=0\\
x_R^i-x_L^i\, +\,  \pi p^+\,\p_i \Phi^{(1)}(x_R)  &= 0   &&\text{at } \sigma=\pi
\end{align}
\end{subequations}
Contrairement au cas de la particule traité ci-dessus, le potentiel à deux corps \eqref{pote} n'est plus harmonique, mais satisfait la relation suivante
\be
\left( \Delta_L + \Delta_R \right) V = \frac{2}{\pi} d_\perp
\ee
où $d_\perp = d-1$ est le nombre de dimensions transverses au c\^one de lumière.
Il est donc possible que ce potentiel admette des extrema stables au lieu de présenter uniquement des points-cols.

\subsubsection{Domaine de stabilité, gradient critique et cordes macroscopiques}

Nous pouvons maintenant mettre en évidence une caractéristique importante de cette configuration de cordes ouvertes, l'existence d'un gradient critique et de cordes macroscopiques. Pour cela, nous étudions la dynamique de notre dip\^ole élastique sous l'effet de potentiels $\Phi^{(a)} = \frac12 h^{(a)}_{ij} x^i x^j$ où nous supposons que $h^{(0)}$ et $h^{(1)}$ commutent. Comme nous l'avons fait à la fin de la sous-section \ref{cchem}, nous définissons $e_a = p^+ h^{(a)}$. L'énergie potentielle dans une direction propre $x^i$ de $e$ s'écrit alors comme une forme quadratique
\be
V^i(x^i_L, x^i_R) = \frac{1}{2\pi} \begin{pmatrix}x^i_L & x^i_R \end{pmatrix} Q^i \begin{pmatrix}x^i_L \\ x^i_R \end{pmatrix}\ ,\quad
Q^i=\begin{pmatrix} 1-\pi e^i_0  & -1 \\ -1 & 1+\pi e^i_1
\end{pmatrix} 
\ee
où $e^i_a$ est la valeur propre correspondante.

\begin{figure}[h]
\begin{center}
\epsfig{file=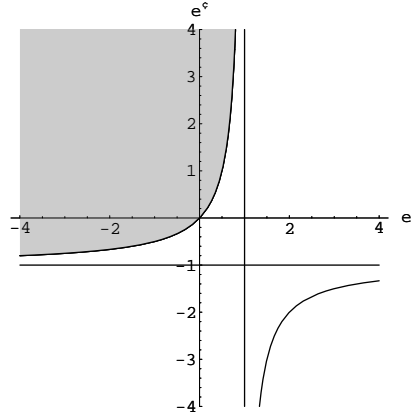,height=5.5cm}
\end{center}
\caption{\label{dsde}\small Diagramme de phase pour le dipole \'elastique dans le plan $(e_0,e_1)$. La r\'egion de stabilit\'e est gris\'ee. Des cordes macroscopiques sont produites sur la s\'eparatrice.}
\end{figure}

Considérons tout d'abord une première direction $x^{i_1}$. Le mouvement est stable dans cette direction si le potentiel $V^{i_1}$ présente un minimum global, c'est-à-dire si les deux valeurs propres de $Q^{i_1}$ sont positives. C'est le cas si le déterminant de $Q^{i_1}$
\be
\pi \delta = \pi (e_1 - e_0) - \pi^2 e_0 e_1
\ee
et la trace de $Q^{i_1}$
\be
\pi \Delta = \pi (e_1 - e_0) + 2
\ee
sont positifs. Nous avons omis l'exposant $i_1$ et introduit les notations de la sous-section \ref{cchem}. Si $\delta >0$ et $\Delta <0$, alors le potentiel présente un maximum global et si $\delta<0$, quel que soit le signe de $\Delta$, le potentiel présente un point-col.
Le diagramme de stabilité est présenté sur la figure \ref{dsde}.

\`A ce stade, nous pouvons faire les remarques suivantes.
\begin{itemize}
\i Nous retrouvons les différents cas que nous avons distingués lors de la quantification de la corde ouverte coupl\'ee \`a des ondes planes. L'origine de l'instabilité est cinématique, elle correspond à une configuration de champs donnée, non à un instabilité de la configuration.
\i Si la corde avait une tension nulle, le domaine de stabilité serait défini par $e_1 >0$, $e_0<0$, c'est-à-dire que chacune des extrémités de la corde serait plongée dans un puits électrostatique. L'énergie de liaison élastique étend ce domaine aux quadrants $e_0 e_1 >0$.
\i Si l'on considère $n$ directions $i_1, \ldots i_n$ simultanément, il n'est pas possible de choisir les gradients $e_0^{i_k}$, $e_1^{i_k}$ tous dans le domaine de stabilité et remplir en même temps la condition d'harmonicité \eqref{cPhi} sur les potentiels $\Phi^{(a)}$, qui se traduit par la condition suivante sur les gradients $\sum_{k=1}^n (e_0^{i_k}, e_1^{i_k}) = (0,0)$. Autrement dit, l'état fondamental de la corde a une polarisabilité \emph{positive}.
\i Le choix d'une configuration où $h^{(0)}$ et $h^{(1)}$ ne commutent pas ne semble pas améliorer le problème de stabilité \cite{Durin:2003gj}.
\end{itemize}

Nous considérons à nouveau une seule direction. La courbe $\delta =0$ correspond au cas où l'une des deux valeurs propres de $Q$ est nulle. Pour la partie de cette courbe qui forme la frontière du domaine de stabilité, le potentiel présente une vallée dégénérée, qui correspond à des cordes de taille arbitraire dans la direction considérée
\be
X = - (e_0 \s -1) x_L
\ee
La longueur de la corde selon $x$ est $\lvert \pi e_0 x_L \rvert$. L'interprétation physique est la suivante. Lorsque $e_0$ est positif ou $e_1$ est négatif, le potentiel électrostatique décro\^it quadratiquement selon $x_L$ et/ou $x_R$ et l'énergie de liaison de la corde non relativiste cro\^it comme le carré de la longueur. Il existe donc une valeur critique des gradients $e_0$, $e_1$ pour laquelle les deux forces électrostatique et de tension s'annulent quelle que soit la taille de la corde. Les cordes peuvent alors être macroscopique. Si l'on modifie les gradients de fa\c con à franchir la ligne de stabilité, le vide devient instable, des cordes d'échelle quantique sont étirées sur des distances infinies et déchargent le condensateur qui génère la configuration de champ, ce qui ramène les valeurs de gradients dans la zone de stabilité. Ce phénomène est très similaire à la présence, en électrodynamique de Born-Infeld ou pour les cordes ouvertes chargées en présence d'un champ électrique \cite{Fradkin:1985qd,Burgess:1986dw}, d'une valeur critique de ce champ, résultat de la compétition entre un potentiel électrique qui écarte les extrémités de la corde et qui cro\^it linéairement avec la longueur de celle-ci et une énergie de liaison qui cro\^it aussi linéairement.

\subsubsection{Stabilisation de la configuration, piège de Penning, piège de Paul}

Nous allons résumer les résultats qui montrent que cette stabilisation est bien possible, comme dans le cas de la particule.
Nous commen\c cons avec le piège de Penning. Par souci de simplicité, nous considérons la configuration suivante, avec des champs d'onde plane dans deux directions $x$ et $y$ et stabilisée par un champ magnétique $B$,
\be
\Phi^{(a)} = \frac12 h^{(a)} \left(x^2 + y^2 - \frac{2}{1+b^2} z^2 \right)
\ ,\quad B=b\; dx\wedge dy
\ee
En présence du champ magnétique, la métrique effective pour les cordes ouvertes est modifiée, ce qui explique le facteur $1/(1+b^2)$.

\begin{figure}[h]
\begin{center}
\epsfig{file=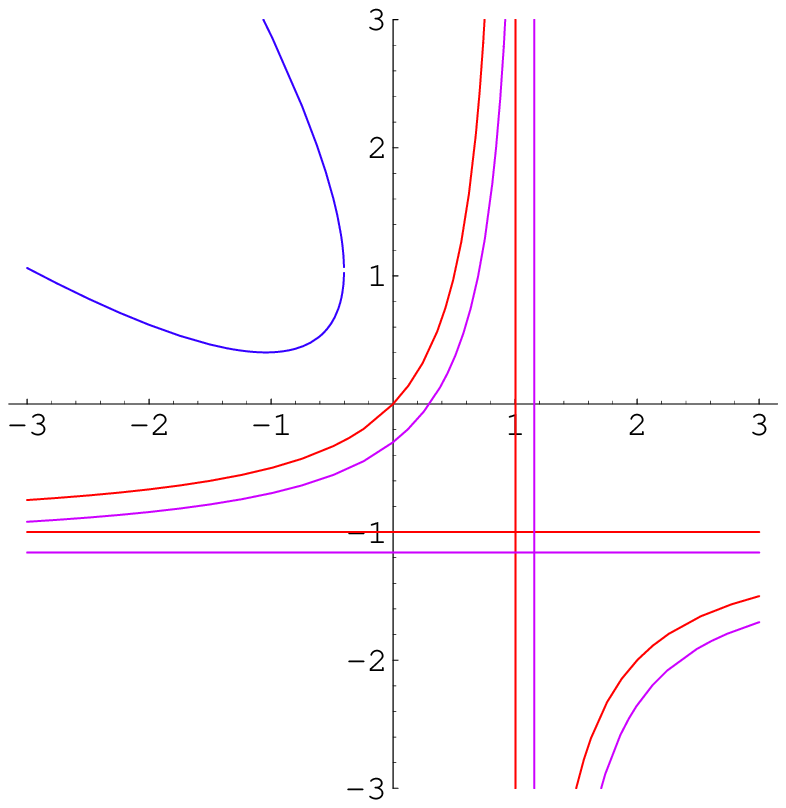,height=4.2cm}
\epsfig{file=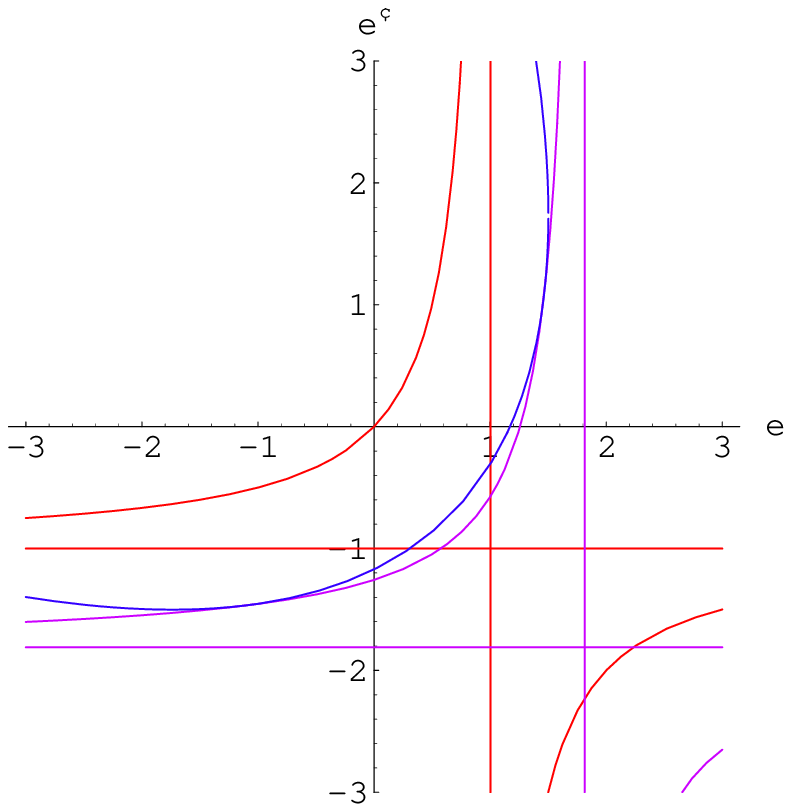,height=4.2cm}
\epsfig{file=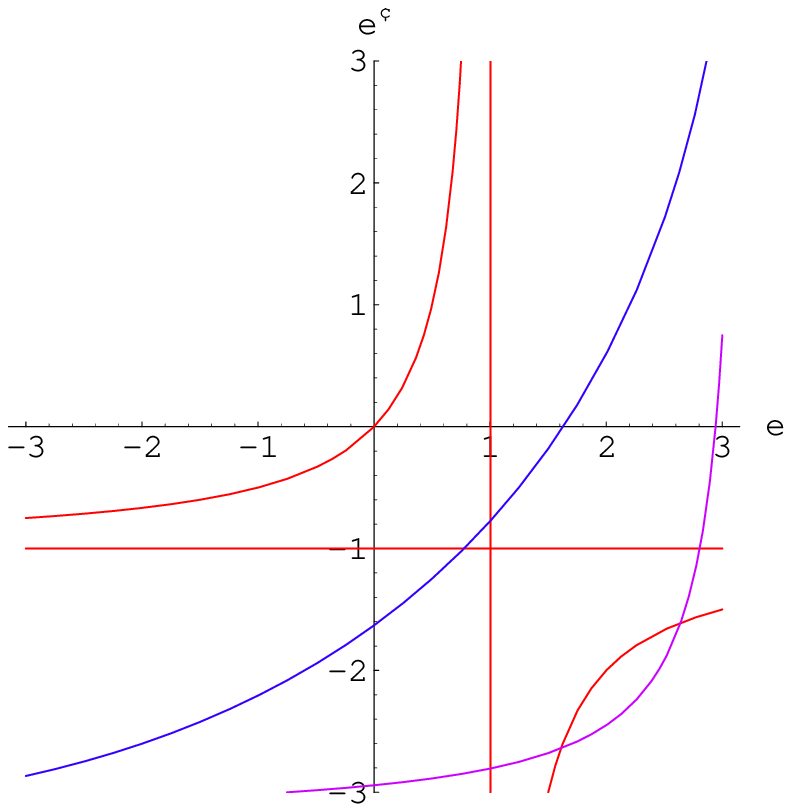,height=4.2cm}
\end{center}
\caption{\label{phasediagmag}\small Diagramme de phase pour la corde ouverte dans un potentiel quadrupolaire dans le plan $(e_0,e_1)$ avec un champ magn\'etique de faible (\`a gauche), moyenne (au centre) et de forte (\`a droite) intensit\'e. La limite du domaine de stabilit\'e est en violet, la ligne de production des cordes macroscopiques, en rouge et la ligne de d\'eg\'en\'erescence accidentelle, en bleu.}
\end{figure}

Nous n'entrerons pas dans les détails de l'analyse (voir \cite{Durin:2003gj}, annexe \ref{a1}).
Pour la direction $z$, le domaine de stabilité est défini par
\bse
\label{trnob}
\begin{gather}
\left(1+b^2 +2e_0\right)\left(1+b^2-2e_1\right) - \left(1+b^2\right)^2 > 0 \\
\left(1+b^2\right) + e_0-e_1 >0 \label{trnobupperleft}
\end{gather}
\ese
Pour chacune des directions $x$ et $y$, le domaine de stabilité $\delta > 0$, $\Delta >0$ est élargi au domaine défini par
\bse
\label{condmag}
\begin{gather}
\left(1+ b^2 - e_0 \right) \left( 1+ b^2 + e_1 \right) - 1  > 0, \\
e_0 < 1+ b^2, \label{condmagupperleft}
\end{gather}
\ese
qui correspond au quadrant supérieur gauche de la courbe dessinée en violet sur la figure \ref{phasediagmag}.
Il est maintenant possible de trouver $e_0$ et $e_1$ vérifiant \eqref{trnob} et \eqref{condmag} de sorte que le mouvement soit stable dans les trois directions $x$, $y$ et $z$.

Pour le piège de Paul, l'intensit\'e des champs \'electromagn\'etiques des ondes planes est modulée par une fonction périodique $H(t)$ de période $2\pi$ de la fa\c con suivante (nous avons choisi la configuration la plus simple)
\be
\Phi(x^+)= \frac12 \left[ h + f H\left(\frac{2\pi}{T}x^+\right) \right] ( x^2 -y^2)
\ee
L'équation du mouvement \emph{pour une particule} s'écrit, après reparamétrisation de $x^+$
\be 
\label{mathieu}
 \ddot{X} + [\w^2 + \alpha^2 H(t)]\, X = 0
\ee
À nouveau, nous invitons le lecteur à se référer à \cite{Durin:2003gj} et à l'article original \cite{vdpol} pour les détails de l'analyse. Le diagramme de stabilité, représenté sur la figure \ref{dsPaul}, montre l'existence d'une région stable dans le domaine $\w^2 < 0$. Il est possible de choisir l'amplitude $f$ et la période $T$ de la modulation pour que $(\w^2, \a^2)$ et $(-\w^2, -\a^2)$ soient tous deux dans le domaine de stabilité et ainsi d'obtenir un mouvement stable pour les deux directions $x$ et $y$.

\begin{figure}[h]
\hfill\epsfig{file=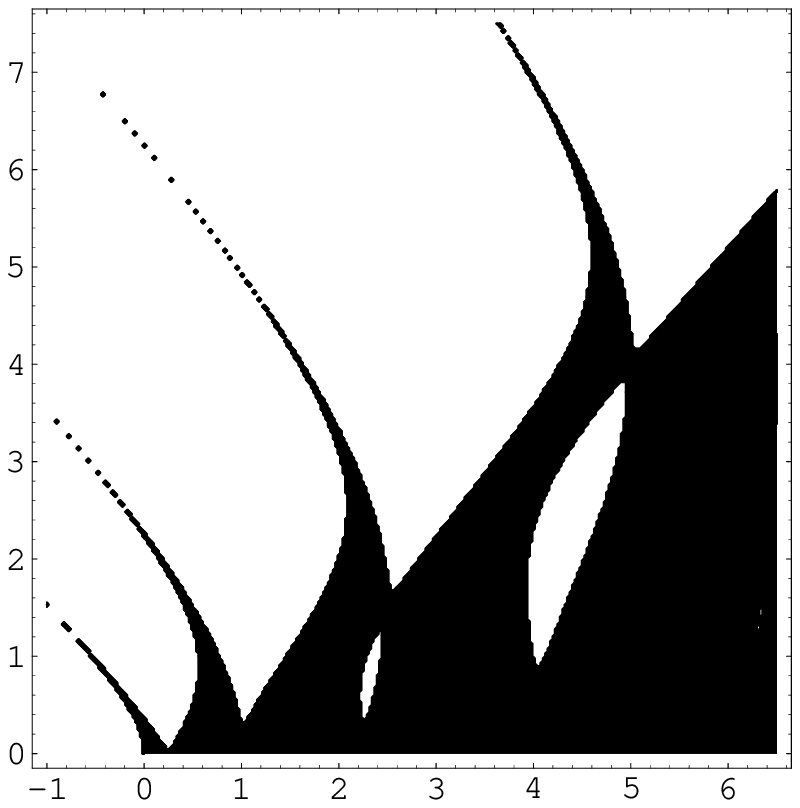,height=4.5cm}
\hfill\epsfig{file=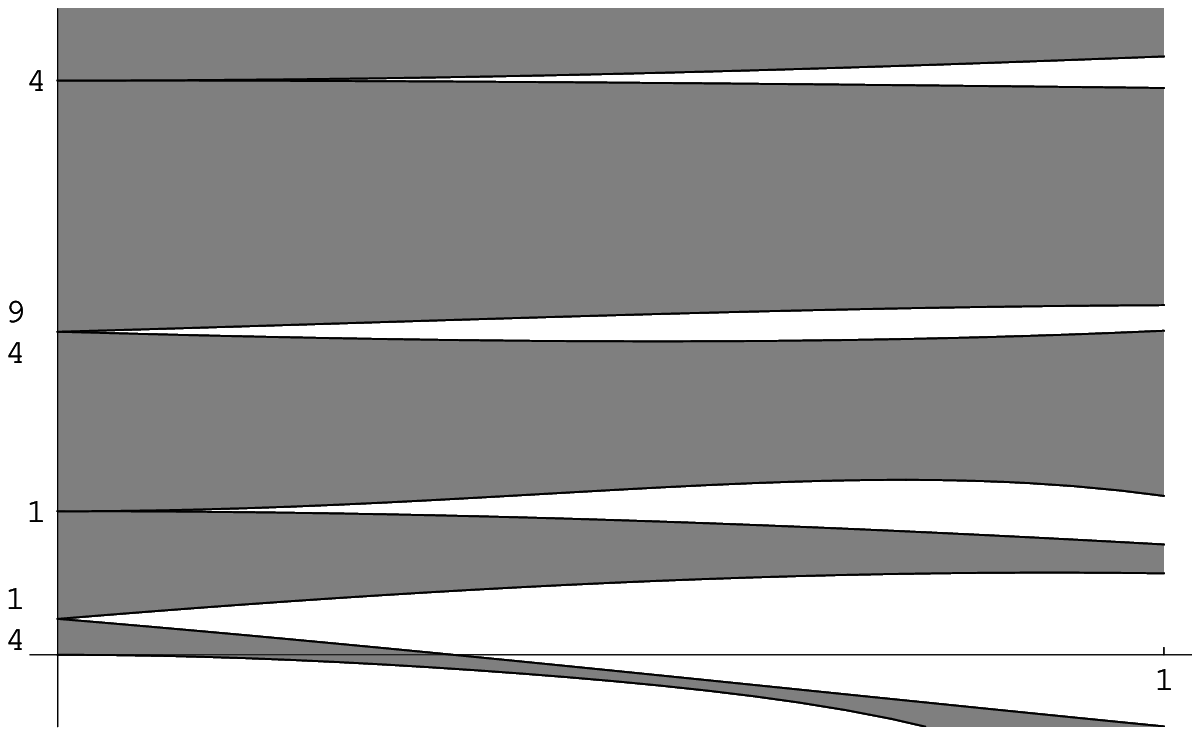,height=4.7cm}
\hfill
\caption{\l{dsPaul}\small À gauche : diagramme de stabilité dans le plan $(\omega^2,\alpha^2)$ pour un oscillateur harmonique dont la fréquence est modulée par un signal rectangulaire. $\omega$ est proportionnel à $h$, $\alpha$ à $f$. La région stable est ombrée. Elle s'étend dans la zone habituellement instable $\om^2<0$. À droite : zoom sur la région où $\alpha^2\ll 1$, les axes sont échangées. Notons enfin que des instabilités apparaissent lorsque $\omega$ est demi-entier, dès que l'amplitude de modulation est infinitésimale.}
\end{figure}

Le cas qui nous intéresse est celui du dip\^ole élastique. L'équation du mouvement s'écrit, en supposant que la même modulation est appliquée aux deux extrémités de la corde,
\be
\begin{pmatrix} \ddot{x}_L \\ \ddot{x}_R \end{pmatrix}
+ \frac{1}{\pi}
\left[ \begin{pmatrix} 1- \pi p^+h^{(0)} \quad -1\quad \\ \quad -1 \qquad 1+ \pi p^+ h^{(1)} \end{pmatrix} - p^+ f H\left(\frac{2\pi}{T} x^+\right) \begin{pmatrix} 1 & 0\\0 & 1 \end{pmatrix} \right]
\begin{pmatrix} x_L \\ x_R \end{pmatrix} = 0
\ee
qui peut être réduite à l'équation \eqref{mathieu} pour chacune des directions propres de la matrice\footnote{La matrice $Q$ est la même que celle que nous avons définie plus haut, elle n'inclut pas la partie de modulation.} $Q$ avec
\bse
\begin{align}
\w^2_{\pm} &= \left(\frac{T}{2\pi p^+}\right)^2  \frac{1}{\pi} \left(1+ \frac{e_1-e_0}{2} \pm \sqrt{1+\frac{(e_0+e_1)^2}{4}} \right) \\
\alpha^2 &= \left(\frac{T}{2\pi p^+}\right)^2 f
\end{align}
\ese
Les modes propres $\w^2_\pm$ dans les directions $x$ et $y$ ne sont plus opposés, mais il est possible de choisir la modulation $f$ de fa\c con à stabiliser le mouvement dans les deux directions, au moins pour une tension suffisamment faible.

\subsection{Corde ouverte couplée à des ondes planes \`a profil lin\'eaire~: stabilisation}

Les solutions de corde ouverte dans cette configuration et leur quantification ont été décrites dans les sous-sections \ref{cchem} et \ref{cocheeq}.

L'analyse faite dans la sous-section \ref{decheq} s'applique aux modes zéro de la corde ouverte et montre que le mouvement de celle-ci dans des ondes planes \`a profil lin\'eaire est instable.

Nous pouvons également transposer les mécanismes de piégeage qui ont fait leur preuve pour des particules ponctuelles chargées. Nous ne traiterons que le piège de Penning car d'une part, il nous suffit de trouver un seul dispositif de stabilisation et d'autre part, le piège de Paul introduit en quelque sorte une dépendance en temps parasite. Nous considérons ainsi le piège de Penning à trois dimensions qui a été exposé à la sous-section \ref{decheq}. Nous supposons par souci de simplicité que le champ $B$ et les gradients $e_a$ commutent. L'énergie des modes de la corde ouverte, déterminée par l'équation \eqref{disp}, est donnée en présence du champ magnétique par la relation de dispersion suivante
\be
\label{dispb}
\tan(\pi\omega) = \frac{\w (e_1-e_0) }{(1+b^2)\w^2 \pm b (e_0+e_1)\w + e_0 e_1}
\ee
où les signes $\pm$ correspondent aux polarisations circulaires gauche et droite. Ces polarisations sont échangées si l'on change $\w$ en $-\w$, nous pouvons donc nous concentrer sur le signe $+$. La ligne de production de cordes macroscopiques $\delta = e_1 - e_0 - e_0 e_1 = 0$ n'est pas modifiée puisque le champ magnétique $b$ n'appara\^it que dans des termes qui sont d'ordre supérieur en $\w$ par rapport aux termes qui déterminent cette ligne. Par contre, lorsqu'on atteint $\delta = 0$, une seule racine devient nulle\footnote{Ceci peut être vu en développant la relation de dispersion \ref{dispb} pour $\w$ petit.} et change de signe avec $\delta$. On n'a donc pas d'instabilité associée à $\delta = 0$, mais la création de cordes statiques macroscopiques comme expliqué ci-dessus.

L'instabilité appara\^it pour un champ magnétique critique
\be
b_{\text{c}}^2 = \frac{4 (e_0-1)(3-3e_0+e_0^2)}{3 e_0^2(e_0-2)^2} \delta \ ,
\quad \delta \ll 1
\ee
Malheureusement, il semble impossible de trouver l'expression de la ligne critique pour un $\delta$ arbitraire. Néanmoins, le domaine de stabilité est élargi~: il est maintenant possible de fixer les valeurs de gradients et du champ magnétiques pour stabiliser le mouvement dans toutes les directions, comme pour le dip\^ole élastique.

\subsection{Corde ouverte couplée à des ondes planes \`a profil lin\'eaire~: dépendance en temps}

Nous considérons maintenant le cas d'ondes planes \`a profil lin\'eaire dépendantes du temps. Notons qu'à cause de la symétrie par translation selon $x^-$, il n'y a pas de productions de cordes. Seuls les états avec $p^+ = 0$ contribuent à l'énergie du vide (ou l'amplitude du vide, pour reprendre les termes utilisés au chapitre précédent) et ces états ne sont pas sensibles aux ondes planes. Par contre, une corde, de manière générale, sera excitée en se propageant dans le champ électromagnétique. On parlera donc de production de modes.

\subsubsection{Production de mode, transformation de Bogolioubov}

Nous considérons des champs d'onde plane à support compact le long de $x^+$. Lorsque $x^+ \to \pm \infty$, la corde ouverte est libre, les coordonnées $X^\mu (\t, \s)$ admettent un développement en modes \gl libres\gr\ \eqref{co}. Pour distinguer l'état initial à $x^+ \to -\infty$ de l'état final à $x^+ \to \infty$, nous utiliserons les notations de \eqref{co} $x_0$, $p_0$ et $a_n$ pour les modes initaux et nous noterons respectivement $y_0$, $q_0$ et $b_n$ les modes finaux. Comme la perturbation est linéaire, les modes finaux dépendent linéairement des modes initiaux. Cette relation de dépendance linéaire est représentée par \emph{la matrice de Bogolioubov}
\be 
\l{bglb}
\begin{pmatrix} y_0 \\ q_0 \\ b_m \end{pmatrix} = \begin{pmatrix} \alpha & \beta & A_n \\ \gamma & \delta & B_n \\ \tilde A_m &  \tilde B_m & B_{mn} \end{pmatrix} \begin{pmatrix} x_0 \\ p_0 \\ a_n \end{pmatrix} 
\ee
Pour trouver une expression explicite des coefficients, il nous a fallu procéder à des approximations. Nous invitons le lecteur à se réferer à \cite{Durin:2003gj} pour les approximations adiabatique et brutale, cette dernière étant pertinente pour traiter le cas du piège de Paul.

Nous exposons les résultats de l'approximation de Born, dans laquelle le champ des ondes planes est traité comme une perturbation autour du cas de la corde ouverte libre. Cette approximation est pertinente pour les états très excités de la corde. Nous faisons ainsi l'hypothèse que les gradients $h^{(a)}$ sont du même ordre $\e$. La décomposition des parties droite et gauche des modes excités, $f$ et $g$ s'écrit
\be
\begin{pmatrix} f \\ g \end{pmatrix} = \frac{i}{n} a_n  \begin{pmatrix} 1 \\ 1 \end{pmatrix} e^{-i n \tau}
+ \begin{pmatrix} \delta f \\ \delta g \end{pmatrix}
\ee
où $\delta f$ et $\delta g$ sont d'ordre $\e$.
Les équations du mouvement peuvent être résolues\footnote{Un calcul très similaire a été fait dans l'article \cite{Bachas:2002jg}.}, ce qui permet d'extraire les coefficients $B_{mn}$, $A_n$ et $B_n$. On complète le calcul des coefficients en considérant le développement au premier ordre en $\e$ des parties droite et gauche des modes zéro $x_0 + p_0 \t$. Nous obtenons
\bse
\begin{align}
\alpha &= 1 - \frac1{\pi} \int_{-\infty}^\infty  \left(\tau (e_0-(Te_1))(p^+\tau) - \pi(Te_1)(p^+ \tau)\right)d\tau\ , \\
\beta &= - \frac1{\pi} \int_{-\infty}^\infty  \left(\tau^2 (e_0-(Te_1))(p^+\tau)
-\pi(\pi +2\tau) (Te_1)(p^+\tau) \right) d\tau\ , \\
\gamma &= \frac1{\pi} \int_{-\infty}^\infty  (e_0-(Te_1))(p^+\tau) d\tau\ , \\
\delta &= 1 + \frac1{\pi} \int_{-\infty}^\infty \left( \tau (e_0-(Te_1))(p^+\tau) - \pi(Te_1) (p^+\tau)\right) d\tau\ , \\
\tilde{A}_m &= \frac1{2\pi} \int_{-\infty}^\infty (e_0-(Te_1))(p^+\tau) e^{im\tau} d\tau\ , \\
\tilde{B}_m &=  \frac1{2\pi} \int_{-\infty}^\infty \left(\tau(e_0-(Te_1))(p^+\tau) - \pi(Te_1)(p^+\tau)  \right) e^{im\tau} d\tau\ ,\\
A_n &=  -\frac{2i}{n\pi} \int_{-\infty}^\infty  \left(\tau (e_0-(Te_1))(p^+\tau) - \pi(Te_1)(p^+ \tau)\right)
e^{-in\tau} d\tau\ , \\
B_n &= 
 \frac{2i}{n\pi^2}  \int_{-\infty}^\infty  (e_0-(Te_1))(p^+\tau) e^{-in\tau} d\tau\ , \\
B_{mn} &= \delta_{mn} + 
\frac{i}{\pi n}  
\int_{-\infty}^\infty  (e_0-(Te_1))(p^+\tau) e^{-i(n-m)\tau} d\tau\ ,
\end{align}
\ese
où par exemple $(e_0-(Te_1))(p^+\tau)$ doit être compris comme la fonction $e_0-(Te_1)$ prise en $p^+ \t$. $T$ est l'opérateur de décalage : $(Te_1)(\t) = e_1(\t+\pi)$.

Il est cependant un phénomène dont la transformation de Bogolioubov ne rend pas compte. Il s'agit de la contre-réaction de la corde sur le champ électromagnétique, qui est l'analogue de la contre-réaction qu'on attend d'une corde dans une géométrie dépendente du temps. En effet, le mode $x^i_0$ de la corde est construit comme la somme des modes constants $f_0$ et $g_0$ des parties droite et gauche. La différence $f^i_0-g^i_0$ n'a dans le cas habituel de la corde libre, aucune signification physique. Il peut être en effet interprété\footnote{Il est aussi possible de le voir comme la position de la D-brane T-duale le long de la direction $x$ et de parvenir à la même conclusion.} comme un potentiel de jauge U(1) constant $A = (f^i_0 - g^i_0) dx^i$ et ce degré de liberté superflu est éliminé en fixant la jauge.

Par contre, dans une configuration dépendante du temps, la \emph{variation} de $f_0 - g_0$ entre l'état initial et l'état final possède une signification invariante de jauge : c'est la variation du potentiel du champ de l'onde plane entre ces deux états\footnote{ou encore le déplacement de la D-brane T-duale} $A_i(+\infty) - A_i(-\infty) = \int_{-\infty}^{+\infty} F_{+i} dx^+$. Ce changement dans la configuration électromagnétique induit par la propagation des cordes constitue ainsi un cas d'école de contre-réaction qu'il serait intéressant de transposer au cas d'une géométrie dépendente du temps et asymptotiquement plate. Dans le cadre de l'approximation de Born, cette contre-réaction peut être précisée, en calculant séparément la variation de $f_0$ et de $g_0$. On trouve le résultat suivant ($X$ est la solution de corde ouverte libre)
\be 
\delta f_0-\delta g_0  = - \int_{-\infty}^{\infty} e_0 (p^+ \tau) X(\sigma=0,\tau) d\tau 
\l{crchem}
\ee
qui représente, comme on vient de l'expliquer, une variation de $F_{i+}$.

\section{Cordes fermées dans un orbifold de boost}
\l{article2}

Comme nous l'avons expliqué auparavant, l'orbifold de boost présente une \emph{singularité de type Big Crunch - Big Bang}\footnote{de manière équivalente, nommée \emph{singularité cosmologique} ou \emph{singularité de genre espace}.} et offre la possibilité d'utiliser le formalisme perturbatif de première quantification. On s'attend ainsi à ce que le calcul des observables physiques telles que les amplitudes nous donnent des informations sur la nature et l'évolution de la singularité. Toutefois ces attentes doivent être tempérées par la nature de la singularité présente dans l'orbifold de boost. Par construction, il ne s'agit que d'une singularité topologique et ne peut constituer qu'un cas d'école\footnote{Nous utilisons cette traduction de \gl toy model\gr\ au lieu de la traduction littérale \gl modèle jouet\gr.}. En outre, nous verrons qu'il est assez difficile d'extraire des informations physiques des résultats que nous avons obtenus. Ceux-ci doivent être vus comme une première étape vers un véritable traitement perturbatif de la singularité de type cosmologique. Nous reviendrons sur ce point au chapitre \ref{persp}.

Nous allons d'abord détailler la géométrie d'orbifold de boost, préciser les caractéristiques du mouvement des particules libres dans cette géométrie et en tirer les conséquences pour les modes zéro de la corde, puis, après avoir rapidement exposé les phénomènes de production de particules, nous exposerons le calcul perturbatif de la contre-réaction et des amplitudes pertinentes.

Précisons que les sous-sections \ref{geommisner}, \ref{pcnt} et \ref{ctob} exposent les résultats de \cite{Berkooz:2004re}, qui ont été résumés dans \cite{Durin:2005ix} mais qui ne rentrent pas directement dans le cadre de cette thèse, alors que les sous-sections \ref{am3c}, \ref{am4c} et \ref{cr} présentent les résultats de \cite{Berkooz:2004yy}.

\subsection{Géométrie de l'orbifold de boost}
\l{geommisner}

L'orbifold de boost est plus connu sous le nom d'\emph{espace de Misner} et a été introduit en tant que modèle local de singularité \cite {Misner} dans les espaces-temps Taub-NUT \cite{Taub}. Comme nous l'avons vu dans la sous-section \ref{oboost}, l'espace de Misner peut être obtenu par quotient de l'espace de Minkowski $\R^{1,1}$ par une transformation de boost de rapidité $2\pi \beta$ : $x^\pm \sim e^{\pm 2 \pi \beta} x^\pm$. Il est donc localement plat et la courbure est localisée sur le c\^one de lumière $x^+ x^- = 0$, qui est fixe sous la transformation de boost. L'espace de Misner peut être représenté  comme quatre c\^one (de signature lorentzienne) réunis à leur sommet et qui correspondent aux quatre quadrants de l'espace de recouvrement $\R^{1,1}$. On distingue ainsi deux \emph{régions de Milne}, définies par $x^+ x^- > 0$ et deux \emph{régions de Rindler}, définies par $x^+ x^- < 0$. En effet, si l'on définit d'une part les coordonnées $t$ et $\theta$ pour les régions de Milne et d'autre part les coordonnées $r$ et $\eta$ pour les régions de Rindler
\begin{align}
x^\pm  &= \frac1{\sqrt{2}} t e^{\pm \beta \theta} \label{xpm1} \\
x^\pm  &= \pm \frac1{\sqrt{2}}  r e^{\pm \beta \eta}\ ,
\end{align}
on obtient les métriques suivantes ($ds^2 = -2 dx^+ dx^-$)
\begin{align}
ds^2_{\text{Milne}}  &= -dt^2 + \beta^2 t^2 d\theta^2\\
ds^2_{\text{Rindler}}  &= dr^2 - \beta^2 r^2 d\eta^2 \ .
\end{align}
à cause de l'identification sous le boost, les coordonnées $\theta$ et $\eta$ paramètrent des directions compactes de période $2\pi$.

Ainsi, la région définie par $x^+ < 0$ et $x^- < 0$, notée (P), décrit une cosmologie en contraction. Plus précisément, il s'agit d'un cercle $\theta \in [0, 2\pi]$ de rayon $\lvert\beta t\rvert$ qui se contracte donc linéairement. De même, la région définie par $x^+ > 0$ et $x^- > 0$, notée (F), décrit une cosmologie en expansion. Le rayon du cercle augmente linéairement. Les régions définies par $x^+ < 0$ et $x^- > 0$ d'une part et $x^+ > 0$ et $x^- < 0$ d'autre part, respectivement notées (L) et (R), sont des géométries indépendantes du temps avec un temps $\eta$ compact (de rayon $2\pi$ également). Ces régions de Rindler contiennent des \emph{courbes fermées de genre temps} qui sont habituellement considérées comme un défaut incurable de la géométrie considérée, ou éliminées par contre-réaction gravitationnelle\footnote{C'est la conjecture de protection chronologique défendue par S. Hawking \cite{Hawking:1991nk}.}. Cependant, le formalisme perturbatif est bien défini et nous ferons l'hypothèse que la production de particules due à la dépendance en temps modifie la géométrie de manière adéquate.

\subsection{Particules et cordes non twistées dans l'orbifold de boost}
\l{pcnt}

Comme nous l'avons expliqué dans la sous-section \ref{otypes}, dans un orbifold il faut distinguer les états twistés des états non-twistés. Ces derniers se comportent de façon similaire à des particules ponctuelles. En effet, les modes zéros de la corde fermée décrivent la trajectoire d'une particule ponctuelle, qui dans l'espace de recouvrement est une ligne droite
\be
X_{\text{zéro}}^\pm = x^\pm + p^\pm \t
\ee
Nous définissons une masse de cône de lumière $m^2_{\text{c.l.}} = 2 p^+ p^-$, qui inclut les impulsions transverses et les modes excités de la corde. Le moment cinétique dans la direction compacte est appelé \emph{moment cinétique de boost} $j=x^+ p^- + p^+ x^-$ et est contraint après quantification à être un multiple de $1/\beta$. Une particule massive avec une énergie positive ($p^+ > 0$, $p^- >0$) arrive de l'infini passé (dans la région (P)) à $\t = - \infty$ et atteint l'infini futur (dans la région (F)) à $\t = +\infty$ après s'être aventuré dans les régions de Rindler pendant un temps propre fini. Lorsque la particule, dans la région (P), s'approche du cône de lumière, sa vitesse angulaire $d\theta/dt \sim 1/t$ le long de la direction compacte croît jusqu'à l'infini. On s'attend alors à ce qu'un abondant rayonnement gravitationnel soit émis et puisse induire une contre-réaction importante qui mènerait à la création d'un large trou noir\footnote{Là aussi, nous faisons l'hypothèse que la production de particules est suffisamment importante pour modifier la géométrie avant que cela n'arrive.} \cite{Horowitz:2002mw}. Un observateur dans l'une des régions de Rindler à un instant $\eta$ fixé verrait un nombre infini de copies de la particule émerger et replonger dans la singularité de manière périodique. $j$ décrit alors l'énergie de Rindler de chacune de ces particules, qui se propagent jusqu'à une distance radiale finie $r=\lvert j \rvert /m_{\text{c.l.}}$.

Au niveau quantique, le mouvement du centre de masse d'une corde non twistée sans spin, qui correspond aux modes zéro, est décrit par une fonction d'onde solution de l'équation de Klein-Gordon dans la géométrie de Misner. Après avoir diagonalisé le moment cinétique de boost (la valeur propre est noté également $j$), le mouvement radial est gouverné par l'équation de Schrödinger suivante
\bse
\begin{align}
-\partial_x^2 - m^2 e^{2x} - j^2 &= 0 && t=\pm\sqrt{2x^+ x^-}=e^x \l{pMnt}\\
-\partial_y^2 + m^2 e^{2y} - j^2 &= 0 && r=\pm\sqrt{-2x^+ x^-}=e^y \l{pRnt}
\end{align}
\ese
Nous constatons ce que nous avons évoqué à la sous-section \ref{otypes} : la différence entre les cordes fermées libres et cordes fermées non twistées se situe non pas au niveau du développement en modes qui est identique \eqref{co} (de même que les relations de commutation), mais au niveau semi-classique.

La particule est réfléchie dans les régions de Rindler par un mur type Liouville, exponentiellement croissant, tandis qu'elle est accélérée dans un puits type Liouville dans les régions de Misner. Remarquons que dans les deux cas, l'origine de l'orbifold de boost est  située à distance infinie dans les système de coordonnées canoniquement normalisées, $x$ dans les régions de Milne, respectivement $y$ dans les régions de Rindler. Il est toutefois possible de définir des fonctions d'ondes entrantes (états \emph{in}) et sortantes (états \emph{out}) dans chaque région et de les étendre par continuation analytique à travers les horizons à $x^+ x^- = 0$ pour définir des fontions d'ondes globales.

D'autre part, nous pouvons définir des fonctions d'onde sur tout l'espace de Misner en superposant les images des fonctions d'onde dans l'espace plat. Comme nous l'avons vu à la sous-section \ref{ontwist}, la fonction d'onde s'écrit :
\begin{equation} \label{propoco}
f_{j,m^2,s}(x^+,x^-)=\int_{-\infty}^{+\infty}dv\ 
\exp\left(-i k^+ X^- e^{-v}
- i k^- X^+ e^{v} - i v j - v s \right) \ .
\end{equation}
Elle est caractérisée par la masse $m$, la valeur du moment cinétique de boost $j$ et nous considérons que le spin SO(1,1) dans $\R^{1,1}$, noté $s$, peut ne pas être nul\footnote{De manière équivalente, nous pouvons permettre que $j$ prenne des valeurs complexes.}.
Cette expression définit bien des fonctions d'ondes globales dans toutes les régions, à condition de déformer le contour d'intégration sur $v$ $(-\infty,+\infty) \to (-\infty+i\e,+\infty-i\e)$.

Remarquons en particulier qu'il n'existe pas de production de particule entre le vide entrant (adiabatique) à $t=-\infty$ et le vide sortant (adiabatique) à $t=+\infty$, alors qu'il y en a une entre le vide entrant (adiabatique) à $t=-\infty$ et le vide sortant (conforme) à $t= 0^-$. Ce dernier résultat découle de l'effet d'accélération/décélération infinie $d\theta/dt \sim 1/t$ près de la singularité.

\subsection{Cordes twistées dans l'orbifold de boost}
\l{ctob}

Les cordes twistées ne se comportent pas comme des particules ponctuelles, mais l'étude des modes zéro nous permettra, de la même fa\c con qu'à la sous-section précédente, de comprendre le comportement semi-classique de ces états de cordes. Avant cela, nous pouvons déjà dire, en lisant la condition de périodicité \eqref{corboost} pour $w \neq 0$, que dans les régions de Milne, les états twistés correspondent à des cordes qui s'enroulent $w$ fois autour de la direction compacte de genre espace paramétrée par $\theta$, alors que dans la région de Rindler ils correspondent à des cordes qui s'enroulent autour de la direction compacte de genre temps paramétrée par $\eta$. Si nous acceptons l'existence des boucles temporelles, une corde enroulée autour d'une direction temporelle ne pose pas de problème, il s'agit juste d'une superposition de $w$ cordes statiques ou temporellement périodiques et couvrant des distances radiales infinies (pour une feuille d'univers avec une topologie de cylindre).

Nous nous restreignons ainsi à la partie de la solution \eqref{orboost} comportant seulement les modes zéro\footnote{Dans ce qui suit, nous omettrons l'indice 0 associé à ces modes.}
\be
\l{x0}
 X^{\pm}_0(\t,\s) =
\frac1{\nu} e^{\mp \nu \sigma} \left[ \pm
\alpha^\pm_0 e^{\pm \nu \tau} 
\mp \tilde \alpha_0^\pm e^{\mp \nu \tau}\right]\ .
\ee
Nous invitons le lecteur à se reporter à la sous-section \ref{qobche} pour les détails de la quantification et les conditions de Virasoro. En se restreignant par souci de simplicité à $j=0$, on peut fixer le module des oscillateurs $\a^\pm$ et $\ta^\pm$ à $\mu/\sqrt{2}$. On distingue alors qualitativement deux types de cordes twistées
\begin{itemize}
\i Pour $\alpha^+\tilde\alpha^->0$, on obtient des configurations de \emph{corde courte},
\be
X^\pm_0 (\t,\s)= \frac{\mu\sqrt2}{\nu}\,\sinh(\nu\t)\,
e^{\mp \nu \s}
\ee
Ces cordes s'enroulent autour des cercles de genre espace dans les régions de Milne et se propagent de l'infini passé à l'infini futur (si $\a^+ >0$). Lorsque $j \neq 0$, elles s'étendent dans les régions de Rindler jusqu'à une distance radiale finie  $r_-^2=(M-\bM)^2/(4\nu^2)$, après que la feuille d'univers ait changé de signature.
\item Pour $\alpha^+\tilde\alpha^-<0$, on obtient des configurations de \emph{corde longue},
\begin{equation}
\label{x0t}
X^\pm_0(\sigma,\tau)= \pm\frac{\mu\sqrt2}{\nu}\,\cosh(\nu\tau)\,
e^{\mp \nu \sigma}
\end{equation}
Ces cordes se propagent seulement dans les régions de Rindler et s'enroulent autour du cercle de genre temps. Elles correspondent à des configurations statiques qui s'étendent de l'infini spatial dans (L) ou (R) jusqu'à une distance finie $r_+^2=(M+\tilde M)^2/(4\nu^2)$ et se replient à nouveau vers l'infini spatial.
La figure \ref{cutting} montrent les configurations de corde courte et longue

\begin{figure}[h]
\begin{center}
\epsfig{file=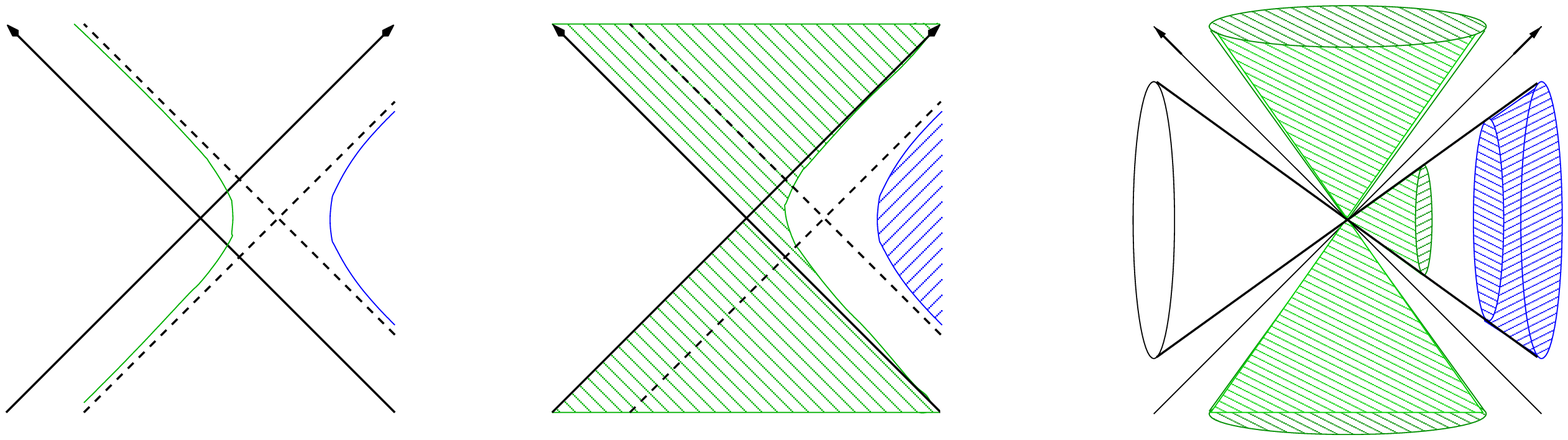,height=3cm}
\caption{\label{cutting}\small Les feuilles d'univers des cordes ferm\'ees (\`a droite) dans l'espace de Misner sont obtenues \`a partir des trajectoires des particules charg\'ees dans l'espace plat en présence d'un champ \'electrique constant (nous en avons soulign\'e l'analogie formelle \`a la sous-section \ref{qobche}). Les cordes courtes (en vert), respectivement longues (en bleu), correspondent aux particules charg\'ees qui franchissent, respectivement ne franchissent pas, l'horizon.}
\end{center}
\end{figure}

\end{itemize}
Contrairement aux orbifolds euclidiens, les cordes twistées ne sont pas localisées près de la singularité.

Dans la sous-section \ref{qobche}, nous avons écrit les équations semi-classiques \eqref{scct1} et \eqref{scct2} correspondant aux modes zéro sans en préciser le sens physique. L'analogie très proche avec la corde ouverte dans un champ électrique suggère d'adapter l'étude de la particule chargée dans un champ électrique au problème de la corde (ouverte ou fermée). C'est de cette fa\c con qu'émerge l'image d'oscillateur harmonique inversé : la trajectoire de la particule chargée sous l'effet du champ électrique peut être ramenée à un mouvement unidimensionel avec un potentiel inversé (harmonique pour un champ électrique uniforme et constant). Ainsi le spectre des états est un continuum d'états de diffusion qui correspondent à des électrons et des positrons réfléchis par le champ électrique et dont le mélange par effet tunnel n'est autre que la production de paires dans le vide. Pour adapter cette image aux cordes twistées, il faut cependant projeter le spectre sur les états invariants sous le boost. Pour cela, il est utile de considérer le problème d'une particule chargée vue par un observateur accéléré dans l'espace de Minkowski, c'est-à-dire un observateur statique dans l'espace de Rindler, car dans cet espace, $j$ est diagonalisé et la projection est alors plus facile à réaliser.

Les particules chargées dans l'espace de Rindler ont été étudiées dans \cite{Gabriel:1999yz} et les résultats, appliqués au cas de la corde fermée dans \cite{Berkooz:2004re}. Les trajectoires classiques sont des hyperboles dans l'espace de Minkowski, décrites en termes des coordonnées de Rindler ($y=e^r,\, \eta$). Pour $j$ fixé, le potentiel gouvernant le mouvement radial s'écrit
\be
\l{schror} V(y) = \frac{M^2 \bM^2}{\nu ^2} - \left( \frac{M^2+\bM^2}{2\nu} - \frac{\nu}{2} r^2 \right)^2
\ee
en terme des paramètres de corde fermée. Contrairement au cas de la particule neutre où nous avions un mur type Liouville (voir \eqref{pRnt}), le potentiel n'est pas borné inférieurement quand $r \to \infty$. Pour $j< M^2/(2\nu)$, condition qui peut être vérifiée par les états twistés non tachyoniques de la corde fermée, les régions asymptotiques $r=0$ et $r=\infty$ sont séparées par une barrière de potentiel (voir figure 3.5.
Les particules à droite ($ r \to \infty$) de la barrière correspondent à des électrons venant de ou retournant vers l'infini spatial de Rindler, alors que pour $j>0$ (respectivement pour $j<0$), les particules à gauche ($r \to 0$) de la barrière correspondent à des positrons (respectivement des électrons) émis ou absorbé par l'horizon de Rindler. L'effet tunnel décrit ainsi à la fois la production de paires dans le champ électrique ($j>0$) et l'émission Hawking de particules chargées par l'horizon ($j<0$).

Quant-au mouvement des particules chargées dans les régions de Milne, il est gouverné, à $j$ fixé, par le potentiel suivant
\be
\l{schrom} V(t) = \frac{M^2 \bM^2}{\nu ^2} - \left( \frac{M^2+\bM^2}{2\nu} + \frac{\nu}{2} t^2 \right)^2
\ee
à comparer à celui correspondant aux particules neutres (états non twistés) \eqref{pMnt}. Le potentiel est maximal et négatif à $t=0$, qui est à une distance infinie dans la coordonnée $x$. Le mouvement classique couvre ainsi tout l'axe temporel $t \in \R$.

Au niveau quantique, l'équation de Klein-Gordon dans les régions de Rindler est équivalente à une équation de Schrödinger dans le potentiel \eqref{schror} pour une énergie nulle. Les solutions peuvent être exprimées en terme de fonctions de Whittaker $M$ et $W$. Les modes \textit{in} et \textit{out} peuvent être définis dans chaque quadrant et prolongés analytiquement à travers les horizons. Par exemple, dans la région (R) de Rindler, les modes entrants à l'infini spatial $r = \infty$ sont représentés par
\be
\l{vinr}
{\cal V}_{in,R}^j = e^{-i j \eta} r^{-1} M_{-i(\frac{j}{2}-\frac{M^2}{2\nu}), -\frac{ij}{2} } ( i \nu r^2 /2 )
\ee
alors que les modes entrants à l'horizon de Rindler $r=0$ sont représentés par
\be
\l{uinr}
{\cal U}_{in,R}^j = e^{-i j \eta} r^{-1} W_{i(\frac{j}{2}-\frac{M^2}{2\nu}), \frac{ij}{2} } ( -i \nu r^2 /2 )
\ee
Les modes sortants peuvent être obtenus simplement par conjugaison de charge
\be
{\cal U}_{out,R}^j (r,\eta) = [ {\cal V}_{in,R}^j (r,-\eta) ]^*\ ,\quad
{\cal V}_{out,R}^j (r,\eta) = [ {\cal U}_{in,R}^j (r,-\eta) ]^*
\ee
Les vides \textit{in} et \textit{out} sont reliés par des transformation de Bogolioubov non triviales. Il y a donc une production de paires, dont les taux dans les régions respectivement de Rindler et de Milne sont
\begin{equation}
\label{prod}
Q_{\text{R.}} =
 e^{-\pi M^2/2\nu}
\frac{|\sinh \pi j|}{\cosh\left[ \pi \tilde M^2/ 2\nu\right]}\ ,\quad
Q_{\text{M.}} =
 e^{-\pi M^2/2\nu}
\frac{\cosh\left[ \pi \tilde M^2/ 2\nu\right]}{|\sinh \pi j|}\ ,
\end{equation}

\begin{center}
\epsfig{file=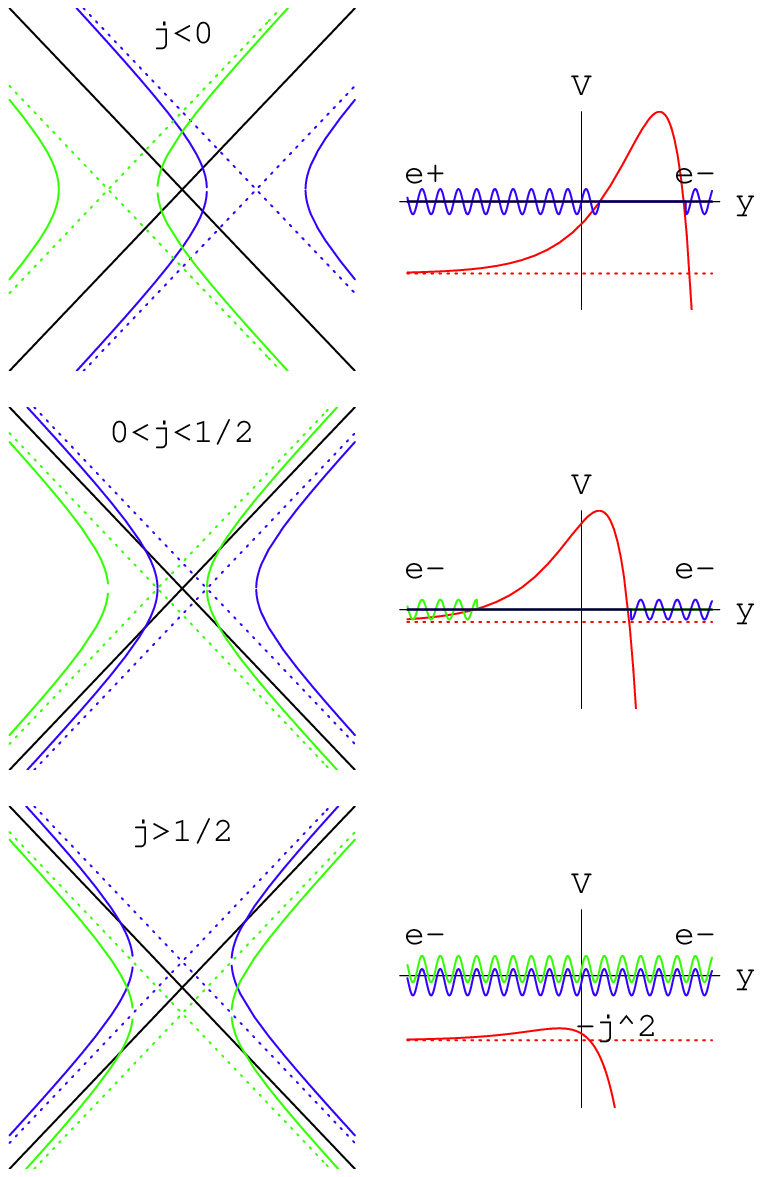,height=14cm}
\end{center}

\begin{quote}
{\sc Fig.} 3.5 --{\small 
\`A gauche : trajectoires classiques d'une particule charg\'ee dans l'espace de Misner (Rindler). $j$ désigne l'\'energie de Rindler ou l'impulsion de Milne et est mesur\'ee en unit\'es de $M^2/\nu$. \`A droite : le potentiel gouvernant le mouvement radial dans la r\'egion de Rindler droite (R), en fonction de la coordonn\'ee standard $y=e^r$.}
\end{quote}

Revenons aux cordes twistées. La fonction d'onde des modes zéro $\a_0^\pm$, $\ta_0^\pm$ est solution de la même équation de Klein-Gordon que la particule chargée, bien que seule la dépendance en $r$ soit pertinente. L'interprétation des solutions est par contre différent : au mouvement d'une particule à droite de la barrière de potentiel dans la région (R) de Rindler correspond une corde longue étirée de l'infini spatial de (R) à une distance radiale finie $r_+$ et repliée sur elle-même vers l'infini. De même, au mouvement d'une particule à gauche de la barrière correspond une corde courte étirée de la singularité à une distance radiale $r_-$. Par effet tunnel, on passe d'un type d'états à l'autre par une évolution en distance radiale imaginaire $r$, évolution qui peut être vue semi-classiquement comme une bande euclidienne étirée entre $r_-$ et $r_+$. Les fonctions d'onde dans les régions de Milne représentent simplement des cordes courtes entrantes ou sortantes à l'infini passé, futur ou près de la singularité.

Le calcul de la production de paires de cordes nécessite en principe de pouvoir définir des vides de seconde quantification, c'est-à-dire choisir une base d'états d'énergie positive et d'états d'énergie négative. Pour les cordes courtes dans les régions de Milne, il est possible de poser une définition claire, mais pour les cordes longues, cela est plus délicat, étant donné qu'elles ont une énergie de Rindler infinie et que cette définition dépend des conditions au bord à l'infini $r= \infty$. Cependant, comme une fonction d'onde globale dans l'espace de Misner peut être écrite comme un état appartenant au produit tensoriel des espaces des états des régions (L) et (R) de Rindler, ces régions sont susceptibles de fournir la formulation la plus naturelle. Ainsi, la dynamique de l'espace cosmologique entier serait décrit comme l'état d'une géométrie indépendante du temps, à ceci près qu'elle comporte des boucles temporelles.
Quoi qu'il en soit, il n'est pas nécessaire de comprendre en détail la définition rigoureuse de la seconde quantification des modes zéro des cordes twistés pour relier les composantes entrantes et sortantes des fonctions d'onde et calculer la production de paires pour des conditions au bord fixées à l'infini spatial de Rindler. En particulier, remarquons que le taux de production $Q_{\text{M.}}$ dans les régions de Milne est infini lorsque le moment cinétique de boost $j$ est nul. Ceci apparaît comme une conséquence de la singularité.

Remarquons finalement que ces résultats peuvent être obtenus en calculant la partie imaginaire de l'amplitude du vide à une boucle. Nous ne détaillerons pas cette approche qui sort du cadre de ce mémoire.

\subsection{Amplitudes à trois cordes}
\l{am3c}

Après avoir étudié les phénomènes physiques associés aux modes zéro de la corde, nous utilisons le formalisme perturbatif pour calculer quelques amplitudes. Nous commen\c cons par préciser ce que nous entendons par amplitude à trois cordes. Sauf si nous précisons le contraire, un état est implicitement le premier état du secteur, c'est-à-dire un état dont le nombre d'occupation des oscillateurs $\a_n$, $\ta_n$, $n>0$ de la corde est nul. Dans la théorie de la corde bosonique, il s'agit de l'état tachyonique, c'est pourquoi nous utiliserons abusivement la même dénomination pour les états twistés (dont l'énergie n'est pas forcément négative).
Nous appellerons \emph{amplitude à trois cordes} les amplitudes comportant deux états twistés tachyoniques et un état non twisté, notées $\langle 1 \rvert\, 0\, \lvert 3 \rangle$ si l'état non twisté est tachyonique et $\langle 1 \rvert\, T^{\mu\nu}\, \lvert 3 \rangle$ si l'état non twisté est le premier état excité (correspondant au dilaton, à la métrique et au champ antisymmétrique $B$). Les états 1 et 3 sont twistés et l'état non twisté est désigné par 2 (on notera par exemple $k_2$ l'impulsion entrante de la corde non twistée correspondante). Nous ne détaillerons pas le cas des amplitudes à trois états twistés, obtenues par continuation analytique des amplitudes obtenues dans une géométrie d'onde plane de Nappi-Witten et calculées dans le formalisme d'un modèle WZW \cite{D'Appollonio:2003dr}. L'analogie formelle entre l'espace de Misner et cette géométrie, ainsi que les expressions des amplitudes, sont données dans \cite{Berkooz:2004yy, Durin:2005ix}. Nous n'entrerons pas plus dans les détails pour ce calcul, car il sort du cadre du travail effectué au cours de la thèse. Nous ne parlerons pas non plus des amplitudes à trois états non twistés, qui ne diffèrent de l'amplitude à trois cordes libres dans l'espace plat que par un facteur $\delta_{-j_1 + j_2 + j_3,0}$ imposant la conservation du moment cinétique de boost.

Rappelons que l'ordre normal d'un opérateur de vertex non twisté dans une amplitude où les deux autres états sont twistés produit un facteur $t=\exp(-k_2^+ k_2^- \Delta)$ avec $\Delta(\nu)=\psi(1+i\nu) + \psi(1-i\nu) -2\psi(1)$. Nous avons exposé les données qui permettent d'en donner une interprétation physique. Comme ce facteur dépend de $w=-\nu/\beta$, il ne peut pas être éliminé par une redéfinition des champs, twistés ou non. Il s'agit plut\^ot du facteur de forme acquis par les états non twistés en présence d'une corde twistée, à cause des fluctuations quantiques de l'état de plus basse \'energie de cette dernière.

En effet la corde twistée polarise les états non twistés, qui forment alors un nuage de taille moyenne $\sqrt{\Delta(\nu)}$. Le comportement à grand twist de $\Delta(\nu)$ est logarithmique
\be
\Delta(\nu) \approx 2 \log \nu - \frac{23}{20} + {\cal O}(\nu^{-2})
\ee
ce qui peut être considérée comme la version T-duale de la croissance de Regge. Ce facteur évoque également un produit non commutatif, comme cela peut être vu plus clairement sur l'expression de l'amplitude à trois cordes.

L'amplitude $\A_0$ entre deux états twistés et un état non twisté tachyonique s'écrit alors, avec les conventions données à la sous-section \ref{aoa}
\be
\A_0 = e^{-k_2^+ k_2^- \Delta} \langle 1 \rvert e^{\xz(1)} e^{\xnf(1)} e^{\xpf(1)} \lvert 3 \rangle \l{a32t}
\ee
où, comme dans la sous-section \ref{ontwist}, nous avons sous-entendu $ik_2$ et l'expression complète de l'opérateur de vertex pour l'état non twisté \eqref{vont}.

L'amplitude se décompose ainsi en une valeur moyenne d'une expression ne comportant que des modes excités de la corde non twistée et en une valeur moyenne d'une expression ne comportant que des modes zéro. La valeur moyenne pour les modes excités est triviale : elle vaut 1. Pour écrire la valeur moyenne pour les modes zéros, que nous noterons également $\A_0$, il faut choisir une représentation pour ces modes. Si l'on choisit la \emph{représentation dans l'espace réel} ou \emph{représentation $x$} pour faire court, les opérateurs $\a^\pm$ et $\ta^\pm$ se mettent sous la forme donnée par \eqref{repx}. On obtient alors, en notant que $\xz(1)$ s'écrit $-i k_2^+ x^- - i k_2^- x^+$,
\begin{multline}
\A_0 (k_2^+, k_2^-) = \delta_{-j_1 + j_2 + j_3,0}\, \delta\left({\textstyle \sum_{i=1}^3} k_i^\perp\right)\\ e^{-k_2^+ k_2^- \Delta(\nu)} \int dx^+ dx^- f_1^*(x^+, x^-) e^{-i k_2^+ x^- - i k_2^- x^+} f_3(x^+, x^-) \l{a32tk}
\end{multline}
dans l'espace des impulsions ou encore
\be
\A_0 (x_2^+, x_2^-)= \delta_{-j_1 + j_2 + j_3,0}\, \delta\left({\textstyle \sum_{i=1}^3} k_i^\perp\right) e^{\Delta(\nu) \p_+ \p_-} f_1^*(x_2^+, x_2^-) f_3(x_2^+, x_2^-)  \l{a32tx}
\ee
avec $\p_\pm = \frac{\p}{\p x_2^\pm}$ et où $\delta_{-j_1 + j_2 + j_3,0}$ provient de l'intégrale sur $v_2$, le paramètre entrant dans la définition de l'opérateur de vertex \emph{intégré}. Le facteur $e^{\Delta(\nu) \p_+ \p_-} f_1^* f_3$ peut être considéré comme un produit non commutatif et constituerait ainsi le pendant de la non-commutativité de $x^+$ et $x^-$ pour une corde ouverte dans un champ électrique \eqref{rcE}.

Les fonctions $f_{1,3}$ doivent être remplacées par les fonctions d'ondes pour les états twistés. On peut \emph{a priori} étendre l'expession \eqref{a32tk} ou \eqref{a32tx} pour des fonctions d'onde d'états hors couche de masse, bien que la pertinence physique du résultat ne soit pas garantie. Par la suite, le facteur $\delta_{-j_1 + j_2 + j_3,0} \delta\left({\textstyle \sum_{i=1}^3} k_i^\perp\right)$ sera parfois sous-entendu.

Pour obtenir plus facilement un résultat explicite pour ces amplitudes à trois cordes, il est plus pratique d'utiliser une autre représentation des modes zéro : la \emph{représentation de mode} ou \emph{représentation $\a$, $\ta$}. Pour cela, nous diagonalisons la moitié des modes zéro (et nous confondons opérateur et valeur propre). Les fonctions d'onde entrantes sont des fonctions $f_{in}(\a^+,\ta^-)$ de $\a^+ \in \R^\eps$ et de $\ta^- \in \R^\beps$. Les états courts correspondent à $\eps \beps = +1$ tandis que les états longs, à $\eps \beps = -1$. Les opérateurs $\a^-$ et $\ta^+$ agissent sur $f_{in}$ de la façon suivante
\be
\a^-=i\nu\p_{\a^+}\ , \quad \ta^+=i\nu\p_{\ta^-}
\ee
De même les fonctions d'onde sortantes sont des fonctions de $\a^-$ et de $\ta^+$  $f_{out}(\a^-,\ta^+)$. Les opérateurs $\a^+$ et $\ta^-$ agissent sur $f_{out}$ de la façon suivante
\be
\a^+=-i\nu\p_{\a^-}\ , \quad \ta^-=-i\nu\p_{\ta^+}
\ee
Les fonctions d'onde sur la couche de masse s'écrivent
\bse
\begin{align}
f_{in}( \a^+,  \ta^- ) &= N_{in} \left(\eps \, \a^+ \right)^{\frac{M^2}{2i\nu} - \frac12} \left( \beps\, \ta^- \right)^{\frac{\bM^2}{2i\nu} - \frac12} \l{foin}\\
f_{out}( \a^-,  \ta^+ ) &= N_{out} \left( \eps \, \a^- \right)^{-\frac{M^2}{2i\nu} - \frac12} \left( \beps\, \ta^+ \right)^{- \frac{ \bM^2}{2i\nu} - \frac12} \l{foout}
\end{align}
\ese
Nous avons volontairement omis de distinguer $\eps$ et $\beps$ entre les états entrants et sortants car nous calculerons des amplitudes où ils sont identiques pour les deux états, par souci de simplicité. à partir du produit scalaire dans la représentation $x$
\be
\langle f_1 \vert f_2 \rangle = \int dx^+ dx^- \, f_1^*(x^+,x^-)~f_2(x^+,x^-)
\ee
on détermine le produit scalaire entre les états $in$ et $out$ dans la représentation $\a$, $\ta$
\be
{}_{out}\langle f_1 \vert f_2 \rangle_{in} = \int\!\! d\ta^+ d\a^- d\a^+ d\ta^-
f_1^*(\ta^+, \a^-) \, e^{\frac{i}{\nu}\left(\a^+ \a^- +  \ta^+ \ta^- \right)} f_2(\a^+, \ta^-)
\l{2pt}
\ee
Pour les expressions des normalisations $N_{in}$ et $N_{out}$, nous invitons le lecteur à se reporter à \cite{Berkooz:2004yy}. La partie de modes zéro de l'amplitude \eqref{a32t} peut s'écrire comme le produit des valeurs moyennes pour la partie de mode gauche et celle de mode droit
\be
\A^0_0 = e^{\frac{k_2^+ k_2^-}{i\nu}}
{}_{out}\langle M_1 \rvert e^{+\frac{i}{\nu} k_2^+ \a^-} e^{-\frac{i}{\nu} k_2^- \a^+} \lvert M_3 \rangle_{in} \;
{}_{out}\langle \bM_1 \rvert e^{+\frac{i}{\nu} k_2^- \ta^+ } e^{-\frac{i}{\nu} k_2^+ \ta^-} \lvert \bM_3 \rangle_{in}
\ee
avec un facteur supplémentaire (le premier) qui provient de l'utilisation de la formule de Baker-Campbell-Hausdorff. Dans cette expression nous avons anticipé le fait que les fonctions d'onde sur la couche de masse $f_1^{out}$, respectivement $f_3^{in}$, s'écrivent comme le produit d'une fonction de $\am$ caractérisée par $M_1$, respectivement de $\ap$ caractérisée par $M_3$, et d'une fonction de $\tap$ caractérisée par $\bM_1$, respectivement de $\tam$ caractérisée par $\bM_3$, comme on peut le constater sur les expressions \eqref{foin} et \eqref{foout}. Il faut finalement calculer les intégrales suivantes
\be
\begin{split}
\A^0_0 =&  \,e^{\frac{k_2^+ k_2^-}{i\nu}} \, \int d\ap d\am \left(\a^- \right)^{\frac{M^1}{2i\nu} - \frac12} e^{\frac{i}{\nu}\left(\a^+ \a^- - k_2^- \a^+ + k_2^+ \a^- \right)}
\left(\a^+ \right)^{\frac{M^3}{2i\nu} - \frac12} \\
& \int d\tam d\tap\left(\ta^+ \right)^{\frac{\bM^1}{2i\nu} - \frac12} e^{\frac{i}{\nu}\left(\ta^+ \ta^- + k_2^- \ta^+ - k_2^+ \ta^-\right)} \left(\ta^- \right)^{\frac{\bM^3}{2i\nu} - \frac12}
\end{split}
\ee
En intégrant d'abord sur $\ap$ et $\tam$  par exemple, on obtient deux intégrales de la forme
\be
\int_0^\infty dt\, t^{a-1} (1+t)^{b-a-1} e^{-zt} = \Gamma(a)\, U(a,b;z)
\ee
où $U$ est la fonction hypergéométrique confluente de Tricomi (voir l'annexe \ref{df}).
On obtient alors pour l'amplitude $\A_0$ le résultat suivant
\begin{multline}
\A_0 = \delta_{-j_1 +j_2 + j_3,0}\,\delta\left({\textstyle \sum_{i=1}^3} k_i^\perp\right)
e^{- k_2^+ k_2^- \tilde \Delta(\nu)}\\
\left(k_2^+\right)^{\mu-1} \left(k_2^-\right)^{\bar \mu-1} U\left(\lambda, \mu, i \frac{k_2^+ k_2^-}{\nu}\right) U\left(\bar \lambda, \bar \mu, i\frac{k_2^+ k_2^-}{\nu}\right)
\label{a32tcmplt}
\end{multline}
Les fluctuations des modes zéro et excités produisent une non-localité caractérisée par
\be 
\tilde \Delta(\nu) = \Delta(\nu) - \frac1{i\nu} = \psi(i\nu)+\psi(1-i\nu) - 2 \psi(1)
\ee
Les paramètres\footnote{Le paramètre $\mu$ ne doit pas être confondu avec la masse de cône de lumière $\mu$.} des fonctions $U$ sont
\be
\begin{aligned}
\lambda &= \frac12+\frac{M_3^2}{2i\nu} & \bar\lambda &= \frac12+\frac{\bM_3^2}{2i\nu} \\
\mu &= 1+\frac{M_3^2-M_1^2}{2i\nu} & \bar\mu &= 1+\frac{\bM_3^2-\bM_1^2}{2i\nu}
\end{aligned} \l{Uparam}
\ee

Nous considérons à présent $\A_1$, l'amplitude entre deux états twistés tachyoniques et un état non twisté de masse nulle. Nous définissons une amplitude $\A_1^{\mu\nu}$ correspondant à la polarisation $\e_{\mu\nu}$ de l'état non twisté
\be
\A_1 = \e_{\mu\nu} \A_1^{\mu\nu}
\ee
et qui s'écrit
\be
\A_1^{\mu\nu} = {}_{out}\langle f_1 \rvert \V_1^{\mu\nu} (1) \lvert f_3 \rangle_{in}
\ee
où l'expression de $\V_1^{\mu\nu}$ est donnée par \eqref{ovsym} et correspond à un état d'impulsion $k_2$.

Nous nous concentrerons sur les amplitudes $\A_1^{\ep\bep}$, $\A_1^{\ep j}$ et $\A_1^{i\bep}$. $\ep = \pm$, $\bep = \pm$ de façon indépendante ; $i$ et $j$ sont des indices correspondant à des directions transverses au cône de lumière . Les amplitudes $\A_1^{ij}$ sont obtenues comme le produit de l'amplitude \ref{a32tcmplt} et de l'amplitude \ref{a3g} avec les exposants $ij$. Considérons tout d'abord $\A_1^{\ep\bep}$. Cette amplitude s'écrit
\be
\A_1^{\ep\bep} = \delta\left({\textstyle \sum_{i=1}^3}  k_i^\perp\right)\, e^{ -\kp_2 \km_2 \Delta(\nu)} {}_{in}\langle f_1 \rvert T_{k_2}^{\ep, \bep} \lvert f_3 \rangle_{out}
\ee
où
\be
T_{k_2}^{\ep, \bep} = 
\left(\ta^{\bep}  +\bep i\nu k_2^\bep - \frac12 k_2^\bep  \right) e^{-i\km_2 \xzp - i\kp_2 \xzm}
\left(\a^\ep -\ep i\nu k_2^\ep + \frac12 k_2^\ep  \right)
\ee
Les termes $ i\nu k_2$ provient de l'ordre normal des modes excités, alors que les termes $\frac12 k_2$ apparaissent lorsqu'on fait commuter les modes zéro avec l'exponentielle en position centrale. Notons qu'ici la position des modes zéro convient à la représentation $x$, où les opérateurs $\a$ et $\ta$ sont représentés par les différentielles covariantes \eqref{repx}. L'amplitude, dans l'espace des positions, s'écrit alors
\begin{multline}
\A_1^{\ep\bep}(x_2^+, x_2^-) = \delta_{-j_1 + j_2 + j_3,0} \delta\left({\textstyle \sum_{i=1}^3} k_i^\perp \right) \\
e^{\Delta(\nu) \p_+ \p_-} \left(\overline{\nabla}_{\bep} -\bep i\nu \p_\bep + \frac12 \p_\bep  \right)  \left(\overline{\nabla}_{\ep} +\ep i\nu \p_\ep - \frac12 \p_\ep\right) f_1^* f_3
\end{multline}
On peut également adopter la représentation de mode, qui permet d'obtenir des expressions en terme de fonctions hypergéométriques $U$. Dans ce cas, l'ordre des modes zéro est différent selon les signes de $\ep$ et $\bep$
\begin{align}
T^{++} &= t \left(\tap -\frac12 (1-2 i\nu) \kp_2 \tun \right) {\cal E} \left(\ap +\frac12(1-2 i\nu) \kp_2\un \right)\\
T^{+-} &= t {\cal E} \left(\tam +\frac12 (1-2 i\nu) \km_2 \tun\right)  \left(\ap +\frac12(1-2 i\nu) \kp_2\un \right)\\
T^{-+} &= t \left(\tap -\frac12 (1-2 i\nu) \kp_2\tun \right)  \left(\am -\frac12(1-2 i\nu) \km_2\un \right) {\cal E}\\
T^{--} &= t \left(\am -\frac12(1-2 i\nu) \km_2\un \right) {\cal E} \left(\tam + \frac12 (1-2 i\nu) \km_2\tun \right)
\end{align}
où 1 et $\tilde 1$ représentent l'opérateur identité (le nombre 1) et servent à distinguer la partie holomorphe 1 et la partie antiholomorphe $\tilde 1$ et ${\cal E} = e^{-i\km_2 \xzp - i\kp_2 \xzm}$.

On peut construire l'expression pour l'amplitude en suivant les règles suivantes
\begin{enumerate}
\i retirer l'exponentielle en position centrale
\i remplacer les opérateurs $\a^\pm$, $\ta^\pm$, 1 et $\tilde 1$ par une fonction appropriée
\begin{itemize}
\i 1 par $(k_2^+)^{\mu-1} U(\lambda, \mu, \frac{i}{\nu} k_2^+ k_2^-)$
\i $\ap$ par $(k_2^+)^{\mu} U(\lambda+1, \mu+1, \frac{i}{\nu} k_2^+ k_2^-)$
\i $\am$ par $(k_2^+)^{\mu-2} U(\lambda, \mu-1, \frac{i}{\nu} k_2^+ k_2^-)$
\i $\tilde 1$ par $(k_2^-)^{\tilde \mu-1} U(\tilde \lambda, \tilde \mu, \frac{i}{\nu} k_2^+ k_2^-)$
\i $\tap$ par $(k_2^-)^{\tilde \mu-2} U(\tilde \lambda, \tilde \mu-1, \frac{i}{\nu} k_2^+ k_2^-)$
\i  $\tam$ par $(k_2^-)^{\tilde \mu} U(\tilde \lambda+1, \tilde \mu+1, \frac{i}{\nu} k_2^+ k_2^-)$
\end{itemize}
où les paramètres des fonctions $U$ sont définis comme précédemment \eqref{Uparam}
\i ajouter aux impulsions un poids exponentiel pour reconstruire la forme invariante de boost de l'opérateur de vertex : $k_2^\pm \to k_2^\pm e^{\mp v_2}$ et ajouter un facteur $e^{-ij_2 v_2}$ dans l'expression
\i ajouter un facteur $e^{(\eps+\beps)v_2}$ à l'expression et intégrer sur $v_2$
\end{enumerate}
Ces deux dernières étapes produisent le terme de conservation du moment cinétique de boost $\delta_{-j_1 + j_2 + j_3,0}$, comme dans les amplitudes précédentes. Un exemple d'amplitude peut être trouvé dans \cite{Berkooz:2004yy}. Notons que ces règles sont les mêmes pour des amplitudes à plus de trois cordes, une fois déterminé l'équivalent de $T$.

Pour les amplitudes $\A_1^{\ep j}$, respectivement $\A_1^{i\bep}$, il suffit d'appliquer les résultats précédents à la partie holomorphe, respectivement antiholomorphe, de l'amplitude et de multiplier par $(k_1-k_3)^i$, respectivement $(k_1-k_3)^j$, facteur qui correspond à l'autre partie. Nous ne détaillerons pas plus cette situation.

\subsection{Amplitudes à quatre cordes}
\l{am4c}

Nous nous intéressons aux amplitudes comportant deux états twistés et deux états non twistés dans le formalisme des opérateurs. Les techniques présentées pour les amplitudes à trois cordes s'appliquent tout aussi bien pour ces amplitudes. Nous obtenons le résultat suivant
\begin{multline}
\label{a4}
 \langle 1 \vert  T(2)  T(3) \vert 4 \rangle = 
\delta_{-j_1 +j_2+j_3+j_4,0} \,\delta\left(\textstyle{\sum_{i=1}^4}
 k_i^\perp \right)\; \int_{-\infty}^{\infty} \!\! dv  ~ e^{i j_3 v}\;\int \! dz d\bar z \\
\lvert 1- z \rvert^{2 k_2^\perp\cdot k_3^\perp}
~\lvert z\rvert^{2 k_3^\perp \cdot k_4^\perp  +k_3^\perp \cdot k_3^\perp -2}
\exp\left[ - \left( k_2^+ k_2^- + k_3^+ k_3^- \right) \tilde \Delta(\nu) \right] \\
\exp\left[
k_2^- k_3^+ e^{-v}
\left( G(z;\nu) +  G(\bar z;-\nu) + \frac{i}{\nu} \bar z^{-i\nu} \right)
+k_2^+ k_3^- e^{v}
\left( G(z;-\nu) +  G(\bar z;\nu) + \frac{i}{\nu} z^{-i\nu} \right)
\right]\\
\left[ 
k_2^+ +k_3^+ e^{-v} z^{i\nu} 
\right]^{\mu-1}
 \left[
k_2^-+k_3^- e^v {\bar z}^{i\nu}
\right]^{\tilde \mu-1} \,
U\left(\lambda, \mu,\frac{i}{\nu} A(v, z)\right) U\left(\tilde\lambda,\tilde
  \mu,\frac{i}{\nu}\tilde A(v, \bar z)\right)
\end{multline}
où $(\lambda,\nu,\tilde\lambda,\tilde\mu)$ ont la même expression \eqref{Uparam} que pour les amplitudes à trois cordes à condition de renuméroter la particule $3$ en $4$. Nous avons défini
\begin{align}
A(v,z) &= (k_2^-+k_3^- e^{v} z^{-i\nu} )(k_2^++k_3^+  e^{-v}  z^{i\nu} ) \\
\tilde A(v,\bar z) &= (k_2^-+k_3^- e^{v} \bar z^{i\nu} )(k_2^++k_3^+  e^{-v}  \bar z^{-i\nu} )
\end{align}
et
\be
G(z;\nu) = \sum_{n=1}^\infty \frac{z^{n+i\nu}}{n+ i\nu} = \frac{z^{1+ i\nu}}{1+ i\nu}\,{}_2 F_1 \left(1+i\nu, 1;2+ i\nu;z\right)
\ee
La valeur moyenne dans \eqref{a4} a été évaluée pour $|z|<1$, mais le même résultat est valable lorsque
$|z|>1$ à cause des propriétés des fonctions hypergeometriques (voir équation \eqref{hypinv} dans l'annexe \ref{df}).

Lorsque $z \to 0$ l'amplitude peut se mettre sous la forme
\begin{multline}
\label{a40}
\langle 1 \vert  T(2)  T(3) \vert 4 \rangle  \to \delta_{-j_1 +j_2+j_3+j_4,0} \delta\left(\textstyle{\sum_{i=1}^4} k_i^\perp \right)\; \int_{-\infty}^{\infty} \!\! dv \,e^{i(j_3-j_1)v} \; \int\! dz d\bar z \\
\lvert z\rvert^{2 k_3^\perp   \cdot k_4^\perp + k_3^\perp \cdot k_3^\perp -2}
\exp\left[ - \left( p_2^+ p_2^-+p_3^+ p_3^- \right) \tilde \Delta(\nu)  \right] \\
(-1)^{\mu+\tilde\mu} \,(k_2^+)^{-\tilde\lambda}\,(k_2^-)^{-\lambda}\,(k_3^+)^{\mu-\lambda-1}\,(k_3^-)^{\tilde\mu-\tilde\lambda-1}\,z^{-\frac12 M_1^2 - \frac{i\nu}{2}}\,
\bar z^{-\frac12 \tM_1^2 - \frac{i\nu}{2}}
\end{multline}
où nous avons utilisé le comportement asymptotique de $U$, donné par \eqref{asymptU}, lorsque $z\to 0 $.

L'amplitude diverge lorsque $j_3 = j_1$ à cause de la propagation de cordes twistées avec un moment cinétique de boost nul dans le canal intermédiaire. Ceci est similaire au problème discuté dans l'article \cite{Berkooz:2002je}, où les amplitudes de diffusion à quatre cordes non twistées à l'ordre des arbres divergent à cause d'un abondant échange de gravitons près de la singularité.

\subsection{Contre-réaction gravitationnelle}
\l{cr}

Il est impossible avec les techniques actuelles de calculer la contre-réaction provoquée par la production quantique de particules et de cordes. Par contre, la réponse linéaire des champs de corde fermée à la pre\'sence d'un condensat classique de cordes twistées peut être calculée dans le formalisme perturbatif. Ajoutons à l'action de la feuille d'univers un condensat d'opérateurs de twist marginaux
\be
\S_\lambda = \int d\t d\s \p X^+ \bp X^- + \lambda_{-w} V_{+w} + \lambda_{+w} V_{-w}
\ee
$w$ indique le secteur twisté mis en jeu. Ceci correspond à la déformation de l'espace de Misner hors du point d'orbifold. Cette déformation est marginale à l'ordre dominant, mais elle induit une fonction à un point non nulle pour les champs non twistés
\be
\l{3p1}
\langle e^{ikX} \rangle_\lambda \sim \lambda_{w} \lambda_{-w}
\langle w |  e^{ikX} | -w \rangle\ ,\quad
\ee
qui doit être annulée en ajoutant à $\S_\lambda$ un terme d'ordre $\lambda^2$ faisant intervenir le champ non twisté. Ce champ est précisément le champ dont la corde twistée avec un opérateur de vertex $V_{\pm w}$ est la source classique. D'autre part, cette même corde twistée est une source pour les états twistés dans les secteurs multiples de $w$
\be
\l{3p2}
\langle V_{-2w} \rangle_\lambda \sim \lambda_{w} \lambda_{w}
\langle w |  V_{-2w} | w \rangle\ ,\quad
\ee

Grâce aux expressions des amplitudes que nous avons données à la sous-section \ref{am3c}, nous pouvons en principe calculer la contre-réaction gravitationnelle (et celle liée aux autres champs) au premier ordre en $\lambda_{w}$, mais nous n'avons pas entrepris ces calculs. Cette procédure permet dans le cas de l'orbifold euclidien de régulariser une singularité conique de type ALE, remplacée par un instanton gravitationnel de type Eguchi Hanson et dont la géométrie est régulière \cite{Adams:2001sv}. Nous n'avons pas répondu à la question de savoir si cela fonctionne aussi pour la singularité de l'espace de Misner, mais des avancées significatives ont été accomplies dans cette direction \cite{Hikida:2005ec, Hikida:2005xa}. Nous en reparlerons au chapitre \ref{persp}.

\section{S-brane et ajustement fin des conditions initiales}
\l{article3}

\subsection{Le concept de S-brane}
\l{Sbranes}

Nous terminons par notre étude de la S-brane de Dirichlet, configuration dépendante du temps particulièrement simple, mais pas aussi simple à comprendre. Nous avons évoqué les problèmes liés à l'existence de S-branes émettant un champ R-R à la sous-section \eqref{sbr}. Nous avons cherché au travers de différentes approche de cerner mieux la nature de cet objet.

Il est nécessaire de préciser \gl S-brane de Dirichlet\gr, car le mot S-brane désigne plusieurs constructions assez différentes en théorie des cordes. Une S-brane désigne de manière générique un défaut topologique localisé sur une hypersurface de genre espace. Cette définition est la contrepartie de la définition première des branes, qui sont localisées sur une hypersurface de genre temps et qui ont été identifiées comme telle dans les différentes théories effectives de supergravité. Le terme \gl S-brane\gr\  a tendance à être appliqué à toute solution dépendante du temps, localisée à un instant donnée (bien que cette localisation puisse être étendue à un intervalle). Le lecteur peut se référer à la revue \cite{Ohta:2004wk} où ces solutions sont examinées dans le contexte de l'inflation.

Les S-branes sont l'un des éléments utilisés pour construire des géométries inflationnaires en théorie des cordes, de même qu'une D-brane sert de source pour les géométries \emph{anti de Sitter} (noté AdS) \cite{Hull:1998vg}. Dans des géométries dépendentes du temps plus générales, elles peuvent être utilisées comme des sondes \cite{Nekrasov:2002kf} ou comme des surfaces de conditions initiales et/ou finales \cite{Nitti:2005ym}.

En théorie des cordes, il existe principalement deux constructions de S-brane. Tout d'abord, nous avons vu dans le chapitre précédent (section \ref{sdbranes}) que la S-brane de Dirichlet\footnote{Elle a été étudiée en premier par les auteurs de \cite{Gutperle:2002ai}.} est obtenue en imposant des conditions au bord de Dirichlet à la coordonnée $X^0$ des cordes ouvertes. Ensuite, il est possible de construire une S-brane sur une D-brane instable. L'évolution d'une D-brane instable est décrite par adjonction d'un terme de bord tachyonique à l'action de corde ouverte. Ce terme correspond à un potentiel en forme de cloche sur lequel le tachyon de la corde ouverte \gl roule\gr\ depuis ou vers le vide de corde fermée (voir \cite{Sen:2004nf} pour une revue du sujet). Une S-brane dans ce contexte est une solution de type kink\footnote{Nous conservons le terme anglais.} temporel sur la D-brane instable. Les deux types de S-branes ne sont pas complètement étrangers l'un à l'autre : dans certains cas, on peut voir une S-brane sur une D-brane instable comme un réseau de S-branes de Dirichlet en temps imaginaire \cite{Gaiotto:2003rm}.

Nous avons vu à la section \ref{brane} que les S-branes de Dirichlet émettant un champ R-R et un champ NS-NS réels n'existaient pas en théorie de type II. Il est donc utile d'évaluer la \gl stabilité spatiale\gr\ des S-branes existantes (qui n'émettent qu'un champ NS-NS), de même qu'on évalue la stabilité temporelle des D-branes en calculant notamment le taux de désintégration d'une paire de D-brane. L'équivalent du taux de désintégration peut être calculé pour une S-brane de Dirichlet et nous exposerons cela à la sous-section \ref{dr}. Nous commençons par exposer le lien entre une configuration de S-brane et une configuration de champ électrique et les enseignements qu'il nous permet de tirer. 

\subsection{S-brane et champ \'electrique surcritique}

D'après la définition première que nous avons donnée des S-branes, on voit que le terme S-brane de Dirichlet s'applique à toute condition au bord de forme
\be
\dpt(X^1 - v X^0) = 0 \text{\ \ ,\ \ } \dps(X^0 - v X^1) = 0
\ee
où $X^1$ est une coordonnée spatiale arbitraire et $\lvert v \rvert > 1$. Pour $\lvert v \rvert < 1$, nous avons à faire à une D-brane et pour $\lvert v \rvert = 1$ il s'agit d'une \emph{brane extrêmale}, c'est-à-dire une D-brane de genre lumière. Une D-brane immobile est obtenue en posant $\lvert v\rvert=0$ tandis qu'une S-brane fixée à un instant précis correspond à $\lvert v\rvert  \to \infty$. Nous considérons seulement le cas de la S-brane $\lvert v \rvert > 1$. Les conditions au bord correspond à la configuration obtenue par T-dualité sur la coordonnée $X^1$ s'écrivent
\be
\dps X^1 - v \dpt X^0 = 0 \text{\ \ ,\ \ } \dps X^0 - v \dpt X^1 = 0
\ee
ce qui, comme on peut le vérifier en se reportant à l'équation \eqref{cbem}, correspond à un champ électrique \emph{surcritique}\footnote{L'action de Born-Infeld \eqref{DBIact}, schématiquement $\sqrt{1-F^2}$, impose \emph{a priori} la valeur limite 1 (un facteur $2\pi\apr$ est inclus dans $F$) à l'intensité du champ électrique $F$. Par ailleurs, des configurations de champs surcritiques ont été récemment étudiées dans \cite{Dorn:2005vg}.} $F_{01} = v$, constant et uniforme, défini sur la surface d'univers d'une D-brane dans les directions $X^0$, $X^1$. Rappelons que par T-dualité sur la coordonnée temporelle $X^0$, on obtiendrait une configuration avec un champ électrique \emph{sous-critique} $F_{01} = \frac1v$ mais de théorie II${}^\star$ \cite{Hull:1998vg}, théorie qui reste mal comprise.

Un tel champ électrique surcritique surmonte la tension des cordes ouvertes qui sont étirées jusqu'à atteindre une longueur infinie. Lorsque $\lvert v\rvert  \to \infty$, la configuration de feuille d'univers qui décrit ce processus est définie par, dans l'approximation de corde rigide,
\begin{align}
X^0 = x^0 + w\s \\
X^1 = x^1 + \frac1F w \t \\
X^\perp = 2\apr p^\perp \tau \l{sbrws}
\end{align}
Nous avons supposé que les deux extremités des cordes ouvertes obéissaient aux mêmes conditions au bord.
Les conditions de Virasoro imposent
\be
w^2 = 4\apr(\apr p_\perp^2 + (N-a)) \frac{F^2}{F^2-1}
\ee
où $a$ est la constante d'ordre normal (1 en théorie bosonique, $\frac12$, respectivement 0, en théorie de supercordes dans le secteur de Neveu-Schwarz, respectivement de Ramond).
Si l'on ajoute les fluctuations à la configuration rigide décrite par \eqref{sbrws}, la feuille d'univers peut être vue comme un processus de nucléation de dipoles qui s'étirent et forment rapidement une corde de longueur infinie (voir figure \ref{stretchos}). Ce processus décharge les plaques du condensateur qui produit le champ électrique et ramène celui-ci à une valeur critique ou sous-critique. Dans la configuration T-duale, on imagine alors que la S-brane est courbée de façon à former une brane extrêmale ou même des paires D-brane, anti-D-brane à partir d'une certaine distance.

\begin{figure}[h]
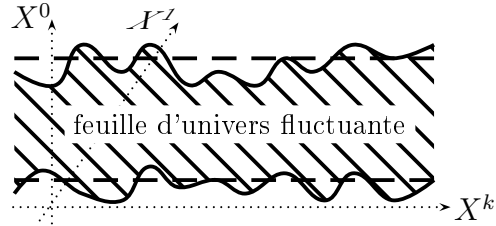

\vspace{2.3cm}
\begin{center}
\rput{0}(-1.8,0){%
  \psset{dotsep=2pt}
  \rput[b]{-35}(0.1,0){%
     \psline[linecolor=black, linestyle=dotted]{->}(0,-.3)(0,3)
     \rput[r]{*0}(0,3.05){\pstilt{55}{\textcolor{black}{$X^1$}}}
  }
  \scalebox{1.8}{%
    \pscustom[linecolor=black]{%
      \pscurve(-.2,.1)(0,.27)(.3,.1)(.6,.05)(.8,.3)(1,.15)(1.3,.2)(1.5,.1)(1.8,.25)(2.1,0.05)(2.3, .25)(2.6,.05)(2.9,.2)   
      \gsave
        \pscurve[liftpen=1](2.9,1.2)(2.6,1.1)(2.3,1.2)(2,1)(1.8,1.1)(1.6,.9)(1.3,1.)(1.1,.9)(.8,1.2)(.6,1)(.3,1.2)(.1,.9)(-.2,1.)
        \fill[fillstyle=vlines]
      \grestore
    }
    \pscurve[linecolor=black]{-}(2.9,1.2)(2.6,1.1)(2.3,1.2)(2,1)(1.8,1.1)(1.6,.9)(1.3,1.)(1.1,.9)(.8,1.2)(.6,1)(.3,1.2)(.1,.9)(-.2,1.)
    \psline[linecolor=black, linestyle=dashed]{-}(-.2,.2)(2.9,.2)
    \psline[linecolor=black, linestyle=dashed]{-}(-.2,1.1)(2.9,1.1)
  }
  \psline[linecolor=black, linestyle=dotted]{->}(-.6,0)(5.3,0)
  \rput[l]{0}(5.35,0){\textcolor{black}{$X^k$}}
  \psline[linecolor=black, linestyle=dotted]{->}(0,-.3)(0,2.5)
  \rput[r]{0}(0,2.55){\textcolor{black}{$X^0$}}
  \rput{0}(2.5,1.1){\psframebox[fillstyle=solid, linestyle=none]{\small feuille d'univers fluctuante}} 
}
\end{center}
\caption{\l{stretchos}\small La feuille d'univers dans l'approximation rigide est délimitée par les lignes en pointillés tandis que sans approximation elle est délimitée par les lignes continues. Si nous suivons les points de la feuille d'univers à $X^0$ constant lorsque $X^0$ augmente, nous constatons l'apparition de dipôles comportant deux charges $e_0$ (nous considérons le bord inférieur de la feuille d'univers). Ces dipôles coalescent rapidement pour former une corde de longueur \gl infinie\gr, qui décharge le condensateur.}
\end{figure}

Ainsi l'image de contre-réaction électrique par création de paires suggère une contre-réaction gravitationnelle induite par la S-brane, ce que nous pouvons examiner en calculant la fonction de partition à une boucle.

\subsection{Amplitude du vide à une boucle, longueur de corrélation de S-brane}
\l{dr}

Précisons un peu la signification physique de la fonction de partition. Pour deux D-branes statiques, l'amplitude du vide à une boucle est égale à l'énergie d'interaction entre les deux D-branes, multipliée par un facteur infini correspondant à la durée infinie de l'interaction. En effet, si l'on considère l'amplitude comme l'amplitude à l'ordre des arbres pour des cordes fermées se propageant entre les deux D-branes, celle-ci s'écrit schématiquement
\be
\cZ = \frac{i}2 \int d^D x\, d^D y\, J_1(x) G_F(x-y) J_2(y) 
= T E_{\text{int}}
\ee
où $G_F$ est le propagateur de Feynman,
\be
J_1(x^0, x^i)=q_1 \delta(x^i-x^i_1)\ ,\quad J_2= q_2 \delta(x^i-x^i_2)
\ee
sont les sources représentants les D-branes au repos et $E_{\text{int.}}$ est l'énergie d'interaction.
Ce résultat en terme d'énergie d'interaction se justifie en quantifiant le long de la direction temporelle $x^0$. Si l'amplitude, et donc cette énergie d'interaction, a une partie imaginaire non nulle, l'état de paire de D-branes a une durée de vie finie : le taux de désintégration est de l'ordre de $1/\im E_{\text{int}} [J]$. Ceci a été appliqué pour calculer le taux de désintégration des D-branes instables dans \cite{Bardakci:2001ck, Craps:2001jp}.
La même interprétation est valable pour les S-branes à condition de considérer l'amplitude comme le résultat d'une quantification le long d'une direction spatiale commune aux deux S-branes, $x^1$ pour la définir. Alors l'amplitude du vide à une boucle correspond au produit entre une longueur infinie d'interaction $L$ et une impulsion d'interaction le long de $x^1$, due aux échanges de cordes fermées entre les deux S-branes.
\be
\cZ = L P_{\text{int}}
\ee
Si l'impulsion d'interaction présente une partie imaginaire non nulle, cela est interprété comme une \gl désintégration\gr\ au cours d'une évolution \gl temporelle\gr\ selon $x^1$, après une distance caractéristique $1/\im P_{\text{int}} [J]$. Nous reviendrons sur cette interprétation après le calcul.

Nous considérons deux S-branes de Dirichlet, fixées  à des instants respectivement $x^0_0$ et $x^0_1$ ainsi qu'à des positions respectivement $x_0^i$ et $x_1^i$. Nous rappelons que $\Delta$ est la séparation entre les S-branes et nous supposerons qu'elle est de genre temps.
\be
\Delta^2 = (x^0_1 - x^0_0)^2 - (x^i_1 - x^i_0)^2
\ee
En appliquant l'expression \eqref{Lbddbr} \`a notre configuration de S-branes, nous pouvons \'ecrire la relation de couche de masse suivante
\be
p_k^2 = \frac{\Delta^2}{4\pi^2 \apr^2} + \frac{a-N}{\apr} \\
\ee
o\`u $k$ d\'esigne les directions parall\`eles \`a la S-brane et $a$ la constante d'ordre normal ($a=\frac12$ dans le secteur NS et $a=0$ dans le secteur R). Nous constatons qu'il existe une infinit\'e d'\'etats tachyoniques et seulement un nombre fini d'\'etats physiques (car nous avons suppos\'e que $\Delta^2 >0$). L'interpr\'etation de ces \'etats tachyoniques n'est pas claire, mais nous pouvons en voir l'effet sur l'amplitude du vide.

La seule différence entre l'amplitude pour les S-branes et celle pour les D-branes se trouve dans la partie concernant les modes zéro. En effet, dans le canal de corde ouverte,
\be
\l{olsbr} \cZ = \frac12 V_{p+1} \int_0^{\infty}\! \frac{dt}{t} \int \!\!
\frac{d^{p+1} p_k}{(2\pi)^{p+1}}\, e^{- t [ 2\pi \apr p_k^2 - 
\Delta^2/(2\pi \apr)] }\, \cZ^* 
\ee
où
\be
\cZ^*  = \frac{ -16 \prod_{n=1}^{\infty} (1+q^{2n})^8 
+ q^{-1} \prod_{n=1}^{\infty} (1+q^{2n-1})^8}
{\prod_{n=1}^{\infty} (1-q^{2n})^8}
\ee
avec  $q = e^{-\pi t}$. Le premier produit infini correspond à la fonction de partition des modes excités des cordes ouvertes dans le secteur de Ramond, le second à la fonction de partition des modes excités dans le secteur de Neveu-Schwarz (les deux sont sans insertion de projecteur). Ces deux termes correspondent, dans le canal de corde fermée à l'échange de cordes fermées dans le secteur NS-NS uniquement. Dans le cas des D-branes émettant un champ R-R, il existe un troisième terme, correspondant dans le canal de cordes ouvertes à la fonction de partition des modes excités dans le secteur de Neveu-Schwarz et dans le canal de cordes fermées à la fonction de partition des modes excités dans le secteur R-R, mais comme nous avons vu qu'il n'existait pas de S-branes émettant de champ R-R en théorie de type II, du moins dans le cadre du formalisme perturbatif, ce terme n'est pas présent dans l'expression \eqref{olsbr}.
L'identité abstruse de Jacobi permet de remplacer les deux termes de produits infinits par le terme du secteur R-R dont l'expression est plus simple, ce qui nous facilitera l'étude ultérieure. Ainsi, l'amplitude s'écrit pour deux S\pbr\ 
\be
\l{olsbrR} \cZ = \frac12 V_{p+1} \int_0^{\infty} \frac{dt}{t} \int
\frac{d^{p+1} p_k}{(2\pi)^{p+1}} e^{- t [ 2\pi \apr p_k^2 - 
\Delta^2/(2\pi \apr)] }  q^{-1} \frac{ \prod_{n=1}^{\infty} (1-q^{2n-1})^8}
{\prod_{n=1}^{\infty} (1-q^{2n})^8} \ .
\ee 
Malgré les apparences, le secteur R-R est bien absent !

Une différence avec l'amplitude pour les D-branes mérite d'être relevée : l'intégrale sur les impulsions $p_k$ correspondant aux directions du volume d'univers de la S-brane est convergente sans qu'on ait besoin de faire une continuation analytique. D'autre part, et il s'agit maintenant d'un point commun, nous avons une divergence ultraviolette quand $t\to 0$, qui correspond à l'échange d'états de masse nulle de corde fermée.

Point crucial, il existe des divergences infrarouges qui s'aggravent lorsque la séparation $\Delta$ augmente. Ces divergences ont en réalité un sens physique bien précis qui est mis en évidence par la régularisation préconisée dans \cite{Marcus:1988vs}. En effet, elles correspondent en fait à une partie imaginaire non nulle pour l'amplitude des S-branes
\be 
\l{imolsbr}
\im \cZ \propto \sum_{n=-1}^{n^*} a_{n} 
\left(\frac{\Delta^2}{2\pi^2\alpha'}-n\right)^{\frac{p+1}2} \ee
où $n^*$ est la partie entière de $\frac{\Delta^2}{2\pi^2\apr}$
et $a_n$ sont les coefficients de Fourier de la forme modulaire suivante
\be
\frac{\vartheta_4^4}{\eta^{12}} 
=q^{-1}\prod_{n=1}^{\infty}\frac{(1-q^{2n-1})^8}{(1-q^{2n})^8}
:=\sum_{n=-1}^{\infty} a_n q^n  \qquad (q=e^{-\pi t})
\ee
Ces coefficients sont donnés par la formule de Rademacher (nous invitons le lecteur à se référer à \cite{Dijkgraaf:2000fq} et aux références qui y sont données). En tronquant la somme de Kloosterman au premier terme nous obtenons\footnote{Le facteur $(-1)^{n+1}$ provient justement de ce premier terme de la somme de Kloosterman.} une très bonne approximation
\be 
a_n = \pi (-1)^{n+1}  n^{-5/2} I_5 \left(2\pi \sqrt{n}\right) \sim (-1)^{n+1} n^{-11/4} e^{2\pi\sqrt{n}}
\ee
Notons en particulier que la série des $a_n$ est alternée. Le comportement de croissance exponentielle de ces coefficients, appelée \emph{croissance de Hagedorn}, permet d'approximer la série de \eqref{imolsbr} par le dernier terme non nul
\be
\l{imapp}
\lvert\im \cZ \rvert \sim \Delta^{-\frac{11}2} 
e^{\Delta\sqrt{2/\alpha'}}
\ee
Le comportement détaillé de la somme en tant que fonction de $\Delta$, notamment ses oscillations est illustré dans la figure \ref{imfig}.

\begin{figure}[h]
\begin{center}
\epsfig{file=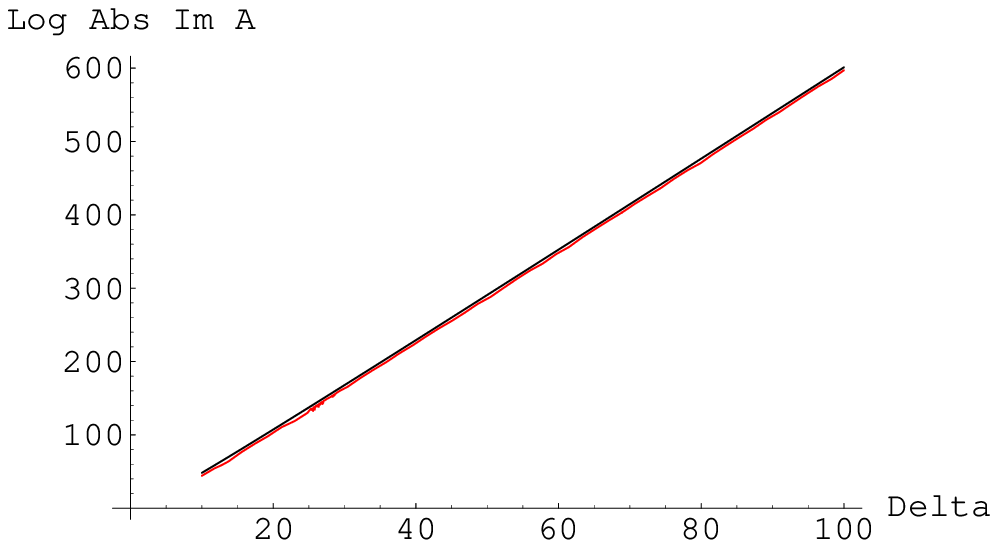, width=6.3cm} \ \ \ \ \ \ \ 
\epsfig{file=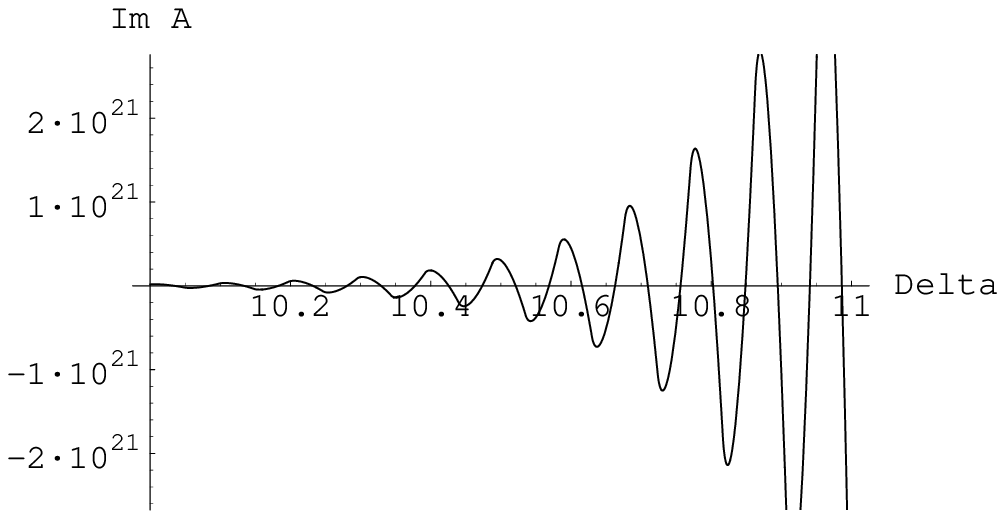, width=6.3cm}
\end{center}
\caption{\l{imfig}\small \`A gauche : $\log \lvert \im \cZ \rvert$
(en rouge) en fonction de $\Delta$. La courbe ne peut \^etre distingu\'ee de son approximation \eqref{imapp} (en noir),
mais ne reproduit pas les oscillations. \`A droite : nous avons grossi l'intervalle $\Delta\in[10,11]$ pour mettre en \'evidence ces oscillations. Nous avons choisi $p=2$ et $\apr = \frac1{2\pi}$ pour les deux courbes.}
\end{figure}

Lorsque $\frac{\Delta^2}{2\pi^2\apr}$ est entier, le dernier terme de la somme s'annule et la somme est dominé par l'avant-dernier terme, dont le signe dépend de la parité de $n^*$. Ceci permet de conclure que l'amplitude s'annule une infinité de fois, ce qui correspond à des paires de S-branes d'extension spatiale infinie qui peuvent se former pour des séparations temporelle dont la valeur est fixée très précisement. Toutefois, en général, la longueur de corrélation est de l'ordre de l'échelle de corde $\sqrt{\apr}$.

La conclusion que nous avons tirée dans notre article \cite{Durin:2005ts} visait à mettre en évidence le caractère hautement ajusté des configurations de S-brane et à rejeter ces configurations comme des candidats génériques à des géométries dépendentes du temps. Malgré ce caractère d'exception, nous pouvons lire ce résultat d'une manière perpendiculaire. Les valeurs de $\Delta$ où la partie imaginaire de $\cZ$ s'annule, que nous noterons $\Delta_n$ donnent la séparation temporelle entre deux S-branes d'extension infinie. Nous pouvons ainsi dire que si une S-brane \gl stable\gr, c'est-à-dire d'extension spatiale infinie, apparaît à l'instant $t$, alors une seconde S-brane \gl stable\gr\ apparaîtra à l'un des instants $t+ \Delta_n$.

%% file: persp.tex

\chapter{Perspectives}
\l{persp}

\section{Résumé des résultats}

L'étude de la configuration des ondes planes électromagnétiques \`a profil lin\'eaire a fait apparaître plusieurs phénomènes.
\begin{itemize}
\i Il existe une valeur critique du gradient à partir de laquelle se développe une instabilité. Les cordes ouvertes deviennent macroscopiques car leur tension est compensée par les champs. Elles s'étirent alors jusqu'à atteindre le condensateur qui crée le champ des ondes planes et le déchargent. Ce phénomène est l'équivalent non relativiste du champ critique en électrodynamique de Born-Infeld
\i La configuration est instable mais il est possible de la stabiliser en ajoutant un champ magnétique constant (ou en modulant de manière paramétrique le gradient des champs électromagnétiques. L'instabilité due au gradient critique n'est pas modifiée par cette stabilisation.
\end{itemize}
Nous avons par ailleurs eu l'occasion de vérifier la dualité de feuille d'univers dans une configuration non triviale et pu calculer, moyennant certaines approximation, les effets de la dépendance en temps, notamment la production de modes. Nous avons enfin mis en évidence la contre-réaction de la corde sur le potentiel de l'onde plane.

L'espace de Misner nous a fourni un cas d'école de singularité cosmologique tout en nous permettant d'utiliser le formalisme perturbatif. Dans le cadre de cette thèse, en nous appuyant sur l'étude de la quantification \cite{Pioline:2003bs} et des modes twistés \cite{Berkooz:2004re}, nous avons étudié les amplitudes à l'ordre des arbres à trois ou quatre points.  Nous avons réussi à calculer les amplitudes faisant intervenir deux états twistés en utilisant le formalisme d'opérateur. Pour cela, il a fallu adapter la définition de l'ordre normal des opérateurs de vertex, ce qui a mis en évidence un étalement des flucuations quantiques des modes excités sur une distance $\Delta(\nu)$. Cette distance croît comme $\log \nu$ lorsque $\nu$ tend vers l'infini et peut ainsi être arbitrairement grande. Cet effet est qualitativement similaire à la géométrie non-commutative.

Le calcul des amplitudes à quatre points permet de mettre en évidence des divergences provoquées par la propagation de cordes twistées avec un moment cinétique de boost nul dans le canal intermédiaire, similaire à ce qui a été constaté dans \cite{Berkooz:2002je} pour les gravitons. Les amplitudes à trois points permettent d'envisager le calcul perturbatif de la contre-réaction gravitationnelle, ce que nous laissons pour un travail futur.

Enfin, nous nous sommes intéressés à la S-brane de Dirichlet. Après avoir montré en détail qu'une S-brane émettant un champ R-R réel n'existait pas en théorie de type II, nous avons étudié les propriétés de \gl stabilité\gr\ des S-branes \gl instables\gr. Nous avons mis en évidence un comportement très particulier dû aux divergences infrarouges de l'amplitude à une boucle de cordes ouvertes tendues entre deux S-branes. En effet, pour certaines valeurs bien précises de la séparation entre les deux S-branes, le système possède une extension spatiale infinie. Ceci permet d'une part de conclure que ces configurations correspondent à un ajustement infini des conditions initiales et d'autre part de prédire l'ensemble des dates auxquelles une S-brane d'extension spatiale infinie peut apparaître si une première S-brane d'extension infinie est déjà apparue à un instant donné (ou inversement, les dates de première apparition d'une S-brane infinie sachant qu'une seconde apparaîtra à un instant donné).

\section{Limites et espoirs du calcul perturbatif}

Nous avons cherché à montrer tout au long de ce mémoire comment le formalisme perturbatif pouvait être exploité pour faire des calculs précis d'observables intéressantes dans des configurations dépendantes du temps. Il apparaît cependant que la complexité formelle de certains résultats limite la possibilité de tirer des conclusions physiques voire ne permet pas de saisir le comportement de certaines configurations.

Toutefois, cet écueil est moins dû à une faiblesse inhérente à cette approche qu'à la difficulté de mener les calculs et d'explorer l'ensemble des pistes possibles. Nous souhaitons clore ce mémoire en illustrant ce propos. En étudiant l'espace de Misner, nous avions espoir que le calcul des amplitudes nous permette de préduire le destin de la singularité, notamment par le calcul perturbatif de la contre-réaction gravitationnelle.

Les articles \cite{Hikida:2005ec,Hikida:2005xa}, s'appuyant notamment sur les travaux précédents \cite{Pioline:2003bs, Berkooz:2004re, Berkooz:2004yy} et dans la même démarche, ont pu, à partir de l'étude des cordes ouvertes et des D-branes dans l'espace de Misner, sonder la condensation du tachyon de corde fermée à l'aide de \emph{D-instantons} et construire un modèle effectif. Ils ont ainsi montré qu'après la condensation, $\beta$ avait diminué, \gl élargissant\gr\ ainsi l'espace de Misner. Bien que l'analyse faite à partir des D-instantons ne soit valable que sur une distance de l'ordre de la longueur de la corde, ce résultat est très encourageant et apporte un premier élément de réponse quant-au destin de la singularité.

D'autre part, dans une approche diff\'erente \cite{McGreevy:2005ci, She:2005mt}, la structure de l'espace-temps \`a l'endroit o\`u classiquement la singularit\'e est attendue a pu être \'eclaircie~: dans \cite{She:2005qq}, une analyse de mod\`ele de matrice sugg\`ere que la singularit\'e n'est pas une description correcte du vide en pr\'esence des paires de cordes enroul\'ees produites par effet tunnel (voir sous-section \ref{ctob}) et qu'il faut la remplacer par un espace-temps non commutatif. Cette nouvelle structure est interpr\'et\'ee \cite{McGreevy:2005ci, She:2005qq} comme l'apparition d'une \'echelle de temps $t_c$ de l'ordre de $\sqrt{\apr \log w}$, \'egale au comportement \`a grand $w$ du facteur de polarisation $\sqrt{\Delta(\nu)}$ de la corde twist\'ee (voir sous-section \ref{am3c}), en de\c ca de laquelle la notion de temps n'existe pas.

%% file: df.tex
\chapter{Définitions et formules utiles \l{df}}

Nous r\'eunissons ici les d\'efinitions et relations qui nous ont \'et\'e utiles au cours de nos calculs et qui ne sont pas souvent utilis\'ees dans d'autres contextes. Ainsi, nous n'aborderons pas les fonctions modulaires, dont les propri\'et\'es utiles sont par exemple donn\'ees dans l'appendice de \cite{Kiritsis:1997hj}.

\section{La fonction hypergéométrique $U$ et les fonctions de Whittaker $M$ et $W$}
\l{unfun}

\subsection{D\'efinition des fonctions}

Nous nous inspirons de la présentation faite dans \cite{slater} et nous en adoptons les conventions.
Il y a quatre fonctions appelées fonctions hypergéométriques confluentes : la fonction de Kummer ${}_1 F_1$, sa fonction associée $U$ et les deux fonctions de Whittaker $M_{k,m} (x)$ et $W_{k,m} (x)$. La fonction ${}_1 F_1 (a,b,x)$ est la solution la plus simple de l'\'equation de Kummer, un cas particulier d'\'equation diff\'erentielle hyperg\'eom\'etrique g\'en\'eralis\'ee,
\be
x y'' + (b-x) y' - a y =0 \l{kummer}
\ee
o\`u $y$ est une fonction de $x$, $y'$ et $y''$ ses d\'eriv\'ees premi\`ere et seconde, $a$ et $b$ des constantes. La solution ${}_1 F_1 (a,b,x)$ est d\'efinie par la s\'erie enti\`ere suivante, lorsque $b \not \in \{0,-1,-2,\ldots\}$
\be
{}_1 F_1 (a,b,x) = \sum_{n=0}^\infty \frac{(a)_n\, x^n}{(b)_n\, n!}
\ee
o\`u la notation $(a)_n$ d\'esigne $a (a+1) \ldots (a+n-1)$.

La fonction hyperg\'eom\'etrique confluente de Tricomi $U$ est une autre solution de l'\'equation \eqref{kummer} d\'efinie par
\be
U(a,b,x) = \frac{\Gamma(1-b)}{\Gamma(1+a-b)} {}_1 F_1 (a,b,x) + \frac{\Gamma(b-1)}{\Gamma(a)} x^{1-b} {}_1 F_1 (1+a-b,2-b,x)
\ee
$U$ est une fonction multivalu\'ee de $x$, la branche principale est choisie comme \'etant celle qui se trouve dans le plan complexe de $x$ coup\'e le long la partie n\'egative de l'axe r\'eel. Notons \'egalement que $U$ est d\'efinie pour $b$ nul ou entier n\'egatif, par exemple \`a l'aide de l'expression suivante,
\be
U(a,b,x) = \frac{\pi}{\sin(\pi b)} \left( \frac{{}_1 F_1 (a,b,x)}{\Gamma(1+a-b)\Gamma(b)} - \frac{x^{1-b} {}_1 F_1 (1+a-b,2-b,x)}{\Gamma(a) \Gamma(2-b)} \right)
\ee

Les fonctions de Whittaker peuvent \^etre d\'efinies par les expressions suivantes
\bse
\begin{align}
M_{k,m} (x) &= x^{\frac12+m}\, e^{-\frac12 x} {}_1 F_1 (\frac12+m-k,1+2m,x) \\
W_{k,m} (x) &= x^{\frac12+m}\, e^{-\frac12 x} U (\frac12+m-k,1+2m,x)
\end{align}
\ese
\subsection{Relations utiles concernant $U$}

La fonction $U$ admet la repr\'esentation int\'egrale suivante
\be
U(a,b,x) = \frac1{\Gamma(a)} \int_0^\infty e^{-xt} t^{a-1} (1+t)^{b-a-1} dt
\ee
avec $\re a >0$ et $\re x >0$. Dans nos calculs, la condition sur $a$ est satisfaite, mais $x \in i\R$ . Il est possible n\'eanmois de faire tourner le contour d'un angle $\phi$ et de prendre la limite $\phi \to 0$ :
\be
U(a,b,x) = \frac1{\Gamma(a)} \int_0^{\infty\, e^{i\phi}} e^{-xt} t^{a-1} (1+t)^{b-a-1} dt
\ee
avec $-\frac{\pi}2 < \phi +\arg x <\frac{\pi}2$.

Le premier th\'eor\`eme de Kummer s'\'ecrit pour $U$
\be
U (a,b,x) = x^{1-b}\, U(1+a-b, 2-b, x)
\ee

Enfin, le comportement asymptotique de $U$ lorsque $x$ est
\be
U(a,b,x) \sim x^{-a} \l{asymptU}
\ee
lorsque $\re a>0$.

\section{La fonction hypergéométrique $\hypF$}

La fonction $\hypF(a,b;c;x)$ est une solution de l'\'equation diff\'erentielle hyper\'geom\'etrique suivante, connue aussi sous le nom d'\'equation Fuchsienne de type g\'en\'eral,
\be
x(1-x) y'' + [c+(1+a+b)x] y' - ab\, y =0
\ee
Nous avons adopt\'e les m\^emes notations que dans la Section pr\'ec\'edente \ref{unfun}

Parmi les relations de transformation principales, nous mentionnons celle-ci
\begin{multline}
\hypF(a,b;c;x) = \frac{\Gamma(b-a) \Gamma(c)}{\Gamma(b) \Gamma(c-a)} (-x)^{-a} \hypF\left(a,a-c+1;a-b+1;\frac1x\right)\\+ \frac{\Gamma(a-b) \Gamma(c)}{\Gamma(a) \Gamma(c-b)} (-x)^{-b} \hypF\left(b,b-c+1;b-a+1;\frac1x\right) \l{hypinv}
\end{multline}

La fonction $\hypF(a,b;c;x)$ admet le développement en série entière suivant
\be
\hypF(a,b;c;x) = \sum_{n=0}^\infty \frac{(a)_n (b)_n}{(c)_n}\, \frac{x^n}{n!}
\ee
En particulier, nous avons la relation suivante, qui intervient à plusieurs reprises dans nos calculs,
\be
G(z;v) = \sum_{n=1}^\infty \frac{z^{n+i\nu}}{n+i\nu} = \frac{z^{1+i\nu}}{1+i\nu}\, \hypF(1+i\nu, 1; 2+i\nu;z)
\ee
où la première égalité définit la notation $G(z;\nu)$

Dans la limite où $z\to 1$, nous avons le comportement asymptotique suivant
\be
G(z;\nu) = -\log (1-z) +\psi(1) -\psi(1+i\nu) + \O (z-1)
\ee
où $\psi$ est la fonction digamma.

\section{La fonction digamma $\psi$}

La fonction digamma, notée $\psi$, est la dérivée logarithmique de la fonction $\Gamma$, ou encore
\be
\psi(x) = \frac1{\Gamma(x)} \frac{d\Gamma (x)}{dx}
\ee
En particulier, $\psi(1)$ est la constante d'Euler $\gamma$.

\section{Relations utiles}

\subsection{Resommation de Poisson}
Soit $f(x)$ une fonction admettant une transformée de Fourier $\F[f](\w)$ définie par
\be \F[f](\w) = \frac1{2\pi} \int_{-\infty}^\infty dx f(x)\, e^{i \w x} \ .\ee
On a alors
\be \sum_{n \in \Z} f(2 \pi n) = \sum_{n \in \Z} \F[f](n) \ee
En particulier pour $f(x) = e^{a x}$
\be \sum_{n \in \Z} e^{2 \pi a n} = \sum_{n \in \Z} \delta(a+n) \ee
D'autres formules intéressantes peuvent être obtenues en choisissant pour $f$ une gaussienne (voir \cite{Kiritsis:1997hj}, Appendix A).

\subsection{Une formule de trigonométrie hyperbolique}

Nous mentionnons pour référence la formule de trigonométrie suivante
\be \arctanh(x) = \frac12 \ln \left\lvert\frac{1+x}{1-x} \right\rvert \ee